**UNIVERSITA' DEGLI STUDI "G. d'Annunzio" Chieti-Pescara**

**SCUOLA SUPERIORE "G. d'Annunzio"**

*DOTTORATO DI RICERCA IN HUMAN SCIENCE CON CURRICULUM IN ECONOMICS AND STATISTICS*

*CICLO XXXVI*

# MODELLING FINANCIAL RETURNS WITH MIXTURES OF GENERALIZED NORMAL DISTRIBUTIONS

**Dipartimento di Scienze Filosofiche, Pedagogiche ed Economico-Quantitative**

**Settore Scientifico Disciplinare SECS-S/01 Statistica**

Dottorando                    Coordinatore
Dott. Pierdomenico Duttilo     Prof. Marco Di Marzio

Tutor
Prof. Stefano Antonio Gattone

Anni Accademici 2020/23

# MODELLING FINANCIAL RETURNS WITH MIXTURES OF GENERALIZED NORMAL DISTRIBUTIONS

*To my parents*

# Contents

# List of Figures

# List of Tables

**Abstract:** This manuscript presents an investigation into the analysis of financial returns using mixture models, focusing on mixtures of generalized normal distributions (MGND) and their extensions. The study addresses several critical issues encountered in the estimation process and proposes innovative solutions to enhance accuracy and efficiency. In Chapter 2, the focus lies on the MGND model and its estimation via expectation conditional maximization (ECM) and generalized expectation maximization (GEM) algorithms. A thorough exploration reveals a degeneracy issue when estimating the shape parameter. Several algorithms are proposed to overcome this critical issue. Chapter 3 extends the theoretical perspective by applying the MGND model on several stock market indices. A two-step approach is proposed for identifying turmoil days and estimating returns and volatility. Chapter 4 introduces constrained mixture of generalized normal distributions (CMGND), enhancing interpretability and efficiency by imposing constraints on parameters. Simulation results highlight the benefits of constrained parameter estimation. Finally, Chapter 5 introduces generalized normal distribution-hidden Markov models (GND-HMMs) able to capture the dynamic nature of financial returns. This manuscript contributes to the statistical modelling of financial returns by offering flexible, parsimonious, and interpretable frameworks. The proposed mixture models capture complex patterns in financial data, thereby facilitating more informed decision-making in financial analysis and risk management.

*Questo studio presenta un'indagine sull'analisi dei rendimenti finanziari attraverso modelli di mistura, nello specifico misture di distribuzioni normali generalizzate (MGND) e loro estensioni. Lo studio affronta diverse questioni critiche riscontrate nel processo di stima e propone soluzioni innovative per migliorarne l'accuratezza e l'efficienza. Nel Capitolo 2, l'attenzione si focalizzata sul modello MGND e la stima dei parametri mediante gli algoritmi ECM e GEM. Un'esplorazione approfondita rivela un problema di degenerasi nella stima del parametro di forma. Vengono proposti vari algoritmi per superare questo problema. Il Capitolo 3 estende la prospettiva teorica applicando il modello MGND a diversi indici azionari. Viene proposto un approccio a due step per identificare i giorni di turbolenza e stimare rendimenti e volatilità. Il Capitolo 4 introduce modelli di mistura vincolati di distribuzioni normali generalizzate (CMGND), migliorando interpretabilità ed efficienza. Infine, il Capitolo 5 introduce gli hidden-Markov models con distribuzioni normali generalizzate (GND-HMM) in grado di catturare la natura dinamica dei rendimenti finanziari. Il presente lavoro contribuisce alla modellazione statistica dei rendimenti finanziari offrendo frameworks flessibili, parsimoniosi e interpretabili. I modelli di mistura proposti catturano gli andamenti complessi dei rendimenti finanziari, facilitando così le decisioni in ambito di gestione del rischio.*

# Introduction | 1

In statistics, the challenge of decomposing an empirical distribution into several components has a long history. We are approaching the 130th anniversary of Karl Pearson's study (K. Pearson 1894) in which he introduced the method of moments to estimate a two-component mixture of normal distributions. The innovative Pearson's work gave rise to a vast literature on mixture models with applications in many fields like agriculture, astronomy, biology, economics, engineering, finance, genetics, imaging, marketing and others (G. J. McLachlan and Peel 2000).

In finance, the idea to fit the distribution of returns with mixture models goes back to Press 1967, Praetz 1972, Mandelbrot 1973, P. K. Clark 1973, and Ball and Torous 1983. It is well known that the empirical return distribution of most financial asset presents significant departures from 'normality'. Specifically, it tends to be a leptokurtic distribution with marked fat tails and a peak around the mean value. Therefore, from a **statistical perspective** '*mixture models may be motivated as models which imply leptokurtosis*' (Kaehler and Marnet 1994). For example, Kon 1984 introduced discrete mixtures of normal distributions to approximate kurtosis and skewness of several US stocks and stock indices.

From a **theoretical perspective** mixture models can be motivated by the practical interpretability of financial returns dynamics. Usually, financial markets are characterized by two market periods: stable and tumultuous. The latter are caused by aggregate events like economic and geopolitical events, corporate announcements or changes in market sentiment. Since mixture components could be different with respect to their variances, a daily return belongs to the stability period if it belongs to the mixture component with the lowest variance. Similarly, a daily return belongs to the turmoil period if it belongs to the mixture component with the highest variance. As a result, '*market participants expect distributions to be drawn from some possibilities*' (Bellalah and Lavielle 2002).

However, normal mixtures impose a priori specific constraints on the shape of the return distribution since mixture components are normal distributions. Mixtures of Student-t distributions and mixtures of generalised normal distributions can overcome this critical issue thanks to the flexibility provided by the additional shape parameter.

This work is about the analysis of financial returns with mixtures of generalized normal distributions (MGND). The MGND model is a natural generalization of the normal mixture model which controls both the peakedness and the tail behaviour with the additional shape parameters. This manuscript provides a comprehensive investigation into the use of mixture models, particularly MGND, in the analysis of financial returns, addressing both theoretical and practical aspects.

After this brief introductory chapter, we turn in **Chapter 2** to the definition of financial asset returns and a literature review about financial returns modelling, including a description of the principal heavy-tailed distributions and mixture models. Next, Chapter 2 focuses on the MGND model and the maximum likelihood estimation via expectation conditional maximization (ECM) and generalized expectation maximization (GEM) algorithms. The study highlights a degeneracy issue of the shape parameter. The ECM algorithm may have some convergence issues when the shape parameter ($v_k$) is higher than 2, since variance and kurtosis are nearly constant and vary slowly for high $v_k$. Several solutions are provided to overcome this critical issue (algorithms 2-6 in Appendix A). The best solution is represented by algorithm 5 in Appendix A which performs a parameter estimation via GEM algorithm[1] with the backtracking line-search and an additional $v_k$ step-size. To conclude, numerical experiments are performed in order to evaluate the estimation performance of the proposed algorithms and the goodness-of-fit performance of the MGND model.

1: The GEM algorithm aims to improve, rather than maximizes, the likelihood function in every M-step.

**Chapter 3** focuses on the theoretical perspective of the MGND model through two empirical studies. The first is focused on ESG[2] indices, while the second on the G7 stock markets[3] . A two step approach is proposed. Firstly, turmoil days are objectively detected employing a two-component MGND model fitted on financial asset returns. Secondly, returns and volatility of several market indices are estimated with generalized autoregressive conditional heteroskedasticity (GARCH) models including the detected turmoil days as exogenous dummy variables. In the specific case of the MGND model, a daily return is classified as a stable day if it belongs to the stable component, i.e. the mixture component with the highest shape parameter/lowest scale parameter. Similarly, a daily return is classified as a turmoil day if it belongs to the turmoil component, i.e. the mixture component with the lowest shape parameter/highest scale parameter. Conventionally, a lower shape parameter/higher scale parameter means a thicker tail and then a higher standard deviation, while a higher shape parameter/lower scale parameter means a thinner tail and then a lower standard deviation. Thus, the inclusion of these dummy variables improves the estimates as well as the interpretation of the results.

2: Environmental, Social and Governance.

3: Canada, France, Germany, Italy, Japan, UK and US.

**Chapter 4** introduces constrained mixture of generalized normal distributions (CMGND). Constraints are imposed on the location and/or scale and/or shape parameters. In the two-component case taking all possible combinations of these constraints into consideration would result to a 7-model family. The implementation of such constraints not only solves the numerical degeneracy of the log-likelihood but also enhances the interpretability and the parametric efficiency[4] of the final solution. The MLE of the parameters is obtained via the GEM algorithm. Additionally, since the iteration equations for some of the parameters are non-linear,

4: The use of similar constraints in the mixture models clearly leads to more parsimonious models as the number of unconstrained parameters increases linearly with the number of components.

the Newton-Raphson method is employed. The proposed methodology is tested on simulated data across different scenarios such as symmetric models (mixtures with common location parameters) and models with the same kurtosis (i.e. mixtures with common shape parameter). In order to test the null hypothesis '*constraints hold*', an appropriate likelihood ratio test (LRTS) is implemented. The LRTS performance is evaluated by a power analysis. Furthermore, an empirical application and a comparative study is performed on financial data.

**Chapter 5** introduces GND-hidden Markov models (GND-HMMs) with unconstrained and constrained parameters. GND-HMMs improve the MGND and CMGND models by allowing temporal dependence with a Markov chain process for the state variable. Specifically, a first-order Markov chain governs the draws of the states from the mixture components. In this way, it is possible to improve the model ability to capture the dynamic nature of the state variable distinguishing stable and turmoil periods. Chapter 5 covers issues like autocorrelation, heteroskedasticity, Markov chain, hidden Markov model, parameter estimation via direct optimization, state prediction, forecast distributions, local-global decoding, simulation study and numerical experiments. The latter aims to answer to the following questions: Who best fits financial returns between unconstrained and constrained GND-HMMs? and then who best fits financial returns between independent GND mixture models and GND-HMMs together with the features exploited?

Table 1.1 summarizes the models covered in this manuscript together with the features exploited.

| Model | Chapter | Leptokurtosis | Temporal dependence | Parsimony | Interpretability |
|---|---|---|---|---|---|
| MGND | Chapter 2-3 | Captured | Not fully captured | Not allowed | Stable/turmoil days |
| CMGND | Chapter 4 | Captured | Not fully captured | Allowed | Stable/turmoil days |
| GND-HMMs | Chapter 5 | Captured | Captured | Allowed | Stable/turmoil periods |

**Table 1.1:** Models by chapters and features exploited.

In relation to the contents of this manuscript:

▶ Section 3.2 in Chapter 3 has been published in AStA journal (Duttilo, Gattone, and Iannone 2023);

▶ Section 3.3 in Chapter 3 has been presented at the conference '*ASA 2023 | Statistics, Technology and Data Science for Economic and Social Development*', 6-8 September 2023, Bologna;

▶ Chapter 4 is under review. In addition, it has been presented at two conferences:

  ● '*SIS 2023 | Statistical Learning, Sustainability and Impact Evaluation*', 21-23 June 2023, Ancona (Duttilo, Kume, and Gattone 2023);
  ● '*COMPSTAT 2023 | 25th International Conference on Computational Statistics*', 22-25 August 2023, London (Duttilo, Gattone, and Kume 2023).

▶ An empirical application on Italian electricity prices of GND-HMMs (Chapter 5) has been presented at the '*SIS 2024 | The 52nd Scientific Meeting of the Italian Statistical Society*', 17-20 June 2024, Bari (Duttilo, Bertolini, et al. 2024).

# Mixture of generalized normal distribution to fit stock returns

# 2

## 2.1 Financial asset returns

'*Returns are changes in price expressed as a fraction of the initial price*' (Ruppert 2011). Assuming no dividends or interests, the net return of an asset over the holding period $[t-1, t]$ is

$$R_t = \frac{P_t}{P_{t-1}} - 1 = \frac{P_t - P_{t-1}}{P_{t-1}},$$

where $R_t$ is the net return at time $t$, $P_t$ is the price of an asset at time $t$ and $P_{t-1}$ is the price at time $t-1$. $P_t - P_{t-1}$ is the revenue ($P_t > P_{t-1}$) or loss ($P_t < P_{t-1}$) obtained during the holding period $[t-1, t]$. As a result, the gross return of an asset at time $t$ is given by

$$\frac{P_t}{P_{t-1}} = 1 + R_t. \tag{2.1}$$

Log returns are calculated with the natural log difference approach

$$r_t = \log(1 + R_t) = \log\left(\frac{P_t}{P_{t-1}}\right) = \log(p_t) - \log(p_{t-1}), \tag{2.2}$$

where $r_t$ denotes the log return of an asset a time $t$, $p_t$ is the log price of an asset at time $t$, while $p_{t-1}$ is the log price at time $t-1$. Many companies decide to distribute dividends to shareholders as a return on invested capital. Similarly bonds pay interests. The adjusted log return of an asset at time $t$ is defined as follows

$$r_t = \log(1 + R_t) = \log\left(\frac{P_t + D_t}{P_{t-1}}\right) = \log(P_t + D_t) - \log(P_{t-1}), \tag{2.3}$$

where $D_t$ is the amount of dividends or interests at time $t$.

Percentage log returns (used in this manuscript) are obtained multiplying the equation 2.2 by 100

$$r_t = \log(1 + R_t) = \log\left(\frac{P_t}{P_{t-1}}\right)100. \tag{2.4}$$

Figure 2.1 shows the percentage log returns of the Euro Stoxx 50 Index[1] (STOXX50E) for the period from 02 April 2007 to 15 August 2023 (16 years). The returns series is featured by two important high volatility periods (volatility clusters) which occur during the global financial crisis (2007-2008) and the Covid-19 crisis (2020-2021).

It is well known that financial returns are non-normal, with fat-tails, high-kurtosis and different degrees of skewness (Massing and Ramos

1: The Euro Stoxx 50 Index is a stock market index representing the performance of 50 large blue-chip companies from Eurozone countries. It serves as a key benchmark for the Eurozone equity market.

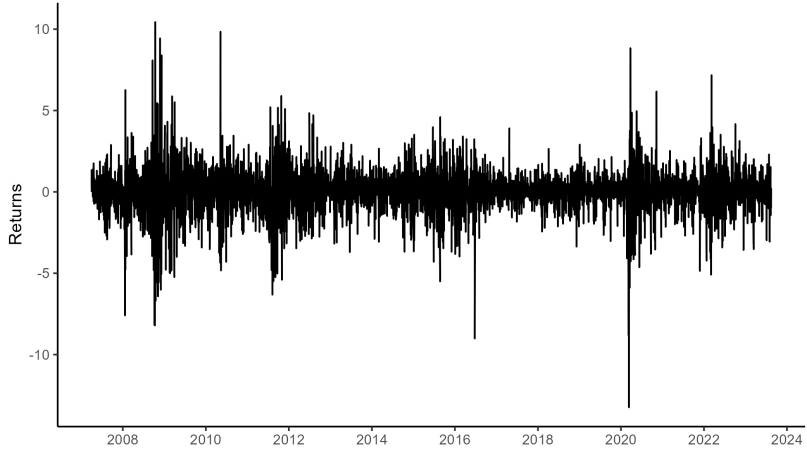

2021). For example, Figure 2.4 shows the density plot and the normal QQ plot for the daily returns of the STOXX50E. The density plot highlights that the normal distribution (dashed black line) cannot capture the density around the origin and in the tails. The normal QQ plot shows the presence of fat-tails: both ends of the QQ plot deviate from the straight line while its center follows the straight line.

Another way to test the normality assumption[2] is the Jarque-Bera test which compares the sample skewness (Sk) and kurtosis (Kur) to their values under the normality assumption. The statistic test is

$$\text{JB} = \frac{\text{T}}{6}\left(\widehat{\text{Sk}}^2 + \frac{1}{4}\left(\widehat{\text{Kur}} - 3\right)^2\right), \quad \text{with } T \text{ as number of observations.} \quad (2.5)$$

The null and alternative hypothesis are defined as follows

▶ $H_0$: the sample comes from a normal distribution, i.e. Sk = 0 and Kur = 3;
▶ $H_1$: the sample comes from a non-normal distribution.

Under $H_0$, JB is asymptotically distributed as a chi-square random variable with 2 degrees of freedom. The normally assumption $H_0$ is rejected if the *p-value* < 0.05.

Sample skewness and kurtosis are indices that describe the shape of a probability distribution. Specifically, the sample skewness (Figure 2.2) *'measures the degree of asymmetry'* (Ruppert 2011)

$$\widehat{\text{Sk}} = \frac{1}{\text{T}}\sum_{t=1}^{T}\left(\frac{r_t - \bar{r}}{\sqrt{s^2}}\right)^3, \quad (2.6)$$

where $\bar{r} = \frac{1}{T}\sum_{t=1}^{T} r_t$ is the sample mean and $s^2 = \frac{1}{T-1}\sum_{t=1}^{T}(r_t - \bar{r})^2$ is the sample variance, with:

▶ $\widehat{\text{Sk}} = 0$ the distribution is symmetric;
▶ $\widehat{\text{Sk}} > 0$ the distribution is positive skewed (or right skewed) with a longer tail on the right;
▶ $\widehat{\text{Sk}} < 0$ the distribution is negative skewed (or left skewed) with a longer tail on the left.

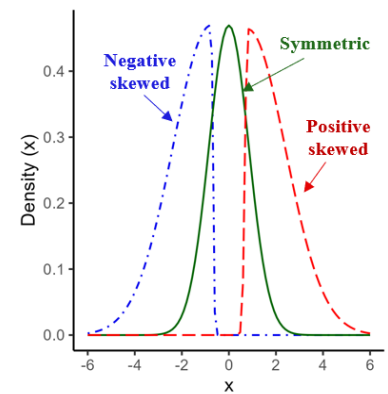

On the other hand, the sample kurtosis (Figure 2.3) indicates '*the extent to which probability is concentrated in the center and especially the tails of the distribution*' (Ruppert 2011)

$$\widehat{\text{Kur}} = \frac{1}{T} \sum_{t=1}^{T} \left( \frac{r_t - \bar{r}}{\sqrt{s^2}} \right)^4,$$ (2.7)

with:

- ► $\widehat{\text{Kur}} = 3$ the distribution is mesokurtic (normal);
- ► $\widehat{\text{Kur}} > 3$ the distribution is leptokurtic, more pointed than normal and with heavier tails;
- ► $\widehat{\text{Kur}} < 3$ the distribution is platykurtic, more flattened than normal.

Table 2.1 show the descriptive statistics and the Jarque-Bera test statistic for the daily returns of the STOXX50E. Daily returns are not normally distributed because $H_0$ is rejected and the descriptive statistics highlight negative skewness and high-kurtosis.

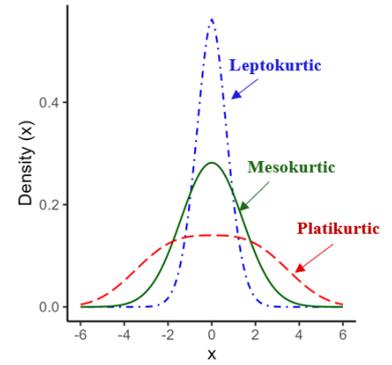

**Figure 2.3:** General shape of kurtosis.

**Table 2.1:** Descriptive statistics summary - Euro Stoxx 50 Index

|          | Mean  | Median | Std   | Sk     | Kur    | JB Test |
|----------|-------|--------|-------|--------|--------|---------|
| STOXX50E | 0.001 | 0.042  | 1.436 | -0.288 | 10.326 | 9249*   |

*Notes.*\* indicates a *p*-value≤ 0.05.

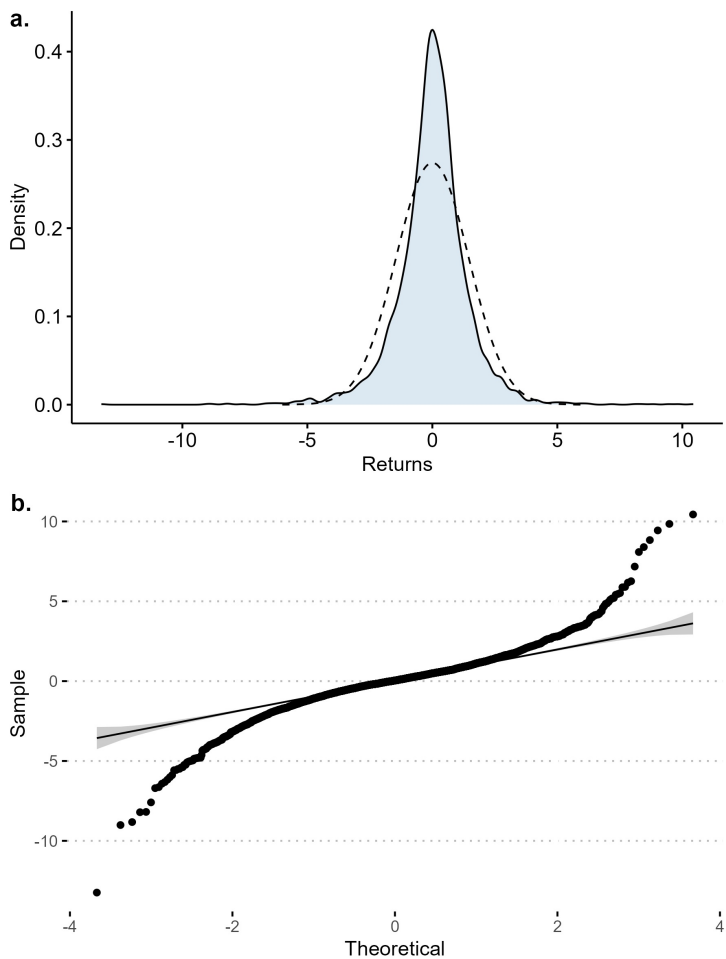

**Figure 2.4:** Density plot and normal QQ plot - Euro Stoxx 50 Index. Panel (a) provides the density plot and panel (b) the normal QQ plot.

However, the classic portfolio theory (Markowitz 1952), the capital asset pricing model-CAPM (Sharpe 1964; Lintner 1965; Mossin 1966) and the Black-Scholes option pricing model (Black and Scholes 1973) rely on the normality assumption. Moreover, the normality assumption can also lead to errors in the estimate of the *Value at Risk* (VaR), *Expected Shortfall* (ES) and other risk measures based on the unconditional returns distribution (Prakash, Sangwan, and Singh 2021). According to Bellalah and Lavielle 2002 '*the identification of a given stochastic process that matches the observed distribution allows one to build pricing models which give prices that fit market prices*'. Therefore, as the distributional form of returns series plays a key role, it is crucial to investigate which distribution gives the best fit in a large class of parametric models.

## 2.2 Literature review

From an historical point of view, the normality assumption was introduced by Bachelier 1900 who proposed a white noise process for distributions of price changes (Peiró 1994). The differences $P_{t+T} - P_t$ are assumed independent and normally distributed[3] with zero mean and variance proportional to the differencing interval $T$ (independent to the price level $P_t$). This assumption was commonly accepted until 1960, although some researchers provided empirical evidence about the non-normal behaviour of returns (Divisia 1927; Mills 1927; Mitchell 1938).

In response to these empirical facts Mandelbrot 1963 recommended the stable Paretian family of distributions. Since this family has unknown[4] density functions, the logarithm of the characteristic function is

$$\log f(t) = i\delta t - |ct|^{\alpha}[1 + i\beta(t/|t|)\tan(\pi\alpha/2)],$$

where $t$ is a real number, $i = \sqrt{-1}$, $\delta$ is the location parameter, $\gamma$ is the scale parameter, $\beta$ is the skewness parameter and $\alpha$ is the characteristic exponent or stability parameter. Of the four parameters defining the family, $\alpha$ is the most important because controls the size of the tails of the distribution. It can take any value in the range $0 < \alpha < 2$, the normal distribution is yielded with $\alpha = 2$, whereas the Cauchy distribution is yielded with $\alpha = 1$. However, the family has undefined variance for $\alpha < 2$, and undefined mean for $\alpha \le 1$. The Mandelbrot's stable Paretian hypothesis '*for distributions of price changes in speculative series*' (Fama 1963) occurs when $1 < \alpha < 2$. In other words, price changes intermediate between a Cauchy and a normal distribution.

While Fama 1963 and Fama and Roll 1968 further supported the Mandelbrot's Paretian hypothesis, Praetz 1972 suggested the Student-t distribution, as an alternative to overcome the three main drawbacks of stable Paretian distributions, i.e. infinite variance, unknown density functions and inefficient estimation methods. The Student-t distribution has a finite variance for degrees of freedom ($\nu$) higher than two and lies also between the Cauchy ($\nu = 1$) and normal ($\nu \to +\infty$) distributions. Findings showed that the Student-t distribution provides a better fit of the Sydney Share Price Indices compared to the stable Paretian and normal distributions.

3: A random variable $X$ is said to have a normal distribution if its probability density function is given by

$$f_N(x|\mu, \sigma) = \frac{1}{\sigma\sqrt{2\pi}} \exp\left\{ -\frac{1}{2}\left(\frac{x-\mu}{\sigma}\right)^2 \right\},$$
(2.8)

with $-\infty < x < \infty$, mean $-\infty < \mu < \infty$ and standard deviation $\sigma > 0$.

4: The stable Paretian family of distributions does not admit a closed expression of density functions, see Gnedenko and A. N. Kolmogorov 1968.

Blattberg and Gonedes 1974 and Upton and Shannon 1979 reached similar conclusions.

Hsu 1982 recommended the use of the exponential power distribution or generalized normal distribution (GND)[5] introduced by Box and Tiao 1973. It is able to model a large variety of statistical behaviours thanks to the additional shape parameter which controls both the peakedness and tail behaviour of the density function. Gray and French 1990 demonstrated that the GND fits the return distribution of the S&P500 Composite Index better than the student-t and logistic distributions.

Eberlein and Prause 2002 recommended the Generalized Hyperbolic (GH) distribution which contains many other models as special or limiting cases like the normal inverse Gaussian distribution introduced by Barndorff-Nielsen 1997.

The innovative Mandelbrot's work gave rise to a vast literature which analyses and compares several probability distributions, other recommendations include the variance gamma distribution (Madan and Seneta 1990), the hyperbolic distribution (Eberlein and Keller 1995), the Meixner distribution (Schoutens 2002), the generalized lambda distribution and skewed t distribution (C. Corlu G. and A. Corlu 2014).

Another important strand of literature fits financial returns with finite mixtures of distributions. These models are widely used to analyse complex distributions of data (G. J. McLachlan, S. X. Lee, and S. I. Rathnayake 2019) since K. Pearson 1894 first introduced a mixture of two normal distributions[6] .

The idea to fit the return distribution with mixture models goes back to Press 1967, Praetz 1972, Mandelbrot 1973, P. K. Clark 1973, and Ball and Torous 1983. Since rare events like bad news can cause turbulence in financial markets, it is possible to distinguish two stock market periods: stable and tumultuous. A daily return belongs to the stability period if it belongs to the mixture component with the lowest variance. Similarly, a daily return belongs to the turmoil period if it belongs to the mixture component with the highest variance. As a result, '*market participants expect distributions to be drawn from some possibilities*' (Bellalah and Lavielle 2002). This is one of the main difference with respect to heavy-tailed distributions which assume a unique distribution.

Kon 1984 introduced a discrete mixture of normal distributions to approximate the excess kurtosis and skewness of returns distributions for several New York Stock Exchange indices as well as for 30 Dow-Jones stocks. However, mixtures of normals with different parameters were compared with Student-t distributions. Findings show that discrete mixture of normal distributions have '*more descriptive validity than the Student-t distribution*' (Kon 1984).

Peiró 1994 analysed and compared the distribution of stock returns of six stock markets. Several distributions were estimated: Stable Paretian, normal, logistic, Student-t, exponential power, discrete mixture of two normals and discrete mixture of two normals with the same expectation. Findings highlight that all markets present a non-normal distribution well represented by the Student-t distribution.

5: The GND is also known as the generalized Gaussian distribution or generalized error distribution. Furthermore, different parametrizations are available (Box and Tiao 1973; Nadarajah 2005; Ruppert 2011). The acronym GND and the parametrization of Nadarajah 2005 are adopted in this manuscript.

6: For a brief history of mixture models see G. J. McLachlan and Peel 2000.

In 1996, *RiskMetrics* proposed the normal mixture model to fit the heavy-tails of some returns distributions. Other experimental studies include Bellalah and Lavielle 2002, Behr and Pötter 2009 and Liyan Han and Zheng 2019.

Normal mixtures modelling is a well-known method used for most applications in a wide range of fields, but datasets characterized by non-normal features require more flexible and powerful tools. Hence, non-normal model-based methods have attracted the attention of researchers (S. Lee and G. Mclachlan 2013).

Heterogeneous mixture distributions have been proposed as an alternative to normal mixture models. They involve combinations of different types of probability distributions, as opposed to homogeneous mixtures, where the components are of the same distribution type (Shin, Ouarda, and T. Lee 2016; Imran, Sheraz, and Dedu 2022).

Markus Haas and Paolella 2006 proposed a discrete mixture of a normal and Laplace distribution[7] to fit the distribution of 25 daily stock return series. Results showed that the proposed mixture model provides an accurate description of the return distributions over-performing the Laplace distribution and generalized error distribution.

J. Li 2012 proposed a discrete mixture of a normal and Cauchy distribution for financial risk management. Findings highlight that the normal distribution underestimates the tails of returns, while the Cauchy distribution overestimates them. By contrast, the Normal-Cauchy mixture model improves the goodness of fit performance and avoids serious VaR violations in the backtesting. S. Zhang et al. 2020 and K. Jiang et al. 2022 proposed several time-varying versions of the Normal-Cauchy mixture model.

However, heterogeneous mixture distributions impose a priori specific constraints on the shape of the return distribution since for example, mixture components are a normal, Laplace or Cauchy distribution. Mixtures of Student-t (MSTD) distributions and mixtures of generalised normal distributions (MGND) can overcome this critical issue thanks to the flexibility provided by the additional shape parameter.

Massing, Puente-Ajovín, and Ramos 2020 introduced three univariate finite mixtures of Student-t with two and three components[8] . The degrees of freedom where fixed a priori in order to obtain a more feasible estimation. Specifically, the proposed mixture models allow three different scenarios with small, moderately high and high degree of freedoms. These models have been applied to study the distribution of the log-growth rates of cities' sizes of four European countries and the USA. In a subsequent work, Massing and Ramos 2021 presented a comparative study for 78 stock indices from 70 countries with hourly and daily return series. Findings show a good performance of the Student-t mixtures with respect to three goodness-of-fit measures: Kolmogorov–Smirnov statistic, Anderson–Darling statistic and Akaike information criterion. On the other hand, the Bayesian information criterion heavily penalizes the parameters of the proposed mixture models preferring heavy-tailed distributions like the Student-t and normal inverse Gaussian.

Mixtures of generalized normal distributions (MGND) *'have the flexibility required to fit the shape of the data better than the Gaussian mixture model'* (T. M.

7: Originally, this model was proposed by Jones and G. J. Mclachlan 1990 and Kanji 1985 in a engineering context.

8: Multivariate mixtures of Student-t where introduced by G. J. McLachlan and Peel 1998. The R *teigen* package allow to estimate univariate and multivariate finite mixtures of Student-t (Andrews, McNicholas, and Subedi 2011; Andrews, Wickins, et al. 2018).

Nguyen, Jonathan Wu, and H. Zhang 2014). They have been successfully applied in computer vision and pattern recognition problems. For example, Bazi, Bruzzone, and Melgani 2006 and Allili, Bouguila, and Ziou 2008 applied univariate MGND for image processing. The estimation of the parameters was performed via the maximum likelihood estimation (MLE), and the expectation-maximization (EM) algorithm. Since the system to resolve the updating equation of the shape parameter is heavily non-linear, the numerical optimization based on Newton-Raphson method is used. As an alternative, Mohamed and Jaïdane-Saïdane 2009 estimated the shape parameter by exploiting the analytical relationship between the shape parameter and kurtosis. T. M. Nguyen, Jonathan Wu, and H. Zhang 2014 proposed a univariate bounded generalized Gaussian mixture model defining a bounded support region in ℝ for each component. Recently, Wen et al. 2022 studied a two-component mixture of univariate GND and proposed an expectation conditional maximization (ECM) algorithm for parameter estimation. In particular, it was found that for the modelling purposes of the S&P 500 and Shanghai Stock Exchange Composite Index (SSEC), such mixtures outperform those constructed by simply using mixtures of normals.

Empirical studies that investigate the distribution of returns involve several assets:

▶ stocks and stock indices (Kon 1984; Gray and French 1990; Peiró 1994; Aparicio and Estrada 2001; Behr and Pötter 2009; Göncü, Karahan, and Kuzubaş 2016; Massing 2019; Massing and Ramos 2021);

▶ exchange rates (C. Corlu G. and A. Corlu 2014; Saralees Nadarajah and Chan 2015; Nasir, Sheraz, and Dedu 2022);

▶ cryptocurrencies (J. Chu, Nadarajah, and Chan 2015; Y. Zhang et al. 2019).

Another branch of the literature employs hidden Markov models to characterize the distribution of returns. Kaehler and Marnet 1994 employed mixtures of normal distributions and normal-HMMs to fit exchange-rates and their empirical dynamics: leptokurtosis and heteroskedasticity. It was pointed out that mixtures of normal distributions effectively capture the leptokurtosis, while normal-HMMs successfully capture both leptokurtosis and heteroskedasticity. Guidolin and Timmermann 2006 demonstrated that HMMs perform effectively in measuring Value at Risk (VaR) at a monthly frequency but these results cannot be extended to daily returns. The latter strongly exhibit non-normal features. Haas 2009 performed a VaR backtesting for several HMMs and independent mixture model considering the daily returns of major European stock markets. Both normal and Student-t mixtures are considered. Findings reveal that the univariate two-state Student-t-HMM demonstrates the best overall performance.

## 2.3 Parametric distributions

A parametric probability distribution is an *'abstract mathematical form'* (Wilks 2011) able to describe variations in a set of data. They are divided in discrete distributions and continuous distributions. While the former describe discrete random variables that can take only specific values, the latter describe continuous random variables that can take any real number in a specific range. However, *'it is convenient and not too inaccurate to represent as continuous those variables that are continuous conceptually but reported discretely'*(Wilks 2011). For example, financial returns are discrete observations when are treating as samples from continuous distributions.

Parametric distributions are characterized by at least three advantages (Wilks 2011):

▶ **Compactness**, the probability distribution reduce large data sets into few distribution parameters;
▶ **Smoothing and interpolation**, the probability distribution represent sampling variations of real data;
▶ **Extrapolation**, the probability distribution fit unobserved behaviour, i.e. observations outside the dataset range.

The mathematical form of a parametric distribution is determined by particular entities called parameters. Commonly, they are classified in location, scale and shape parameters (Ruppert 2011).

▶ The **location parameter** quantifies the central tendency and shifts the parametric distribution without affecting its shape and scale.
▶ The **scale parameter** determines the dispersion of the parametric distribution.
▶ The **shape parameter** is not influenced by changes of the location and scale parameters. It is used to quantify the skewness or tail weight.

An important task of data analysis is finding *parsimonious* models that fit the data well without an excessive number of parameters, i.e. the well known bias/variance trade-off. A model with few parameters does not fit the data well, while a model with too many parameters is source of over-fitting for the prediction of new observations (Ruppert 2011).

The rest of the Chapter 2 is organized as follows. Section 2.4 analysed some important heavy-tailed distributions: the Laplace distribution, the Cauchy distribution, the Student-t distribution, the generalized hyperbolic distribution and the generalized normal distribution. In Section 2.5, mixture models are treated, namely the normal mixture model, heterogeneous mixtures and the Student-t mixture model (MSTD). Mixture of generalized normal distribution (MGND) are studied in Section 2.6. After analysing its moments, 6 algorithms are proposed for estimating the parameters. Section 2.7 has a dual target. Firstly, a simulation study is performed to find the best algorithm to fit the parameters of the MGND model. Specifically, algorithm 5 in the Appendix A is the best one, as it fit well across multiple simulated scenarios. Secondly, a comparative analysis is performed for the 52 stocks of the Euro Stoxx 50 Index using the MGND model and the models introduced in sections 2.4-2.5. Several goodness-of-fit measures are used to determine the best fit: the Kolmogorov–Smirnov statistic, the Anderson–Darling statistic, the Akaike information criterion and the Bayesian information criteria. Findings show that the AIC, KS and AD, mainly select the generalized hyperbolic distribution, while the MGND and MSTD models are often selected as second choice. The student-t distribution is the most selected according to the BIC.

## 2.4 Heavy-tailed distributions

### 2.4.1 Laplace distribution

The Laplace distribution (or double exponential distribution) is a continuous probability distribution named for Pierre Simon Laplace (1749–1827). A random variable X has a Laplace distribution if its probability density function (pdf) is

$$f_L(x|\phi, b) = \frac{1}{2b} \exp\left(-\frac{|x - \phi|}{b}\right),$$ (2.9)

with $-\infty < x < +\infty$, location parameter $-\infty < \phi < +\infty$ and scale parameter $b > 0$ (Kozubowski and Nadarajah 2010). The scale parameter $b$ controls the tail behaviour.

The $n$th moments of the equation 2.9 are given by

$$E(X^n) = \frac{1}{2b}\left[\sum_{k=0}^{n} \binom{n}{k}\left\{1 + (-1)^k\right\}\phi^{n-k}b^{k+1}k!\right].$$ (2.10)

As a result, the variance is defined as $\text{VAR}(X) = 2b^2$.

Figure 2.5 shows a comparison between the Laplace and normal densities with the same mean and variance. The former is more peaked around the center and has much fatter tails than the normal density. Figure 2.6 shows the Laplace pdf for $\phi = 0$ and different $b$ values.

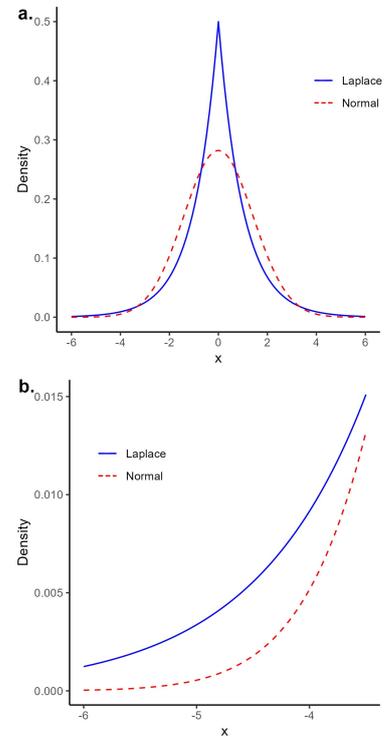

**Figure 2.5:** Panel (a) provides a comparison between the Laplace density and the normal density with the same mean and variance. Panel (b) provides the left tails of both densities.

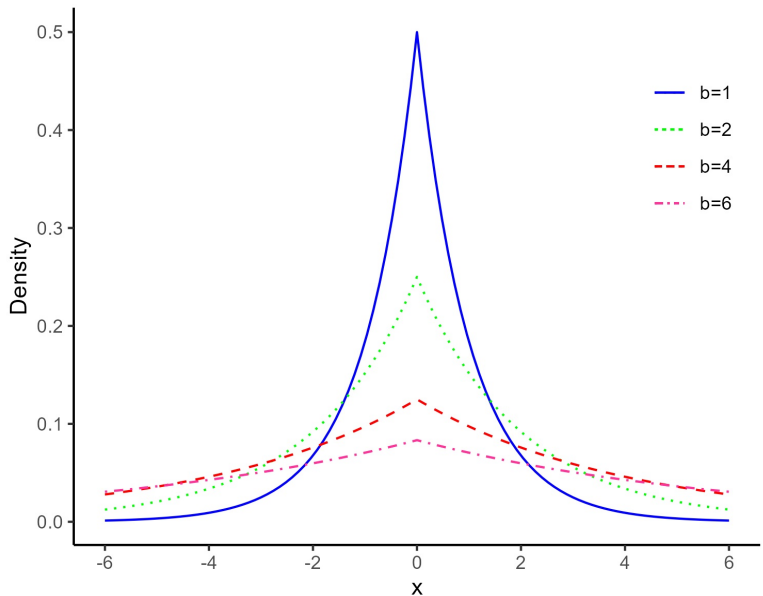

**Figure 2.6:** Laplace density for $\phi = 0$ and different $b$ values.

The maximum likelihood estimators of $\phi$ and $b$ in equation 2.9 are $\hat{\phi} = \text{median}(x_1, x_2, ..., x_n)$ and the mean absolute deviation from the median $\widehat{b} = \frac{1}{n}\sum_{i=1}^{n}|x_i - \widehat{\phi}|$. In this manuscript, the R function *eLaplace* in the *ExtDist* package is used to obtain the estimates of the parameters.

### 2.4.2 Cauchy distribution

The Cauchy distribution is a continuous probability distribution named for Augustin Cauchy (1789-1857). A random variable X has a Cauchy distribution if its pdf is

$$f_C(x|\delta, l) = \frac{1}{\pi l}\left(1 + \left(\frac{x-\delta}{l}\right)^2\right)^{-1},$$ (2.11)

with $-\infty < x < +\infty$, location parameter $-\infty < \delta < +\infty$ and scale parameter $l > 0$. The distribution is symmetric, the median is $\delta$ and the upper and lower quartiles are $\mu \pm l$. However, equation 2.11 has undefined moments. Nadarajah 2011 introduced a truncated version of the Cauchy distribution to overcome this critical issue.

Figure 2.7 shows a comparison between the Cauchy density and the normal density with the same scale and centred to 0. The Cauchy density is less pointed around the center than the normal density, but the tails of the Cauchy density are heavier than the tails of the normal density. Figure 2.8 shows the Cauchy pdf for $\delta = 0$ and different $l$ values.

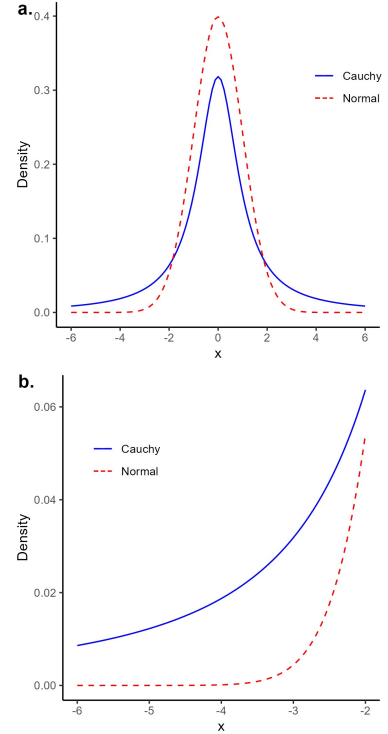

**Figure 2.7:** Panel (a) provides a comparison between the Cauchy density and the normal density with the same scale and centred to 0. Panel (b) provides the left tails of both densities.

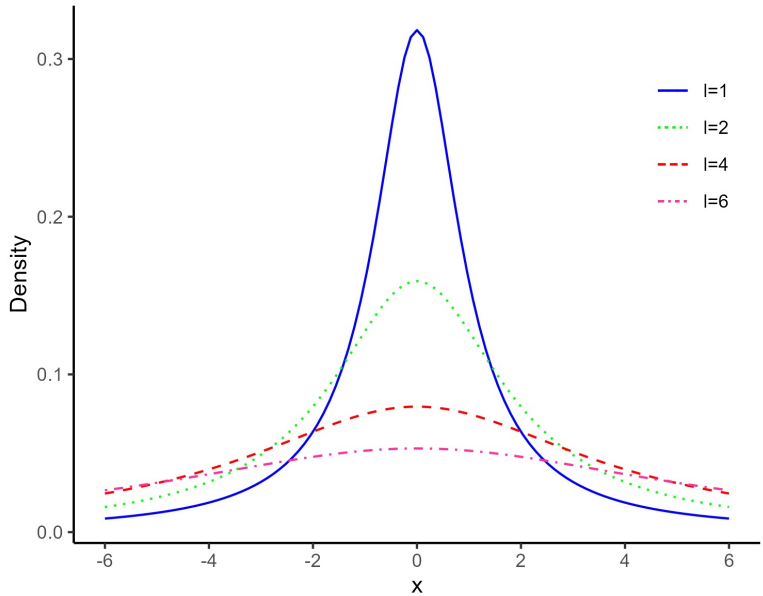

**Figure 2.8:** Cauchy density for $\delta = 0$ and different $l$ values.

The MLE of the parameters is difficult as it requires to find the roots of a high degree polynomial (Ferguson 1978). The Newthon-Raphson method or the Nelder-Mead approach (S. Zhang et al. 2020) can be used as an alternative. In this manuscript, the R function *mlcauchy* in the *univariateML* package is used to obtain the estimates of the parameters.

### 2.4.3 Student-t distribution

The Student-t distribution (or t-distribution) is a family of continuous probability distributions introduced by Willieam S. Gosset (1908) with the work on '*the probable error of a mean*' published under the name 'Student' (Ahsanullah, Kibria, and Shakil 2014). The t-density function is

$$f_{St}(x|\mu, \sigma, \nu) = \frac{\Gamma\left(\frac{\nu+1}{2}\right)}{\Gamma\left(\frac{\nu}{2}\right)\sqrt{\pi\nu\sigma^2}}\left(1 + \frac{1}{\nu}\left(\frac{x-\mu}{\sigma}\right)^2\right)^{-\frac{\nu+1}{2}}, \tag{2.12}$$

with $-\infty < x < +\infty$, location parameter $-\infty < \mu < +\infty$, scale parameter $\sigma > 0$ and shape parameter $\nu > 0$ (degrees of freedom). The gamma function $\Gamma(\cdot)$ is defined as follows

$$\Gamma(t) = \int_0^\infty x^{t-1}\exp(-x)dx, \qquad t > 0. \tag{2.13}$$

For $\nu = 1$ the Student-t distribution reduces to the Cauchy distribution, while for $\nu \to \infty$ the Student-t distribution converges to the normal distribution (Ahsanullah, Kibria, and Shakil 2014). Figure 2.10 shows how the shape of the t-distribution varies as the degrees of freedom increase.

All moments of order $n < \nu$ are finite, $E(X) = \mu$ for $\nu > 1$ and $VAR(X) = \frac{\nu\sigma^2}{\nu-2}$ for $\nu > 2$. As a result, $E(X)$ does not exist for $\nu < 1$ and $VAR(X)$ is infinite for $\nu \leq 2$. Both the kurtosis and the weight of the tails increase as $\nu$ gets smaller. The kurtosis is infinite for $\nu = 4$, otherwise is equal to $Kur(X) = 3 + \frac{6}{\nu-4}$. Figure 2.9 shows a comparison between the normal density and the t-density with $\nu = 5$. Both densities have $\mu = 0$ and $\sigma = 1$. The t-density is more peaked around the center and has much fatter tails than the normal density.

With $\mu = 0$ and $\sigma = 1$ equation 2.12 reduces to the standardized t-distribution

$$f_{St}(x|\mu, \sigma, \nu) = \frac{\Gamma\left(\frac{\nu+1}{2}\right)}{\Gamma\left(\frac{\nu}{2}\right)\sqrt{\pi\nu}}\left(1 + \frac{1}{\nu}x^2\right)^{-\frac{\nu+1}{2}}. \tag{2.14}$$

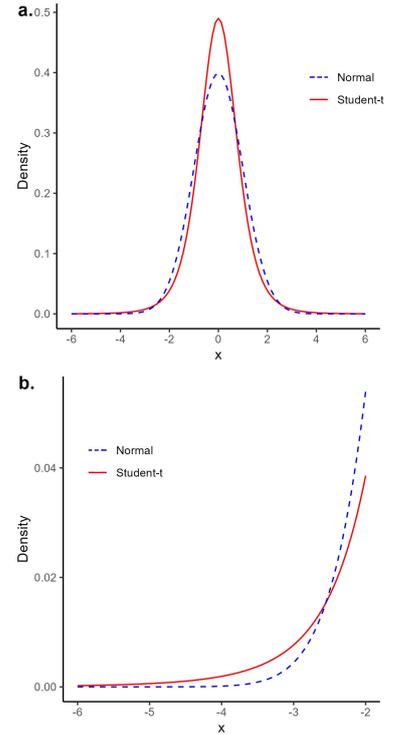

**Figure 2.9:** Panel (a) provides a comparison between the Student-t density with $\nu = 5$ and the normal density. Both densities have $\mu = 0$ and $\sigma = 1$. Panel (b) provides the left tails of both densities.

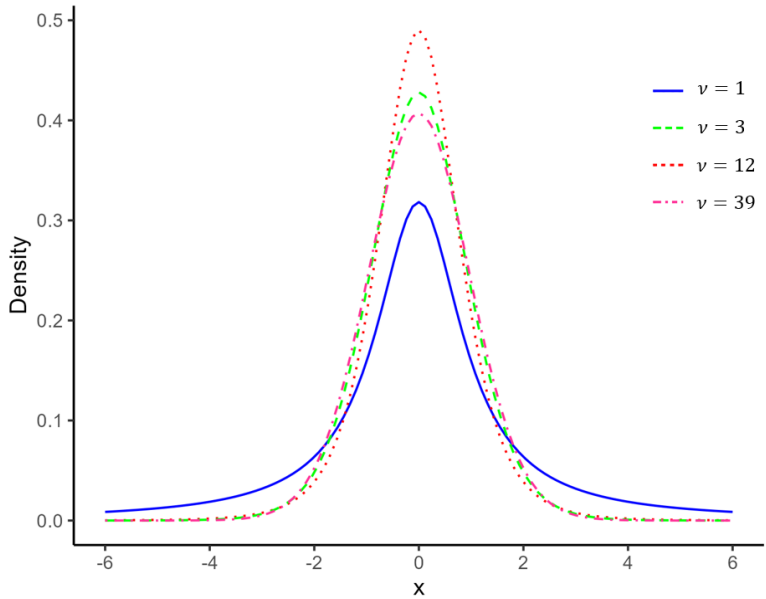

**Figure 2.10:** Student-t density for $\mu = 0$, $\sigma = 1$ and different $\nu$ values.

In this manuscript, the R function *optim* is applied to obtain the ML estimates of the parameters with the Nelder-Mead approach (Nelder and Mead 1965).

### 2.4.4 Generalized hyperbolic distribution

The generalized hyperbolic (GH) distribution introduced by Barndorff-Nielsen 1997 is a powerful model to fit financial returns (X. Jiang, Nadarajah, and Hitchen 2023).The pdf of the GH distribution is

$$
\begin{aligned}
f_{GH}(x|\lambda, \alpha, \beta, \delta, \mu) = & C[\delta^2 + (x-\mu)^2]^{\frac{\lambda}{2} - \frac{1}{4}} K_{\lambda - \frac{1}{2}}\left(\alpha\sqrt{\delta^2(x-\mu)^2}\right) \\
& \times \exp\{\beta(x-\mu)\},
\end{aligned} \tag{2.15}
$$

with $-\infty < x < +\infty$, location parameter $-\infty < \mu < +\infty$, scale parameter $\delta \leq 0$, shape parameters $-\infty < \lambda < +\infty$, $-\infty < \alpha < +\infty$ and skewness parameter $-\infty < \beta < +\infty$. The normalizing constant $C$ is defined as follows

$$
C = \frac{\sqrt{\alpha^2 - \beta^2}^{\lambda}}{\sqrt{2\pi}\alpha^{\lambda - \frac{1}{2}}\delta^{\lambda} K_{\lambda}\left(\delta\sqrt{\alpha^2 - \beta^2}\right)},
$$

and $K_\nu(\cdot)$ is the modified Bessel function of the second kind

$$
K_\nu(x) = \begin{cases} \frac{\pi \csc(\pi\nu)}{2}[I_{-\nu}(x) - I_\nu(x)], & \text{if } \nu \in \mathbb{Z}, \\ \lim_{\mu \to \nu} K_\mu(x), & \text{if } \nu \notin \mathbb{Z}, \end{cases}
$$

where $I_\nu(\cdot)$ is the modified Bessel function of the first kind

$$
I_\nu(x) = \sum_{k=0}^{\infty} \frac{1}{\Gamma(k + \nu + 1)k!}\left(\frac{x}{2}\right)^{2k+\nu}.
$$

The $n$th moments of the equation 2.15 are given by

$$
E(X^n) = \frac{\sqrt{\alpha^2 - \beta^2}^{\lambda}}{K_\lambda \delta\sqrt{\alpha^2 - \beta^2}} \frac{\partial^n}{\partial t^n} \frac{\exp\{\mu t\} K_\lambda \delta\sqrt{\alpha^2 - (\beta + t)^2}}{[\alpha^2 - (\beta + t)^2]^{\frac{\lambda}{2}}}. \tag{2.16}
$$

Figure 2.11 show a comparison among the GH density $f_{GH}(1, 0, 1, 0, -0.5)$, t-density $f_{St}(0, 1, 5)$, and normal density $f_N(0, 1)$. The GH density is more peaked around the center and has much fatter tails than Student-t and normal densities. The GH distribution contains many distributions as limiting or particular cases:

▶ hyperbolic distribution for $\lambda = 1$;
▶ normal inverse Gaussian distribution for $\lambda = \frac{1}{2}$;
▶ the variance gamma distribution for $\delta = 0$;
▶ Student-t distribution with $\nu$ degrees of freedom for $\lambda = -\frac{\nu}{2}$, $\alpha = 0$, $\beta = 0$ and $\delta = \sqrt{\nu}$;
▶ normal distribution for $\delta \to \infty$ and $\frac{\delta}{\alpha} \to \sigma^2$;
▶ GH skew Student-t distribution for $\lambda = -\frac{\nu}{2}$ and $\alpha \to |\beta|$.
▶ generalized inverse Gaussian distribution for $\alpha\delta^2 \to \chi$, $\alpha - \beta = \frac{\psi}{2}$ and $\mu = 0$.

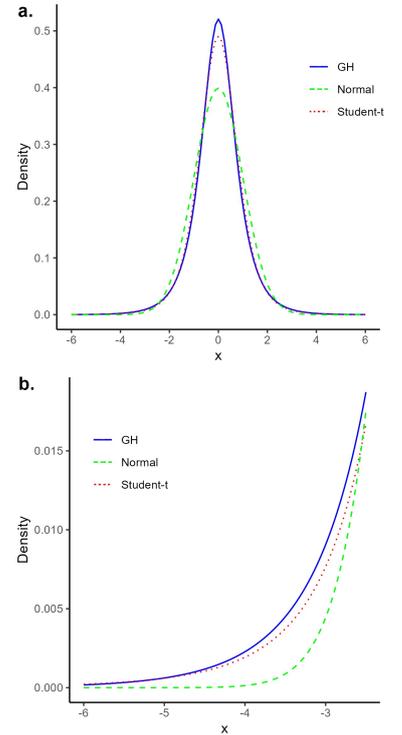

**Figure 2.11:** Panel (a) provides a comparison among the GH density $f_{GH}(1, 0, 1, 0, -0.5)$, t-density $f_{St}(0, 1, 5)$, and normal density $f_N(0, 1)$. Panel (b) provides the left tails.

The likelihood function has a complicated form because depends on five parameters (Massing and Ramos 2021). To find a local maximum with the MLE it is possible to use the numerical ML estimation algorithm of Breymann and Lüthi 2013, the Newthon-Raphson method (Wuertz et al. 2023) or the Nelder-Mead approach (Nelder and Mead 1965; Massing and Ramos 2021). In this manuscript, the R function *ghFit* in the *fBasics* package is used to obtain the estimates of the parameters via the Newthon-Raphson method.

### 2.4.5 Generalized normal distribution

The generalized normal distribution (GND) is '*a natural generalization of the normal distribution*' (Nadarajah 2005). A random variable $X$ is said to have the GND with location $\mu$, scale $\sigma$ and shape $\nu$ if its pdf is given by

$$f_{GND}(x|\mu, \sigma, \nu) = \frac{\nu}{2\sigma\Gamma(1 / \nu)} \exp\left\{-\left|\frac{x - \mu}{\sigma}\right|^{\nu}\right\}, \qquad (2.17)$$

with $-\infty < x < \infty$, $-\infty < \mu < \infty$, $\sigma > 0$, $\nu > 0$. Figure 2.13 shows the probability density function of the GND for $\mu = 1$, $\sigma = 1$ and different shape values. The shape parameter $\nu$ controls both the peakedness and tail weights. If $\nu = 1$ the GND reduces to the Laplace distribution and if $\nu = 2$ it coincides with the normal distribution. It is noticed that $1 < \nu < 2$ yields an 'intermediate distribution' between the normal and the Laplace distribution. As limit cases, for $\nu \to \infty$ the distribution tends to a uniform distribution, while for $\nu \to 0$ it will be impulsive (Nadarajah 2005; Bazi, Bruzzone, and Melgani 2006; Dytso et al. 2018). Figure 2.12 shows a comparison among the GND density and the Student-t density. The GND density is featured by a sharp peak at the center that decreases as $\nu$ increases. By contrast, the Student-t density is smoother in the center. Panel (b) shows that the GND has heavier tails than the Student-t distribution. However, the sharpness and the tails heaviness depend on the shape parameter value (Ruppert 2011).

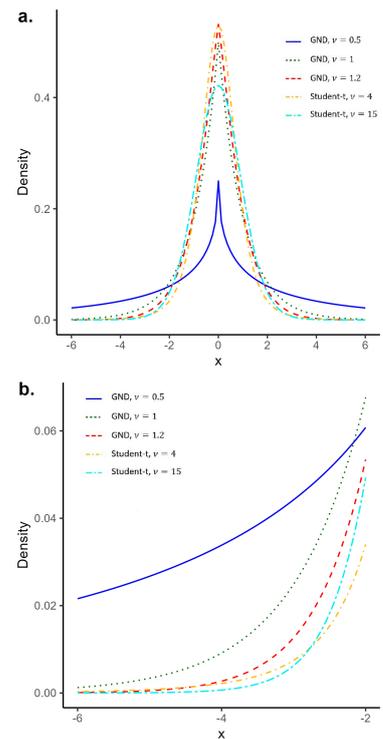

**Figure 2.12:** Panel (a) provides a comparison among GND and Student-t densities. Panel (b) provides the left tails.

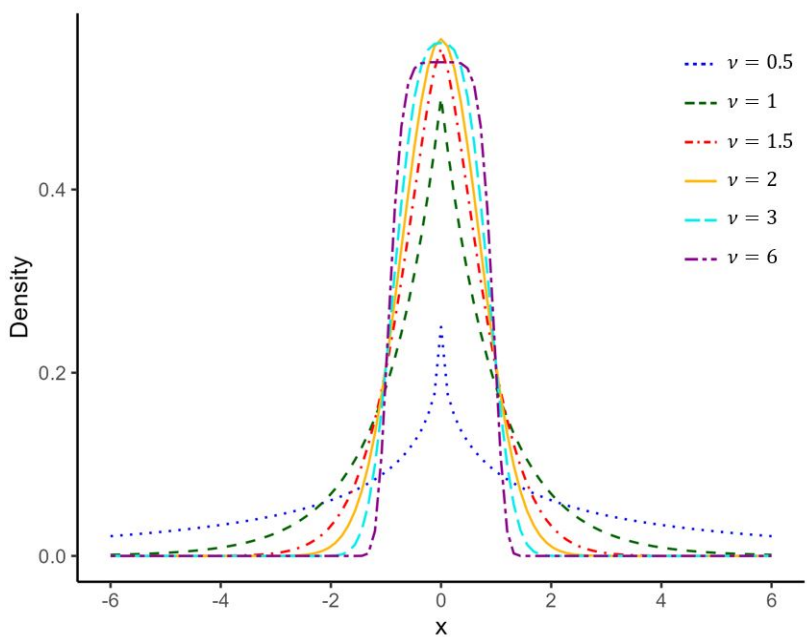

**Figure 2.13:** GND densities for $\mu = 1$, $\sigma = 1$ and different shape values.

The standardized GND is defined as follows

$$f_{GND}(z|\mu, \sigma, \nu) = \frac{\nu \exp{-|z|^\nu}}{2\Gamma(1/\nu)} \qquad \text{with} \qquad Z = \frac{X - \mu}{\sigma}. \quad (2.18)$$

The cumulative density function (cdf) is defined as follows

$$F_{GND}(x|\mu, \sigma, \nu) = \frac{1}{2} + \text{sign}(x - \mu) \frac{1}{2\Gamma(1/\nu)} \gamma\left(\frac{1}{\nu}, \left|\frac{x - \mu}{\sigma}\right|^\nu\right), \quad (2.19)$$

where $\gamma(s, x) = \int_0^x t^{s-1} \exp(-t) dt$ is the lower incomplete gamma function.

The $n$th moments of the equation 2.17 are given by

$$E(X^n) = \frac{\mu^n \sum_{k=0}^n \binom{n}{k} (\sigma/\nu)^k \{1 + (-1)^k\} \Gamma((k+1)/\nu)}{2\Gamma(1/\nu)}. \quad (2.20)$$

Specifically, the first four moments are

$$\begin{aligned}
E(X) &= \mu, \\
E(X^2) &= \mu^2 + \frac{\sigma^2 \Gamma(3/\nu)}{\Gamma(1/\nu)}, \\
E(X^3) &= \mu^3 + \frac{3\mu\sigma^2 \Gamma(3/\nu)}{\Gamma(1/\nu)}, \\
E(X^4) &= \mu^4 + \frac{6\mu^2\sigma^2 \Gamma(3/\nu)}{\Gamma(1/\nu)} + \frac{\sigma^4 \Gamma(5/\nu)}{\Gamma(1/\nu)}.
\end{aligned} \quad (2.21)$$

The $n$th central moments are obtained as follows

$$E[(X - \mu)^n] = \frac{\sigma^n \{1 + (-1)^n\} \Gamma((n+1)/\nu)}{2\Gamma(1/\nu)} \quad (2.22)$$

Mean, variance, skewness and kurtosis obtained from equation 2.22 are

$$\begin{aligned}
E(X) &= \mu, \\
\text{VAR}(X) &= \frac{\sigma^2 \Gamma(3/\nu)}{\Gamma(1/\nu)}, \\
\text{Sk}(X) &= 0, \\
\text{Kur}(X) &= \frac{\Gamma(1/\nu)\Gamma(5/\nu)}{\Gamma(3/\nu)^2}.
\end{aligned} \quad (2.23)$$

As a result, both variance and kurtosis turn on the shape parameter $\nu$. Figure 2.14 shows that variance and kurtosis decrease with respect to $\nu$. What is interesting is that both variance and kurtosis decrease very slowly beyond $\nu = 2$. Table 2.2 shows some values of variance and kurtosis of the GND for $\sigma = 1$.

Varanasi and Aazhang 1989 and Nadarajah 2005 explored two methods for parameters estimation: the method of moments and the MLE. In any case, the estimates do not have a closed form[9] and must be obtained numerically by applying the Newton-Raphson method or the Nelder–Mead approach (Nelder and Mead 1965). The later is a direct optimization

**Table 2.2:** Values of variance and kurtosis of the GND (equation 2.23) versus $\mu = 1$, $\sigma = 1$ and $\nu = 0.5, ..., 40$.

| $\nu$ | VAR$(X)$ | Kur$(X)$ |
|---|---|---|
| 0.5 | 120 | 25.2 |
| 1 | 2 | 6 |
| 1.5 | 0.74 | 3.76 |
| 2 | 0.5 | 3 |
| 3 | 0.37 | 2.42 |
| 10 | 0.31 | 1.88 |
| 20 | 0.32 | 1.82 |
| 40 | 0.33 | 1.81 |

9: This topic will be explored in more detail in Section 2.6.

method which performs a numerical maximization of the log-likelihood function

$$(\widehat{\mu}, \widehat{\sigma}, \widehat{\nu}) \leftarrow \underset{\mu, \sigma, \nu}{\mathrm{argmax}} \; n \log \left\{ \frac{\nu}{2\sigma\Gamma(1/\nu)} \right\} - \sum_{i=1}^{n} \left| \frac{(x_i - \mu)}{\sigma} \right|^{\nu}.$$

In this manuscript, the R function optim is applied to obtain the estimates of the parameters with the Nelder-Mead approach (Nelder and Mead 1965).

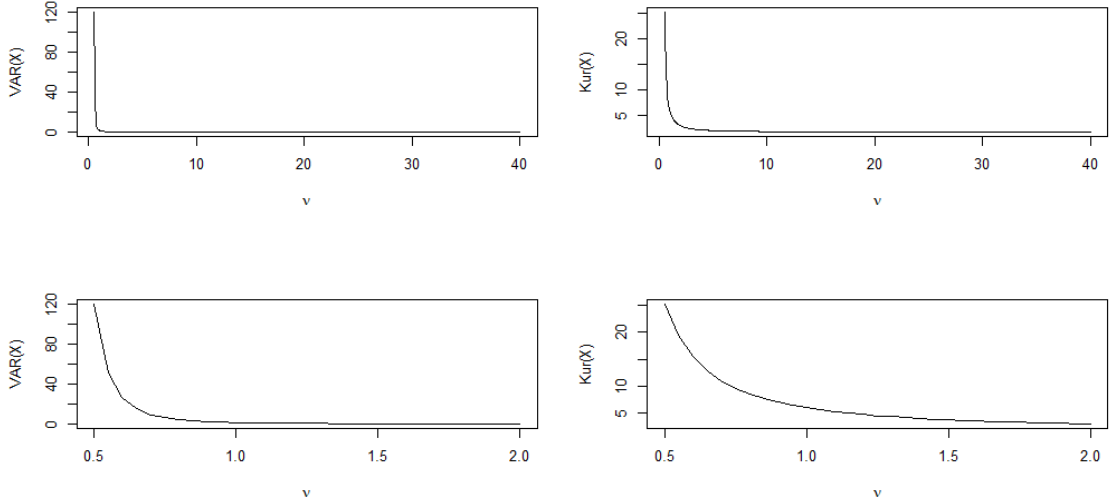

**Figure 2.14:** Variation of variance and kurtosis of the GND versus $\mu = 1$, $\sigma = 1$ and $\nu = 0.5, ..., 40$.

Two other parametrizations of the GND can be found in the literature. The exponential power (EP) distribution introduced by Box and Tiao 1973

$$f_{EP}(x|\mu, \sigma, \nu) = \frac{\exp\left( -\frac{1}{2} \left( \frac{|x-\mu|}{\sigma} \right) \right)^{\frac{2}{1+\nu}}}{2^{\frac{3+\nu}{2}} \sigma \Gamma\left( \frac{3+\nu}{2} \right)} \tag{2.24}$$

with $-\infty < x < \infty$, $-\infty < \mu < \infty$, $\sigma > 0$, $\nu > 0$. For $\nu = 1$ the EP distribution reduces to the Laplace distribution, and for $\nu = 0$ it coincides with the normal distribution. As limit case, for $\nu \to -1$ the distribution tends to a uniform distribution.

The other parametrization is used to model the standardized errors of GARCH models, it is the standardized generalized error distribution (GED) (D. B. Nelson 1991; Ruppert 2011)

$$f_{GED}(x|\nu) = k(\nu) \exp\left\{ -\frac{1}{2} \left| \frac{y}{\lambda_\nu} \right|^2 \right\}, \quad \text{with } -\infty < y < \infty, \tag{2.25}$$

where $k(\nu)$ and $\lambda_\nu$ are constants given by

$$k(\nu) = \frac{\nu}{\lambda_\nu 2^{1+1/\nu} \Gamma(\nu^{-1})},$$

$$\lambda_\nu = \left\{ \frac{2^{-2/\nu}\Gamma(\nu^{-1})}{\Gamma(3/\nu)} \right\}^{1/2}.$$

While for $\nu = 1$ the generalized error distribution reduces to the Laplace distribution, for $\nu = 2$ it coincides with the normal distribution. As limit cases, for $\nu \to 0$ the tail weights will be more extreme than those of the Laplace distribution, while for $\nu \to +\infty$, a uniform distribution is yielded.

## 2.5 Mixture models

### 2.5.1 Normal mixture model

A finite mixture of univariate normal (MN) distributions with $K$ components is given by the marginal distribution of the random variable $X$

$$f_{MN}(x|\theta) = \sum_{k=1}^{K} \pi_k f_k(x|\mu_k, \sigma_k), \qquad (2.26)$$

where $f_k(\cdot) = f_N(\cdot)$ and $-\infty < x < \infty$. The set of all mixture parameters is given by $\theta = \{\pi_k, \mu_k, \sigma_k, k = 1, ..., K\}$ belonging to the parameter space $\Theta = \{\theta : 0 < \pi_k < 1, \sum_{k=1}^{K} \pi_k = 1, -\infty < \mu_k < \infty, \sigma_k > 0, k = 1, ..., K\}$ with $\dim(\theta) = p$. The two-component MN, $(K = 2)$ is given by

$$f_{MN}(x|\theta) = \pi_1 f_1(x|\mu_1, \sigma_1) + \pi_2 f_2(x|\mu_2, \sigma_2), \quad \text{with } \dim(\theta) = 5. \quad (2.27)$$

In this manuscript, the R function *Mclust* of the package *mclust* is applied to obtain the estimates of the parameters.

### 2.5.2 Heterogeneous mixture models

**Normal-Laplace mixture model**

A finite mixture of a normal distribution and a Laplace distribution (MNL) is given by the marginal distribution of the random variable $X$

$$f_{MNL}(x|\theta) = \pi_1 f_1(x|\mu, \sigma) + \pi_2 f_2(x|\phi, b). \qquad (2.28)$$

where $f_1(\cdot) = f_N(\cdot)$, $f_2(\cdot) = f_L(\cdot)$ and $-\infty < x < \infty$. The set of all mixture parameters is given by $\theta = \{\pi_k, \mu, \sigma, \phi, b, k = 1, 2\}$ belonging to the parameter space $\Theta = \{\theta : 0 < \pi_k < 1, \sum_{k=1}^{2} \pi_k = 1, -\infty < \mu < \infty, \sigma > 0, -\infty < \phi < \infty, b > 0, k = 1, 2\}$ with $\dim(\theta) = 5$. In this manuscript, the EM algorithm and the R function *optim* are applied to obtain the estimates of the parameters.

**Normal-Cauchy mixture model**

A finite mixture of a normal distribution and a Cauchy distribution (MNC) is given by the marginal distribution of the random variable $X$

$$f_{MNC}(x|\theta) = \pi_1 f_1(x|\mu, \sigma) + \pi_2 f_2(x|\delta, l). \qquad (2.29)$$

where $f_1(\cdot) = f_N(\cdot)$, $f_2(\cdot) = f_C(\cdot)$ and $-\infty < x < \infty$. The set of all mixture parameters is given by $\theta = \{\pi_k, \mu, \sigma, \delta, l, k = 1, 2\}$ belonging to the parameter space $\Theta = \{\theta : 0 < \pi_k < 1, \sum_{k=1}^{2} \pi_k = 1, -\infty < \mu < \infty, \sigma > 0, -\infty < \delta < \infty, l > 0, k = 1, 2\}$ with $\dim(\theta) = 5$. In this manuscript, the EM algorithm and the R function *optim* are applied to obtain the estimates of the parameters (Y. Zhang et al. 2019; S. Zhang et al. 2020).

### 2.5.3 Student-t mixture model

A finite mixture of univariate Student-t (MSTD) distributions with $K$ components is given by the marginal distribution of the random variable $X$

$$f_{MSTD}(x|\theta) = \sum_{k=1}^{K} \pi_k f_k(x|\mu_k, \sigma_k, \nu_k), \qquad (2.30)$$

where $f_k(\cdot) = f_{St}(\cdot)$ and $-\infty < x < \infty$. The set of all mixture parameters is given by $\theta = \{\pi_k, \mu_k, \sigma_k, \nu_k, k = 1, ..., K\}$ belonging to the parameter space $\Theta = \{\theta : 0 < \pi_k < 1, \sum_{k=1}^{K} \pi_k = 1, -\infty < \mu_k < \infty, \sigma_k > 0, \nu_k > 0, k = 1, ..., K\}$ with $\dim(\theta) = p$. The two-component MSTD ($K = 2$) is given by

$$f_{MSTD}(x|\theta) = f_1(x|\mu_1, \sigma_1, \nu_1) + f_2(x|\mu_2, \sigma_2, \nu_2), \quad \text{with } \dim(\theta) = 7.$$
$$(2.31)$$

This model nests several mixtures of distributions as its sub-models by varying the degrees of freedom $\nu_k$. Equation 2.31 reduces to:

▶ Normal mixture model for $\nu_1 \to +\infty$ and $\nu_2 \to +\infty$;
▶ Cauchy mixture model for $\nu_1 = \nu_2 = 1$;
▶ Normal-Cauchy mixture model for $\nu_1 \to +\infty$ and $\nu_2 = 1$;
▶ Normal-Student's mixture model for $\nu_1 \to +\infty$ and $\nu_2 > 0$;
▶ Cauchy-Student's mixture model for $\nu_1 = 1$ and $\nu_2 > 0$.

In this manuscript, the R function *teigen* of the package *teigen* (Andrews, Wickins, et al. 2018) is used to obtain the estimates of the parameters.

## 2.6 Mixture of generalized normal distributions

A finite mixture of univariate generalized normal distributions (MGND) with $K$ components is given by the marginal distribution of the random variable $X$

$$f_{MGND}(x|\theta) = \sum_{k=1}^{K} \pi_k f_k(x|\mu_k, \sigma_k, \nu_k), \qquad (2.32)$$

where $f_k(\cdot) = f_{GND}(\cdot)$ and $-\infty < x < \infty$. The set of all mixture parameters is given by $\theta = \{\pi_k, \mu_k, \sigma_k, \nu_k, k = 1, ..., K\}$ belonging to the parameter space $\Theta = \{\theta : 0 < \pi_k < 1, \sum_{k=1}^{K} \pi_k = 1, -\infty < \mu_k < \infty, \sigma_k > 0, \nu_k > 0, k = 1, ..., K\}$ with $\dim(\theta) = p$. The two-component MGND ($K = 2$) is given by

$$f_{MGND}(x|\theta) = \pi_1 f_1(x|\mu_1, \sigma_1, \nu_1) + \pi_2 f_2(x|\mu_2, \sigma_2, \nu_2), \quad \text{with } \dim(\theta) = 7.$$
$$(2.33)$$

This model nests several mixtures of distributions as its sub-models by varying the shape parameter $\nu_k$. Equation 2.33 reduces to:

► Normal mixture model for $\nu_1 = \nu_2 = 2$;
► Laplace mixture model for $\nu_1 = \nu_2 = 1$;
► Normal-Laplace mixture model for $\nu_1 = 2$ and $\nu_2 = 1$;
► Normal-GND mixture model for $\nu_1 = 2$ and $\nu_2 > 0$;
► Laplace-GND mixture model for $\nu_1 = 1$ and $\nu_2 > 0$.

### 2.6.1  Moments

From equation 2.22, the $k$th component mean, variance, standard deviation and kurtosis are defined as follows

$$
\begin{aligned}
\mathrm{E}_k &= \mu_k, \\
\mathrm{VAR}_k &= \frac{\sigma_k^2 \Gamma(3/\nu_k)}{\Gamma(1/\nu_k)}, \\
\mathrm{Std}_k &= \sqrt{\frac{\sigma_k^2 \Gamma(3/\nu_k)}{\Gamma(1/\nu_k)}}, \\
\mathrm{Kur}_k &= \frac{\Gamma(1/\nu_k)\Gamma(5/\nu_k)}{\Gamma(3/\nu_k)^2}.
\end{aligned}
\tag{2.34}
$$

The $n$th moments (Nadarajah 2005; Wen et al. 2022) of the equation 2.32 are given by

$$
\mathrm{E}(X^m) = \sum_{k=1}^{K} \pi_k \mu_k^m \sum_{i=0}^{m} \binom{m}{i} \left(\frac{\sigma_k}{\nu_k}\right)^m \frac{\{1 + (-1)^m\}\Gamma((m+1)/\nu_k)}{2\Gamma(1/\nu_k)}, \tag{2.35}
$$

for $K = 2$ (equation 2.33), the first four moments are

$$
\begin{aligned}
\mathrm{E}(X) &= \pi_1 \mu_1 + \pi_2 \mu_2, \\
\mathrm{E}(X^2) &= \pi_1 \left[ \mu_1^2 + \frac{\sigma_1^2 \Gamma(3/\nu_1)}{\Gamma(1/\nu_1)} \right] + \pi_2 \left[ \mu_2^2 + \frac{\sigma_2^2 \Gamma(3/\nu_2)}{\Gamma(1/\nu_2)} \right], \\
\mathrm{E}(X^3) &= \pi_1 \left[ \mu_1^3 + \frac{3\mu_1 \sigma_1^2 \Gamma(3/\nu_1)}{\Gamma(1/\nu_1)} \right] + \pi_2 \left[ \mu_2^3 + \frac{3\mu_2 \sigma_2^2 \Gamma(3/\nu_2)}{\Gamma(1/\nu_2)} \right], \\
\mathrm{E}(X^4) &= \pi_1 \left[ \mu_1^4 + \frac{6\mu_1^2 \sigma_1^2 \Gamma(3/\nu_1)}{\Gamma(1/\nu_1)} + \frac{\sigma_1^4 \Gamma(5/\nu_1)}{\Gamma(1/\nu_1)} \right], \\
&\quad + \pi_2 \left[ \mu_2^4 + \frac{6\mu_2^2 \sigma_2^2 \Gamma(3/\nu_2)}{\Gamma(1/\nu_2)} + \frac{\sigma_2^4 \Gamma(5/\nu_2)}{\Gamma(1/\nu_2)} \right].
\end{aligned}
\tag{2.36}
$$

The $n$th central moments are obtained as follows

$$
\begin{aligned}
\mathrm{E}(X - \mathrm{E}X)^m &= \sum_{k=1}^{K} \pi_k \sum_{i=0}^{m} \binom{m}{i} \left( \mu_k - \sum_{k=1}^{K} \pi_k \mu_k \right)^{(m-i)} \\
&\quad \times \frac{\sigma_k^i \{1 + (-1)^m\}\Gamma((m+1)/\nu_k)}{2\Gamma(1/\nu_k)},
\end{aligned}
\tag{2.37}
$$

for $K = 2$ (equation 2.33) mean, variance, skewness are defined as follows

$$\mathrm{E}(X) = \pi_1\mu_1 + \pi_2\mu_2,$$

$$\mathrm{VAR}(X) = \pi_1\left[\mu_1^2 + \frac{\sigma_1^2\Gamma(3/\nu_1)}{\Gamma(1/\nu_1)}\right] + \pi_2\left[\mu_2^2 + \frac{\sigma_2^2\Gamma(3/\nu_2)}{\Gamma(1/\nu_2)}\right] + \pi_1\pi_2(\mu_1 - \mu_2)^2,$$

$$\mathrm{Sk}(X) = \frac{\pi_1\left[(\mu_1 - \pi_1\mu_1)^3 + 3(\mu_1 - \pi_1\mu_1)\frac{\sigma_1^2\Gamma(3/\nu_1)}{\Gamma(1/\nu_1)}\right]}{\left[(\mu_1 - \pi_1\mu_1)^2\frac{\sigma_1^2\Gamma(3/\nu_1)}{\Gamma(1/\nu_1)}\right]^{\frac{3}{2}}}$$

$$+ \frac{\pi_k\left[(\mu_2 - \pi_2\mu_2)^3 + 3(\mu_2 - \pi_2\mu_2)\frac{\sigma_2^2\Gamma(3/\nu_2)}{\Gamma(1/\nu_2)}\right]}{\left[(\mu_2 - \pi_2\mu_2)^2\frac{\sigma_2^2\Gamma(3/\nu_2)}{\Gamma(1/\nu_2)}\right]^{\frac{3}{2}}},$$

$$\mathrm{Kur}(X) = \frac{\pi_1\left[(\mu_1 - \pi_1\mu_1)^3 + 6(\mu_1 - \pi_1\mu_1)^2\frac{\sigma_1^2\Gamma(3/\nu_1)}{\Gamma(1/\nu_1)}\frac{\sigma_1^4\Gamma(5/\nu_1)}{\Gamma(1/\nu_1)}\right]}{\left(\left[(\mu_1 - \pi_1\mu_1)^2 + \frac{\sigma_1^2\Gamma(3/\nu_1)}{\Gamma(1/\nu_1)}\right]\right)^2}$$

$$+ \frac{\pi_2\left[(\mu_2 - \pi_2\mu_2)^3 + 6(\mu_2 - \pi_2\mu_2)^2\frac{\sigma_2^2\Gamma(3/\nu_2)}{\Gamma(1/\nu_2)}\frac{\sigma_2^4\Gamma(5/\nu_2)}{\Gamma(1/\nu_2)}\right]}{\left(\left[(\mu_2 - \pi_2\mu_2)^2 + \frac{\sigma_2^2\Gamma(3/\nu_2)}{\Gamma(1/\nu_2)}\right]\right)^2}, \tag{2.38}$$

As a result, $\mathrm{VAR}(X)$, $\mathrm{Sk}(X)$ and $\mathrm{Kur}(X)$ depend on all mixture parameters and in particular decrease with respect to $\nu_1$ and $\nu_2$. An example is provided in Figure 2.16 where $\mathrm{VAR}(X)$ and $\mathrm{Kur}(X)$ are computed with respect to $\nu_2 = 0.5, ..., 40$ keeping all other parameters constant. Similar to the non-mixture case, the decrease becomes more slowly and gradual beyond $\nu_k = 2$. For high values of $\nu_k$, $\mathrm{VAR}(X)$ and $\mathrm{Kur}(X)$ are almost constant.

Moreover, $\mathrm{VAR}(X)$ and $\mathrm{Kur}(X)$ increase with respect to $\sigma_2 = 0.5, ..., 40$ keeping all other parameters constant. $\mathrm{Kur}(X)$ increases slowly for high values of $\sigma_2$.

Table 2.3 reports some values of variance and kurtosis versus $\pi_1 = 0.7$, $\mu_1 = 1$, $\mu_2 = 5$, $\sigma_1 = 1$, $\sigma_2 = 3$, $\nu_1 = 2$, and $\nu_2 = 0.5, ..., 40$.

Figures 2.17-2.15 show the functional relationship between $\mathrm{VAR}(X)$, $\mathrm{Kur}(X)$, $\sigma_k$ and $\nu_k$. $\mathrm{Kur}(X)$ is constant when the two mixture components assume the same scale parameter value, while when $\sigma_1 \neq \sigma_2$ the monotonicity of $\mathrm{Kur}(X)$ is irregular.

Wen et al. 2022 explored also the functional relationship between $\mathrm{Sk}(X)$, $\sigma_k$ and $\nu_k$. It was found that when $\sigma_k$ increases $\mathrm{Sk}(X)$ becomes large, while when $\nu_k$ increases $\mathrm{Sk}(X)$ becomes small keeping all other parameters constant.

**Table 2.3:** Values of variance and kurtosis of the MGND (equation 2.38) versus $\pi_1 = 0.7$, $\mu_1 = 1$, $\mu_2 = 5$, $\sigma_1 = 1$, $\sigma_2 = 3$, $\nu_1 = 2$, and $\nu_2 = 0.5, ..., 40$.

| $\nu_2$ | VAR($X$) | Kur($X$) |
|---|---|---|
| 0.5 | 756.91 | 35.26 |
| 1 | 13.51 | 3.33 |
| 1.5 | 5.56 | 0.84 |
| 2 | 4.06 | 0.44 |
| 3 | 3.26 | 0.26 |
| 10 | 2.89 | 0.18 |
| 20 | 2.92 | 0.18 |
| 40 | 2.96 | 0.18 |

$\pi_1 = 0.7, \mu_1 = 1, \mu_2 = 5, \sigma_1 = 1, \sigma_2 = 3$

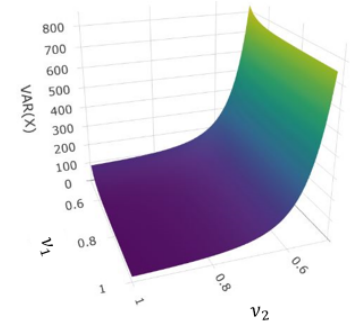

$\pi_1 = 0.7, \mu_1 = 1, \mu_2 = 5, \sigma_1 = 1, \sigma_2 = 3$

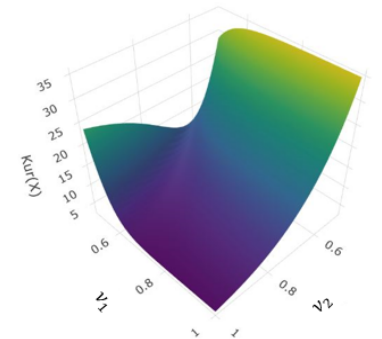

**Figure 2.15:** Functional relationship between variance, kurtosis and $\nu_k$.

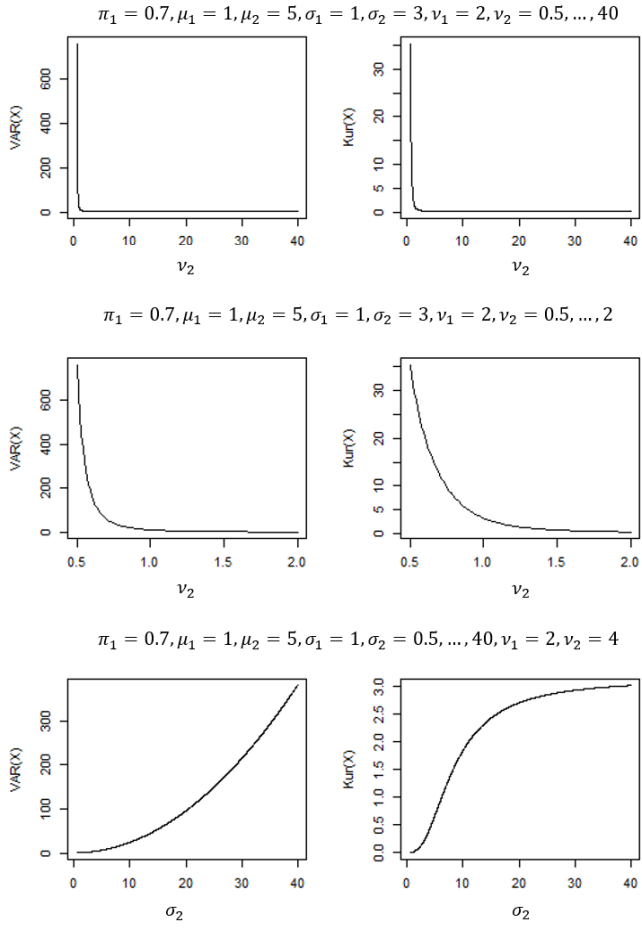

**Figure 2.16:** Variation of variance and kurtosis (equation 2.38) of the MGND versus $\sigma_k$ and $\nu_k$.

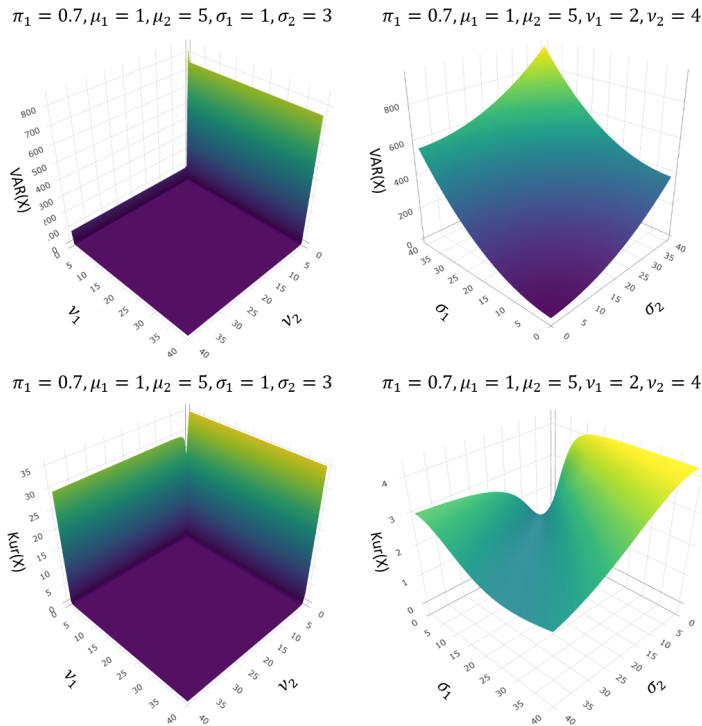

**Figure 2.17:** Functional relationship between variance, kurtosis, $\sigma_k$ and $\nu_k$.

### 2.6.2 Parameter Estimation

The expectation maximization algorithm (EM) is a well-known method to perform the parameter estimation of mixture models (Dempster, Laird, and Rubin 1977). The EM algorithm is divided into two steps: the E-step and the M-step. The E-step computes the expected value of the complete-data log-likelihood, while the M-step maximizes the parameters according to the complete-data log-likelihood (Bishop 2006; Andrews, Wickins, et al. 2018). Wen et al. 2022 proposed an expectation conditional maximization (ECM) algorithm to estimate the parameters of a two-component MGND model. The ECM algorithm (Meng and Rubin 1993) is an extension of the EM which adjusts the M-step into several conditional maximization (CM) steps. The parameter estimation of the MGND model is performed with the ECM algorithm and Newton-Raphson updates.

From equation 2.33 the log-likelihood function is given by

$$\log L(\theta) = \sum_{n=1}^{N} \log \left[ \sum_{k=1}^{K} \pi_k f_k(x_n | \mu_k, \sigma_k, \nu_k) \right]. \tag{2.39}$$

The **E-step** involves computing the following equation

$$Q(\theta, \theta^{(m-1)}) = \sum_{k=1}^{K} \sum_{n=1}^{N} z_{kn}^{(m-1)} \log \left[ \pi_k^{(m-1)} f_k(x_n | \mu_k^{(m-1)}, \sigma_k^{(m-1)}, \nu_k^{(m-1)}) \right], \tag{2.40}$$

where

$$z_{kn}^{(m-1)} = \frac{\pi_k^{(m-1)} f_k(x_n | \mu_k^{(m-1)}, \sigma_k^{(m-1)}, \nu_k^{(m-1)})}{\sum_{k=1}^{K} \pi_k^{(m-1)} f_k(x_n | \mu_k^{(m-1)}, \sigma_k^{(m-1)}, \nu_k^{(m-1)})}. \tag{2.41}$$

The term $z_{kn}^{(m-1)}$ represents the current estimate of the posterior probability or '*responsibility*' (Bishop 2006) at the (*m-1*) iteration, i.e. the probability that the observation $n$ belongs to group $k$ given the current component parameters $\theta^{(m-1)}$.

The **CM-Step** maximizes $Q(\theta, \theta^{(m-1)})$ with respect to $\theta$ to obtain the parameter estimation at the (*m*) iteration and increases the expectation of the log-likelihood function in equation 2.39. In what follows the updating equations are provided.

**Mixture weights** Set $\frac{\partial Q(\theta, \theta^{(m-1)})}{\partial \pi_k} = 0$, then

$$\pi_k^{(m)} = \frac{\sum_{n=1}^{N} z_{kn}^{(m-1)}}{\sum_{k=1}^{K} \sum_{n=1}^{N} z_{kn}^{(m-1)}}. \tag{2.42}$$

**Location parameter** Set

$$
\frac{\partial Q(\theta, \theta^{(m-1)})}{\partial \mu_k} = \frac{v_k^{(m-1)}}{(\sigma_k^{(m-1)})^{v_k^{(m-1)}}} \left( \sum_{x_n \geq \mu_k^{(m-1)}}^{N} z_{kn}^{(m-1)} (x_n - \mu_k^{(m-1)})^{v_k^{(m-1)}-1} \right.
$$
$$
\left. - \sum_{x_n < \mu_k^{(m-1)}}^{N} z_{kn}^{(m-1)} (\mu_k^{(m-1)} - x_n)^{v_k^{(m-1)}-1} \right) = 0,
$$
$$
\text{for } k = 1, 2, ..., K.
$$

Since the previous equation is non-linear, the iterative Newton-Raphson method is applied as follows:

$$
\mu_k^{(m)} = \mu_k^{(m-1)} - \frac{g(\mu_k^{(m-1)})}{g'(\mu_k^{(m-1)})}, \tag{2.43}
$$

where

$$
g(\mu_k^{(m-1)}) = \frac{v_k^{(m-1)}}{(\sigma_k^{(m-1)})^{v_k^{(m-1)}}} \left( \sum_{x_n \geq \mu_k^{(m-1)}}^{N} z_{kn}^{(m-1)} (x_n - \mu_k^{(m-1)})^{v_k^{(m-1)}-1} \right.
$$
$$
\left. - \sum_{x_n < \mu_k^{(m-1)}}^{N} z_{kn}^{(m-1)} (\mu_k^{(m-1)} - x_n)^{v_k^{(m-1)}-1} \right), \tag{2.44}
$$

$$
g'(\mu_k^{(m-1)}) = -\frac{v_k^{(m-1)}}{(\sigma_k^{(m-1)})^{v_k^{(m-1)}}} \left( \sum_{x_n \geq \mu_k^{(m-1)}}^{N} z_{kn}^{(m-1)} (x_n - \mu_k^{(m-1)})^{v_k^{(m-1)}-2} \right.
$$
$$
\times (v_k^{(m-1)} - 1) + \sum_{x_n < \mu_k^{(m-1)}}^{N} z_{kn}^{(m-1)} (\mu_k^{(m-1)} - x_n)^{v_k^{(m-1)}-2} (v_k^{(m-1)} - 1) \right). \tag{2.45}
$$

After some further calculations, the location parameter of the *k-th* component is estimated by the following iteration equation (Wen et al. 2022; Bazi, Bruzzone, and Melgani 2006)

$$
\mu_k^{(m)} = \mu_k^{(m-1)} + \frac{A_k}{B_k} \tag{2.46}
$$

where

$$
A_k = \sum_{x_n \geq \mu_k^{(m-1)}}^{N} z_{kn}^{(m-1)} (x_i - \mu_k^{(m-1)})^{v_k^{(m-1)}-1} - \sum_{x_n < \mu_k}^{N} z_{kn}^{(m-1)} (\mu_k^{(m-1)} - x_n)^{v_k^{(m-1)}-1},
$$

$$
B_k = \sum_{n=1}^{N} z_{kn}^{(m-1)} |x_n - \mu_k^{(m-1)}|^{v_k^{(m-1)}-2} (v_k^{(m-1)} - 1).
$$

**Scale parameter** Set

$$\frac{\partial Q(\theta, \theta^{(m-1)})}{\partial \sigma_k} = \sum_{n=1}^{N} z_{kn}^{(m-1)} \left( -\frac{1}{\sigma_k^{(m-1)}} \right) + \frac{v_k^{(m-1)}}{(\sigma_k^{(m-1)})^{v_k^{(m-1)}+1}}$$

$$\times \sum_{n=1}^{N} z_{kn}^{(m-1)} |x_n - \mu_k^{(m)}|^{v_k^{(m-1)}} = 0,$$

$$\text{for } r = 1, 2, ..., R$$

after some calculations, it is possible to obtain the fixed point update (Wen et al. 2022; Bazi, Bruzzone, and Melgani 2006)

$$\sigma_k^{(m)} = \left[ \frac{v_k^{(m-1)} \sum_{n=1}^{N} z_{kn}^{(m-1)} |x_n - \mu_k^{(m)}|^{v_k^{(m-1)}}}{\sum_{n=1}^{N} z_{kn}^{(m-1)}} \right]^{\frac{1}{v_k^{(m-1)}}}. \tag{2.47}$$

**Shape parameter** Set

$$\frac{\partial Q(\theta, \theta^{(m-1)})}{\partial v_k} = \sum_{n=1}^{N} z_{kn}^{(m-1)} \frac{1}{v_k^{(m-1)}} \left( \frac{1}{v_k^{(m-1)}} \Psi \left( \frac{1}{v_k^{(m-1)}} \right) + 1 \right)$$

$$- \sum_{n=1}^{N} z_{kn}^{(m-1)} \left| \frac{x_n - \mu_k^{(m)}}{\sigma_k^{(m)}} \right|^{v_k^{(m-1)}} \log \left| \frac{x_n - \mu_k^{(m)}}{\sigma_k^{(m)}} \right| = 0,$$

$$\text{for } r = 1, 2, ..., R.$$

Since the above equation is a non-linear equation, the iterative Newton-Raphson method is applied as follows:

$$v_k^{(m)} = v_k^{(m-1)} - \frac{g\left(v_k^{(m-1)}\right)}{g'\left(v_k^{(m-1)}\right)}, \tag{2.48}$$

where

$$g\left(v_k^{(m-1)}\right) = \sum_{n=1}^{N} z_{kn}^{(m-1)} \frac{1}{v_k^{(m-1)}} \left( \frac{1}{v_k^{(m-1)}} \Psi \left( \frac{1}{v_k^{(m-1)}} \right) + 1 \right)$$

$$- \sum_{n=1}^{N} z_{kn}^{(m-1)} \left| \frac{x_n - \mu_k^{(m)}}{\sigma_k^{(m)}} \right|^{v_k^{(m-1)}} \log \left| \frac{x_n - \mu_k^{(m)}}{\sigma_k^{(m)}} \right|, \tag{2.49}$$

$$g'\left(v_k^{(m-1)}\right) = \sum_{n=1}^{N} z_{kn}^{(m-1)} - \frac{1}{\left(v_k^{(m-1)}\right)^2} \left( 1 + \frac{2}{v_k^{(m-1)}} \Psi \left( \frac{1}{v_k^{(m-1)}} \right) \right.$$

$$\left. + \frac{1}{\left(v_k^{(m-1)}\right)^2} \Psi' \left( \frac{1}{v_k^{(m-1)}} \right) \right) - \sum_{n=1}^{N} z_{kn}^{(m-1)} \left| \frac{x_n - \mu_k^{(m)}}{\sigma_k^{(m)}} \right|^{v_k^{(m-1)}}$$

$$\times \left( \log \left| \frac{x_n - \mu_k^{(m)}}{\sigma_k^{(m)}} \right| \right)^2. \tag{2.50}$$

The shape parameter of the *k-th* component is estimated by the following iteration equation (Wen et al. 2022; Bazi, Bruzzone, and Melgani 2006)

$$\nu_k^{(m)} = \nu_k^{(m-1)} - \frac{\sum_{n=1}^{N} z_{kn}^{(m-1)} A_k - \sum_{n=1}^{N} z_{kn}^{(m-1)} B_k}{\sum_{n=1}^{N} z_{kn}^{(m-1)} C_k - \sum_{n=1}^{N} z_{kn}^{(m-1)} D_k}, \qquad (2.51)$$

where

$$A_k = \frac{1}{\nu_k^{(m-1)}} \left( \frac{1}{\nu_k^{(m-1)}} \Psi\left( \frac{1}{\nu_k^{(m-1)}} \right) + 1 \right),$$

$$B_k = \left| \frac{x_n - \mu_k^{(m)}}{\sigma_k^{(m)}} \right|^{\nu_k^{(m-1)}} \log \left| \frac{x_n - \mu_k^{(m)}}{\sigma_k^{(m)}} \right|,$$

$$C_k = -\frac{1}{(\nu_k^{(m-1)})^2} \left( 1 + \frac{2}{\nu_k^{(m-1)}} \Psi\left( \frac{1}{\nu_k^{(m-1)}} \right) + \frac{1}{(\nu_k^{(m-1)})^2} \Psi'\left( \frac{1}{\nu_k^{(m-1)}} \right) \right),$$

$$D_k = \left| \frac{x_n - \mu_k^{(m)}}{\sigma_k^{(m)}} \right|^{\nu_k^{(m-1)}} \left( \log \left| \frac{x_n - \mu_k^{(m)}}{\sigma_k^{(m)}} \right| \right)^2,$$

The digamma $\Psi(1/\nu_k)$ and trigamma $\Psi'(1/\nu)$ functions are

$$\Psi(1/\nu_k) = \frac{\partial \Gamma(1/\nu_k)}{\partial(1/\nu_k)} \log \Gamma(1/\nu_k), \qquad \Psi'(1/\nu) = \frac{\partial^2 \Gamma(1/\nu_k)}{\partial(1/\nu_k)^2} \log \Gamma(1/\nu_k). \qquad (2.52)$$

Algorithm 1 performs the parameter estimation of the MGND model via the ECM algorithm and Newton-Raphson updates (Wen et al. 2022).

---

**Algorithm 1:** Parameter estimation of the MGND model via the ECM algorithm and Newton-Raphson updates.

---

1. **require:** data $x_1, x_2, ..., x_N$.
2. **set the initial estimates:** *k-means* initialization
   minimize $\sum_{k=1}^{K} W(P_k)$, where $P_k$ denotes the set of units belonging to the *k*th cluster and $W(P_k)$ is the within cluster variation.
   $\mu_k^{(m-1)} \leftarrow \text{mean}(P_k), \sigma_k^{(m-1)} \leftarrow \text{std}(P_k),$
   $\nu_k^{(m-1)} \leftarrow$ randomly generated in $[0.5, 3]$
   $\pi_k^{(m-1)} \leftarrow$ randomly generated in $[0, 1], z_{kn}^{(m-1)} \leftarrow$ Eq. 2.41,
   $\epsilon \leftarrow 10^{-5}$
3. **while** $\|\theta^{(m-1)} - \theta^{(m)}\| \le \epsilon$ not convergence **do**
   $\mu_k^{(m)} \leftarrow$ Eq. 2.46, $\sigma_k^{(m)} \leftarrow$ Eq. 2.47, $\nu_k^{(m)} \leftarrow$ Eq. 2.51, $\pi_k^{(m)} \leftarrow$ Eq. 2.42, $z_{kn}^{(m)} \leftarrow$ Eq. 2.41, $\log L(\theta^{(m)}) \leftarrow$ equation 2.39
   **evaluate** $\|\theta^{(m-1)} - \theta^{(m)}\| \le \epsilon$
   $\pi_k^{(m-1)} \leftarrow \pi_k^{(m)}, \mu_k^{(m-1)} \leftarrow \mu_k^{(m)}, \sigma_k^{(m-1)} \leftarrow \sigma_k^{(m)}, \nu_k^{(m-1)} \leftarrow \nu_k^{(m)},$
   $\log L(\theta^{(m-1)}) \leftarrow \log L(\theta^{(m)})$
   **end while**
4. **return** $\theta^{(m)}, \log L(\theta^{(m)})$

---

The Brent method, implemented by the R function *optimize* (Brent 2013), is an alternative to the Newton-Raphson method. It performs a direct optimization by maximizing numerically equation 2.40. The parameter estimation of the MGND model via the ECM algorithm and 'Brent updates' is reported in Algorithm 2 in Appendix A.

These two algorithms may have some convergence issues when $v_k > 2$. As explained in the previous section, variance and kurtosis are nearly constant and vary slowly for high values of $v_k$. A case study with sample size 500 is reported. Data are generated using the R function *rgnorm* of the package *gnorm* with the following parameter setting (Wen et al. 2022) $\mu_1 = \mu_2 = 0$, $\sigma_1 = 2$, $\sigma_2 = 5$, $v_1 = 5$, and $v_2 = 3$. The estimated parameters with algorithms 1-2 differ significantly from their real values (Table 2.5). Figure 2.18 shows the convergence of the parameter estimates with Algorithm 1. The likelihood function is maximized but the final solution is a spurious solution: the degeneracy of $\widehat{v_2} = 33.45$ from its real value $v_2 = 5$ is remarkable.

In order to control the degeneracy of $v_k$, a step size $\alpha$ is introduced as follows

$$v_k^{(m)} = v_k^{(m-1)} - \alpha \frac{g\left(v_k^{(m-1)}\right)}{g'\left(v_k^{(m-1)}\right)} \quad \text{with } \alpha \in [1, 0]. \tag{2.53}$$

The lower the value of $\alpha$ the lower the update provided by $\frac{g\left(v_k^{(m-1)}\right)}{g'\left(v_k^{(m-1)}\right)}$. If $\alpha = 1$, equation 2.53 reduces to equation 2.48. A step size $\alpha$ is introduced also for the location parameter,

$$\mu_k^{(m)} = \mu_k^{(m-1)} - \alpha \frac{g\left(\mu_k^{(m-1)}\right)}{g'\left(\mu_k^{(m-1)}\right)} \quad \text{with } \alpha \in [1, 0]. \tag{2.54}$$

The scale parameter does not require a fixed step size because its update is in closed form. Algorithm 3 in the appendix A performs the parameter estimation via the ECM algorithm and Newton-Raphson updates with a fixed step-size $\alpha$. Results are showed in Table 2.5 for $\alpha = 0.8, 0.5, 0.2, 0.1$. Moving from $\alpha = 0.8$ to $\alpha = 0.1$ the degeneracy of $v$ is gradually reduced. Specifically, the estimates are similar to those provided by algorithm 2 for $\alpha = 0.8$, while the estimate of $v_2$ is too much limited for $\alpha = 0.1$. Good estimates are provided by $\alpha = 0.2$. However, in the next section, it will be shown that the optimal $\alpha$ depends on the simulation scenario and sample size.

In order to automate the choice of the step-size $\alpha$, a backtracking line-search is introduced. To calculate a suitable step-length $\alpha$ it is possible to start from $\alpha = 1$ and check if the likelihood increases otherwise $\alpha$ is decremented by a factor of 0.8. If $\alpha < 0.01$, the search is terminated with no parameter update, i.e. $v_k^{(m)} = v_k^{(m-1)}$. If the search is terminated with $\alpha = 1$, a step size is not required to estimate $v_k^{(m)}$. The backtracking line-search is introduced also for the location parameter update. Since the backtracking line-search framework requires that **CM-steps** increase, rather than maximize, the log-likelihood function, a generalized expectation maximization (GEM) algorithm (G. McLachlan and Krishnan 2008) is proposed (algorithm 4 in the appendix A). Table 2.5 shows that algorithm 4 does not fix the degeneracy of the shape parameters as $v_1 = 33.439$.

Thus, an additional step-size on the shape parameter update is required

$$v_k^{(m)} = v_k^{(m-1)} - \alpha^{(i)} \frac{1}{v_k^{(m-1)}} \frac{g\left(v_k^{(m-1)}\right)}{g'\left(v_k^{(m-1)}\right)} \quad \text{with } \alpha^{(i)} \in [1,0], \qquad (2.55)$$

where the step-size term $\frac{1}{v_k^{(m-1)}}$ controls the magnitude of $v_k^{(m)}$. The higher the value of $v_k^{(m-1)}$ the lower the update provided by $\frac{g\left(v_k^{(m-1)}\right)}{g'\left(v_k^{(m-1)}\right)}$.

Algorithm 5 in the appendix A performs parameter estimation via the GEM algorithm and Newton-Raphson updates with the backtracking line-search and the additional step-size. It solves the degeneracy issue obtaining better estimates with $v_1 = 6.461$ and $v_2 = 2.808$ (Table 2.5). The convergence of parameter estimates is shown in Figure 2.19. It is possible to notice how $v_1$ and $v_2$ converge towards their real values addressing the degeneracy problem encountered with Algorithm 1 as shown in Figure 2.18.

Besides, the parameter estimation via the ECM algorithm and Newton-Raphson updates with adaptive $v$ step-size are explored with Algorithm 6 in the appendix A. It can be argued that the adaptive $v$ step-size is sufficient to guarantee the convergence of the shape parameter (Table 2.5).

In Table 2.5 the estimated variance and kurtosis provided by algorithms 5-6 (3.561 and 5.144) are almost equal to the sample variance and kurtosis (3.556 and 5.059).

In addition, Figure 2.20 shows a comparison among the estimated densities. To enhance the reading, the estimates of algorithms 4 and 6 are omitted because they are similar to those of algorithms 1 and 5, respectively. Panel (a) represents a comparison among the estimated densities and the true density. The estimated density of algorithms 1-2-3 have a central peak that diverges from the true density. Panel (b) shows the estimated and true first mixture component densities. The estimated first mixture component density by Algorithm 5 is close to the true one.

The case study provides a first overview of the algorithm's performance. Algorithm 1-2 do not prevent the degeneracy issue for $v_k > 2$. Algorithm 3 requires experimenting different step-size $\alpha$ to identify the optimal value. Algorithm 4 automates the choice of the step-size but does not prevent the degeneracy issue. Finally, algorithm 5-6 prevent the degeneracy issue thanks to the additional step-size on the shape parameter. However, the case study is not sufficient to identify the best algorithm, a simulation study is performed in the next section.

**Table 2.4:** Algorithms summary.

| Algorithm | Method |
|---|---|
| Algorithm 1 | ECM |
| Algorithm 2 | ECM with optim updates |
| Algorithm 3 | ECM with step-size $\alpha$ |
| Algorithm 4 | GEM with line-search |
| Algorithm 5 | GEM with line-search and adaptive step-size $1/v$ |
| Algorithm 6 | ECM with adaptive step-size $1/v$ |

**Table 2.5:** Case study: parameter estimates of Algorithms 1-6.

|  | $\pi_1$ 0.7 | $\mu_1$ 0 | $\sigma_1$ 2 | $\nu_1$ 5 | $\mu_2$ 0 | $\sigma_2$ 5 | $\nu_2$ 3 | VAR(X) | Kur(X) | $\log L(\theta)$ |
|---|---|---|---|---|---|---|---|---|---|---|
| Algorithm 1 | 0.450 | -0.287 | 1.928 | 33.456 | 0.190 | 1.974 | 1.121 | 3.618 | 6.691 | -984.43 |
| Algorithm 2 | 0.564 | -0.211 | 1.945 | 8.999 | 0.282 | 2.801 | 1.401 | 3.593 | 6.079 | -986.49 |
| Algorithm 3, $\alpha = 0.1$ | 0.612 | -0.182 | 1.691 | 2.408 | 0.343 | 3.099 | 1.496 | 3.576 | 6.154 | -993.22 |
| Algorithm 3, $\alpha = 0.2$ | 0.641 | -0.143 | 1.936 | 6.019 | 0.287 | 3.582 | 1.721 | 3.563 | 5.822 | -987.14 |
| Algorithm 3, $\alpha = 0.5$ | 0.597 | -0.168 | 1.941 | 7.148 | 0.272 | 3.098 | 1.512 | 3.571 | 5.990 | -986.99 |
| Algorithm 3, $\alpha = 0.8$ | 0.574 | -0.195 | 1.947 | 8.998 | 0.273 | 2.893 | 1.436 | 3.587 | 6.034 | -986.54 |
| Algorithm 4 | 0.450 | -0.287 | 1.928 | 33.439 | 0.190 | 1.975 | 1.122 | 3.618 | 6.690 | -984.43 |
| Algorithm 5 | 0.719 | -0.118 | 1.975 | 6.461 | 0.238 | 4.938 | 2.808 | 3.561 | 5.144 | -984.33 |
| Algorithm 6 | 0.719 | -0.118 | 1.975 | 6.460 | 0.238 | 4.937 | 2.806 | 3.561 | 5.144 | -984.33 |

*Notes.* The sample variance and kurtosis are equal to 3.556 and 5.059, respectively.

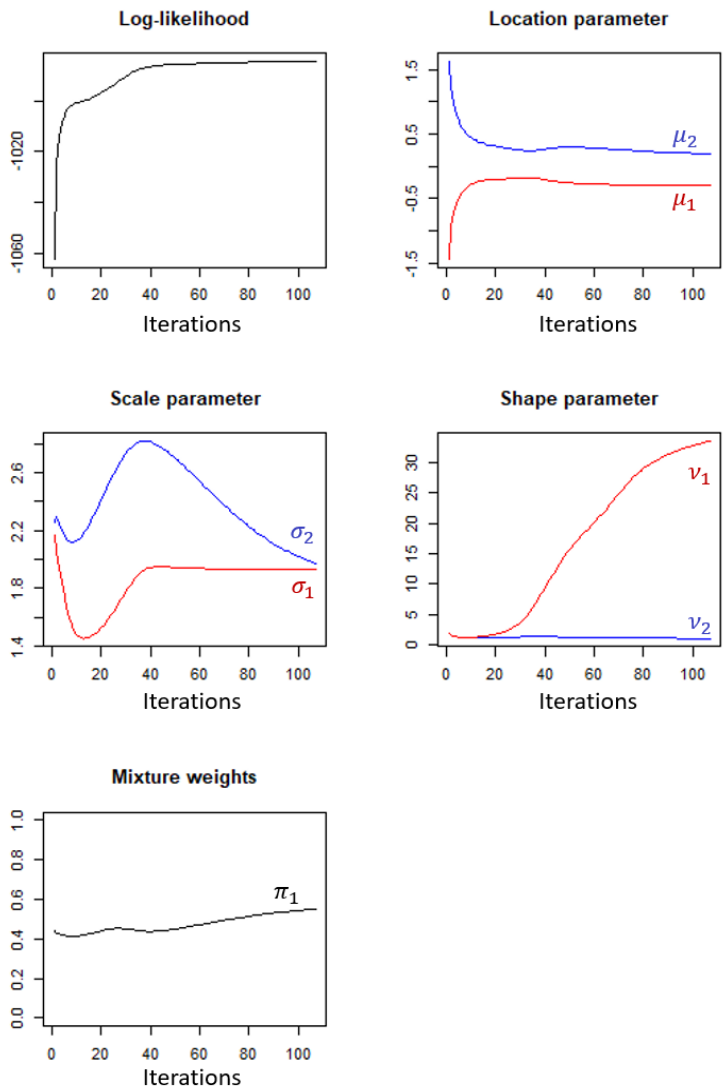

**Figure 2.18:** Convergence of parameter estimates via the ECM algorithm and Newton-Raphson updates for sample size of 500 with $\mu_1 = \mu_2 = 0$, $\sigma_1 = 2$, $\sigma_2 = 5$, $\nu_1 = 5$, and $\nu_2 = 2$.

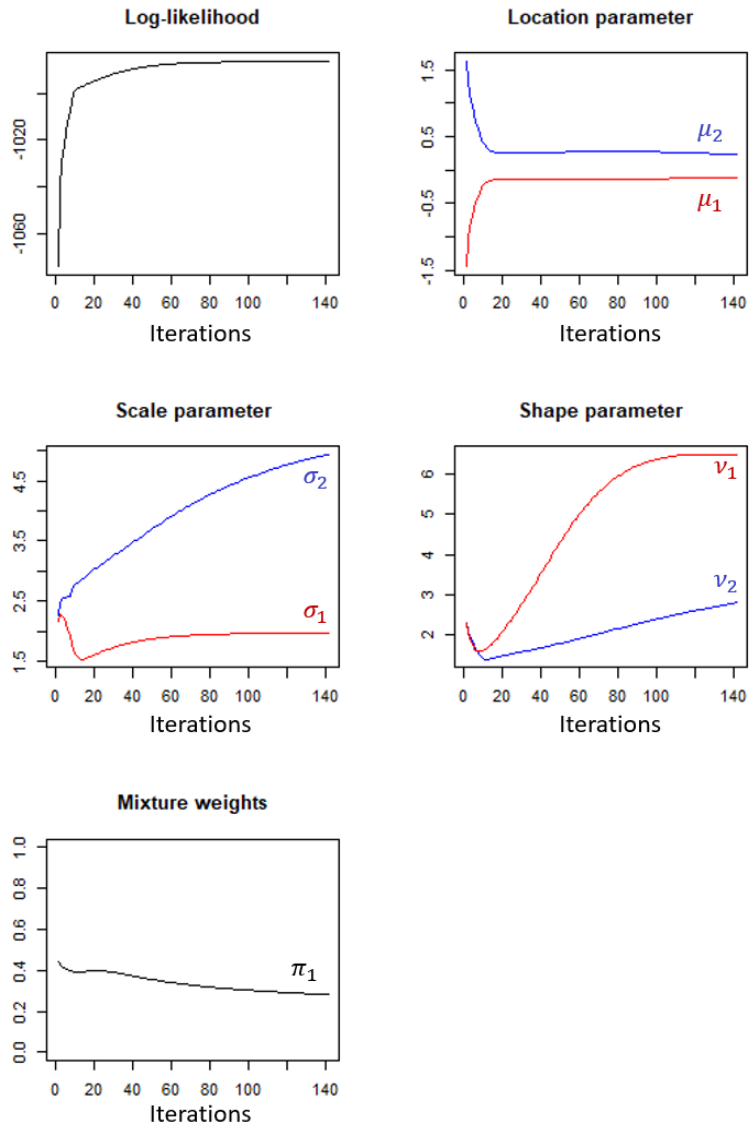

**Figure 2.19:** Convergence of parameter estimates via the GEM algorithm and Newton-Raphson updates with the adaptive $\nu$ step-sizes and backtracking line-search for sample size of 500 with $\mu_1 = \mu_2 = 0$, $\sigma_1 = 2$, $\sigma_2 = 5$, $\nu_1 = 5$, and $\nu_2 = 2$.

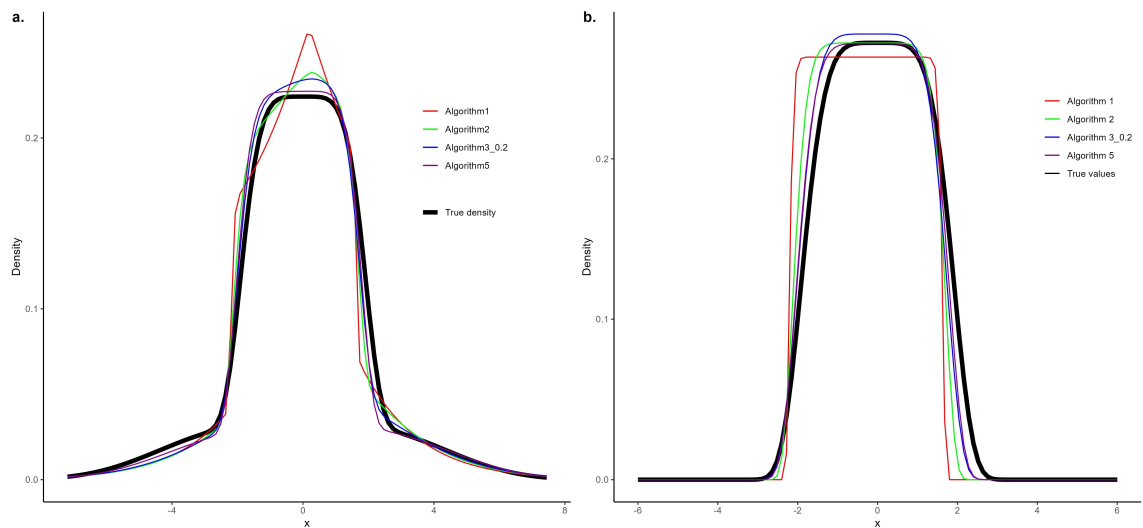

**Figure 2.20:** Case study, comparison among densities. Panel (a): estimated densities and the true density. Panel (b): estimated and true first mixture component densities.

## 2.7 Numerical experiments

### 2.7.1 Simulation study

The main target of the simulation study is to study the performance of algorithms 1-6 across different simulated scenarios, identifying which of them fits the MGND model best.

The simulated scenarios are shown in Table 2.6. Scenario 1 shares the same parameter setting of the previous section's case study. In scenario 2, the predominant mixture component is a normal distribution with $\nu_1 = 2$, while the less frequent mixture component gets $\nu_2 = 4$. In scenario 3, mixture parameters take on values very close to those estimated on financial returns. It is featured by a normally distributed principal component ($\nu_1 = 2$) and a less frequent component with heavy-tails ($\nu_2 = 0.8 < 1$). Scenario 4 has the same parameter setting of scenario 3, except for the location parameter which differs between the two mixture components ($\mu_1 = 1$ and $\mu_2 = 5$).

| | $\pi_1$ | $\mu_1$ | $\sigma_1$ | $\nu_1$ | $\pi_2$ | $\mu_2$ | $\sigma_2$ | $\nu_2$ |
|---|---|---|---|---|---|---|---|---|
| Scenario 1 | 0.7 | 0 | 2 | 5 | 0.3 | 0 | 5 | 3 |
| Scenario 2 | 0.7 | 1 | 1 | 2 | 0.3 | 5 | 3 | 4 |
| Scenario 3 | 0.7 | 0 | 1 | 2 | 0.3 | 0 | 3 | 0.8 |
| Scenario 4 | 0.7 | 1 | 1 | 2 | 0.3 | 5 | 3 | 0.8 |

**Table 2.6:** MGND simulated scenarios

For each scenario $S = 250$ samples with sample size $N = 500, 2000$ are generated using the R function *rgnorm* of the package *gnorm*. Algorithms 1-6 are estimated for each sample.

As a measure of quality estimation, for each parameter of the component mixture the mean square error (MSE) is computed as follows

$$\text{MSE}(\widehat{\theta}) = \frac{1}{S} \sum_{s=1}^{S} (\widehat{\theta}_s - \theta)^2, \qquad (2.56)$$

where $\theta$ is the true parameter value and $\widehat{\theta}_s$ is the estimate of $\theta$ for the $s$-th simulated data. The average of the estimates is also computed for each parameter of the component mixture

$$\text{AVG}(\widehat{\theta}) = \frac{1}{S} \sum_{s=1}^{S} \widehat{\theta}. \qquad (2.57)$$

| Algorithm | Method |
|---|---|
| Algorithm 1 | ECM |
| Algorithm 2 | ECM & optim updates |
| Algorithm 3 | ECM & step-size $\alpha$ |
| Algorithm 4 | GEM & line-search |
| Algorithm 5 | GEM & line-search & $1/\nu$ |
| Algorithm 6 | ECM & $1/\nu$ |

Tables B.1-B.8 in the Appendix B report the simulation results.

Results from scenario 1 (tables B.1-B.2 in Appendix A) validate the conclusions made in the previous section's case study. For $N = 500$, algorithms 2, 3, 5 and 6 provide accurate estimates of all parameters. Algorithm 1 fails to prevent the degeneracy of the shape parameter, with $\text{AVG}(\widehat{\nu_1}) = 18.43$ significantly diverging from the true value $\nu_1 = 5$ (Figure 2.21). Algorithm 2 can be used as an alternative because simulation results do not highlight the degeneracy issue of $\nu_1$. By looking at the different step size $\alpha$ values (algorithm 3) it is possible to observe how the degeneracy of $\nu_1$ increases as $\alpha$ increases from 0.1 to 0.8. As a consequence, the MSE($\widehat{\nu_1}$) increases from 3.91 to 5.40. The line search approach (algorithm 4) does not fully fix the degeneracy issue since $\text{AVG}(\widehat{v_1}) = 18.89$. On the contrary, the adaptive step-size $1/v_k$ yields accurate estimates for all parameters, especially for $v_1$. Both algorithms 5 and 6 yields $\text{AVG}(\widehat{v_1}) = 6.21$. For $N = 2000$, all algorithms obtain a more accurate estimation of $v_1$. For large samples, the degeneracy problem of $v_1$ disappears and therefore Algorithm 1 works as expected. Panel (b) in Figure 2.21 illustrates the similar performance of algorithms 1 and 5, with the former exhibiting a higher number of outliers. Moreover, it is worth noting that the best $\alpha$ value for algorithm 3 is 0.1 for $N = 500$ and 0.5 for $N = 2000$, respectively.

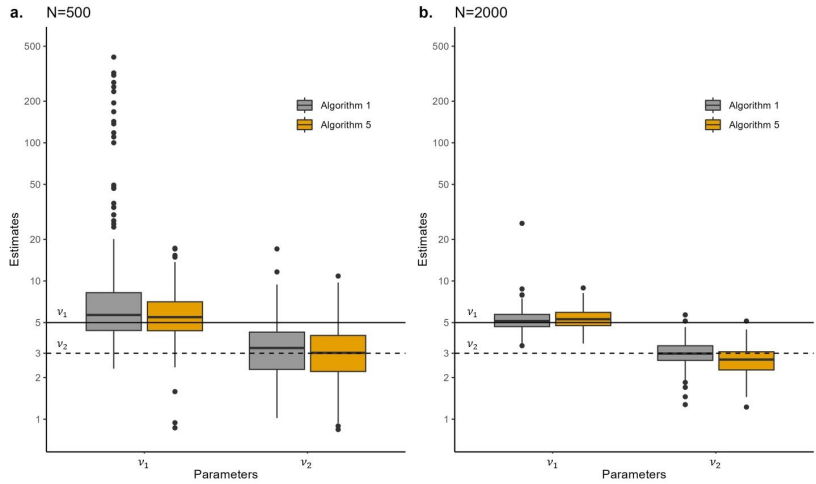

**Figure 2.21:** MGND simulations, scenario 1, boxplot of the shape parameters estimates via algorithms 1 and 5. The y-axis is in $\log_{10}$-scale.

In scenario 2 (tables B.3-B.4 in Appendix A), for $N = 500$, algorithm 1 fails to prevent the degeneracy of $v_2$ yielding $\text{AVG}(\widehat{v_2}) = 10.30$ (Figure 2.22). Algorithms 5 and 6 provide accurate estimates for all parameters. However, algorithm 5 get lower MSE values than algorithm 6 for the less frequent mixture component. For $N = 2000$, all algorithms get better estimates. Also for the scenario 2, the degeneracy problem of $v_2$ disappears and therefore Algorithm 1 works as expected. The best $\alpha$ value for algorithm 3 is 0.1 for $N = 500$ and 0.2 for $N = 2000$.

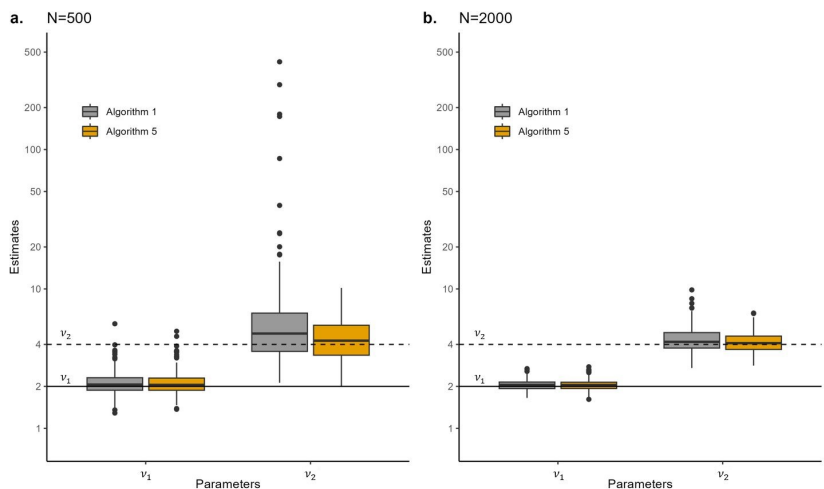

**Figure 2.22:** MGND simulations, scenario 2, boxplot of the shape parameters estimates via algorithms 1 and 5. The y-axis is in $\log_{10}$-scale.

In scenario 3 (Tables B.5-B.6 in Appendix A) mixture parameters take on values very close to those estimated on financial returns. For both sample sizes, algorithm 5 provide more accurate estimates for all parameters compared to the other algorithms. Similar results can be observed in scenario 4 (Tables B.7-B.8 in Appendix A) where $\mu_1 = 1$ and $\mu_2 = 5$.

Figures 2.23 and 2.24 show the MSE dotplots by algorithm, simulated scenario and sample size.

Some interesting considerations arise from the simulation study.

▶ The degeneracy of the shape parameter can be controlled by increasing the sample size and using algorithms 2, 3, 5 and 6;

▶ The degeneracy issue is not observed if $\nu_k \leq 2$ in the population;

▶ Parameters' estimation is heavily affected when the sample size is small;

▶ Algorithm 1 does not prevent the degeneracy issue for $\nu_k > 2$ and small sample sizes;

▶ Algorithm 2 can be used as an alternative but sometimes its estimation underperforms that of algorithms 3-5-6;

▶ Algorithm 3 require experimenting different step-size $\alpha$ to identify the optimal value that depends on the specific scenario and sample size;

▶ Algorithm 4 automates the choice of the step-size but does not prevent the degeneracy issue, its estimation accuracy is better than algorithm 1 when $\nu$ is not very high;

▶ Algorithm 5 automates the choice of the step-size and prevent the degeneracy issue providing accurate estimates for all parameters in multiple scenarios and sample sizes;

▶ Algorithm 6 prevent the degeneracy issue but for some scenarios and sample sizes algorithm 5 get more accurate estimates;

Overall, algorithm 5 is the best-performing algorithm across multiple scenarios and sample sizes. It overcomes the drawbacks of the other algorithms and yields more accurate estimates. In addition, Chapter 4 will demonstrate how Algorithm 5 becomes essential when carrying out the likelihood ratio test.

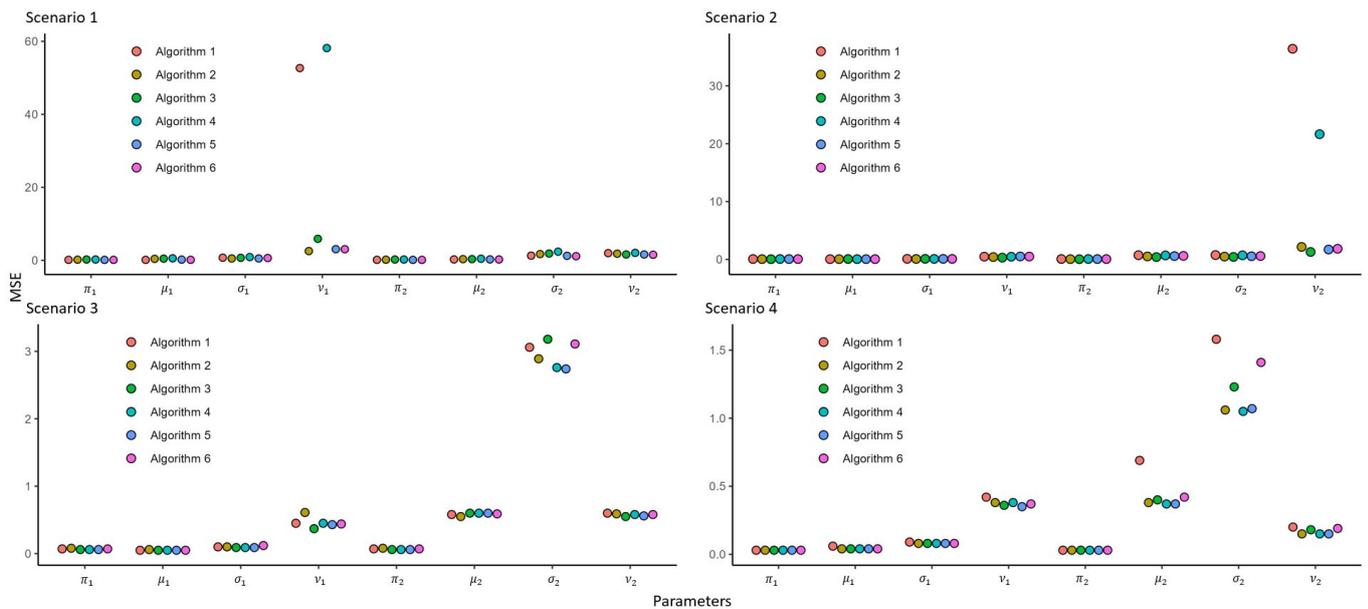

**Figure 2.23:** MGND simulations, $N = 500$, MSE dotplot by algorithm.

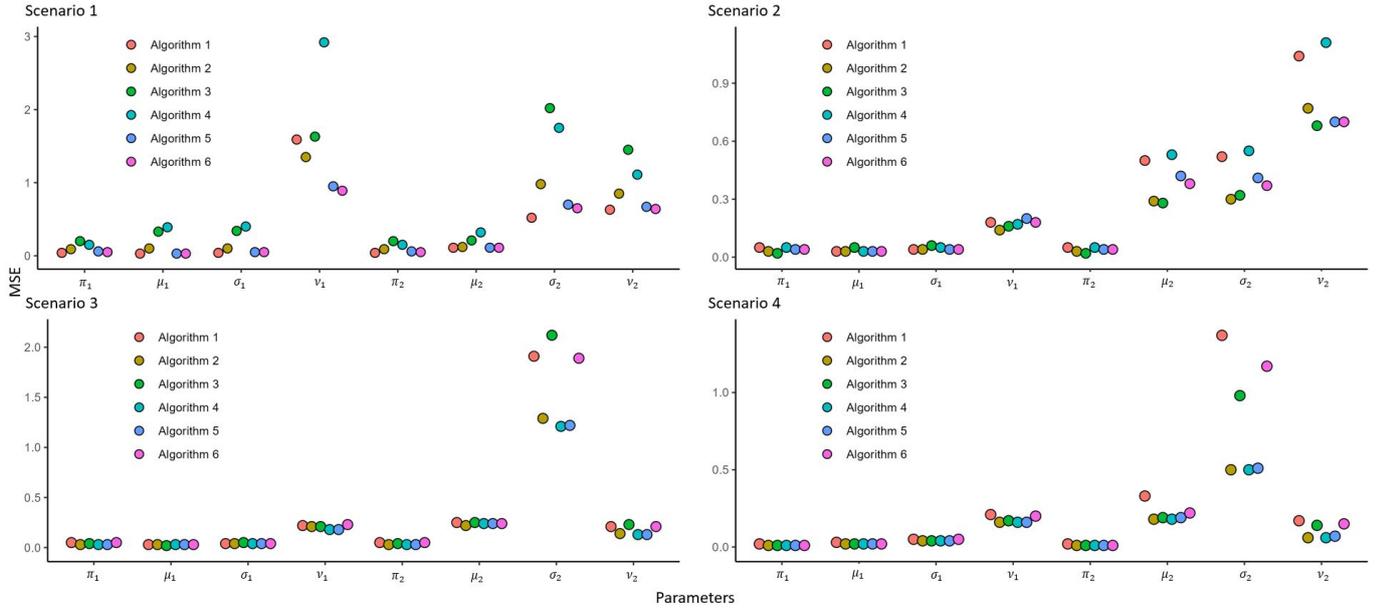

**Figure 2.24:** MGND simulations, $N = 2000$, MSE dotplot by algorithm.

### 2.7.2 Comparative analysis part I

In this section a comparative analysis is performed for the 52 stocks that belong to the STOXX50E using all models introduced in the previous sections, along with different goodness-of-fit measures such as the Akaike information criterion, Bayesian information criterion, Kolmogorov–Smirnov statistic, and Anderson–Darling statistic.

**Data**

Data on daily closing prices of the 52 stocks that belong to the STOXX50E are collected from Yahoo Finance for the period from 01 January 2010 to 30 September 2023 (shorter for stocks entering the STOXX50E later than January 2010). Daily log returns are computed for each stock according to equation 2.4. All descriptive statistics and the JB test are computed for daily log returns for the full period. Table B.9 in Appendix B shows for each stock (alphabetically ordered) the number of observations ($N$), mean, median, standard deviation, skewness, kurtosis, JB statistics, and minimal and maximal values. All stocks have a mean and median close to 0. The standard deviation range from 1.2624 to 2.7371. Returns are positively and negatively skewed with an empirical kurtosis greater than three, indicating fat-tails (leptokurtic distribution). For all indices the JB test rejects the null hypothesis which states that data are normally distributed.

**Models and estimation methods**

All models introduced in the previous sections are implemented to perform a comparative analysis of the 52 stocks. Specifically, models and estimation methods are resumed in Table 4.10. Mixture models with two components are considered because the results obtained for three component are insignificant and irrelevant. Besides, the MGND model is estimated with algorithm 5 as supported by the simulation study in the previous section.

| Model | Equation | Estimation method | R function |
|-------|----------|-------------------|-----------|
| Laplace | 2.9 | MLE | *eLaplace* |
| Cauchy | 2.11 | MLE | *mlcauchy* |
| Student-t | 2.12 | MLE and Nelder-Mead approach | *optim* |
| GH | 2.15 | MLE and Newton-Raphson method | *ghFit* |
| GND | 2.17 | MLE and Nelder-Mead approach | *optim* |
| MN | 2.27 | EM algorithm | *mclust* |
| MNL | 2.28 | EM algorithm and Nelder-Mead updates | *optim* |
| MNC | 2.29 | EM algorithm and Nelder-Mead updates | *optim* |
| MSTD | 2.31 | EM algorithm | *teigen* |
| MGND | 2.33 | Algorithm 5 | |

**Table 2.7:** Comparative analysis part I: models and methods.

**Goodness-of-fit measures**

Model selection is an important step of statistical modelling. Four different goodness-of-fit measures are used to find the best-fit model.

The Akaike information criterion (AIC) introduced by Akaike 1974 is defined by

$$\text{AIC} = 2p - 2\log L(\widehat{\theta}), \tag{2.58}$$

where $p$ is the number of parameters of the model and $\log L(\widehat{\theta})$ is the estimated log-likelihood of the model. On the other hand, the Bayesian information criterion (BIC) introduced by Schwarz 1978 is defined by

$$\text{BIC} = p\log(N) - 2\log L(\widehat{\theta}), \tag{2.59}$$

where $N$ is the number of observations.

The Kolmogorov–Smirnov (KS) statistic (A. Kolmogorov 1933) is defined as follows

$$\text{KS} = \sup_{x \in \mathbb{R}} |F_n(x) - F_m(x, \widehat{\theta})|, \tag{2.60}$$

where:

▶ $F_n(x) = \frac{1}{n}\sum_{i=1}^{n}\mathbb{1}_{(-\infty, x]}(x_i)$ is the empirical cumulative density function for $n$ independent and identically distributed (i.i.d.) ordered observations $X_i$;

▶ $\mathbb{1}_{(-\infty, x]}(x_i)$ is the indicator function equal to 1 if $X_i \leq x$ and equal to 0 otherwise;

▶ $F_m(x, \widehat{\theta})$ is the cumulative density function of the fitted model.

The Anderson–Darling (AD) statistic (T. W. Anderson and Darling 1954) is defined as follows

$$AD = \int_{-\infty}^{\infty} \frac{(F_n(x) - F_m(x, \widehat{\theta}))^2}{F_m(x, \widehat{\theta})(1 - F_m(x, \widehat{\theta}))}dF_m(x, \widehat{\theta}). \tag{2.61}$$

Both AIC and BIC balance the goodness-of-fit with model complexity, namely the number of parameters. AIC penalizes less for model complexity compared to BIC. Therefore, BIC prefers simpler models, providing a stronger penalty for over-fitting. By contrast, KS and AD are employed to compare the empirical distribution of returns with the fitted distribution. As explained by Massing and Ramos 2021 *'the KS statistic better reflects the deviance between the empirical and the fitted distribution close to the center and the AD statistic better reflects the deviance in the tails'* (Massing and Ramos 2021). Goodness-of-fit measures may not consistently converge to the same conclusion since they are based on different decision criteria target. Figure 2.25 shows the guidelines for model selection selection via AIC, BIC, KS and AD.

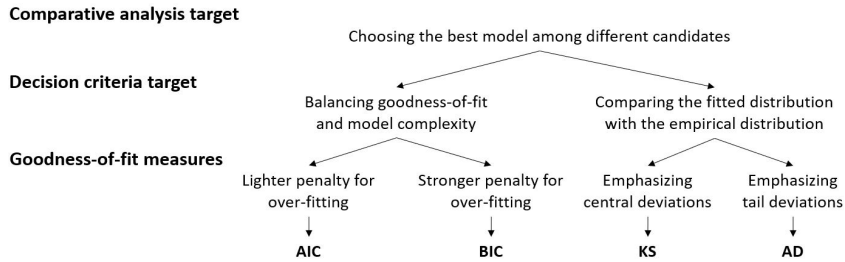

**Figure 2.25:** Guidelines for model selection via AIC, BIC, KS and AD.

### Results

Tables B.10,B.11,B.12 and B.13 in the Appendix B present the AIC, BIC, KS and AD values for daily log-returns over the past 13 years. All models presented here are compared for each stock, and the minimum statistic is highlighted in bold to signify the best fit.

**AIC**  AIC mainly selects the GH and MNL models. The MGND model is selected only 6 times, while the MSTD model is selected 5 times. Panel (a) of Figure 2.26 shows the AIC boxplot for all models ordered by median. The GH, MGND, and MNL models have the lowest median AIC values.

**BIC**  The student-t model is the most often selected model. Panel (b) of Figure 2.26 shows the BIC boxplot for all models ordered by median. The Student-t, GH and MNL report the lowest median BIC values. The BIC prefers simpler models, i.e. models with few parameters.

**KS**  The MSTD, MGND and GH models are consistently selected by KS. Panel (a) of Figure 2.27 shows the KS boxplot of the top four models by median, namely the GH, MSTD, MGND, and STD models.

**AD**  The GH, MGND and MSTD models are consistently selected by AD. Panel (b) of Figure 2.27 shows the AD boxplot of the top four models by median, namely the GH, MGND, MSTD, and STD models.

The GH model is more often selected according to the AIC, KS and AD, while the MGND and MSTD models are often selected as second choice. The student-t model is the most selected according to the BIC. However, if a model is chosen several times as first according to the median value, the second and third model are still competitive as show Figures 2.26 and 2.27.

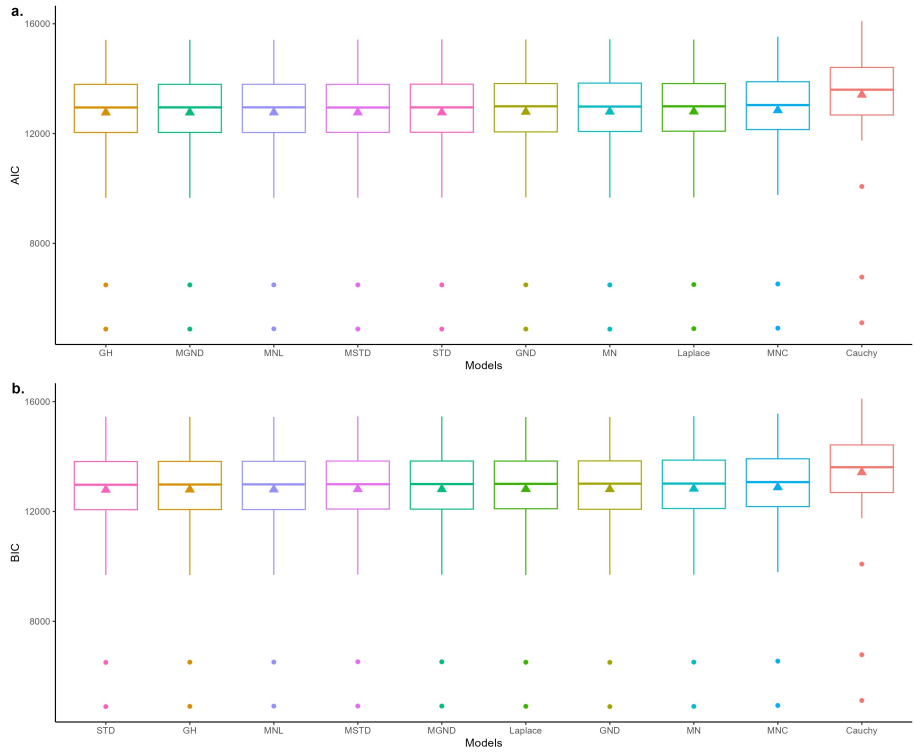

**Figure 2.26:** Comparative analysis part I, AIC and BIC boxplot. Models are ordered by median.

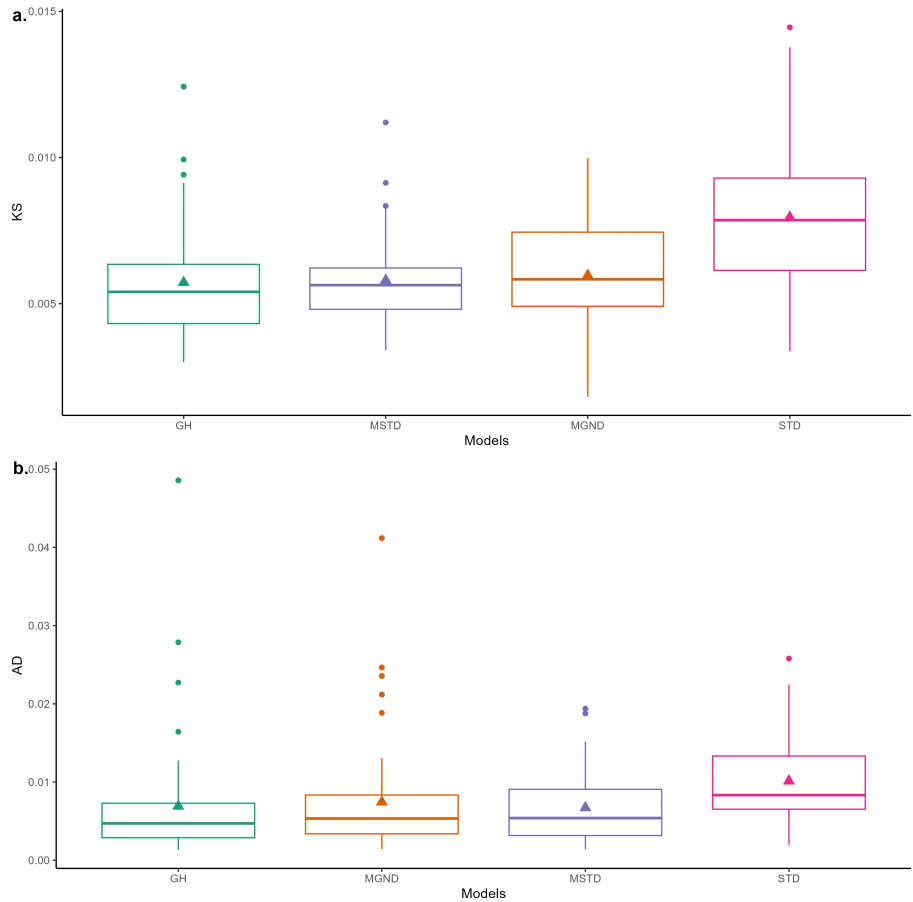

**Figure 2.27:** Comparative analysis part I, KS and AD boxplot. Models are ordered by median.

# Turmoil days identification with MGND models

# 3

## 3.1 Turmoil days identification

In financial research, it is often assumed that there are two distinct trading days to capture market dynamics: stable and turmoil days. By definition, turmoil days are more volatile and less frequent than stable days (Kim and White 2004). Other interesting features are resumed in Table 3.1.

**Table 3.1:** Main features of stable and turmoil days.

| Stable days | Turmoil days |
| --- | --- |
| **Characteristics** | |
| Small and gradual price fluctuations | Large and quick price fluctuations |
| Low uncertainty and market calmness | High uncertainty and market anxiety |
| Low trading volumes | High trading volumes |
| **Causes** | |
| Market stability | Economic and geopolitical events |
| | Corporate announcements |
| | Changes in market sentiment |
| **Impact** | |
| Occasional adjustments of investing strategies | Quickly adjustments of investing strategies |

It is worth noting that the distinction between stable and turmoil days can be subjective[1] . For instance, Figure 3.1 shows the percentage log returns of the STOXX50E. The shadow areas highlight high volatility periods like the global financial crisis, sovereign debt crisis, 2015-2016 stock market sell-off, COVID-19 pandemic and Russia-Ukrainian conflict.

1: An empirical application of subjective turmoil days is provided in Duttilo, Gattone, and Di Battista 2021.

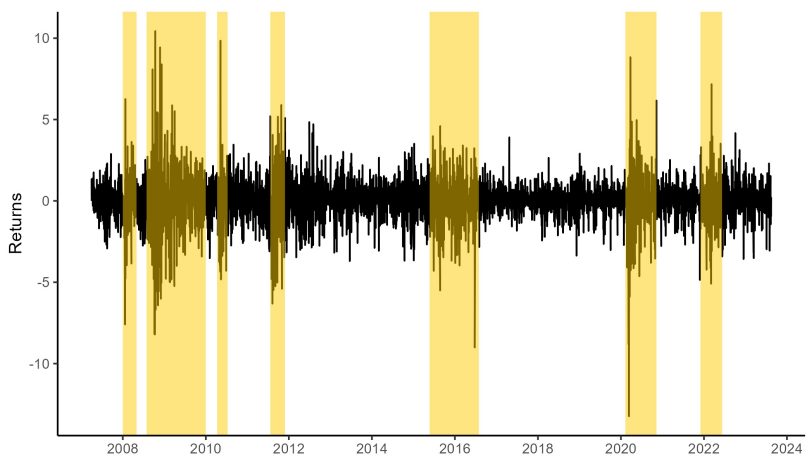

**Figure 3.1:** Percentage log returns - Euro Stoxx 50 Index and crisis periods.

In this chapter turmoil days are objectively detected employing a two-component MGND model fitted on financial asset returns. Specifically, a daily return is classified as a stable day if it belongs to the stable component, i.e. the mixture component with the highest shape parameter/lowest scale parameter. Similarly, a daily return is classified as a turmoil day if it belongs to the turmoil component, i.e. the mixture component with the lowest shape parameter/highest scale parameter. Conventionally, a smaller shape parameter/higher scale parameter means a thicker tail and then a higher standard deviation, while a higher shape parameter/lower scale parameter means a thinner tail and then a lower standard deviation. Shifts from the stable component to the turmoil component are assumed to be due to exogenous market events, i.e. to '*time-ordered shifts*' (Kon 1984).

Kon 1984 assigned daily returns to a specific mixture component through the Naïve Bayes' classifier. Originally, this classification rule was proposed for mixtures of linear models by Kon and Lau 1979, and applied by Christie 1983 and Kon 1983. As explained by Kon 1984 this procedure '*may be particularly useful for efficient markets tests when the estimated data partition can be associated with corresponding public announcements or information signals in accounting numbers released prior to the event*' (Kon 1984). In other words it may be particularly useful to detect important market events and test hypothesis.

The two-component mixture MGND model (equation 2.33) is exploited to classify the daily returns based on the Naïve Bayes' classification rule (Frühwirth-Schnatter 2006) that assigns each return to the class with the highest posterior probability (equation 2.33). The classification rule is defined as follows:

$$\max_k \pi_k f_k(r_t|\theta_k), \tag{3.1}$$

where the selected mixture component $k$ generating the return $r_t$ has the largest posterior probability. Consequently, a decoded variable which is an indicator of market turmoil is obtained by a '*soft assignment*' (Bishop 2006) of observations to classes. In this way, financial turmoil days are objectively detected.

For instance, a two-component MGND model (equation 2.33) is estimated on STOXX50E daily returns for the period from 02 April 2007 to 15 August 2023. Table 3.2 below shows the estimated parameters.

| Stable component | | | | Turmoil component | | | |
|---|---|---|---|---|---|---|---|
| $\pi_1$ | $\mu_1$ | $\sigma_1$ | $\nu_1$ | $\pi_2$ | $\mu_2$ | $\sigma_2$ | $\nu_2$ |
| 0.8787 | 0.0752 | 1.0659 | 1.2456 | 0.1213 | -0.5556 | 2.6515 | 1.2316 |

**Table 3.2:** Estimated parameters of the two-component MGND on STOXX50E daily returns.

The stable component is predominant compared to the turmoil component ($\pi_1 > \pi_2$) which has a negative location parameter ($\mu_2 = -0.5556$) and a higher scale parameter ($\sigma_2 = 2.6515$). The shape parameter of the turmoil component ($\nu_2 = 1.2316$) is slightly lower than the shape parameter of the stable component ($\nu_1 = 1.2456$). As shown in Table 3.3 the stable component has a low standard deviation (Std$_1 = 1.1056$), while the turmoil component has a high standard deviation (Std$_2 = 2.7873$). The kurtosis of turmoil component Kur$_2 = 4.6034$ is slightly higher than the kurtosis of the stable component Kur$_1 = 4.5451$. The estimated overall standard deviation and kurtosis fit well the respective empirical measures.

Panels (a) and (b) in Figure 3.2 show the estimated density and left tail, respectively. It can be seen that one component describes the central and intermediate values of the data, while another component describes the more extreme tail behaviours. Indeed, as shown in panel (c) extreme returns are generated by the turmoil component as its posterior probability is equal or greater than 50%. Panel (d) shows the STOXX50E daily returns and the detected turmoil days (yellow vertical lines).

The use of the MGND model introduces an objective and data-driven way to identify financial market turmoil days. Its flexibility to accommodate different shapes of data distributions improves the accuracy in identifying turmoil days. This departure from the subjective approach provides a more rigorous way for understanding periods of market stress. In addition, it minimizes the risk of subjective bias in the analysis.

Two empirical applications that combine this approach with generalized autoregressive conditional heteroskedasticity (GARCH) models (Bollerslev 1986) will be presented in the next two sections.

**Table 3.3:** Estimated standard deviation and kurtosis of the two-component MGND on STOXX50E

| Stable component | | Turmoil component | | Overall | |
| --- | --- | --- | --- | --- | --- |
| $Std_1$ | $Kur_1$ | $Std_2$ | $Kur_2$ | Std | Kur |
| 1.1056 | 4.5451 | 2.7873 | 4.6034 | 1.4350 | 9.7728 |

**Note.** The empirical standard deviation and kurtosis are equal to 1.436 and 10.326, respectively.

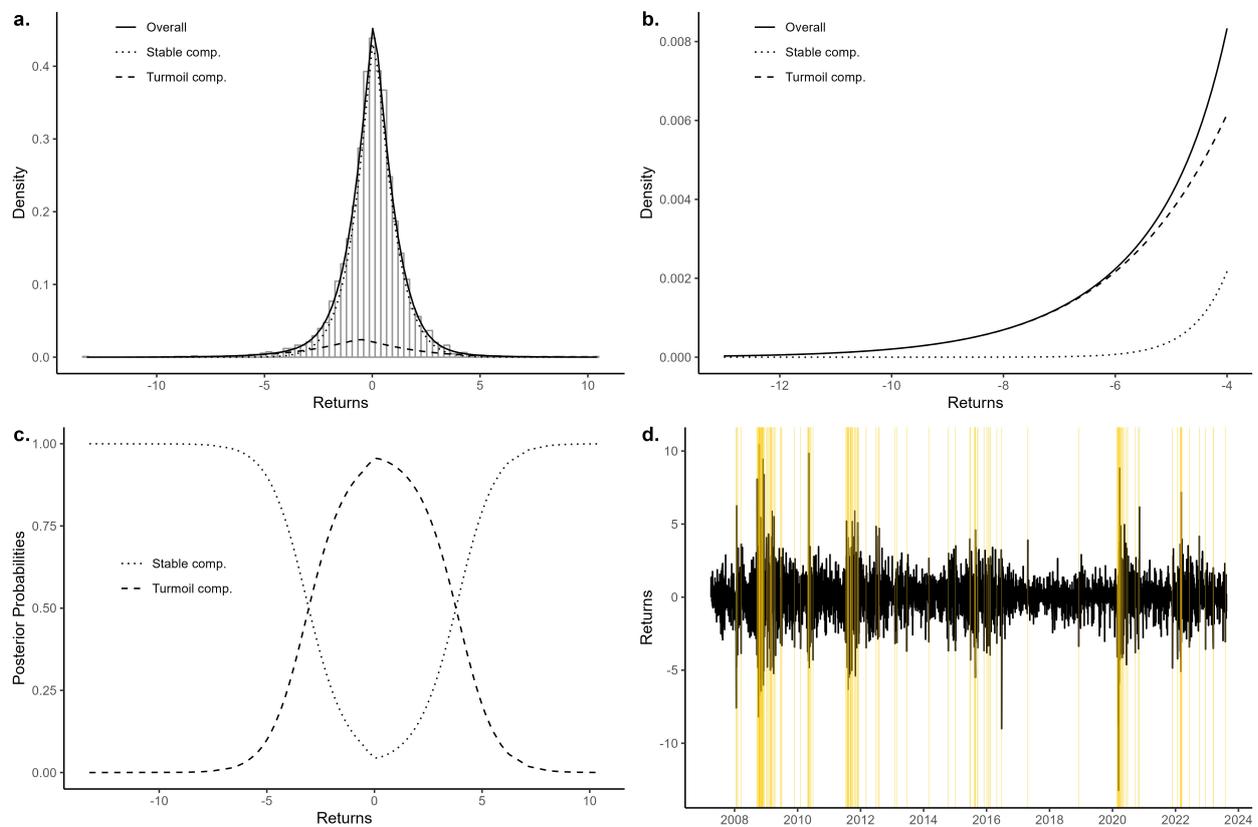

**Figure 3.2:** Estimated two-component MGND on STOXX50E, density - panel (a), left tail - panel (b), posterior probabilities - panel (c) and detected turmoil days - panel (d).

## 3.2 ESG vs traditional investments

The Paris Agreement and the 2030 Agenda for sustainable development demonstrate how the environmental, social and governance (ESG) challenges are widely complex (Friede 2019).

In the last decades, the investors' attention on the ESG factors has been growing, becoming a recurring theme of research in different disciplines (Abate, Basile, and Ferrari 2021). While several scholars claim that ESG investments mean giving up a portion of profits for ethical reasons (Bauer, Derwall, and Otten 2007; Bauer and Smeets 2015; Fich, Harford, and Tran 2015; Arouri and Pijourlet 2017), other scholars state that behind such choices there is much more (Nilsson 2009; Krosinsky and N. Robins 2012; Hemerijck 2018; Kaufer and Steponaitis 2019).

Traditional investors choose their investments by adopting economic-financial criteria, like returns, investment duration, risk aversion, risk premium, liquidity and so on. By contrast, ESG investors combine the economic performance with measurable environmental and social factors (Revelli 2017; Benlemlih and Bitar 2018; Lapanan 2018; Oikonomou, Platanakis, and Sutcliffe 2018; Chatzitheodorou et al. 2019; Rossi et al. 2019; Gomes 2020). The latter '*recognize that the generation of long-term sustainable returns is dependent on stable, well-functioning, and well-governed social, environmental, and economic systems*' (University of Cambridge 2022).

However, traditional and ESG investors assess the market risk considering a mix of economic and socio-political factors or aggregate events that may cause financial turmoil days and deep shocks to stock markets (Zigrand 2014; Berk and DeMarzo 2018; Szczygielski et al. 2021; Duttilo, Gattone, and Di Battista 2021). The dotcom bubble burst (2000-2002), the global financial crisis (2007-2008), the European sovereign debt crisis (2010-2012) and the COVID-19 pandemic are examples of turmoil periods.

This empirical study examines the effect of turmoil periods on the performance and volatility of several Dow Jones Sustainability Indices (DJSIs) and compares them with their respective market benchmarks (traditional indices). For this purpose, two different tools are employed: a two-component MGND model and an exponential GARCH-in-Mean (EGARCH-M) model with exogenous dummy variables.

The MGND model is applied to fit the return distribution of the MSCI All-Country World Equity Index (MSCIW). This index represents the performance of large and mid-cap stocks of 23 developed and 24 emerging markets. Financial market turmoil days are objectively detected by applying the Naïve Bayes' classifier to the mixture model results.

The EGARCH-M model is commonly used to predict financial returns and their volatility (Hoti, McAleer, and Pauwels 2007). The model is applied on several ESG and non-ESG indices including an exogenous dummy variable denoting the turmoil days as detected by the MGND model.

The entire analysis provides insights on potential differences between ESG indices and their traditional market benchmarks investigating the impact of stable and turmoil days on the conditional mean and volatility

This section is based on *"Mixtures of generalized normal distributions and EGARCH models to analyse returns and volatility of ESG and traditional investments"* by Duttilo P., Gattone S.A. and Iannone B., originally published in AStA Advances in Statistical Analysis, and is licensed under a Creative Commons Attribution 4.0 International License. Open access funding provided by Università degli Studi G. D'Annunzio Chieti Pescara within the CRUI-CARE Agreement. DOI: https://doi.org/10.1007/s10182-023-00487-7.

and other interesting aspects of financial returns such as risk premium, leverage effect and volatility persistence.

These indices comprise the following markets: Global, US, Europe (EU) and emerging markets (EM). Hence, Global and regional DJISIs are selected: Dow Jones Sustainability World Index, Dow Jones Sustainability U.S. Composite Index, Dow Jones Sustainability Europe Index, Dow Jones Sustainability Emerging Markets. Likewise, their respective traditional market benchmarks are collected: Dow Jones Global Index, Dow Jones Industrial Average Index, Dow Jones Europe Index, Dow Jones Emerging Markets Index. The time period taken for the study is from 4 January 2016 to 30 September 2022.

### 3.2.1 Literature review

There is a considerable amount of research that analyses the ESG equity indices by comparing their financial performance with traditional equity benchmarks with many different methods, techniques, dates, and variables. However, there is no absolute consensus on the fact that ESG portfolios are less volatile than market benchmark portfolios (Ouchen 2022). Cunha et al. 2020 distinguished three different kinds of results about the financial performance of ESG equity indices: positive, neutral and mixed. The latter refers to studies that '*found one or more positive, neutral or negative return-risk performance findings*'. In order to highlight different methods and techniques used in this research field, the existing literature has been grouped into three main methodological framework: portfolio performance measures, Markov-switching (MS) models and GARCH models.

The first methodological framework is featured by works that applied portfolio performance measures such as Sharpe ratios, Treynor ratios, Jensen's alphas (Schröder 2007; Collison et al. 2008; Consolandi et al. 2009; Belghitar, E. Clark, and Deshmukh 2014; Lean and D. K. Nguyen 2014; Cunha et al. 2020).

The second methodological framework stands out for the use of MS models to analyse the performance of ESG indices (Shunsuke, Tatsuyoshi, and Akimi 2012; Ouchen 2022).

In the last methodological framework, GARCH models are employed to analyse the conditional mean and volatility of daily returns on ESG indices. Usually the conditional analysis is supported by the unconditional analysis which uses different methodologies such as the portfolio performance measures.

Employing GARCH models, Hoti, McAleer, and Pauwels 2007 analysed the conditional volatility of some sustainability and ethical indices: Ethibel, ASPI Eurozone, Calvert Social Index, Ethical Index and FSTE4Good (Global, USA, UK and Europe). Results showed differences in the volatility persistence in the short and long run and in the leverage effect. Lean and D. K. Nguyen 2014 analysed the performance and volatility of DJISIs for the Global and three regional markets over the period 2004 to 2013. The conditional analysis performed via the EGARCH model highlighted that the 2008 global financial crisis had a substantial impact on both return and volatility of sustainable investments. Similarly, Ang 2015 explored the behaviour of the DJSI Korea. Results showed that both return and volatility of Korea ESG portfolio are less affected by the 2008 crash. Using mean-variance testing and GARCH models, Sudha 2015 compared the performance and volatility of the S&P ESG India Index with two market benchmarks, i.e. the Nifty and the S&P CNX 500. Although the volatility clustering featured all three indices, it was found that S&P ESG India Index was less volatile compared to the Nifty. Jain, Sharma, and Srivastava 2019 explored whether ESG investments offer better financial returns than the market benchmarks in developed and emerging markets. Applying GARCH models, the study concluded that the US large-cap ESG index (TRESGUS) provided the highest return and a suitable level of risk. Sabbaghi 2022 analysed the impact of good and bad news on the volatility of ESG firms employing the MSCI indices and the GARCH framework. Results showed the presence of the leverage effect for ESG firms, i.e. bad news increase volatility by a larger amount than good news.

### 3.2.2 Data

The MSCIW index is selected to detect stable and turmoil days. This global traditional index provides a good representation of the global market and regional ones by the performance of large and mid cap stocks of 23 developed and 24 emerging markets.

The DJSI Index Family tracks the stock performance of leading sustainability-driven companies which stand out in terms of economic, environmental and social criteria (Hoti, McAleer, and Pauwels 2007; S&PGlobal 2022). This index family was launched in 1999 as the first global sustainability benchmark. The DJSI Index Family is a '*best-in-class*' benchmarks because only the top ranked companies in the S&P Global ESG Scores are selected for inclusion in the Index Family (S&PGlobal 2022). Its composition is reviewed annually and rebalanced quarterly.

Data on daily closing prices of traditional and ESG indices are collected as shown in Table 3.4. The time period taken for the study is from January 4, 2016 to September 30, 2022. To underline the COVID-19 pandemic scenario, the considered period does not include the global financial crisis (2007-2009) and the sovereign debt crisis (2010-2011) periods. Daily returns of all equity indices under study are calculated with the natural log difference approach in equation 2.4.

| Index | Ticker | Type | Area |
|---|---|---|---|
| MSCI All-Country World Equity Index | MSCIW | Traditional | Global |
| Dow Jones Global Index | W1DOW | Traditional | Global |
| Dow Jones Sustainability World Index | W1SGI | ESG | Global |
| Dow Jones Industrial Average | DJUS | Traditional | US |
| Dow Jones Sustainability U.S. Composite Index | AASGI | ESG | US |
| Dow Jones Europe | E1DOW | Traditional | EU |
| Dow Jones Sustainability Europe Index | DJSEUR | ESG | EU |
| Dow Jones Emerging Markets Index | W5DOW | Traditional | EM |
| Dow Jones Sustainability Emerging Markets | DJSEMUP | ESG | EM |

**Table 3.4:** Selected traditional and ESG indices.

**Source.** S&P Global, Yahoo Finance, Investing.com (October 20, 2022).

### 3.2.3 Exponential GARCH-in-Mean model with exogenous dummy variables

Originally, the work of Engle 1982 introduced the autoregressive conditional heteroscedastic (ARCH) model giving rise to a vast literature and variety of models. Next, in order to capture the risk premium, Engle, Lilien, and R. P. Robins 1987 extended the ARCH model '*to allow the conditional volatility to be a determinant of the mean*', it was called ARCH-in-mean model or ARCH-M. Subsequently, D. B. Nelson 1991 introduced the exponential ARCH model to overcome some limitations of the ARCH model. The EGARCH-M model is a generalization of the exponential ARCH model. These last two asymmetric models are able to capture the leverage effect, an important stylised fact of financial time series. Commonly, the leverage effect occurs when negative returns increase volatility by a larger amount than positive returns (Francq and Zakoian 2019).

In this study the conditional mean and volatility equations are modelled with the EGARCH-M model. In both equations, an exogenous dummy variable is included to take into account the state of the financial market at time $t$. Specifically, the dummy variable $\text{TURMOIL}_t$ assumes the value of 1 during turmoil days, otherwise it is equal to 0, i.e. during stability days. In this way, it is possible to describe the impact of stable and turmoil days on conditional mean and volatility of equity indices. The EGARCH(1,1)-in-Mean model with exogenous dummy variables is specified as follows:
**Conditional mean equation**

$$r_t = \mu + m_1 \text{TURMOIL}_t + \phi_1 r_{t-1} + \lambda h_t + \epsilon_t, \tag{3.2}$$

**Conditional volatility equation**

$$\ln(h_t^2) = \omega + v_1 \text{TURMOIL}_t + \alpha_1 z_{t-1} + \gamma_1(|z_{t-1}| - E[|z_{t-1}|]) + \beta_1 \ln(h_{t-1}^2)$$
$$\text{where} \quad z_t = \frac{\epsilon_t}{\sqrt{h_t^2}} \sim \text{Skewed-GND}(0, 1, \nu, s). \tag{3.3}$$

In equation (3.2), $r_t$ and $\epsilon_t$ indicate the returns and error terms of equity index at time $t$, respectively. Besides, $\mu$ is the constant term. The coefficient $m_1$ determines the impact of turmoil days on the conditional mean. If $m_1$ is negative and statistically significant, turmoil days cause a reduction of the conditional mean. To capture the autocorrelation of returns (i.e. the linear relationship between lagged values of returns time series), the conditional mean equation includes a stationary first-order autoregressive process AR(1) like Hoti, McAleer, and Pauwels 2007. The coefficient $\phi_1$ measures the time link between $r_t$ and $r_{t-1}$. Following Engle, Lilien, and R. P. Robins 1987 the conditional mean equation also includes the risk premium coefficient $\lambda$. If $\lambda > 0$ and statistically significant, returns are positively related to their conditional standard deviation ($h_t$).

In equation (3.3), $\ln(h_t^2)$ denotes the natural logarithm of the conditional volatility, $\omega$ is the constant term, $\alpha_1$ captures the sign effect and $\gamma_1$ the size effect, while $\beta_1$ is the GARCH effect and the volatility persistence. $E[|z_{t-1}|]$ is the expected value of the absolute standardized residual. The coefficient $v_1$ determines the impact of turmoil days on the conditional volatility. If $v_1$ is positive and statistically significant, turmoil days cause an increasing of the conditional volatility. In order to better describe leptokurtosis and fatter tails of returns, the standardized residuals $z_t$ are modelled using the skewed-GND distribution with mean 0, variance 1, $v$ and $s$ as shape and skewness parameters, respectively.

The EGARCH(1,1)-M model with exogenous dummy variables is estimated for all indices under study through the R's *rugarch* package (Ghalanos 2022).

### 3.2.4 Results of exploratory data analysis

Table 3.5 presents the basic statistics of traditional and ESG indices. The mean of traditional and ESG indices is almost the same for Global and US markets, while the mean of ESG indices is slightly higher than the mean of traditional indices for EU and EM markets. More importantly, in the EU market the traditional index turns out to have a higher standard deviation than its ESG counterpart. In the other markets, traditional and ESG indices have approximately the same standard deviation. All indices show negative skewness and excess kurtosis. In general, ESG indices show lower skewness and kurtosis than non-ESG indices with only two exceptions. In the EU market the level of skewness is approximately the same, while the ESG index has a higher kurtosis and skewness than the traditional one in the EM market.

| | Mean | | Stdev | | Skew | | Kur | |
| | Trad. | ESG | Trad. | ESG | Trad. | ESG | Trad. | ESG |
|---|---|---|---|---|---|---|---|---|
| Global | 0.0166 | 0.0182 | 0.9718 | 0.9815 | -1.3752 | -1.3488 | 18.5939 | 17.9680 |
| US | 0.0303 | 0.0337 | 1.2115 | 1.2305 | -1.0819 | -0.9107 | 23.0781 | 17.9696 |
| EU | -0.0098 | 0.0042 | 1.2266 | 1.0289 | -1.3371 | -1.3499 | 16.8315 | 15.6846 |
| EM | 0.0070 | 0.0127 | 0.9860 | 1.0262 | -0.8100 | -1.0156 | 6.1659 | 9.2857 |

**Table 3.5:** Basic statistics of traditional and ESG indices.

Table B.14 shows the results of some preliminary statistical hypothesis tests. According to the Jarque-Bera (JB) test the daily returns are not Normally distributed. The Augmented Dickey-Fuller (ADF) test and Phillips-Perron (PP) test show that the null hypothesis of unit root can be rejected and all indices are stationary in their first difference at 1% significance. The ARCH-LM test confirm the presence of ARCH effect and heteroscedasticity because the null hypothesis of no ARCH effect is rejected at 1% significance.

### 3.2.5 Turmoil days identification

Table 3.11 shows the estimated parameters of the two-component MGND on MSCIW daily returns.

| Stable component | | | | Turmoil component | | | |
| $\pi_1$ | $\mu_1$ | $\sigma_1$ | $v_1$ | $\pi_2$ | $\mu_2$ | $\sigma_2$ | $v_2$ |
|---|---|---|---|---|---|---|---|
| 0.7280 | 0.1244 | 0.5735 | 1.1019 | 0.2720 | -0.1197 | 0.4523 | 0.6977 |

**Table 3.6:** Estimated parameters of the two-component MGND on MSCIW daily returns.

As mentioned in Section 3.1, stable and turmoil components could be identified on the basis of the shape parameter. Firstly, the stable component is predominant compared to the turmoil component as $\pi_1 > \pi_2$. Secondly, the estimated MGND model is bi-modally asymmetric with $\mu_1 = 0.1244 > \mu_2 = -0.1197$. Thirdly, the tails of the stable component intermediate between the Laplace and Normal distribution $1 < \nu_1 = 1.1019 < 2$. On the other hand, tails of the turmoil component are more extreme than those of the Laplace distribution because $\nu_2 = 0.6977 < 1$. A smaller shape parameter means a thicker tail and then a higher standard deviation. Conversely, a higher shaper parameter means a thinner tail and then a lower standard deviation. Table 3.7 shows that the standard deviation of the stable component $\text{Std}_1 = 0.6978$ is less than the standard deviation of turmoil component $\text{Std}_2 = 1.4300$. The kurtosis of turmoil component $\text{Kur}_2 = 11.1379$ is higher than the kurtosis of the stable component $\text{Kur}_1 = 5.2652$ which have a 'mild kurtosis'. The combination of the kurtosis of both components fit the largest kurtosis Kur= 16.0891 of daily returns.

| Stable component | | Turmoil component | | Overall | |
|---|---|---|---|---|---|
| $\text{Std}_1$ | $\text{Kur}_1$ | $\text{Std}_2$ | $\text{Kur}_2$ | Std | Kur |
| 0.6978 | 5.2652 | 1.4300 | 11.1379 | 0.9605 | 16.0891 |

**Table 3.7:** Estimated standard deviation and kurtosis of the two-component MGND model on MSCIW.

Figure 3.3 shows that the two-component MGND model well describes the heavy-tailed and leptokurtic characteristics of the MSCIW daily returns. Moreover, it confirms the presence of negative skewness given by a longer tail on the left. Figure 3.4 illustrates the MSCIW daily returns. Grey vertical lines identify turmoil days detected by the two-component MGND model and the Naïve Bayes' classifier. It can be seen that the COVID-19 pandemic constitutes the most turbulent period in terms of timing and return fluctuations.

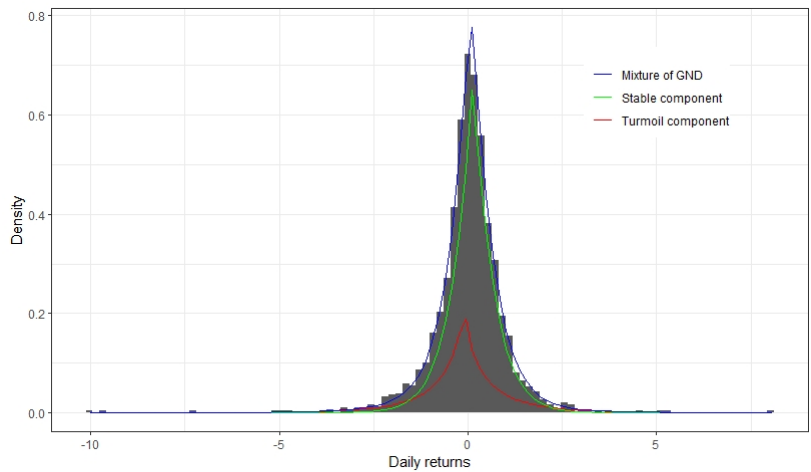

**Figure 3.3:** Estimated density of daily returns on MSCIW.

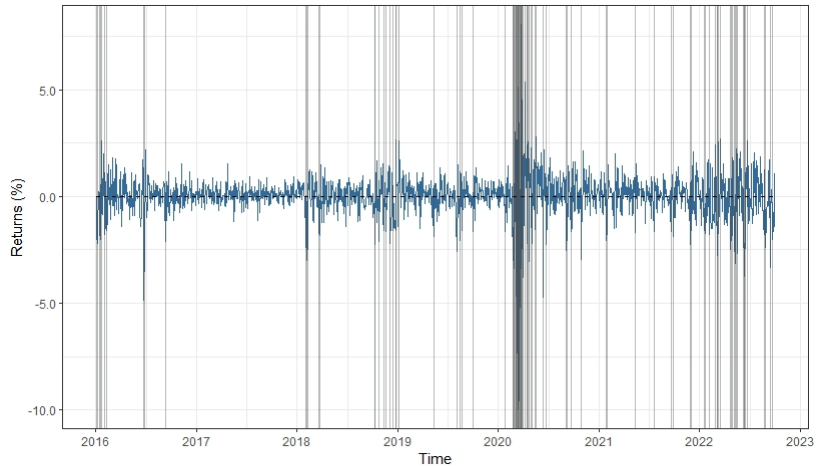

**Figure 3.4:** Daily returns on MSCIW index and turmoil days (grey vertical lines).

The AIC and BIC are applied to compare the goodness of fit performance and complexity among the normal distribution, GND, two-component MGND model, mixture of two Gaussian distributions and mixture of a Gaussian and a Laplace distribution. The latter two are nested models of the finite mixture of GND when the shape parameter $\nu_k$ takes on specific values. Table B.15 shows that both the AIC and BIC support the use of the two-component MGND model to fit the daily return distribution of MSCIW.

### 3.2.6 Results of conditional mean equation

Table 3.8 shows the estimated coefficients of the conditional mean equation. The constant $\mu$ is negative and statistically significant for all indices, except for the DJSEMUP index. Turmoil days decrease the conditional mean because the coefficient $m_1$ is negative and statistically significant for all indices. Figure 3.5 provides an easily interpretable view of the turmoil days impact on the conditional mean by index type and market. The magnitude of the impact is higher in the US market, which is followed by the Global, EU and EM markets. Results showed no significant differences between ESG and non-ESG indices except for the EU market where the conditional mean of the traditional index is more impacted than the conditional mean of the ESG index.

The AR(1) coefficient $\phi_1$ is statistically significant for all indices except for the DJUS. In addition, it is positive for the Global and EM indices but negative for the US and EU indices. These different signs may be related to the spurious consequences of the non-synchronous trading among the index's component stocks (Campbell, Lo, and MacKinlay 1997; Koutmos 1997; Basher, Hassan, and Islam 2007).

Moreover, the results reveal a statistically significant coefficient of the risk premium for all indices. The ESG indices have a higher risk premium than traditional indices in the EU and EM markets.

| Index | $\mu$ | $m_1$ | $\phi_1$ | $\lambda_1$ |
|---|---|---|---|---|
| W1DOW | -0.1047** | -2.0329** | 0.1083** | 0.3435** |
| W1SGI | -0.1159** | -1.9621** | 0.0865** | 0.3352** |
| DJUS | -0.1275** | -2.4445** | -0.0361 | 0.3586** |
| AASGI | -0.1425** | -2.5841** | -0.0756** | 0.3746** |
| E1DOW | -0.1008** | -1.8971** | -0.0569** | 0.1998** |
| DJSEUR | -0.1413** | -1.5315** | -0.0525* | 0.2771** |
| W5DOW | -0.0917** | -1.0966** | 0.1641** | 0.1903** |
| DJSEMUP | -0.1844 | -1.0750** | 0.1675** | 0.3159* |

**Table 3.8:** Estimated conditional mean of AR(1)-EGARCH-M model.

**Note.** * and ** denote p-value significance at the 5% and 1% levels, respectively.

### 3.2.7 Results of conditional volatility equation

Table 3.9 illustrates the estimated coefficients of the conditional volatility equation. The constant $\omega$ is negative and statistically significant for all indices. Turmoil days increase the conditional volatility because the coefficient $v_1$ is positive and statistically significant for all indices. According to Figure 3.6 the magnitude of the impact varies by index type and market. It is higher in the US market, followed by the Global, EU and EM markets. A more pronounced impact on the conditional volatility has been found for the traditional indices in the Global and EM markets.

Coefficients $\alpha_1$ and $\beta_1$ are statistically significant highlighting the presence of ARCH and GARCH effects for all indices. Additionally, the conditional volatility is characterized by a negative leverage effect given by the negative sign of $\alpha_1$ and the statistical significance of $\gamma_1$. Generally, the leverage effect is lower for ESG indices compared to their market benchmark, except for the US and EM markets which exhibit a reverse pattern.

**Table 3.9:** Estimated conditional volatility of AR(1)-EGARCH-M model.

| Index | $\omega$ | $v_1$ | $\alpha_1$ | $\gamma_1$ | $\beta_1$ | $\nu$ | $s$ |
|---|---|---|---|---|---|---|---|
| W1DOW | -0.0850** | 0.4574** | -0.1153** | 0.1404** | 0.9391** | 1.7114** | 1.0115** |
| W1SGI | -0.0728** | 0.3956** | -0.1185** | 0.1275** | 0.9398** | 1.5935** | 1.0005** |
| DJUS | -0.0764** | 0.5450** | -0.1020** | 0.1527** | 0.9268** | 1.5195** | 1.0117** |
| AASGI | -0.0738** | 0.5584** | -0.1098** | 0.1640** | 0.9242** | 1.5978** | 0.9906** |
| E1DOW | -0.0327** | 0.3595** | -0.1023** | 0.1942** | 0.9348** | 1.2096** | 0.9230** |
| DJSEUR | -0.0564** | 0.3459** | -0.1616** | 0.1046** | 0.9323** | 1.3622** | 0.8717** |
| W5DOW | -0.0601** | 0.4020** | -0.1020** | 0.1348** | 0.9105** | 1.6336** | 0.9065** |
| DJSEMUP | -0.0489** | 0.3364** | -0.0520** | 0.1699** | 0.9227** | 1.4946** | 0.9104** |

**Note.** ** denotes p-value significance at the 1% level.

Volatility shocks have '*long memory*' because the volatility persistence ($\beta_1$ coefficient) is close to 1 for all indices. However, it is necessary to consider the estimated (conditional) volatility and its distribution for risk assessment purposes. Figure 3.7 shows the estimated volatility by index type and market. The graphical examination suggests some interesting considerations.

**Global**   The difference in terms of estimated volatility between the ESG and traditional index is slight. During the financial crisis due to the COVID-19 pandemic, the traditional index exhibits a highest peak of estimated volatility (8.64) compared to the ESG index (7.02).

**US** The estimated volatility of the ESG and traditional index is approximately the same. During the financial crisis due to the COVID-19 pandemic, the ESG index exhibits the same peak of estimated volatility (11.84) compared to the traditional index (11.82).

**Europe** The estimated volatility of the ESG and traditional index is different because the latter is higher than the former. During the financial crisis due to COVID-19 pandemic, the traditional index exhibits a highest peak of estimated volatility (7.17) compared to the ESG index (6.38). In addition, the time series highlight other two pronounced volatility peaks: the first occurs during the Brexit vote (June 2016), while the second occurs during the Russia-Ukraine War (February 2022).

**Emerging Markets** The estimated volatility of the ESG and traditional index is approximately the same for almost the entire study period. During the COVID-19 pandemic, the ESG index exhibits a slightly highest peak of estimated volatility (5.72) compared to the traditional index (5.35).

These findings are also confirmed by the distribution of the estimated volatility in Figure 3.8. It is possible to infer that stable periods are characterized by low volatility, while turmoil days by high volatility, for both index type (in line with the assumptions in Section 2.6). Specifically, the median volatility of both index type is approximately the same during stable periods (dashed lines). In contrast, the median volatility of traditional indices is higher than that of ESG indices during turmoil days (straight lines), 1.43 against 1.34.

### 3.2.8 Diagnostic test results

The Skewed-GND effectively captures the leptokurtosis and skewness of the standardized error terms (as represented by Equation 3.3), as evidenced by the statistical significance of coefficients $v$ and $s$ across all indices (Table 3.9). Additionally, Figures 3.9 and 3.10 demonstrate that the Skewed-GND distribution closely approximates both the empirical density and quantiles of the standardized residuals.

In order to ascertain the absence of autocorrelation and heteroscedasticity in the standardized residuals of the models, the Weighted Ljung-Box test and the Weighted ARCH-LM test are performed. Table B.16 illustrates that the standardized residuals are not affected by both the serial correlation and heteroskedasticity because all test statistics have p-values> 0.05.

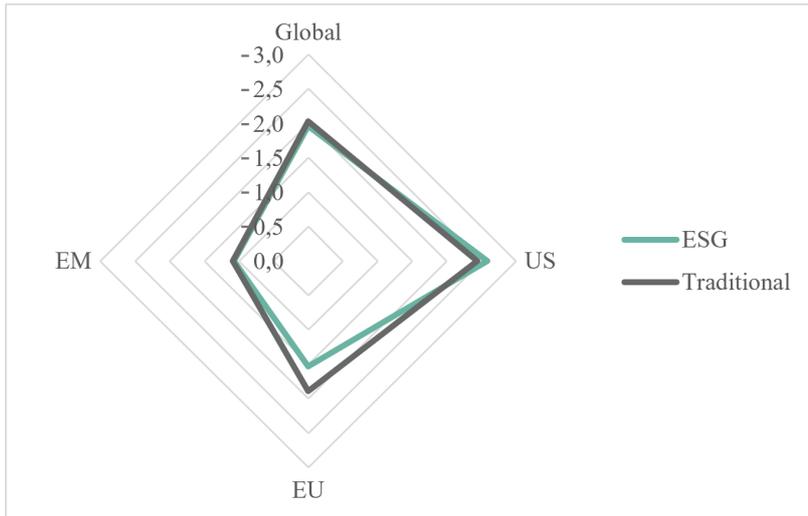

**Figure 3.5:** Impact of turmoil days on the conditional mean ($m_1$ coefficient) by index type and market.

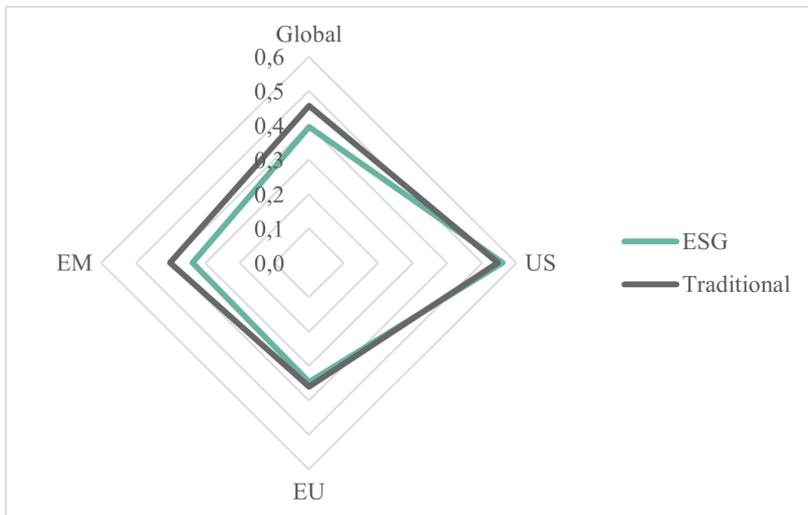

**Figure 3.6:** Impact of turmoil days on the conditional volatility ($v_1$ coefficient) by index type and market.

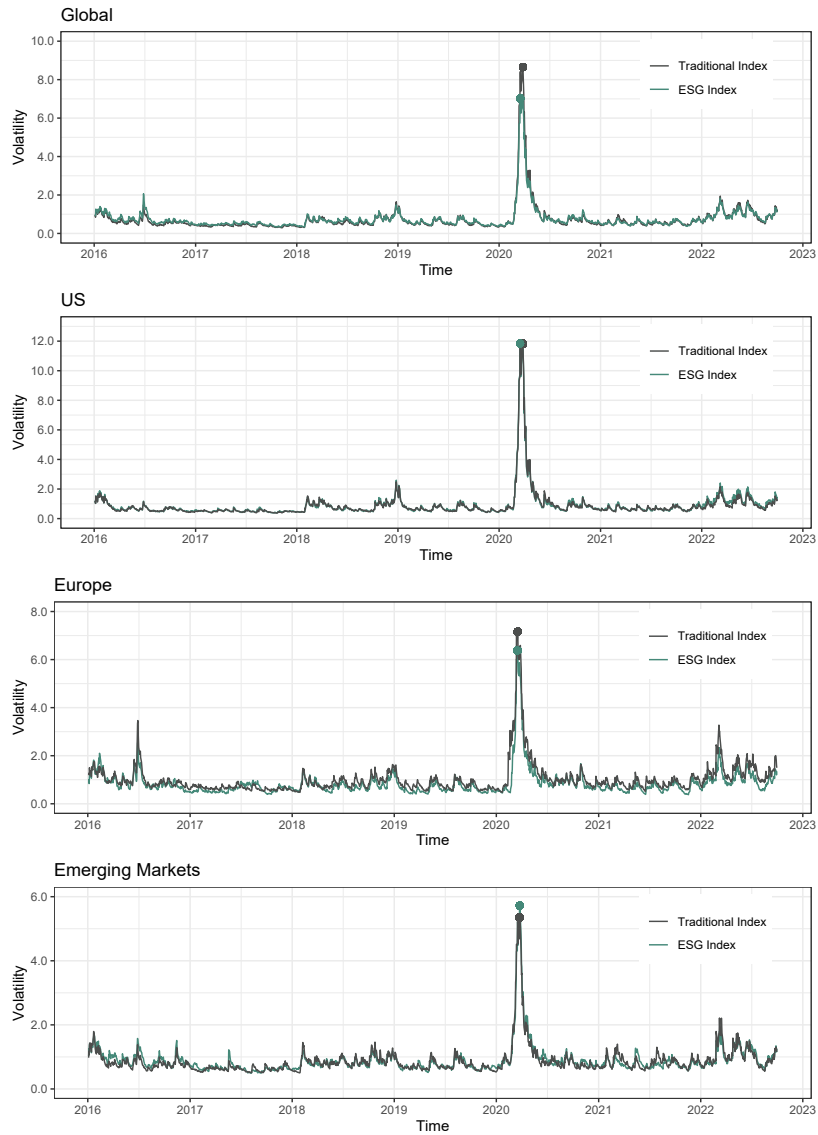

**Figure 3.7:** Estimated conditional volatility by index type and market.

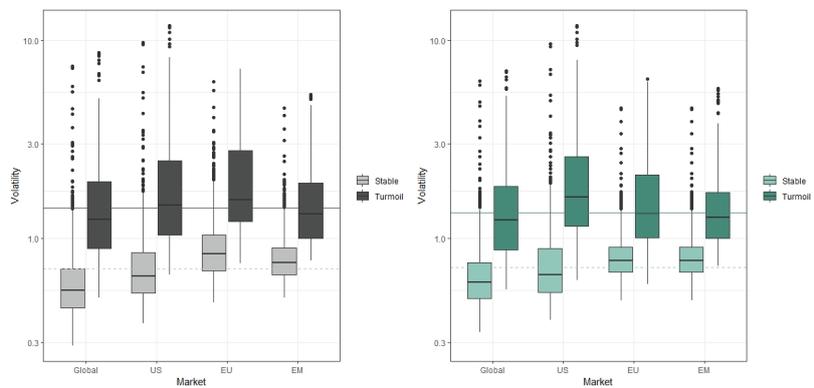

**Figure 3.8:** Distribution of the estimated conditional volatility by index type and market.

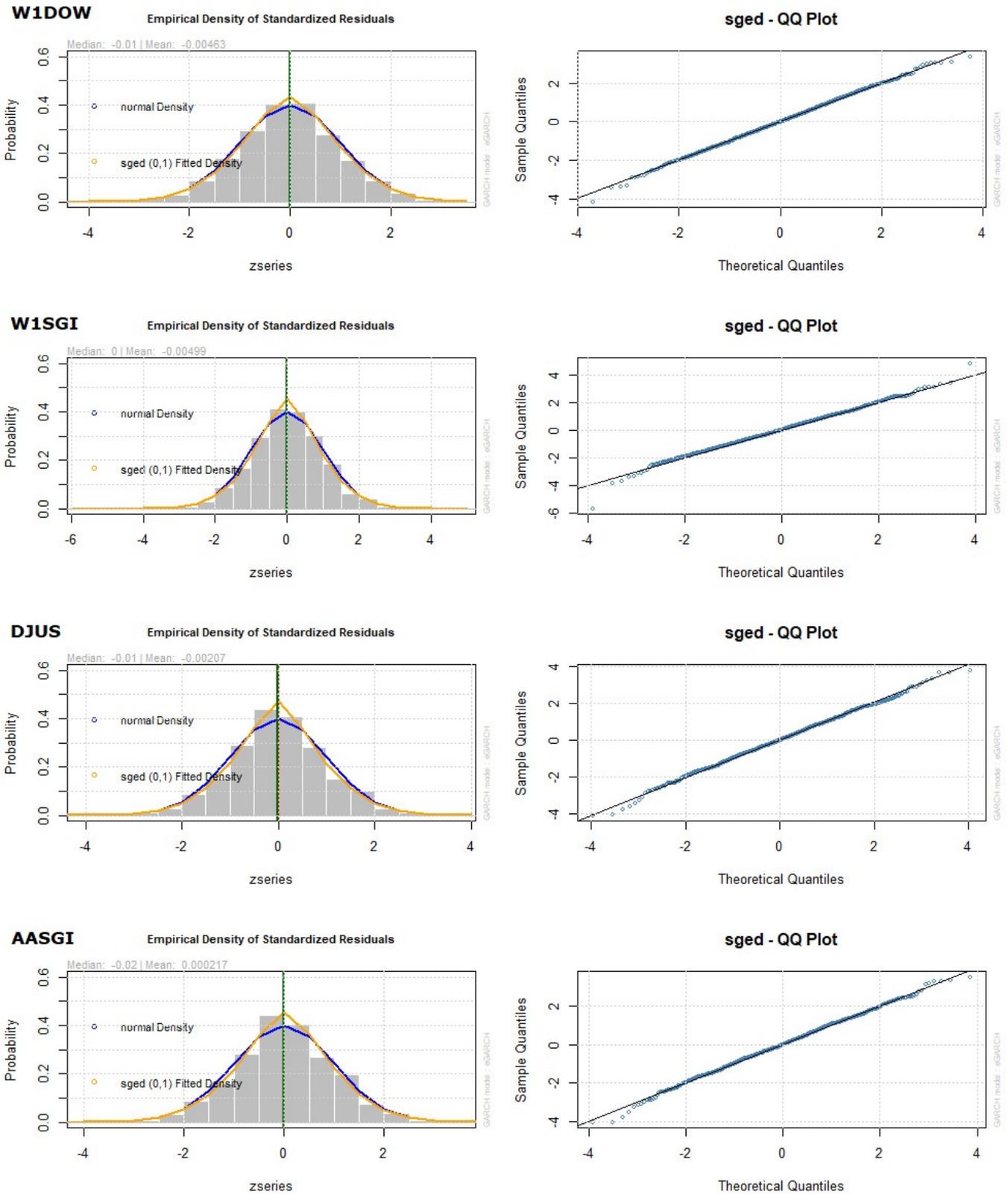

**Figure 3.9:** Densities and QQ-plot of standardized residuals (Part 1).

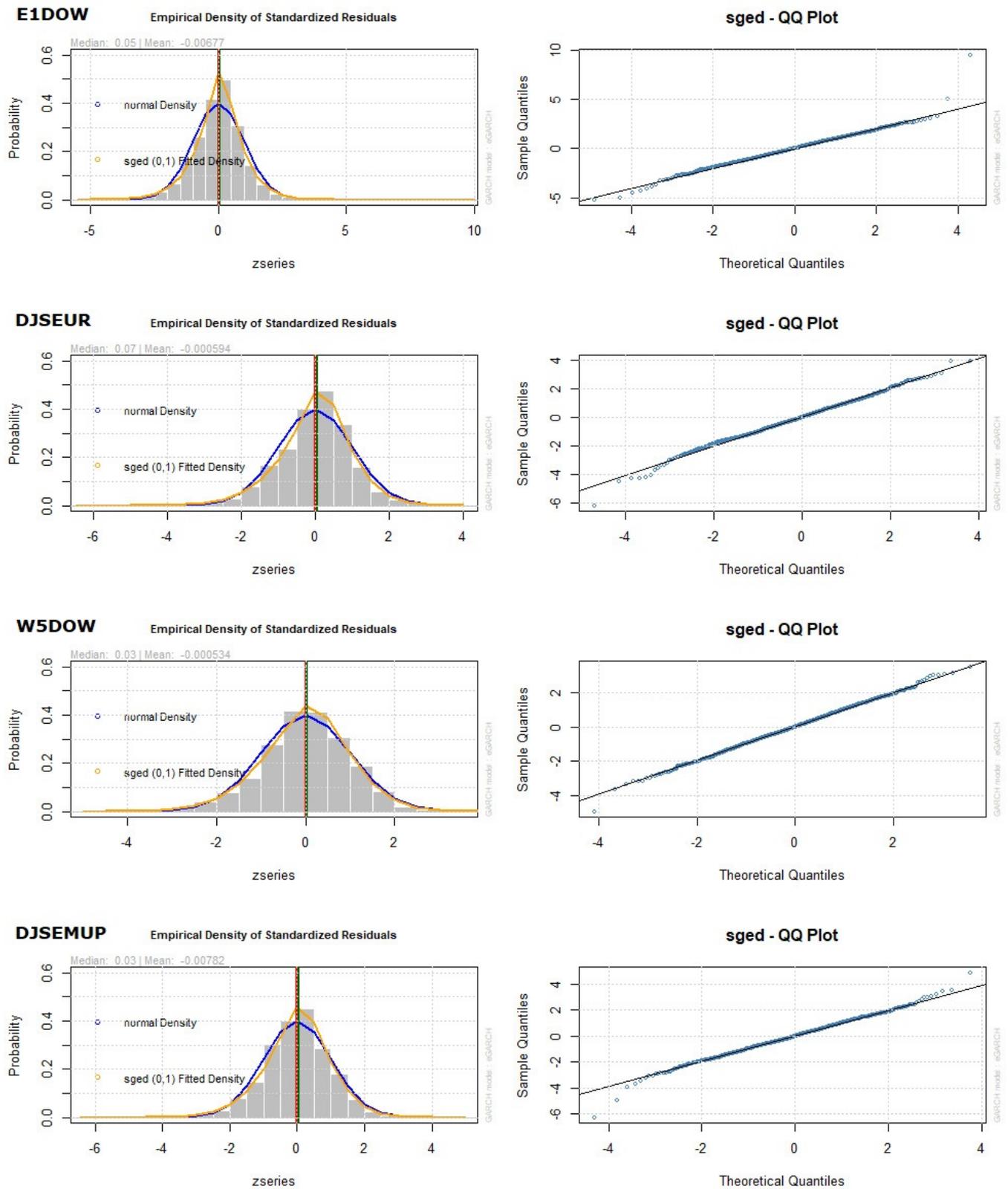

**Figure 3.10:** Densities and QQ-plot of standardized residuals (Part 2).

### 3.2.9 Final remarks

This study adds to the literature as it investigates the reactions of ESG and traditional indices to turmoil days.

Firstly, a finite mixture of two GND is estimated on the MSCIW index returns. According to the initial hypotheses (Section 3.1), findings show that the two-component MGND well describes the heavy-tailed and leptokurtic characteristics of the daily returns. Similarly to Kim and White 2004 and Wen et al. 2022, the estimated mixture contains a '*predominant*' stable component and a '*relatively rare*' turmoil component. The former is characterized by a high shape parameter that implies a thinner tail and then a lower standard deviation, while the latter is featured by a small shape parameter that implies a thicker tail and then a higher standard deviation. Consequently, the mixture model results are exploited to objectively detect financial market turmoil days.

Secondly, turmoil days are included as exogenous dummy variables in the EGARCH-in-Mean model to capture their impact on the conditional mean and conditional volatility of several ESG and non-ESG Dow Jones indices. Results show that the return-risk performance as well as the impact of turmoil days on return and volatility vary among ESG and traditional indices, and across markets. Mixed return-risk findings (Cunha et al. 2020) are obtained similarly to Hoti, McAleer, and Pauwels 2007, Lean and D. K. Nguyen 2014, and Ang 2015.

Whereas the European ESG index results to be less volatile than its traditional market benchmark, the estimated conditional volatility is approximately the same in the other markets.

The turmoil days impact on the conditional mean and volatility of both index type shows a similar pattern: the most affected market is US followed by the Global, EU and EM markets. However, the conditional mean of ESG indices is less affected in the EU market, while they are more resilient in the Global and EM markets in terms of conditional volatility.

Other interesting key aspects concern the risk-premium and leverage effect. ESG indices have a higher risk premium than traditional indices in the EU and EM markets. As identified by Sabbaghi 2022, results confirm the presence of the leverage effect for ESG indices. The leverage effect is lower for ESG indices compared to their market benchmark, except for the US market and EMs which exhibit a reverse pattern.

Findings are important to equity portfolio managers, investors, policy-makers, academics and anyone who decide to encourage ESG investments. In order to contain the risk during turmoil days, portfolio managers may apply rebalancing strategies placing higher weight on the less volatile asset. Besides, they may obtain more investment information about decisions, in terms of diversifying the risks considering also the turmoil period impact, risk premium and leverage effect (Markowitz 1952). Lastly, corporate executives yet shall use it to benchmark their own results against peers and track news as well.

It is important to note that the results obtained depend on the definition of the turmoil period (Section 3.1). For example, considering market specific turmoil days, larger differences between ESG and non-ESG indices might be observed. Additionaly, the Naïve Bayes' classification rule makes the assignment without considering the temporal dependence of the returns. It would be interesting to explore a dynamic time-varying estimation of the mixture model.

To conclude, another extension of this work could be the study of the evolution of correlations among ESG and traditional indices focusing on the impact of turmoil days and spillover effects among markets.

## 3.3 An overview of the G7 stock markets

Considered as economic barometers, stock markets rapidly reflect the impact of important events on traded stocks (Kartal, Ertugrul, and Ulussever 2022). These events can lead to financial turmoil days, often associated with tail risk events, i.e. extreme and unexpected market movements that fall in the tails of a probability distribution (Bollerslev and Todorov 2011). Thus, analysing the features of turmoil days can contribute to a better understanding of tail risk, allowing for improved risk management strategies to mitigate the impact of rare market fluctuations. The 2015-2016 stock market sell-off, the US-China trade war, the COVID-19 pandemic, and the Russian-Ukrainian conflict, are cited as examples (Awan et al. 2021; So, Mak, and A. Chu 2022; Izzeldin et al. 2023). Several contributions (Yousef 2020; Shehzad, Xiaoxing, and Kazouz 2020), point out that the stock market volatility increased over crisis periods leading to a sudden and unpredictable stock market behaviour.

Yousef 2020 stated that the COVID-19 pandemic increased the volatility of G7 stock markets. Similar results were obtained by Chaudhary, Bakhshi, and Gupta 2020, Awan et al. 2021, Mobin et al. 2022 and Basuony et al. 2022. Shehzad, Xiaoxing, and Kazouz 2020 investigated the impact of the global financial crisis, the US-China trade war and the COVID-19 pandemic on several G7 stock markets. Findings highlight that Asian markets were less affected by the COVID-19 pandemic compared to European and US markets.

Common approaches to assess the impact of specific events on stock markets include the use of subjectively identified dummy variables (Yousef 2020; Chaudhary, Bakhshi, and Gupta 2020; Shehzad, Xiaoxing, and Kazouz 2020; Duttilo, Gattone, and Di Battista 2021), sub-periods (Liyan Han and Zheng 2019), and Markov-switching GARCH models (Ardia et al. 2019; Ouchen 2022).

This work implements a two-step approach (Duttilo, Gattone, and Iannone 2023) to study the impact of global turmoil days (or global tail risk events) on the G7 stock markets returns and volatility. Firstly, turmoil days are objectively detected with a two-component mixture of generalised normal distributions (MGND) on the Dow Jones Global Index (W1DOW) returns. The MGND model is able to fit the returns distribution and its non-normal features like skewness and excess kurtosis (Wen et al. 2022; Duttilo, Gattone, and Iannone 2023). Returns are generated by turmoil days if they belong to the mixture component with the lowest shape parameter which accounts an extreme tail behaviour. The resulting Naïve Bayes' classification is encoded in three different dummy variables:

pre-COVID-19 turmoil days (2016-2019), COVID-19 turmoil days (2020-2021) and Russian-Ukrainian conflict turmoil days (2022-2023). Secondly, returns and volatility are estimated with an AR(1)-NAGARCH(1,1)-M model for each G7 stock market including the three exogenous dummy variables defined in the first step.

The following questions are addressed: Do turmoil days have a significant impact on returns and volatility of G7 indices? Which market has been most affected by turmoil days? and in what period?

### 3.3.1 Data

The study is performed considering the returns of the daily closing prices of the W1DOW and stock market indices of the G7 countries, namely GSPTSE (Canada), FCHI (France), DAX (Germany), FTSE.MIB (Italy), N225 (Japan), FTSE (UK) and S&P500 (US). The time period taken for the study is from January 4, 2016 to June 30, 2023. Daily returns of all indices under study are calculated with the natural log difference approach in equation 2.4.

### 3.3.2 Nonlinear-asymmetric GARCH-in-Mean model with exogenous dummy variables

Hentschel 1995 developed the family GARCH model which nests the most popular symmetric and asymmetric GARCH models. Notable among these is the nonlinear-asymmetric GARCH (NAGARCH) model proposed by Engle and Ng 1993. This model allows to capture the leverage effect by a right-shift of the news impact curve (NIC)*. In this work, the AR(1)-NAGARCH(1,1)-M model with exogenous dummy variables is specified as follows:

**Conditional mean equation**

$$r_{i,t} = \mu + \eta_1 \text{pre-COVID-19} + \eta_2 \text{COVID-19} + \eta_3 \text{WAR} + \phi_1 r_{i,t-1} + \lambda_1 h_{i,t}^2 + \epsilon_{i,t},$$
(3.4)

**Conditional volatility equation**

$$h_{i,t}^2 = \omega + \delta_1 \text{pre-COVID-19} + \delta_2 \text{COVID-19} + \delta_3 \text{WAR} + \alpha_1 h_{i,t-1}^2 (z_{i,t-1} - \gamma_1)^2 + \beta_1 h_{i,t-1}^2$$
$$\text{with } z_{i,t} = \frac{\epsilon_{i,t}}{\sqrt{h_{i,t}^2}} \sim \text{SGND}(0, 1, \nu, s).$$
(3.5)

In equation (3.4) $\mu$ is the constant term. The coefficients $\eta_1$, $\eta_2$, $\eta_3$ determine the impact of turmoil days on the conditional mean during the pre-COVID-19 period, COVID-19 pandemic and Russian-Ukrainian conflict, respectively. Consequently, pre-COVID-19, COVID-19 and WAR are three dummy variables which assume the value of 1 during turmoil days, otherwise they are equal to 0 (stability days). $\phi_1$ is the first-order autoregressive coefficient. Based on the ideas of Engle, Lilien, and R. P.

---

* The NIC associates past returns shocks to current volatility. The leverage effect occurs when negative past returns shocks increase volatility by a larger amount than positive returns shocks (Engle and Ng 1993).

Robins 1987, Eq. (3.4) includes the risk premium coefficient $\lambda_1$. $\epsilon_{i,t}$ indicates the error terms of stock index $i$ at time $t$.

In equation (3.5), $h_{i,t}^2$ denotes the conditional volatility of stock index $i$ at time $t$. $\omega$ is the constant term. The coefficients $\delta_1$, $\delta_2$, $\delta_3$ determine the impact of turmoil days on the conditional volatility during the pre-COVID-19 period, COVID-19 pandemic and Russian-Ukrainian conflict, respectively. The dummy variables are defined as in equation (3.4). The parameters $\alpha_1$ and $\beta_1$ capture the ARCH and GARCH effects. Since $\alpha_1 > 0$ and $h_{i,t-1}^2 > 0$ the leverage effect is observed if $\gamma_1 > 0$. $z_{i,t}$ is the standardized residual of stock index $i$ at time $t$. Standardized residuals are modelled with a skewed generalised normal distribution (SGND) to capture non-normal features of returns. The AR(1)-NAGARCH(1,1)-M model with exogenous dummy variables is estimated by the R's *rugarch* package for all G7 indices (Ghalanos 2022).

### 3.3.3 Results

Table 3.10 shows descriptive statistics and some preliminary statistical hypothesis tests. The negative skewness, excess kurtosis and JB test stated that G7 indices are not normally distributed. Moreover, the ADF tests highlight that all indices are stationary in their first difference, while the ARCH-LM tests confirm the presence of ARCH effects and heteroscedasticity.

**Table 3.10:** Descriptive statistics and statistical tests results of G7 stock markets.

| G7 Index | Mean | SD | Skewness | Kurtosis | JB | ADF | ARCH-LM |
|----------|------|------|----------|----------|---------|----------|---------|
| GSPTSE | 0.0223 | 1.0358 | -1.7955 | 43.1601 | 132726** | -12.12** | 746.26** |
| FCHI | 0.0259 | 1.2213 | -1.1107 | 14.2981 | 14819** | -12.66** | 336.76** |
| DAX | 0.0203 | 1.2511 | -0.7521 | 13.2396 | 12568** | -12.18** | 363.94** |
| FTSE.MIB | 0.0208 | 1.4705 | -1.9337 | 22.8831 | 38116** | -12.30** | 187.36** |
| N225 | 0.0212 | 1.2520 | -0.2189 | 4.6937 | 1574** | -12.66** | 220.26** |
| FTSE | 0.0044 | 1.0448 | -1.0696 | 15.2159 | 16711** | -12.62** | 451.59** |
| S&P500 | 0.0430 | 1.2124 | -0.8726 | 16.6982 | 19951** | -12.86** | 736.31** |

**Note.** * denotes p-value significance at the 1% level of statistical hypothesis tests: Jarque-Bera test (JB), Augmented Dickey-Fuller test (ADF) and ARCH-LM test.

Table 3.11 shows the estimated two-component MGND on W1DOW returns. The stable component is predominant compared to the turmoil component which has a negative location parameter and a higher scale parameter. In addition, the shape parameter of the turmoil component is lower than the shape parameter of the stable component. As a result the standard deviation of the turmoil component (2.0339) is higher than the standard deviation of the stable component (0.7119). These findings are also confirmed in panel (a) of Figure 3.11 where it is evident as the turmoil component (dashed line) captures the extreme tail behaviour of W1DOW returns. Panel (b) shows the W1DOW daily returns and the detected turmoil days (gray vertical lines).

**Table 3.11:** Estimated parameters of the two-component MGND on W1DOW.

| Stable component | | | | Turmoil component | | | |
|------|------|------|------|------|------|------|------|
| $\pi_1$ | $\mu_1$ | $\sigma_1$ | $\nu_1$ | $\pi_2$ | $\mu_2$ | $\sigma_2$ | $\nu_2$ |
| 0.8900 | 0.0781 | 0.6737 | 1.2262 | 0.1100 | -0.3097 | 1.7347 | 1.1313 |

Table 3 shows the estimated coefficients of the conditional mean and variance equations. The coefficients $\alpha_1$ and $\beta_1$ are positive and statistically significant for all G7 indices, underlining the presence of ARCH and GARCH effects. In addition, the leverage effect ($\gamma_1$) and risk premium ($\lambda_1$) are statistically significant for all indices, except for the N255 (Japan) where the risk premium term is low and not significant. The SGND suitably captures the non-normal features of daily returns as the coefficients $v$ and $s$ are statistically significant for all indices. Turmoil days decrease the conditional mean ($\eta_i < 0$ for $i = 1, 2, 3$) and increase the conditional volatility ($\delta_i > 0$ for $i = 1, 2, 3$). Figure 3.12 provides an easy view of the turmoil days impact on the conditional mean and variance. The magnitude of impact vary across periods (pre-COVID-19, COVID-19 and Russia-Ukrainian conflict) and among indices. With the only exception of N225 (Japan), pre-COVID-19 turmoil days have a low impact on the conditional mean and variance. On the other hand, COVID-19 turmoil days greatly affect the conditional mean and variance compared to the other two periods. Specifically, whereas S&P500 (US) and DAX (Germany) are the most affected indices in term of conditional mean, the impact on the conditional volatility is almost homogeneous among G7 indices. The impact of the Russia-Ukrainian conflict turmoil days on the conditional mean is lower than the impact of pre-COVID-19 turmoil days, except for S&P500 (US) and DAX (Germany). By contrast, the impact on the conditional volatility is approximately the same, except for N225 (Japan).

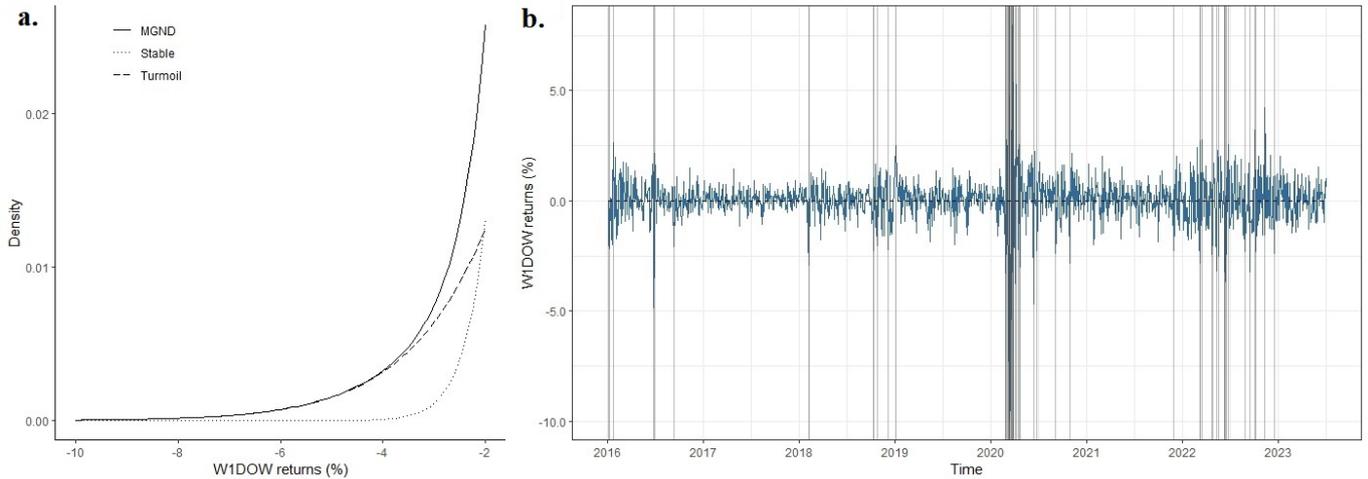

**Figure 3.11:** Left tail of estimated MGND model panel (a) and turmoil days panel (b).

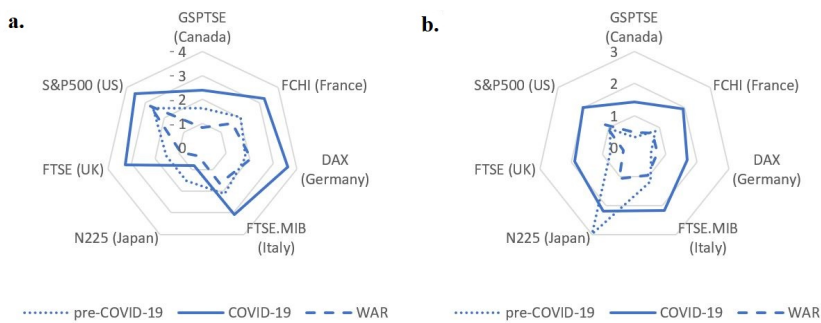

**Figure 3.12:** Turmoil days impact on the conditional mean - panel (a) and variance - panel (b) by G7 stock market.

### 3.3.4 Final remarks

This study adds to the literature as it investigates the reactions of G7 indices to turmoil days in three different periods: pre-COVID-19 (2016-2019), COVID-19 (2020-2021) and Russia-Ukrainian conflict (2022-2023). Turmoil days are objectively detected with a two-component MGND on the W1DOW returns. Then, returns and volatility are estimated with the AR(1)-NAGARCH(1,1)-M model including the detected turmoil days as exogenous dummy variables. Findings highlight that the impact of turmoil days vary across markets and periods. Specifically, COVID-19 turmoil days greatly affect the conditional mean and variance compared to the other two periods. In contrast, the impact of Russia-Ukrainian conflict turmoil days on the conditional volatility is almost equal to the impact of pre-COVID-19 turmoil days, except for N225 (Japan). A possible extension of this work could be a comparison between the two-step approach and Markov-Switching GARCH models which identify stable and turmoil regimes directly within the GARCH equations (Ardia et al. 2019).

**Table 3.12:** Estimated coefficients of the AR(1)-NAGARCH(1,1)-M model.

|  | GSPTSE | FCHI | DAX | FTSE.MIB | N225 | FTSE | S&P500 |
|---|---|---|---|---|---|---|---|
| Conditional mean equation | | | | | | | |
| $\mu$ | 0.0198 | -0.0065 | -0.0607** | -0.0064 | 0.0430 | -0.0608** | 0.0218 |
| $\eta_1$ | -1.6350** | -2.0333** | -1.8392** | -2.1130** | -1.5196* | -1.5262** | -2.6136** |
| $\eta_2$ | -2.3978** | -3.2706** | -3.6107** | -3.0733** | -0.8107 | -3.2512** | -3.5712** |
| $\eta_3$ | -0.8332** | -1.5956** | -2.0588** | -1.6348** | -0.3717 | -0.8998** | -2.7547** |
| $\phi_1$ | 0.0181 | -0.0077 | -0.0281* | -0.0341 | -0.0163 | -0.0421* | -0.0383 |
| $\lambda_1$ | 0.0483* | 0.0677** | 0.1149** | 0.0466* | 0.0025 | 0.1264** | 0.0769** |
| Conditional volatility equation | | | | | | | |
| $\omega$ | 0.0169** | 0.0458** | 0.0570** | 0.0765** | 0.1121** | 0.0528** | 0.0250** |
| $\delta_1$ | 0.3262** | 0.8267** | 0.5294** | 1.1566* | 2.9796** | 0.7684** | 0.9547** |
| $\delta_2$ | 1.4338** | 1.9458** | 1.6791** | 2.1641** | 2.1865** | 1.8920** | 2.0159** |
| $\delta_3$ | 0.4726** | 0.7307** | 0.7363** | 0.9548** | 1.0635* | 0.3341* | 1.1547** |
| $\alpha_1$ | 0.0866** | 0.0561** | 0.0419** | 0.0671** | 0.0827** | 0.0809** | 0.1053** |
| $\beta_1$ | 0.7954** | 0.6316** | 0.6395** | 0.7131** | 0.6877** | 0.7287** | 0.7387** |
| $\gamma_1$ | 0.8968** | 2.1652** | 2.4852** | 1.5542** | 1.2111** | 1.1511** | 1.0493** |
| $\nu$ | 1.7202** | 1.3568** | 1.2578** | 1.3669** | 1.4119** | 1.3200** | 1.4822** |
| $s$ | 0.8391** | 0.8974** | 0.8982** | 0.8863** | 0.9544** | 0.8838** | 0.9663** |

Note. ** and * denote p-value significance at the 1% and 5%, respectively

# Constrained mixture of generalized normal distribution to fit stock returns

# 4

## 4.1 Study framework

Finite mixtures of distributions have been widely used to analyse complex distributions of data (G. J. McLachlan, S. X. Lee, and S. I. Rathnayake 2019) since K. Pearson 1894 first introduced a two-component normal mixture model. Normal mixture modelling is a well-known method used for most applications in a wide range of fields (biology, social sciences, etc.). However, datasets characterized by non-normal features like high skewness, leptokurtosis and heavy tails require more flexible tools. Hence, non-normal model-based methods (S. Lee and G. Mclachlan 2013) received increasing attention over the years, the MGND and MSTD models (Chapter 2) can be cited as examples.

Several approaches to estimate mixture models are available in the literature, including the method of moments or the maximum likelihood approach. The MLE can be obtained via the EM algorithm (Dempster, Laird, and Rubin 1977). It is well known that MLE of normal mixture models could be problematic if the parameters are not appropriately constrained. Indeed, in some cases, both for univariate and multivariate data the resulting likelihood is unbounded (G. J. McLachlan and Peel 2000). For this reason, the maximum likelihood estimation of normal mixture models is problematic for both univariate and multivariate data. For the univariate case the problem occurs in the presence of one fitted component having a very small estimate of the variance, i.e., a few data points relatively close together (Biernacki and Chrétien 2003). As a result, the likelihood function either increases intolerably high iteratively (degeneracy) or leads to spurious solutions[1] , i.e., with a high likelihood but also with a high bias. Existing methods to overcome this drawback are based on the seminal work of Hathaway 1986 who, in the univariate case, imposed a lower bound to the ratios of the scale parameters. Similarly, in the multivariate case, the lower bound is imposed on the eigenvalues of each pair of covariance matrices (Banfield and Raftery 1993; Celeux and Govaert 1995; Rocci, Gattone, and Di Mari 2018).

In order to handle the degeneracy of the likelihood function, in the pathological context of univariate normal mixtures, Quandt and Ramsey 1978 and Chauveau and Hunter 2013 consider imposing linear constraints on both mean and variance parameters. Similarly, Andrews, McNicholas, and Subedi 2011, Andrews, Wickins, et al. 2018 and Massing and Ramos 2021 presented a family of univariate mixtures of Student-t while requiring the variances and the degrees of freedom remain equal across mixture components.

The use of such constraints in the mixture models clearly leads to more parsimonious models as the number of unconstrained parameters increases linearly with the (potentially high) number of components. More specifically for the univariate MGND, the unknown number of [1]: An example of spurious solutions has been introduced in the case study of Chapter 2 for the estimation of the MGND model via algorithm 1.

parameters is $4K - 1$ where $K$ is the number of mixture components. For large $K$ that can be an issue while in some circumstances the nature of the constraints could also stem from the underlying data domain.

To the best of our knowledge, none of the existing studies implement equality constraints on the parameters of the univariate MGND. The following proposal involves exploring constrained mixtures for generalized normal distributions (CMGND), where parameters are allowed to be equal across any subset of mixture components. Since the GND is a generalization of the normal distribution, it is straightforward to observe that the aforementioned degeneracy in MLE for normal mixtures also holds for MGND. Furthermore, the implementation of such constraints not only solves the numerical degeneracy of the log-likelihood but also enhances the interpretability and parametric efficiency of the final solution.

Constraints are imposed on the location and/or scale and/or shape parameters. In the two-component case ($K = 2$) taking all possible combinations of these constraints into consideration would result to a 7-model family. The MLE of the parameters is obtained via the GEM algorithm. Additionally, since the iteration equations for some of the parameters are non-linear, the iterative Newton-Raphson method is employed.

The proposed methodology is tested on simulated data across different scenarios such as symmetric models (mixtures with common location parameters) and models with the same kurtosis (i.e. mixtures with common shape parameter). In order to test the null hypothesis *'constraints hold'*, an appropriate likelihood ratio test (LRTS) is implemented. The LRTS performance is evaluated by a power analysis.

Furthermore, an empirical application and a comparative study (part II) is performed on financial data with constrained models.

## 4.2 Constrained mixture of generalized normal distribution

A finite constrained mixture of generalized normal distribution (CMGND) with K components is given by the marginal distribution of the random variable $X$

$$f_{CMGND}(x|\theta) = \sum_{k=1}^{K} \pi_k f_k(x|\mu_k, \sigma_k, \nu_k) \tag{4.1}$$

where $f_k(\cdot) = f_{GND}(\cdot)$, and the set of all mixture parameters is given by $\theta = \{\pi_k, \mu_k, \sigma_k, \nu_k, k = 1, ..., K\}$ belonging to the parameter space $\Theta = \{\theta : 0 < \pi_k < 1, \sum_{k=1}^{K} \pi_k = 1, \mu_k \in \mathbb{R}, \sigma_k > 0, \nu_k > 0, k = 1, ..., K\}$ with $\dim(\theta) = p$. Let define a partition of the set $\{1, 2, ..., K\}$ as $C_1, C_2, ..., C_R$ such that $C_1 \cup C_2 \cup \cdots \cup C_R = \{1, 2, ..., K\}$ and $C_i \cap C_j = \emptyset$ $\forall_{i,j} = 1, 2, ..., K$. The parameter space $\Theta$ can be constrained by imposing equality constraints such as $\mu_k = \mu_r$ and/or $\sigma_k = \sigma_r$ and/or $\nu_k = \nu_r$ for all $k \in C_r$. For K = 2, Table 4.1 displays the 7-model family of CMGND together with the MGND model. By constraining the parameter space ($p$) the number of parameters ($p$) of each model is gradually reduced as shown in the last column of Table 4.1.

**Table 4.1:** 7-model family of CMGND and MGND models. 'C' denotes constrained, 'U' denotes unconstrained.

| Model | $\mu_k$ | $\sigma_k$ | $\nu_k$ | $p$ |
|---|---|---|---|---|
| MGND | U | U | U | 7 |
| CMGND-CUU | C | U | U | 6 |
| CMGND-UCU | U | C | U | 6 |
| CMGND-UUC | U | U | C | 6 |
| CMGND-CCU | C | C | U | 5 |
| CMGND-CUC | C | U | C | 5 |
| CMGND-UCC | U | C | C | 5 |
| CMGND-CCC | C | C | C | 4 |

## 4.3 Parameter estimation

In this subsection a GEM algorithm with Newton-Raphson updates is introduced to perform the parameter estimation of the CMGND model. Since the Newton-Raphson updates will not necessarily increase the log-likelihood, a backtracking line-search is included to ensure the monotonicity of the algorithm in the update of the unconstrained and constrained parameters. In particular to calculate a suitable step-length $\alpha$ it is possible to start from $\alpha = 1$ and check if the likelihood increases otherwise $\alpha$ is decremented by a factor of 0.8. If $\alpha < 0.01$, the search is terminated with no parameter update. Furthermore, an adaptive step-size is also included to avoid the degeneracy issue[2] in the update of the unconstrained and constrained shape parameter.

From equation 4.1 the log-likelihood function is given by

$$\log L(\theta) = \sum_{n=1}^{N} \log \left[ \sum_{k=1}^{K} \pi_k f_k(x_n | \mu_k, \sigma_k, \nu_k) \right]. \tag{4.2}$$

The **E-step** involves computing the following equation

$$Q(\theta, \theta^{(m-1)}) = \sum_{k=1}^{K} \sum_{n=1}^{N} z_{kn}^{(m-1)} \log \left[ \pi_k^{(m-1)} f_k(x_n | \mu_k^{(m-1)}, \sigma_k^{(m-1)}, \nu_k^{(m-1)}) \right], \tag{4.3}$$

where

$$z_{kn}^{(m-1)} = \frac{\pi_k^{(m-1)} f_k(x_n | \mu_k^{(m-1)}, \sigma_k^{(m-1)}, \nu_k^{(m-1)})}{\sum_{k=1}^{K} \pi_k^{(m-1)} f_k(x_n | \mu_k^{(m-1)}, \sigma_k^{(m-1)}, \nu_k^{(m-1)})}. \tag{4.4}$$

The term $z_{kn}^{(m-1)}$ represents the current estimate of the posterior probability or '*responsibility*' (Bishop 2006) at the $(m-1)$-th iteration, i.e. the probability that the observation $n$ belongs to group $k$ given the current component parameters $\theta^{(m-1)}$.

The **M-Step** maximizes $Q(\theta, \theta^{(m-1)})$ with respect to $\theta$ to obtain the parameter estimation at the $m$-th iteration and increases the expectation of the log-likelihood function in equation 4.2. In what follows we provide the updating equations for the CMGND model[3].

**Mixture weights**    Set $\frac{\partial Q(\theta, \theta^{(m-1)})}{\partial \pi_k} = 0$, then

$$\pi_k^{(m)} = \frac{\sum_{n=1}^{N} z_{kn}^{(m-1)}}{\sum_{k=1}^{K} \sum_{n=1}^{N} z_{kn}^{(m-1)}}. \tag{4.5}$$

**Location parameter**    In order to obtain the iteration equation for the constrained location parameter at any partition $r \in 1, 2, ...R$, let $\mu_k = \mu_r$ for all $k \in C_r$, then for $r = 1, 2, ..., R$

$$\frac{\partial Q(\theta, \theta^{(m-1)})}{\partial \mu_r} = \sum_{k \in C_r} \left[ \frac{\nu_k^{(m-1)}}{(\sigma_k^{(m-1)})^{\nu_k^{(m-1)}}} \left( \sum_{x_n \geq \mu_r^{(m-1)}}^{N} z_{kn}^{(m-1)} (x_n - \mu_r^{(m-1)})^{\nu_k^{(m-1)}-1} \right. \right.$$
$$\left. \left. - \sum_{x_n < \mu_r^{(m-1)}}^{N} z_{kn}^{(m-1)} (\mu_r^{(m-1)} - x_n)^{\nu_k^{(m-1)}-1} \right) \right] = 0.$$

Since the above equation is non-linear, the iterative Newton-Raphson method is applied as follows:

$$\mu_r^{(m)} = \mu_r^{(m-1)} - \alpha^{(i)} \frac{g(\mu_r^{(m-1)})}{g'(\mu_r^{(m-1)})}, \qquad (4.6)$$

where

$$g(\mu_r^{(m-1)}) = \sum_{k \in C_r} \left[ \frac{\nu_k^{(m-1)}}{(\sigma_k^{(m-1)})^{\nu_k^{(m-1)}}} \left( \sum_{x_n \geq \mu_r^{(m-1)}}^{N} z_{kn}^{(m-1)} (x_n - \mu_r^{(m-1)})^{\nu_k^{(m-1)}-1} \right. \right.$$
$$\left. \left. - \sum_{x_n < \mu_r^{(m-1)}}^{N} z_{kn}^{(m-1)} (\mu_r^{(m-1)} - x_n)^{\nu_k^{(m-1)}-1} \right) \right],$$
$$\qquad (4.7)$$

$$g'(\mu_r^{(m-1)}) = \sum_{k \in C_r} \left[ -\frac{\nu_k^{(m-1)}}{(\sigma_k^{(m-1)})^{\nu_k^{(m-1)}}} \left( \sum_{x_n \geq \mu_r^{(m-1)}}^{N} z_{kn}^{(m-1)} (x_n - \mu_r^{(m-1)})^{\nu_k^{(m-1)}-2} \right. \right.$$
$$\left. \left. \times (\nu_k^{(m-1)} - 1) + \sum_{x_n < \mu_r^{(m-1)}}^{N} z_{kn}^{(m-1)} (\mu_r^{(m-1)} - x_n)^{\nu_k^{(m-1)}-2} (\nu_k^{(m-1)} - 1) \right) \right].$$
$$\qquad (4.8)$$

After some further calculations, equation 4.6 is shown to be as follows

$$\mu_r^{(m)} = \mu_r^{(m-1)} + \alpha^{(i)} \frac{\sum_{k \in C_r} A_k}{\sum_{k \in C_r} B_k}, \qquad (4.9)$$

where

$$A_k = \sum_{x_n \geq \mu_r^{(m-1)}}^{N} z_{kn}^{(m-1)} (x_n - \mu_r^{(m-1)})^{\nu_k^{(m-1)}-1} - \sum_{x_n < \mu_r^{(m-1)}}^{N} z_{kn}^{(m-1)} (\mu_r^{(m-1)} - x_n)^{\nu_k^{(m-1)}-1},$$

$$B_k = \sum_{n=1}^{N} z_{kn}^{(m-1)} |x_n - \mu_r^{(m-1)}|^{\nu_k^{(m-1)}-2} (\nu_k^{(m-1)} - 1).$$

**Scale parameter** In order to obtain the iteration equation for the constrained scale parameter at any partition $r \in 1, 2, ... R$, let $\sigma_k = \sigma_r$ for all

$k \in C_r$, then for $r = 1, 2, ..., R$

$$\frac{\partial Q(\theta, \theta^{(m-1)})}{\partial \sigma_r} = \sum_{k \in C_r} \left[ \sum_{n=1}^{N} z_{kn}^{(m-1)} \left( -\frac{1}{\sigma_r^{(m-1)}} \right) \right.$$
$$\left. + \frac{\nu_k^{(m-1)}}{(\sigma_r^{(m-1)})^{\nu_k^{(m-1)}+1}} \sum_{n=1}^{N} z_{kn}^{(m-1)} \left| x_n - \mu_k^{(m)} \right|^{\nu_k^{(m-1)}} \right] = 0.$$

The iterative Newton-Raphson method is applied as:

$$\sigma_r^{(m)} = \sigma_r^{(m-1)} - \alpha^{(i)} \frac{g(\sigma_r^{(m-1)})}{g'(\sigma_r^{(m-1)})}, \qquad (4.10)$$

where

$$g(\sigma_r^{(m-1)}) = \sum_{k \in C_r} \left[ \sum_{n=1}^{N} z_{kn}^{(m-1)} \left( -\frac{1}{\sigma_r^{(m-1)}} \right) + \frac{\nu_k^{(m-1)}}{(\sigma_r^{(m-1)})^{\nu_k^{(m-1)}+1}} \right.$$
$$\left. \times \sum_{n=1}^{N} z_{kn}^{(m-1)} \left| x_n - \mu_k^{(m)} \right|^{\nu_k^{(m-1)}} \right], \qquad (4.11)$$

$$g'(\sigma_r^{(m-1)}) = \sum_{k \in C_r} \left[ \sum_{n=1}^{N} z_{kn}^{(m-1)} \frac{1}{(\sigma_r^{(m-1)})^2} + \nu_k^{(m-1)} \left( -\nu_k^{(m-1)} - 1 \right) \right.$$
$$\left. \times (\sigma_r^{(m-1)})^{-\nu_k^{(m-1)}-2} \sum_{n=1}^{N} z_{kn}^{(m-1)} \left| x_n - \mu_k^{(m)} \right|^{\nu_k^{(m-1)}} \right]. \qquad (4.12)$$

**Shape parameter** In order to obtain the iteration equation for the constrained shape parameter at any partition $r \in 1, 2, ... R$, let $\nu_k = \nu_r$ for all $k \in C_r$, then for $r = 1, 2, ..., R$

$$\frac{\partial Q(\theta, \theta^{(m-1)})}{\partial \nu_r} = \sum_{k \in C_r} \left[ \sum_{n=1}^{N} z_{kn}^{(m-1)} \frac{1}{\nu_r^{(m-1)}} \left( \frac{1}{\nu_r^{(m-1)}} \Psi \left( \frac{1}{\nu_r^{(m-1)}} \right) + 1 \right) \right.$$
$$\left. - \sum_{n=1}^{N} z_{kn}^{(m-1)} \left| \frac{x_n - \mu_k^{(m)}}{\sigma_k^{(m)}} \right|^{\nu_r^{(m-1)}} \log \left| \frac{x_n - \mu_k^{(m)}}{\sigma_k^{(m)}} \right| \right] = 0.$$

Since the above equation is a non-linear equation, the iterative Newton-Raphson method is applied as:

$$\nu_r^{(m)} = \nu_r^{(m-1)} - \alpha^{(i)} \frac{1}{\nu_r^{(m-1)}} \frac{g(\nu_r^{(m-1)})}{g'(\nu_r^{(m-1)})}, \qquad (4.13)$$

where

$$g(\nu_r^{(m-1)}) = \sum_{k \in C_r} \left[ \sum_{n=1}^{N} z_{kn}^{(m-1)} \frac{1}{\nu_r^{(m-1)}} \left( \frac{1}{\nu_r^{(m-1)}} \Psi \left( \frac{1}{\nu_r^{(m-1)}} \right) + 1 \right) \right.$$
$$\left. - \sum_{n=1}^{N} z_{kn}^{(m-1)} \left| \frac{x_n - \mu_k^{(m)}}{\sigma_k^{(m)}} \right|^{\nu_r^{(m-1)}} \log \left| \frac{x_n - \mu_k^{(m)}}{\sigma_k^{(m)}} \right| \right], \qquad (4.14)$$

$$g'\left(v_r^{(m-1)}\right) = \sum_{k \in C_r}\left[\sum_{n=1}^{N} z_{kn}^{(m-1)} - \frac{1}{\left(v_r^{(m-1)}\right)^2}\left(1 + \frac{2}{v_r^{(m-1)}}\Psi\left(\frac{1}{v_r^{(m-1)}}\right)\right.\right.$$

$$\left.\left. + \frac{1}{\left(v_r^{(m-1)}\right)^2}\Psi'\left(\frac{1}{v_r^{(m-1)}}\right)\right)\right) - \sum_{n=1}^{N} z_{kn}^{(m-1)}\left|\frac{x_n - \mu_k^{(m)}}{\sigma_k^{(m)}}\right|^{v_r^{(m-1)}} \qquad (4.15)$$

$$\times \left(\log\left|\frac{x_n - \mu_k^{(m)}}{\sigma_k^{(m)}}\right|\right)^2\right].$$

The ratio $1/v_r^{(m-1)}$ in equation 4.13 represents the adaptive step size of the shape parameter like in equation 2.55 for the MGND model.

Algorithm 7 in the Appendix A performs the parameter estimation of the CMGND model via equations 4.6, 4.10, 4.13 for constrained parameters, as well as equations 2.47, 2.54, 2.55 for unconstrained parameters. The latter are estimated as in algorithm 5.

## 4.4 Model selection

Using the nested structure of the proposed constrained mixtures models it is possible to build the likelihood ratio test (LRTS) for testing $H_0 : \theta \in \Theta_0$ vs $H_1 : \theta \in \Theta$, where $\Theta_0$ and $\Theta$ are the constrained and unconstrained parameter spaces. Specifically, $H_0$ states that the *constrained model is true*, while $H_1$ states that the *unconstrained model is true*. The test statistic is defined as follows

$$\Lambda = 2(\log L(\widehat{\theta}) - \log L(\widehat{\theta}_0)) \quad \text{with } \Lambda \sim \chi^2(d), \qquad (4.16)$$

where: $\log L(\widehat{\theta})$ and $\log L(\widehat{\theta}_0)$ are the unconstrained and constrained estimated log-likelihood, respectively; $d$ are the degrees of freedom equal to the difference in the number of parameters between the two mixture models.

The LRTS is performed via algorithm 7 because the ECM algorithm[4] does not guarantee $\Lambda > 0$. On the contrary, the GEM algorithm ensures $\log L(\widehat{\theta}) > \log L(\widehat{\theta}_0)$ because increases, rather than maximizes, the log-likelihood function in every **M-step** (G. McLachlan and Krishnan 2008). Figure 4.1 shows the $\Lambda$ values obtained with the ECM and GEM algorithms for $S = 250$ replications, sample size $N = 250, 500, 1000, 1500, 2000$ and the following hypothesis system:

$$H_0 = \pi_1 = 0.7, \mu_1 = 1, \sigma_1 = 1, v_1 = 2, \pi_2 = 0.3, \mu_2 = 1, \sigma_2 = 3, v_1 = 4;$$
$$H_1 = \pi_1 = 0.7, \mu_1 = 1, \sigma_1 = 1, v_1 = 2, \pi_2 = 0.3, \mu_2 = 5, \sigma_2 = 3, v_1 = 4.$$

It appears that the GEM algorithm is preferred as it provides non-negative $\Lambda$ values consistent to the expected $\chi^2(1)$ distribution, whereas the ECM algorithm yields negative $\Lambda$ values for $N = 250, 500, 1000$.

To conclude, model selection can be also approached in terms of goodness-of-fit measures (G. J. McLachlan and Peel 2000) like the AIC, BIC, KS and AD defined in Chapter 2 Section 2.7.2.

4: During the simulation study the unconstrained model has been initialized with the solutions provided by the constrained model. Nevertheless, the ECM algorithm did not guarantee $\log L(\widehat{\theta}) > \log L(\widehat{\theta}_0)$.

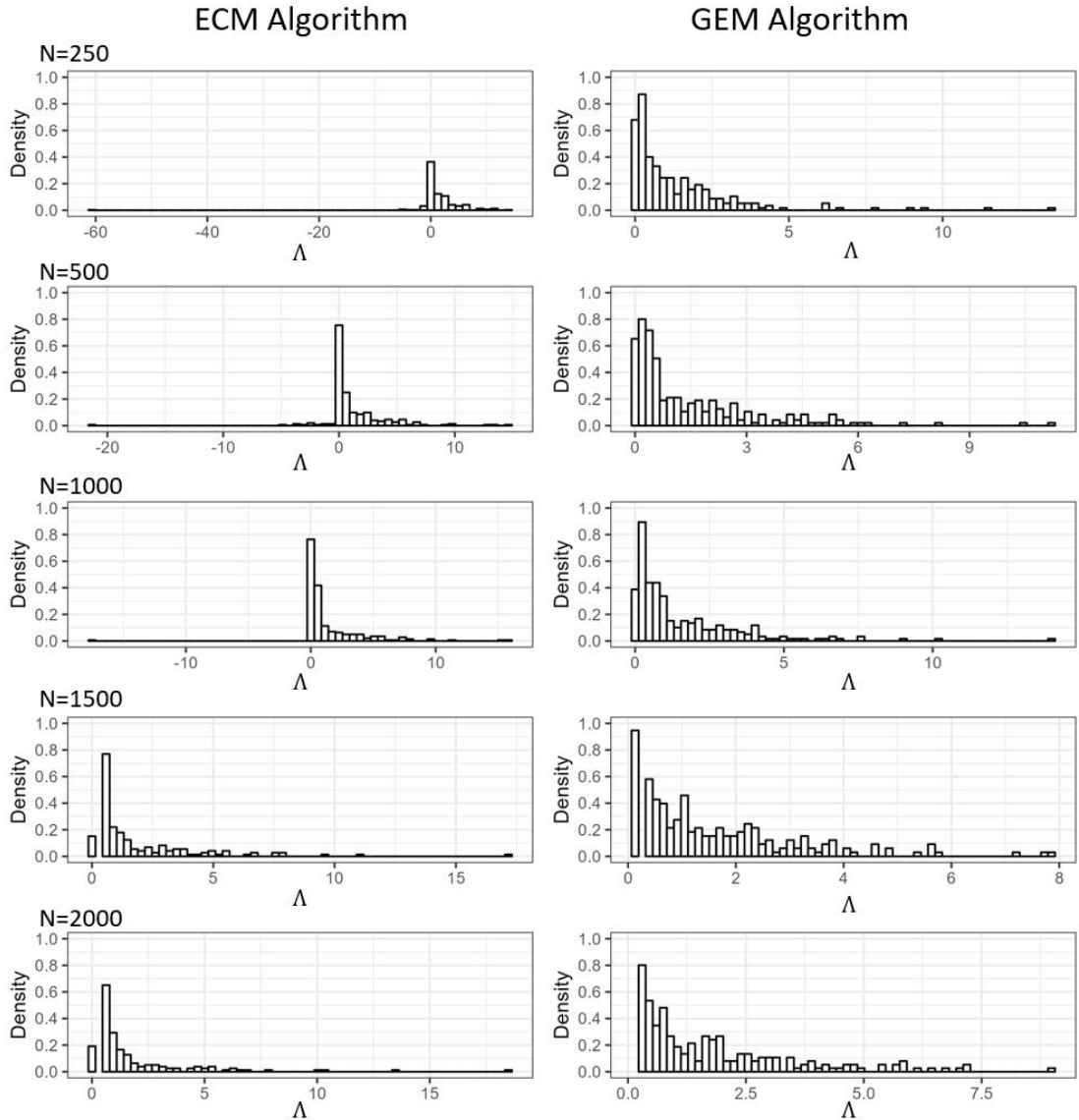

**Figure 4.1:** LRTS case study, $\Lambda$ values estimated via ECM and GEM algorithms.

## 4.5  Numerical experiments

### 4.5.1  Power analysis

The aim of this section is to evaluate the statistical power and the empirical level of the null hypothesis $H_0$: *Constraints hold*. For this purpose, the LRTS detailed in Section 4.4 is applied to perform the power analysis (Cohen 1992; Agresti and Franklin 2009; Cohen 2013). The key parts of the power analysis are:

1. Nominal Significance level ($\alpha$) - the predetermined level at which a researcher sets the probability of rejecting the null hypothesis when it is true, commonly set at 0.05;

2. Empirical Level - the significance level observed in the study based on the collected data;

3. Sample size ($N$);

4. Effect size - the magnitude of an outcome present in the population;

5. Statistical power $(1 - \beta)$ - the probability of rejecting the null hypothesis when it is false.

Table 4.2 resumes the four possible results of the LRTS given $H_0$: *Constraints hold*.

|  | Model selection decision | |
| --- | --- | --- |
| $H_0$: *Constraints hold* | Do not reject $H_0$ | Reject $H_0$ |
| True | Correct decision | Type I error ($\alpha$) |
| False | Type II error ($\beta$) | Correct decision ($1 - \beta$) |

The parameters of the simulated models under $H_0$ and $H_1$ are displayed in Table 4.3. The power analysis is performed for $S = 250$ replications, sample size $N = 250, 500, 1000, 1500, 2000$ and a nominal significance level $\alpha = 0.05$. The LRTS statistic $\Lambda$ is asymptotically $\chi^2(d)$-distributed with $d = 1$ for CUU, UCU, and UUC, $d = 2$ for CCU, CUC, and UCC and $d = 3$ for CCC.

|  | CMGND ($H_0$) | | | | | | | MGND ($H_1$) |
| --- | --- | --- | --- | --- | --- | --- | --- | --- |
| $\theta$ | CUU | UCU | UUC | CCU | CUC | UCC | CCC | UUU |
| $\pi_1$ | 0.7 | 0.7 | 0.7 | 0.7 | 0.7 | 0.7 | 0.7 | 0.7 |
| $\mu_1$ | 1 | 1 | 1 | 1 | 1 | 1 | 1 | 1 |
| $\sigma_1$ | 1 | 3 | 1 | 3 | 3 | 3 | 4 | 1 |
| $\nu_1$ | 2 | 2 | 4 | 2 | 4 | 4 | 4 | 2 |
| $\pi_2$ | 0.3 | 0.3 | 0.3 | 0.3 | 0.3 | 0.3 | 0.3 | 0.3 |
| $\mu_2$ | 1 | 5 | 5 | 1 | 1 | 5 | 1 | 5 |
| $\sigma_2$ | 3 | 3 | 3 | 3 | 3 | 3 | 3 | 3 |
| $\nu_2$ | 4 | 4 | 4 | 4 | 4 | 4 | 4 | 4 |

Figure 4.2 shows empirical levels and powers for a nominal 5% level. Results show that the power reaches 100% for all sample sizes and models except for the UUC model which displays a low power for small sample sizes but reaching 100% at $N = 2000$. In this case, it is possible to notice the smallest effect size among all those considered. The UUC model vs UUU model differ only in terms of the shape parameter in one mixture component equal to 4 and 2, respectively, being all the other parameters the same. Figure 4.4 confirms that the UUC model (purple line) and UUU model (pink line) have a very similar density compared to the others.

The empirical level is around or lower the nominal value of 5% for all the models but the CUU reporting an empirical level always greater than 5% and sometime more close to 10%.

The power analysis is performed also for $\pi_1 = 0.5$, results are shown in Figure 4.3. The statistical power and empirical level are almost the same when $\pi_1$ reduces from 0.7 to 0.5. A notable exception is the CUU model as the empirical level improves for $N = 500, 1000, 1500$.

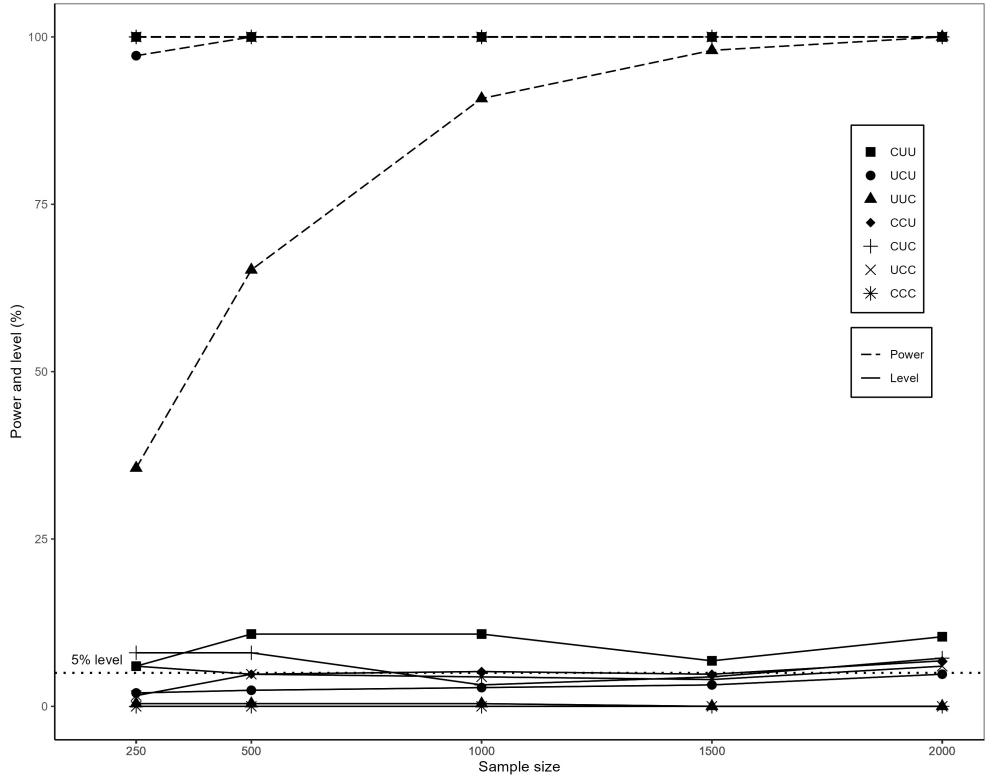

**Figure 4.2:** LRTS empirical levels (solid line) and powers (dashed line) for nominal levels of 5% (dotted line) for the 7-model family of CMGND with $\pi_1 = 0.7$.

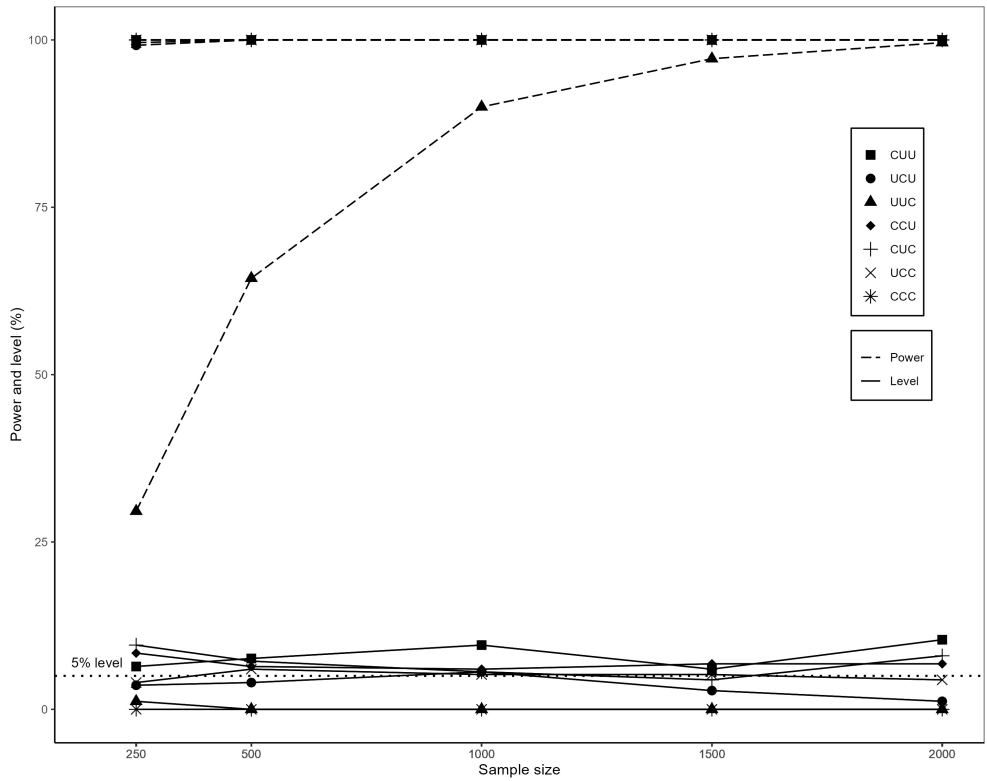

**Figure 4.3:** LRTS empirical levels (solid line) and powers (dashed line) for nominal levels of 5% (dotted line) for the 7-model family of CMGND with $\pi_1 = 0.5$.

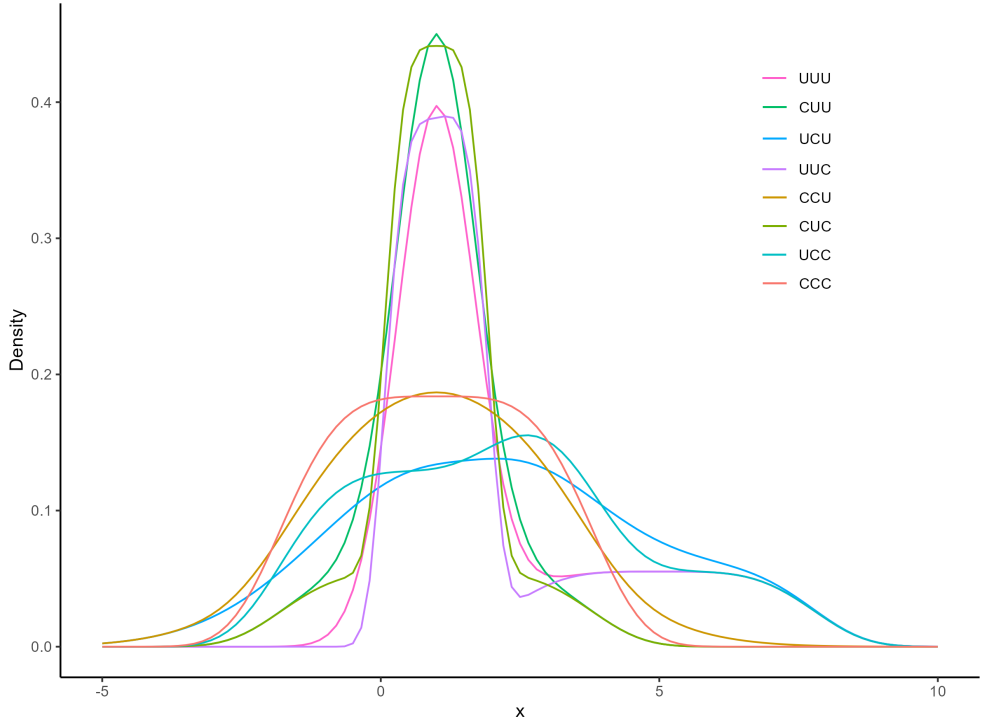

**Figure 4.4:** Power analysis, density of the simulated models.

### 4.5.2 Simulation Study

The aim of this simulation study is to study the performance of the MGND and CMGND models estimated via algorithms 6-7, in order to identify which of them fit the simulated scenarios best[5] .

We simulate from the 7-model family with parameters shown in Table 4.4. The simulation study is repeated also for $\pi_1 = \pi_2 = 0.5$.

For each model $S = 250$ samples with sample sizes $N = 250, 500, 1000, 1500, 2000$ are generated using the R function *rgnorm* of the package *gnorm*. For each sample, 'constrained' and 'unconstrained' estimates are computed.

As a measure of quality estimation, for each parameter of the component mixture, the Percent Root Mean Squared Error is computed as follows:

$$PRMSE\left(\hat{\theta}\right) = 100 \frac{\sqrt{\frac{1}{S} \sum_{s=1}^{S}(\hat{\theta}_s - \theta)^2}}{\theta} \tag{4.17}$$

where $\theta$ is the true parameter value and $\hat{\theta}_s$ is the estimate of $\theta$ for the $s$-th simulated data.

Tables B.17-B.20 in the Appendix B report the PRMSE obtained over the 250 simulations runs. A summary of the results for $\pi_1 = 0.7$ is displayed in figure 4.5 where the difference between the PRMSE of the CMGND and MGND estimates (the PRMSE gain) is reported across all the models, parameters and sample sizes. A similar plot for the case $\pi_1 = \pi_2 = 0.5$ is reported in Figure 4.6. Findings are quite similar except for the CUU model which reports a lower PRMSE gain[6]  in the estimation of $\sigma_1$, $\nu_1$ and $\sigma_2$.

5:  The GEM algorithm (algorithm 7) aims to improve, rather than maximizes (algorithm 6), the likelihood function in every M-step.

6:  It may be due to the high overlap ratio of the two distributions when $\pi_1 = 0.7$ decreases to $\pi_1 = 0.5$.

| θ | CMGND | | | | | | |
|---|---|---|---|---|---|---|---|
| | CUU | UCU | UUC | CCU | CUC | UCC | CCC |
| $\pi_1$ | 0.7 | 0.7 | 0.7 | 0.7 | 0.7 | 0.7 | 0.7 |
| $\mu_1$ | 1 | 1 | 1 | 1 | 1 | 1 | 1 |
| $\sigma_1$ | 1 | 3 | 1 | 3 | 1 | 3 | 3 |
| $\nu_1$ | 2 | 2 | 4 | 2 | 4 | 4 | 4 |
| $\pi_2$ | 0.3 | 0.3 | 0.3 | 0.3 | 0.3 | 0.3 | 0.3 |
| $\mu_2$ | 1 | 5 | 5 | 1 | 1 | 5 | 1 |
| $\sigma_2$ | 3 | 3 | 3 | 3 | 3 | 3 | 3 |
| $\nu_2$ | 4 | 4 | 4 | 4 | 4 | 4 | 4 |

**Table 4.4:** 7-model family simulated parameters.

For all models and sample sizes the CMGND estimates show a higher degree of accuracy with respect to the MGND ones. There is only one exception: for $\sigma$ in one component when $S = 250$ in the CUU model. The gain generally decreases when the sample size increases. For example, for $N = 250$ (with $\pi_1 = 0.7$) the PRMSE gain of $\mu_1$ and $\nu_1$ of the UCU model decreases from 30.20 and 27.71 to 12.84 and 3.77 by moving from $N = 250$ to $N = 2000$.

As expected, higher gains are observed when estimating the constrained parameters, i.e. the parameters with the same value across the components. However, for the unconstrained MGND model, the presence of constraints also deteriorates the estimates of the 'free' remaining parameters, i.e. parameters not constrained to be equal across the components. For instance, for $N = 250$ (with $\pi_1 = 0.7$) the PRMSE of $\mu_1$ and $\nu_1$ of the UCU model are equal to 74.81 and 89.36 for the unconstrained model, and 44.61 and 61.65 for the constrained model.

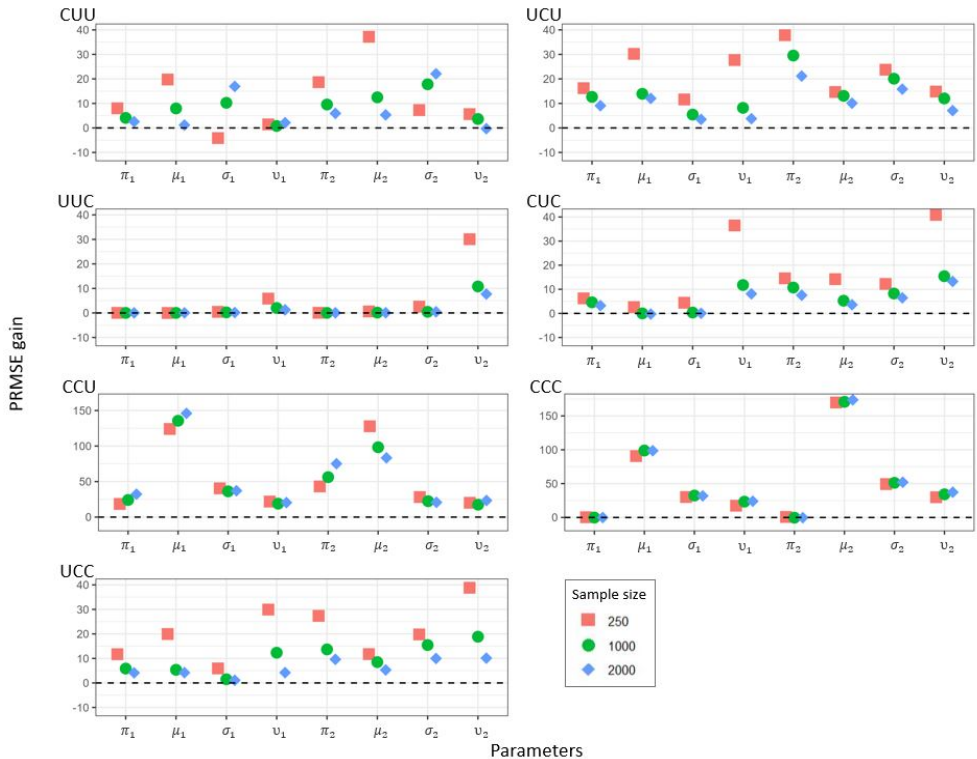

**Figure 4.5:** CMGND vs MGND: PRMSE gain with $\pi_1 = 0.7$.

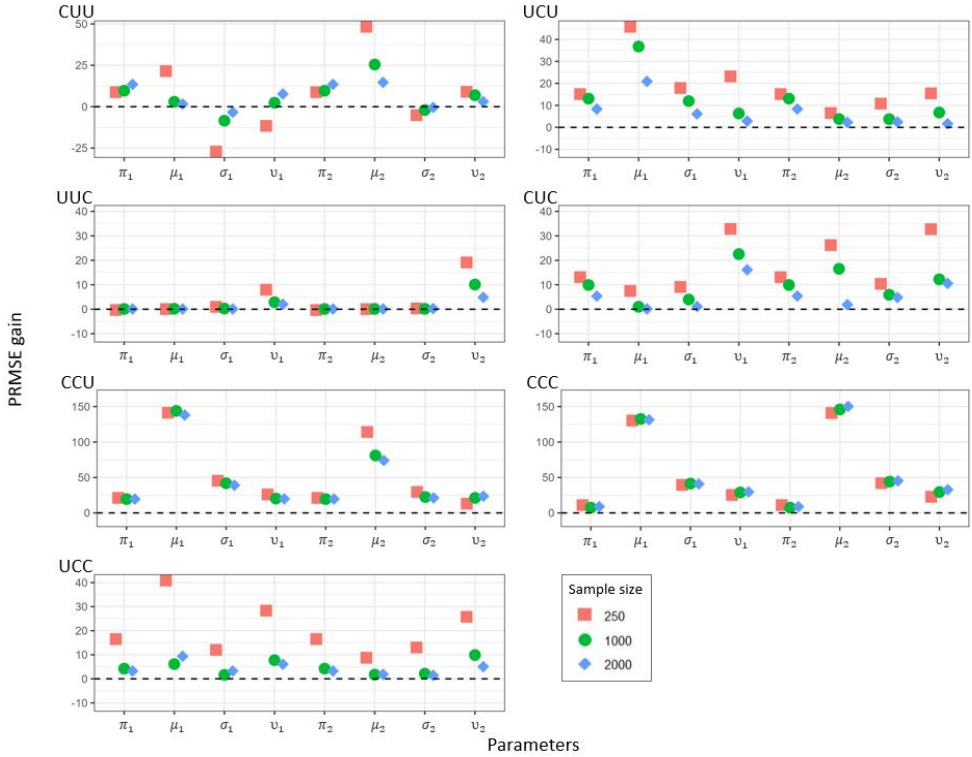

**Figure 4.6:** CMGND vs MGND: PRMSE gain with $\pi_1 = 0.5$.

### 4.5.3 An empirical application

An empirical application on returns of S&P500[7] and STOXX50E indices is presented in this section. Data on daily closing prices of the two indices are collected from Yahoo Finance. The time period taken for the study is from April 2, 2007 to August 15, 2023. Daily returns are calculated with the natural log difference approach in equation 2.4.

Figure 4.7 shows the daily returns of the S&P500 and STOXX50E highlighting two periods of high volatility: the global financial crisis (2007-2008) and Covid-19 crisis (2020-2021). Besides, Table 4.5 demonstrates that daily returns of the S&P500 and STOXX50E are not normally distributed because they are characterised by very high kurtosis and negative skewness.

7: The Standard & Poor's 500 (S&P 500) is a stock market index that measures the performance of 500 large companies listed on stock exchanges in the United States. It is one of the most widely followed equity indices and is considered a representative benchmark for the overall performance of the U.S. stock market.

|  | Mean | Median | Std | Skew | Kur | JB Test |
|---|---|---|---|---|---|---|
| S&P500 | 0.028 | 0.067 | 1.295 | -0.510 | 14.8150 | 24183* |
| STOXX50E | 0.001 | 0.042 | 1.436 | -0.288 | 10.3261 | 9249* |

*Notes.* * indicates a *p*-value $\leq 0.05$.

**Table 4.5:** Descriptive statistic summary of S&P500 and STOXX50E.

After running the MGND and the 7-model family of CMGND, the LRTS and BIC are used to select the best model and compare the goodness-of-fit performance (G. J. McLachlan and S. Rathnayake 2014; Wen et al. 2022). According to the LRTS and BIC, the CMGND-UUC results to be the preferred model for both dataset (Tables 4.8-4.9). The common shape parameter is equal to $\nu = 1.0368$ for the S&P500 while the STOXX50E reports an estimate of $\nu = 1.1935$. Both values confirm the feature of daily

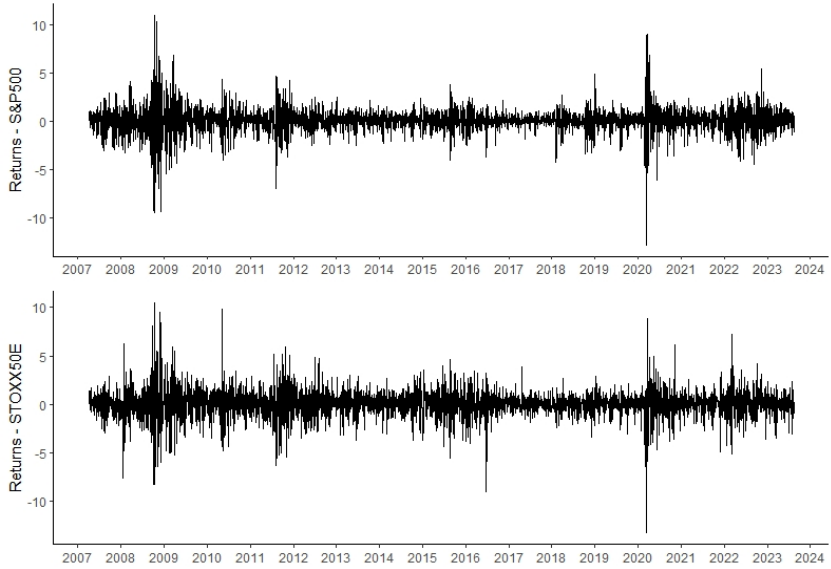

returns distribution to have thinner peaks and heavier tails compared to the normal distribution.

Figure 4.8 shows the estimated density and posterior probability of the LRTS and BIC preferred models. It can be seen that one component describes the central and intermediate values of the data, while another component describes the more extreme tail behaviours. Indeed, as shown in Table 4.6, the first component of both models is featured by low standard deviation ($\sigma_1 = 0.9520$ for S&P500 and $\sigma_1 = 1.1265$ for STOXX50E) while the second component captures the extreme movements through a high standard deviation ($\sigma_2 = 2.8320$ for S&P500 and $\sigma_2 = 2.7810$ for STOXX50E). Both mixture components of the CMGND-UUC have the same kurtosis. The estimated overall standard deviation and kurtosis are almost equal to the empirical ones.

|  | 1-st comp. | | 2-nd comp. | | Overall | |
|---|---|---|---|---|---|---|
|  | Std$_1$ | Kur$_1$ | Std$_2$ | Kur$_2$ | Std | Kur |
| S&P500 | 0.9520 | 5.7057 | 2.8320 | 5.7057 | 1.2977 | 15.4880 |
| STOXX50E | 1.1265 | 4.7734 | 2.7810 | 4.7734 | 1.4321 | 9.8002 |

**Table 4.6:** Estimated standard deviation and excess kurtosis of the two-component CMGND-UUC.

Finally, Table 4.7 compares the BIC of the CMGND-UUC with the BIC of several competing mixture models obtained with the R's packages *Mclust* (Scrucca et al. 2016) and *teigen* (Andrews, McNicholas, and Subedi 2011; Andrews, Wickins, et al. 2018): MN model, mixture of normals with common variance (MN-UC), MSTD model, and constrained mixtures of student-t (CMSTD)[8] . The CMGND-UUC remains the best model over the competitors for both indices. Interestingly, the UUC model, i.e. a model where the shape parameter is constrained to be equal across the components, results to be the preferred one also among the constrained mixtures of student-t.

8: Defined in Appendix C.

**Table 4.7:** Goodness-of-fit performance of the CMGND-UUC, CMN and CMSTD. In bold the best model according to the BIC.

| Index | CMGND-UUC | MN | MN-UC | CMSTD-UUU | CMSTD-UCU | CMSTD-UUC | CMSTD-UCC |
|---|---|---|---|---|---|---|---|
| S&P500 | **12324.02** | 12481.74 | 13863.91 | 12338.09 | 12345.54 | 12330.57 | 12337.24 |
| STOXX50E | **13730.35** | 13809.50 | 14655.45 | 13740.20 | 13756.78 | 13730.73 | 13748.46 |

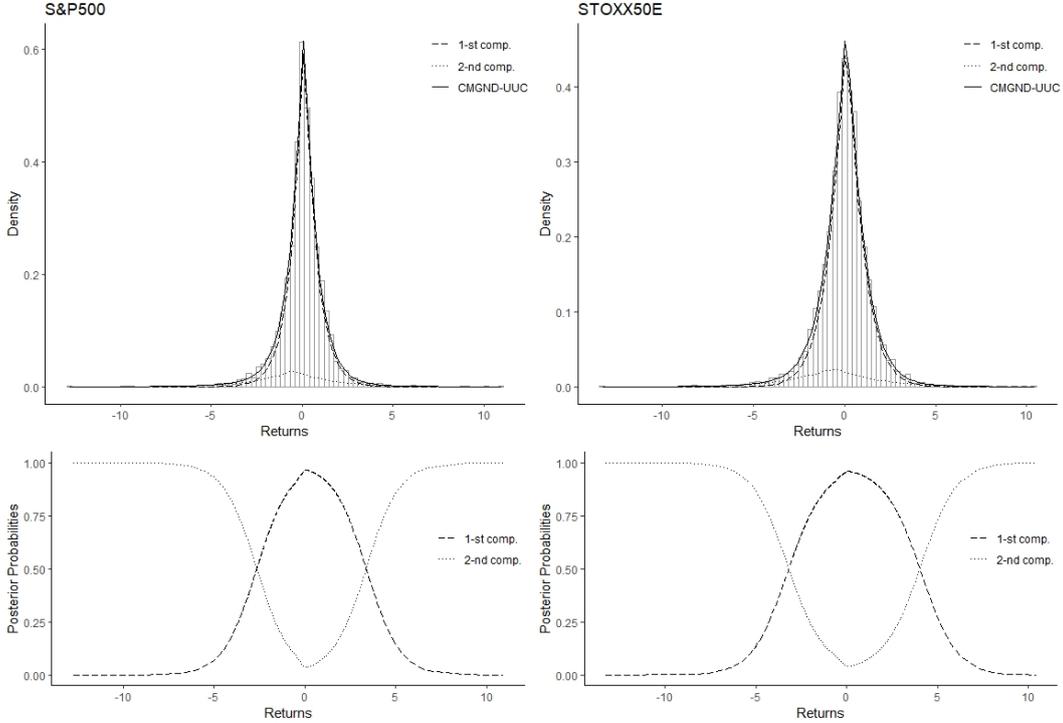

**Figure 4.8:** Density (up) and posterior probability plots (down) of the estimated CMGND-UUC (S&P500) and CMGND-UUC (STOXX50E).

**Table 4.8:** Estimated MGND and 7-model family of CMGND for daily returns on S&P500.

| $\theta$ | MGND | CMGND | | | | | | |
|---|---|---|---|---|---|---|---|---|
| | | CUU | UCU | UUC | CCU | CUC | UCC | CCC |
| $\pi_1$ | 0.7318 | 0.8191 | 0.5468 | 0.8975 | 0.6010 | 0.1295 | 0.8701 | 0.5081 |
| $\mu_1$ | -0.0063 | 0.0672 | -0.0384 | 0.0904 | 0.0672 | 0.0661 | 0.0159 | 0.0670 |
| $\sigma_1$ | 0.6256 | 0.7183 | 0.7812 | 0.7139 | 0.7348 | 2.0648 | 0.5625 | 0.5988 |
| $\nu_1$ | 0.7895 | 1.0767 | 0.8264 | 1.0368 | 1.1404 | 1.0635 | 0.8007 | 0.8230 |
| $\pi_2$ | 0.2682 | 0.1809 | 0.4532 | 0.1025 | 0.3990 | 0.8705 | 0.1299 | 0.4919 |
| $\mu_2$ | 0.2862 | 0.0672 | 0.1829 | -0.6373 | 0.0672 | 0.0661 | 0.3052 | 0.0670 |
| $\sigma_2$ | 0.7946 | 1.5124 | 0.7812 | 2.1239 | 0.7348 | 0.7282 | 0.5625 | 0.5988 |
| $\nu_2$ | 1.7319 | 0.9485 | 1.4629 | 1.0368 | 0.7783 | 1.0635 | 0.8007 | 0.8230 |
| $\log L(\hat{\theta})$ | -6136.32 | -6141.97 | -6138.96 | -6137.04 | -6142.68 | -6141.91 | -6154.71 | -6160.42 |
| $\mathrm{LRTS_{UUU}}$ | | 11.31* | 5.28* | 1.44 | 12.71* | 11.19* | 36.77* | 48.20* |
| $\mathrm{LRTS_{CUU}}$ | | | | | 1.40 | 1 | | 36.89* |
| $\mathrm{LRTS_{UCU}}$ | | | | | 7.43* | | 31.48* | 42.92* |
| $\mathrm{LRTS_{UUC}}$ | | | | | | 9.74* | 35.33* | 46.77* |
| BIC | 12330.91 | 12333.89 | 12327.86 | **12324.02** | 12326.97 | 12325.45 | 12351.03 | 12354.14 |

*Notes.* * indicates a *p*-value≤ 0.05. In bold the best model according to the BIC.

**Table 4.9:** Estimated MGND and 7-model family of CMGND for daily returns on STOXX50E.

| $\theta$ | MGND | CMGND | | | | | | |
| | | CUU | UCU | UUC | CCU | CUC | UCC | CCC |
|---|---|---|---|---|---|---|---|---|
| $\pi_1$ | 0.8787 | 0.2125 | 0.6926 | 0.8864 | 0.6598 | 0.8518 | 0.4113 | 0.5883 |
| $\mu_1$ | 0.0752 | 0.0374 | 0.1208 | 0.0756 | 0.0428 | 0.0334 | 0.3848 | 0.0400 |
| $\sigma_1$ | 1.0659 | 1.9058 | 1.1214 | 1.0310 | 1.1455 | 1.0816 | 0.7805 | 0.9154 |
| $\nu_1$ | 1.2456 | 1.0887 | 1.3789 | 1.1935 | 1.4229 | 1.2915 | 0.8829 | 0.9556 |
| $\pi_2$ | 0.1213 | 0.7875 | 0.3074 | 0.1136 | 0.3402 | 0.1482 | 0.5887 | 0.4117 |
| $\mu_2$ | -0.5556 | 0.0374 | -0.1858 | -0.6124 | 0.0428 | 0.0334 | -0.1133 | 0.0400 |
| $\sigma_2$ | 2.6515 | 1.0658 | 1.1214 | 2.5453 | 1.1455 | 2.6920 | 0.7805 | 0.9154 |
| $\nu_2$ | 1.2316 | 1.3025 | 0.8720 | 1.1935 | 0.8924 | 1.2915 | 0.8829 | 0.9556 |
| $\log L(\hat{\theta})$ | -6840.14 | -6845.36 | -6841.18 | -6840.22 | -6845.84 | -6845.83 | -6855.66 | -6863.81 |
| LRTS$_{UUU}$ | | 10.45* | 2.09 | 0.16 | 11.40* | 11.39 | 31.02* | 47.34* |
| LRTS$_{CUU}$ | | | | | 0.95 | 0.94 | | 36.90* |
| LRTS$_{UCU}$ | | | | | 9.31* | | 28.93* | 45.25* |
| LRTS$_{UUC}$ | | | | | | 11.23* | 30.87* | 47.18* |
| BIC | 13738.52 | 13740.64 | 13732.29 | **13730.35** | 13733.28 | 13733.26 | 13752.90 | 13760.90 |

*Notes.* * indicates a *p*-value$\leq 0.05$. In bold the best model according to the BIC.

### 4.5.4 Comparative analysis part II

In this section a comparative analysis is performed for the 52 stocks that belong to the STOXX50E using the 7-model family of univariate two-component CMGND and the best models of the comparative analysis part I (Section 4.5.4), i.e. Student-t, GH, MGND and MSTD models. Additionally, the CMSTD model (Appendix C) is also considered. The goodness-of-fit measures are those used in the comparative analysis part I (Section 2.7.2), namely AIC, BIC, KS and AD.

**Data**

Data on daily closing prices of the 52 stocks that belong to the STOXX50E are collected from Yahoo Finance for the period from 01 January 2010 to 30 September 2023 (shorter for stocks entering the STOXX50E later than January 2010). Daily log returns are computed for each stock according to equation 2.4. All descriptive statistics and the JB test are computed for daily log returns for the full period. Table B.9 in Appendix B shows for each stock (alphabetically ordered) the number of observations ($N$), mean, median, standard deviation, skewness, kurtosis, JB statistics, and minimal and maximal values. All stocks have a mean and median close to 0. The standard deviation range from 1.2624 to 2.7371. Returns are positively and negatively skewed with an empirical kurtosis greater than three, indicating fat-tails (leptokurtic distribution). For all indices the JB test rejects the null hypothesis which states that data are normally distributed.

**Models and estimation methods**

Models and estimation methods are resumed in Table 4.10. Mixture models with two components are considered because the results obtained for three component are insignificant and irrelevant. Moreover, the best constrained model within the 7-model family is chosen on the basis of the considered goodness-of-fit measure[9] . Tables B.21-B.24 in Appendix B compare the two-component MGND and CMGND models for each stock and for each goodness-of-fit measure. Table 4.11 reports the percentage of times (%) a constrained model is selected by a goodness-of-fit measure, i.e. it shows a comparison within the 7-model family. AIC and BIC prefer models with two constrained parameters (CCU,CUC,UCC), while KS and AD prefer models with one constrained parameter (CUU,UCU,UUC). AIC and BIC are parsimonious compared to KS and AD (Figure 2.25). Table 4.12 show the percentage of times (%) the MGND and CMGND models are selected by a goodness-of-fit measure, i.e. it shows a comparison between MGND and CMGND models. The MGND model is preferred only by KS and AD since they do not consider the model complexity. The AIC prefers the MGND model only 3.85% of the time, while the BIC never prefers it.

| Model | Equation | Estimation method | R function |
|---|---|---|---|
| Student-t | 2.12 | Nelder-Mead approach | *optim* |
| GH | 2.15 | Newton-Raphson method | *ghFit* |
| MSTD | 2.31 | EM algorithm | *teigen* |
| CMGND | 2.31 | EM algorithm | *teigen* |
| MGND | 2.33 | Algorithm 5 | |
| CMGND | 4.1 | Algorithm 7 | |

**Table 4.10:** Comparative analysis part II: models and methods.

| | CMGND | | | | | | |
|---|---|---|---|---|---|---|---|
| | CUU | UCU | UUC | CCU | CUC | UCC | CCC |
| AIC | 7.69 | 26.92 | 5.77 | 13.46 | 38.46 | 7.69 | |
| BIC | 3.85 | | | 11.54 | 69.23 | 9.62 | 5.77 |
| KS | 15.38 | 28.85 | 30.77 | 9.62 | 7.69 | 7.69 | |
| AD | 7.69 | 38.46 | 25.00 | 21.15 | 5.77 | 1.92 | |

**Table 4.11:** Percentage of times (%) a constrained model is selected by a goodness-of-fit measure.

| | MGND | CMGND | | | | | | |
|---|---|---|---|---|---|---|---|---|
| | | CUU | UCU | UUC | CCU | CUC | UCC | CCC |
| AIC | 3.85 | 5.77 | 26.92 | 5.77 | 13.46 | 36.54 | 7.69 | |
| BIC | | 3.85 | | | 11.54 | 69.23 | 9.62 | 5.77 |
| KS | 21.15 | 13.46 | 21.15 | 21.15 | 7.69 | 7.69 | 7.69 | |
| AD | 26.92 | 5.77 | 26.92 | 17.31 | 17.31 | 3.85 | 1.92 | |

**Table 4.12:** Percentage of times (%) the MGND and CMGND models are selected by a goodness-of-fit measure.

### Results

Tables B.25-B.28 in the Appendix B present the AIC, BIC, KS and AD values for daily log-returns over the past 13 years. All models are compared for each stock, and the minimum statistic is highlighted in bold to signify the best fit.

**AIC**   AIC mainly selects the CMGND model (22 times). Panel (a) of Figure 4.9 shows the AIC boxplot for all models ordered by median. The CMSTD, MSTD, GH, and CMGND models have the lowest median AIC values. However, if a model is chosen several times as first, the other models are still competitive because the median values are very close to each other. The CMGND model has the lowest average AIC value.

**BIC**   The student-t model is the most often selected model (44 times), while the CMGND model is selected 6 times. Panel (b) of Figure 4.9 shows the BIC boxplot for all models ordered by median. The Student-t, GH, CMSTD and CMGND report the lowest median BIC values. The BIC prefers simpler models, i.e. models with few parameters. The CMGND and GH models have almost the same average BIC value.

**KS**   The MGND and CMGND models are consistently selected by KS (30 times). Panel (c) of Figure 4.9 shows the KS boxplot for all models ordered by median. The CMGND, CMSTD, GH report the lowest median values.

**AD**   The MGND and CMGND models are consistently selected by AD (28 times). Panel (d) of Figure 4.9 shows the AD boxplot for all models ordered by median. The GH, CMGND and MGND report the lowest median values.

The CMGND model is more often selected according to the AIC, KS and AD, while the GH, MGND and CMSTD models are often selected as second choice. The student-t model is the most selected according to the BIC.

In summary, a significant modelling flexibility is achieved by allowing certain parametric equalities among the different GND components. The proposed 7-model family of univariate two-component CMGND is a good candidate for modelling financial data. The empirical application and the comparative analysis part II suggest that returns can be modelled by the CMGND model as the MGND model is too complex. By adding constraints the CMGND model obtains better results in terms of goodness-of-fit performance. In addition, it is worth noting that mixture models offer a practical interpretability of financial returns by distinguishing periods of high and low volatility as shown in Chapter 3. To conclude, the CMGND and CMSTD models combine flexibility, parsimony, and interpretability showing a clear advantage over non-mixture models.

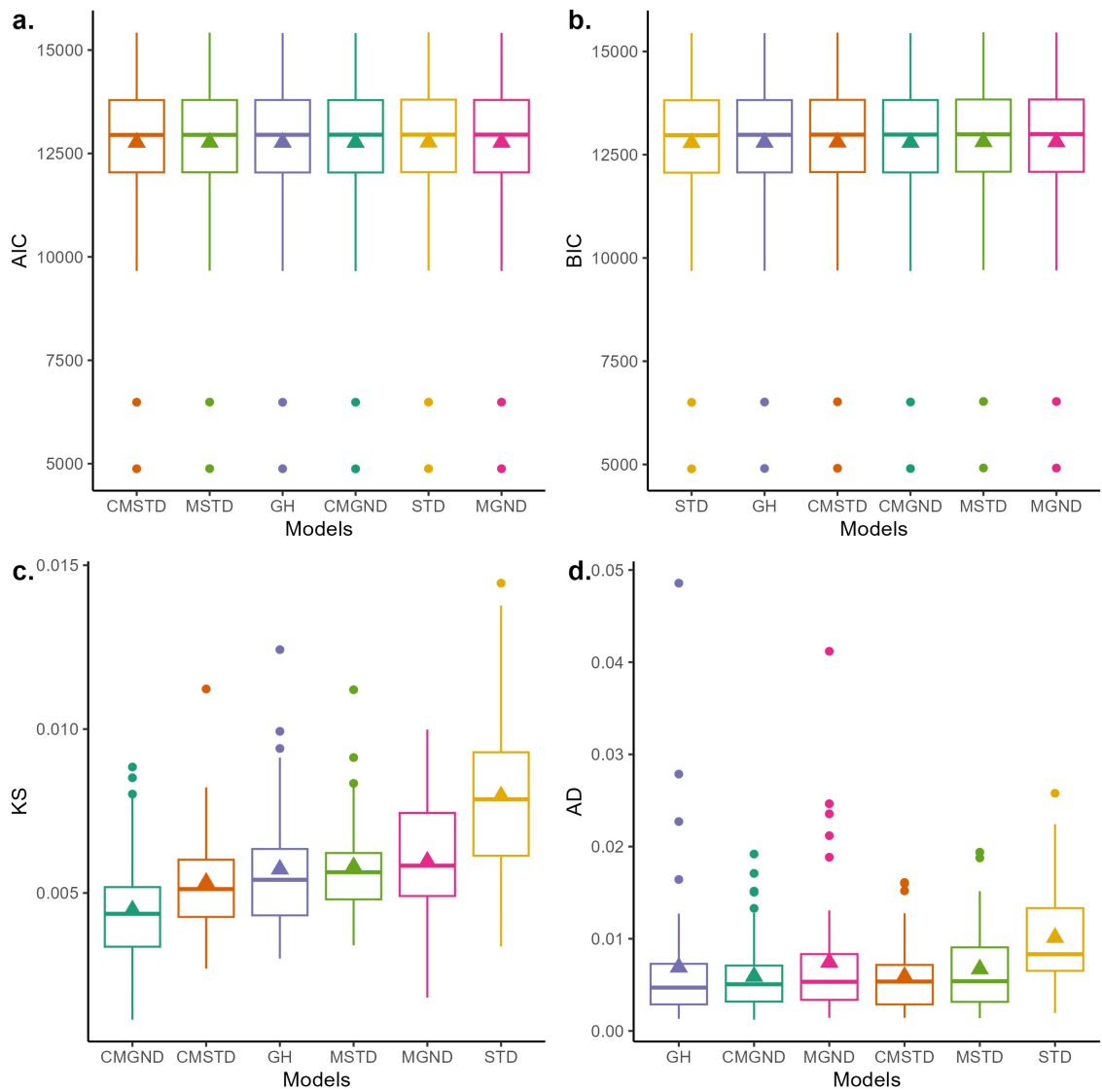

**Figure 4.9:** Comparative analysis part II, AIC, BIC, KS and AD boxplots. Models are ordered by median.

# GND-hidden Markov models to fit stock returns

## 5.1 Study framework

It is well known that the empirical return distribution of most financial asset presents significant departures from 'normality'. Specifically, it tends to be a leptokurtic distribution with marked fat tails and a peak around the mean value. Most notably, daily returns are not independently distributed over time.

Firstly, daily returns exhibit autocorrelation, meaning that returns are correlated over time. According to R. M. Anderson et al. 2012, autocorrelation can be attributed to four main sources.

1. **Non-synchronous trading**: irregular trading times of financial assets causing delays in the transaction reporting (Fisher 1966; Scholes and Williams 1977; Llorente et al. 2002).
2. **Partial price adjustment**: transactions that do not fully reflect all available information indicating a slow adjustment (R. M. Anderson et al. 2012; Boulatov, Hendershott, and Livdan 2013).
3. **Bid-ask bounce**: trading prices are biased toward the bid or ask price (R. M. Anderson et al. 2012).
4. **Time-varying risk premium**: changes in the investor's compensation for bearing risk over time (Duttilo, Gattone, and Di Battista 2021; Duttilo, Gattone, and Iannone 2023).

Secondly, as already identified by Mandelbrot 1963, daily returns exhibit volatility clustering, *'large changes (in price) tend to be followed by large changes-of either sign-and small changes tend to be followed by small changes'* (Mandelbrot 1963). They display small fluctuations during stable periods and extreme fluctuations during turmoil periods (Haas and Pigorsch 2009). In statistical terms, this means heteroskedasticity, i.e. non-constant variance over time (Kaehler and Marnet 1994). Daily squared returns ($r_t^2$) can be viewed as a *'noisy approximation of realised volatility'* (Hol 2003). Figure 5.1 shows the daily squared returns on the STOXX50E for the period from 02 April 2007 to 15 August 2023. There are important volatility clusters that occur during the 2008 global financial crisis, the 2015-2016 stock market sell-off, the Covid-19 pandemic and the Russian-Ukrainian conflict.

A common method to assess serial dependence is to examine the autocorrelation function (ACF) or perform the Ljung-Box test with the following hypotheses:

$$\begin{cases} H_0 : \rho_1 = \rho_2 = ... = \rho_{(M)} = 0 \text{ (serial indipendence)} \\ H_1 : \rho_m \neq 0 \text{ for some } m = 1, ..., M, \text{ (serial dependence)} \end{cases}$$

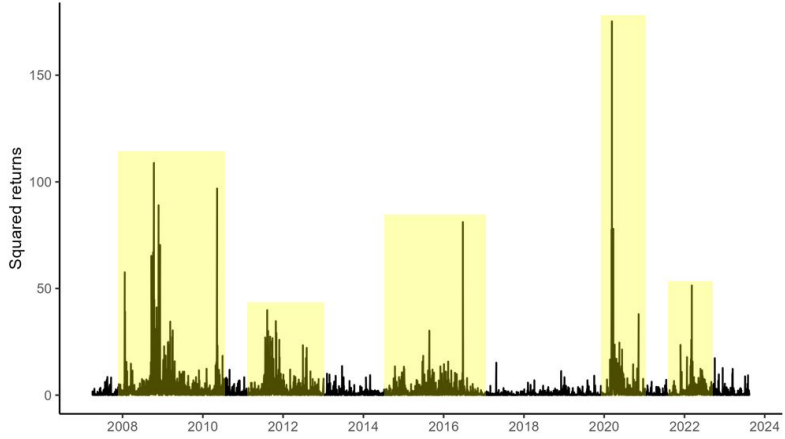

where $\rho_m$ is the autocorrelation function at lag $m$. The Ljung-Box statistic is defined as follows:

$$Q_x(M) = T(T+2) \sum_{m=1}^{M} \frac{\widehat{\rho_m^2}}{T-m}. \tag{5.1}$$

where $T$ is the time series length. Under $H_0$, $Q_x(M)$ has asymptotically a $\chi^2$ distribution with $M$ degrees of freedom. Figure 5.2 reports the ACF and the $Q_x(20)$ statistic. Both the ACF and $Q_x(20)$ statistic confirm the presence of serial dependence.

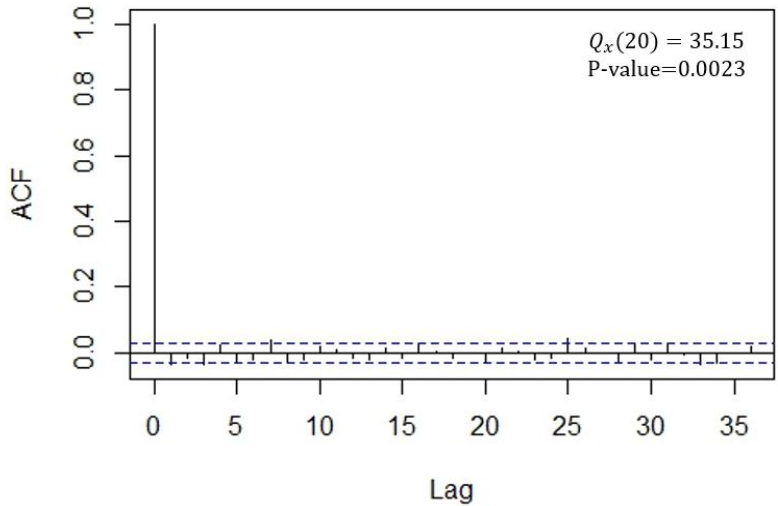

However, as explained by Kaehler and Marnet 1994, '*in the presence of heteroscedasticity, the Bartlett standard errors of the ACF are downward biased and the Ljung-Box statistic is upward biased, i.e. we would reject the $H_0$ of independence too often*' (Kaehler and Marnet 1994). For example, Table B.29 in Appendix B shows that $H_0$ is rejected for 57.69% of the stocks that belong to the STOXX50E. In order to quantify the heteroskedasticity, the Ljung-Box statistic can be applied also for squared data (McLeod and W. K. Li 1983; Kaehler and Marnet 1994). The Ljung-Box test for squared returns is denoted with $Q_{xx}(M)$. Table B.29 in Appendix B shows that there is an extreme strong serial dependence of volatility for all stocks that belong to the STOXX50E.

As a result heteroskedasticity cannot be neglected by the simplified assumptions of i.i.d. returns (Traut and Schadner 2023). Independent mixture models, like the MGND and CMGND models, are compatible with the strong stylized fact of leptokurtosis due to their flexibility in handling a large class of statistical distributions. On the contrary, they cannot fully capture heteroskedasticity. Since mixture components are mainly different with respect to their variances and do not account for temporal dependence (Kaehler and Marnet 1994), independent mixture models can only capture turmoil days and not volatility clusters, i.e. turmoil periods (Chapter 3).

To examine the dynamic properties of these models it is possible to apply a test of serial dependence on the state sequence. Let $S_t$ is the unobservable state variable at time $t$ which determines from which component a realization is drawn. Assuming two mixture components, if $S_t$ = 'turmoil' there will be a drawing from the turmoil component at time $t$, while if $S_t$ = 'stable' there will be a drawing from the stable component at time $t$. For independent mixture model, $S_t$ can be estimated with the Naïve Bayes' classification rule in equation 3.1 (Frühwirth-Schnatter 2006). The (two-sided) runs test (Wald and Wolfowitz 1940; Barton and David 1957; Bujang and Sapri 2018) is applied on the estimated state sequence $S_t$ for $t = 1, ..., T$ to examine its randomness or independence. The null and alternative hypothesis are defined as follows:

$$\begin{cases} H_0 : \text{The state sequence is determined by a random process;} \\ H_1 : \text{The state sequence is not determined by a random process.} \end{cases}$$

A run represents a sequence where a single state $S_t$ is repeated one or more times. The number of runs correspond to the number of times the state changes. For large samples ($T > 20$), the test statistic is defined by an approximation of the normal distribution

$$z = \frac{r - \mu_r + c}{\sigma_r} \tag{5.2}$$

where $r$ is the number of runs, $\mu_r$ is the expected number of runs and $\sigma_r$ is the standard deviation of the number of runs. The constant $c$ is set to 0.5 if $r < \mu_r$ and to $-0.5$ if $r > \mu_r$. The expected number of runs and its standard deviation are defined as follows:

$$\begin{aligned} \mu_r &= \frac{2n_1 n_2}{n_1 + n_2} + 1 \\ \sigma_r &= \sqrt{\frac{(2n_1 n_2)(2n_1 n_2 - n_1 - n_2)}{(n_1 + n_2)^2 (n_1 + n_2 - 1)}} \end{aligned} \tag{5.3}$$

where $n_1$ and $n_2$ are the number of observations in the two states ($n_1 + n_2 = T$). At the 5% significance level, $H_0$ is rejected if the test statistic has an absolute value greater than 1.96.

Figure 5.3 shows the detected turmoil days for the STOXX50E with the two-component CMGND-UUC model[1]. In addition, the estimated coefficients of the runs test are reported at the bottom of the figure. $H_0$ is rejected at the 5% significance level, the estimated sequence $S_t$ is not determined by a random process. There is positive dependence in the states because there are less runs than expected under $H_0$ as $r < \mu_r$

1: For the STOXX50E, the two-component CMGND-UUC model is the best model in the 7-model family (Chapter 4).

(Kaehler and Marnet 1994).

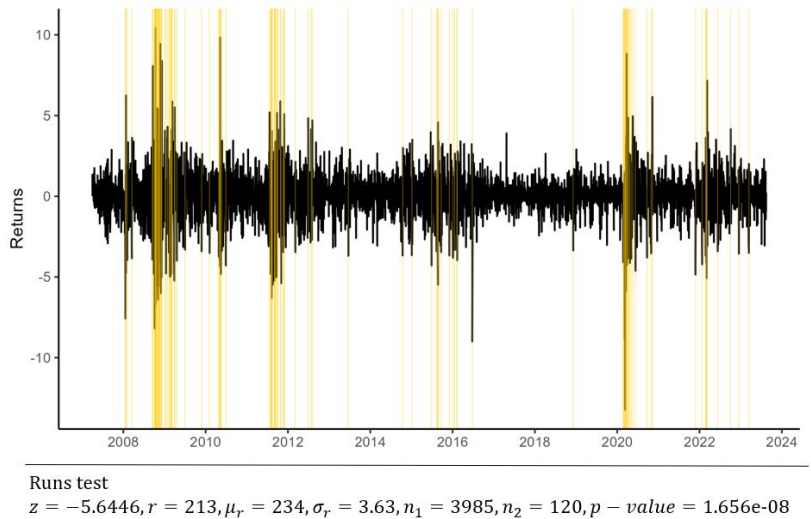

Runs test
$z = -5.6446, r = 213, \mu_r = 234, \sigma_r = 3.63, n_1 = 3985, n_2 = 120, p-value = 1.656e\text{-}08$

**Figure 5.3:** Detected turmoil days for the STOXX50E with the two-component CMGND-UUC model.

Therefore one notable limitation of the independent mixture model is its inability to fully account for heteroskedasticity. Hidden Markov models (HMMs) are a powerful statistical tool used to introduce dynamic elements when modelling financial returns in order to capture the above-mentioned unconditional and conditional returns properties.

HMMs are also known as '*hidden Markov process*' (Ephraim and Merhav 2002), '*Markov-dependent mixture*' (Le et al. 1992), '*Markov-switching model*', '*models subject to Markov regime*', '*Markov mixture model*' or '*latent Markov model*' (Bartolucci 2011). Originally, HMMs were introduced for the first time by Baum and Petrie 1966 with their landmark paper titled '*Statistical Inference for Probabilistic Functions of Finite State Markov Chains*'. Important subsequent studies were Baum, Petrie, et al. 1970 and Lindgren 1978.

Hamilton J.D. extended HMMs to analyse various economic phenomena, including interest rates, exchange rates, and the business cycle (Hamilton 1988; Hamilton 1989; Hamilton 1990; Hamilton 1991; Hamilton 1993; Engel and Hamilton 1990). Turner, Startz, and C. R. Nelson 1989 is another pioneering study who introduced a variety of models in which the variance of a portfolio's excess return depends on a state variable generated by a first-order Markov process.

Kaehler and Marnet 1994 employed mixtures of normal distributions and normal-HMMs to fit exchange-rates and their empirical dynamics: leptokurtosis and heteroskedasticity. It was pointed out that mixtures of normal distributions effectively capture the leptokurtosis, while normal-HMMs successfully capture both leptokurtosis and heteroskedasticity.

Guidolin and Timmermann 2006 demonstrated that HMMs perform effectively in measuring Value at Risk (VaR) at a monthly frequency but these results cannot be extended to daily returns. The latter strongly exhibit non-normal features.

Haas 2009 performed a VaR backtesting for several HMMs and independent mixture model considering the daily returns of major European stock markets. Both normal and Student-t mixtures are considered. Findings reveal that the univariate two-state Student-t-HMM demonstrates the best overall performance.

In the upcoming section, an extension of the MGND and CMGND models is presented in order to remove the heteroskedasticity deficiency. This improvement involves the introduction of a Markov chain process for the state variable. Specifically, a first-order Markov chain governs the draws of the states from mixture components. In this way, it is possible to improve the model ability to capture the dynamic nature of the state variable.

## 5.2 GND-hidden Markov models

### 5.2.1 Markov chain

Independent mixture models assume observations to be independent of each other. In time series analysis, the assumption of independence among observations is unrealistic as they exhibit temporal dependence. The mixture mechanism should be set in a way that emphasizes the probability of choosing the same component distribution at time $t + 1$ as the one selected at time $t$. In such situations, Markov chains becomes essential.

A sequence of discrete random variables $\{S_t : t \in \mathbb{N}\}$ is a Markov chain if for all $t \in \mathbb{N}$ it is satisfied the Markov property

$$\Pr(S_{t+1}|S_t, ..., S_1) = \Pr(S_{t+1}|S_t). \tag{5.4}$$

The future state $S_{t+1}$ depends only on its current state $S_t$ and not on the state sequence that preceded it (Figure 5.4), i.e. '*the past and the future are dependent only through the present*' (Zucchini, MacDonald, and Langrock 2017).

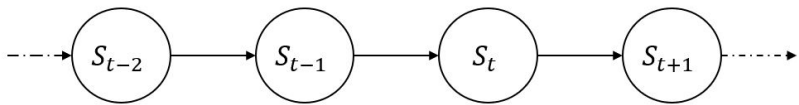

A Markov chain is featured by important quantities called transition probabilities:

$$\gamma_{ij}(t) = \Pr(S_{c+t} = j|S_c = i). \tag{5.5}$$

The Markov chain is homogeneous if the transition probabilities do not depend on $c$, namely, they are constant over time

$$\gamma_{ij}(t) = \gamma_{ij} \text{ for all } t. \tag{5.6}$$

A Markov chain with $K$ states (or regimes) can be defined by a $K \times K$ transition probability matrix (t.p.m.) defined as follows

$$\mathbf{\Gamma} = \begin{pmatrix} \gamma_{11} & \cdots & \gamma_{1K} \\ \vdots & \ddots & \vdots \\ \gamma_{K1} & \cdots & \gamma_{KK} \end{pmatrix}, \text{ with } \gamma_{ij} \in [0, 1] \text{ for all } ij \text{ and } \sum_{k=1}^{K} \gamma_{ij} = 1. \tag{5.7}$$

**Figure 5.4:** Markov property.

A Markov chain with t.p.m. $\Gamma$ have stationary distribution $\pi$ if $\pi\Gamma = \pi$ and $\pi\mathbf{1}' = 1$. The vector $\pi$ is subject to $\sum_{k=1}^{K} \pi_k = 1$ with $\pi_k > 0$. Each element $\pi_k$ represents the probability of the chain being in state $k$ when the chain is in a state described by $\pi$. A stationary Markov chain starts in its stationary distribution $\pi$ and the process continues to have this unconditional distribution at all subsequent time points. It is possible to demonstrate that $\pi$ is a stationary distribution of the Markov chain with t.p.m. $\Gamma$ if

$$\pi(\mathbf{I}_K - \Gamma + U) = \mathbf{1} \tag{5.8}$$

where $I_k$ is the $K \times K$ identity matrix, $U$ is the $K \times K$ matrix of ones and $\mathbf{1}$ is a row vector of ones.

**Example: stable and turmoil days**

Consider a sequence of stable and turmoil days over 21 trading days:

> stable, stable, stable, stable, stable, stable, stable, stable, turmoil,
>
> turmoil, turmoil, turmoil, turmoil, turmoil, turmoil, stable, stable,
>
> stable, stable, stable, stable.

A two-state Markov chain model for these data assumes that tomorrow's market state $S_{t+1}$ is stochastically determined by today's market state $S_t$ with $2 \times 2$ t.p.m.

$$\Gamma = \begin{pmatrix} \gamma_{11} & \gamma_{12} \\ \gamma_{21} & \gamma_{22} \end{pmatrix} = \begin{pmatrix} \text{Pr(stable} \rightarrow \text{stable)} & \text{Pr(stable} \rightarrow \text{turmoil)} \\ \text{Pr(turmoil} \rightarrow \text{stable)} & \text{Pr(turmoil} \rightarrow \text{turmoil)} \end{pmatrix}.$$

The t.p.m. can be estimated as follows

$$\widehat{\Gamma} = \begin{pmatrix} \frac{\text{trans. stable}\rightarrow\text{stable}}{\text{trans. from stable}} & \frac{\text{trans. stable}\rightarrow\text{turmoil}}{\text{trans. from stable}} \\ \frac{\text{trans. turmoil}\rightarrow\text{stable}}{\text{trans. from turmoil}} & \frac{\text{trans. turmoil}\rightarrow\text{turmoil}}{\text{trans. from turmoil}} \end{pmatrix} = \begin{pmatrix} \frac{12}{13} & \frac{1}{13} \\ \frac{1}{7} & \frac{6}{7} \end{pmatrix} \approx \begin{pmatrix} 0.92 & 0.08 \\ 0.14 & 0.86 \end{pmatrix}.$$

The vector $\pi$ is obtained as follows

$$\begin{aligned} \widehat{\pi} &= \frac{1}{\widehat{\gamma}_{12} + \widehat{\gamma}_{21}} (\widehat{\gamma}_{21}, \widehat{\gamma}_{12}) \\ &= \frac{1}{0.08 + 0.14} (0.14, 0.08) \\ &= (0.64, 0.36). \end{aligned}$$

In general, with a $K \times K$ t.p.m., $\pi$ can be estimated by solving equation 5.8 with the R function *solve*.

## 5.2.2 Hidden Markov model

A $K$-state HMM $\{X_t : t \in \mathbb{N}\}$ is a dependent mixture model with a doubly stochastic process in discrete time. Firstly, an unobserved state process $S_1, S_2, ..., S_T$ taking value in $\{1, ..., K\}$ satisfying the Markov property. Secondly, an observed state-dependent process $X_1, X_2, ..., X_T$ in which the distribution of $X_t$ depends only on the current state $S_t$ and not on previous states or observations (Figure 5.5).

The distribution of $S_t$ depends on $S_{t-1}$. Besides, each state has a different distribution. If the Markov chain is in state $i$ at time $t$, the probability

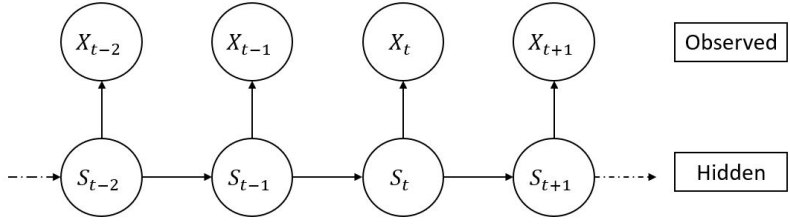

mass function of $X_t$ is

$$f_k(x) = Pr(X_t = x | S_t = k) \text{ for } k = 1, 2, ..., K. \tag{5.9}$$

The unconditional or marginal distribution of $X_t$ is given by

$$f(x_t) = \sum_{k=1}^{K} \pi_k f_k(x_t). \tag{5.10}$$

The distributions $f_k$, $k = 1, ..., K$, are called state-dependent distributions of the model. Thus, an HMM is a dependent mixture model: the distribution selected at time $t - 1$ does affect the distribution that will be selected at time $t$. Table 5.1 resume the main differences and similarities between independent mixture models and HMM.

| Independent mixture models | HMM |
|---|---|
| **Similarities** | |
| $K$ component distributions | |
| $S_t$ selects which component distribution is active at time $t$ | |
| Each state has a different distribution | |
| **Differences** | |
| The distribution of $S_t$ **does not depend** on $S_{t-1}$ | The distribution of $S_t$ **does depend** on $S_{t-1}$ |

Figure 5.6 provides an example of the process generating returns in a two-state HMM. The Markov chain follows the path: turmoil, stable, stable, stable, turmoil, stable. State-dependent distributions are shown in the middle, while returns are generated from the active distributions (Zucchini, MacDonald, and Langrock 2017).

### 5.2.3 GND-hidden Markov model

The Markov chain (equations 5.4 and 5.7) together with the CMGND model (equation 4.1) gives the GND-hidden Markov model (GND-HMM). For K=2, a two-state GND-HMM is obtained with parameter vector $\theta = \{\gamma_{11}, \gamma_{22}, \pi_1, \mu_1, \mu_2, \sigma_1, \sigma_2, \nu_1, \nu_2\}$. Constraints are imposed on $\mu_k$, $\sigma_k$ and $\nu_k$ to be equal across the two state-dependent distributions: $\mu_k = \mu$, $\sigma_k = \sigma$, $\nu_k = \nu$, for $k = 1, 2$. Table 5.2 displays the two-state GND-HMM with constrained and unconstrained parameters. By constraining the parameter space the number of parameters ($p$) of each model is gradually reduced as shown in the last column of Table 5.2.

The parameter estimation is performed with the direct optimization method (Zucchini, MacDonald, and Langrock 2017). The likelihood function of an HMM is given by

$$L_T = \Pr(\boldsymbol{X}^{(T)} = \boldsymbol{x}^{(T)}) = \boldsymbol{\pi} \boldsymbol{P}(x_1) \boldsymbol{\Gamma} \boldsymbol{P}(x_2) \dots \boldsymbol{\Gamma} \boldsymbol{P}(x_T) \boldsymbol{1}', \tag{5.11}$$

| Model | $\mu_k$ | $\sigma_k$ | $\nu_k$ | $p$ |
|---|---|---|---|---|
| GND-HMM-UUU | U | U | U | 9 |
| GND-HMM-CUU | C | U | U | 8 |
| GND-HMM-UCU | U | C | U | 8 |
| GND-HMM-UUC | U | U | C | 8 |
| GND-HMM-CCU | C | C | U | 7 |
| GND-HMM-CUC | C | U | C | 7 |
| GND-HMM-UCC | U | C | C | 7 |
| GND-HMM-CCC | C | C | C | 6 |

where $\boldsymbol{\pi}$ is the initial distribution of $S_1$ and $\boldsymbol{P}(x)$ is the $K \times K$ diagonal matrix with diagonal element equal to $f_k(x_t)$, the $k$-th state-dependent distribution. The estimation is performed with the unconstrained R optimizer *nlm* (Zucchini, MacDonald, and Langrock 2017). However, in the case of a GND-HMM there are three relevant constraints to consider for the likelihood maximization:

▶ the scale parameter $\sigma_k$ of the state-dependent distribution must be non negative for $k = 1, ..., K$;
▶ the shape parameter $\nu_k$ of the state-dependent distribution must be non negative for $k = 1, ..., K$;
▶ the rows of the t.p.m. $\boldsymbol{\Gamma}$ must add to 1 with non-negative elements $\gamma_{ij}$.

According to Zucchini, MacDonald, and Langrock 2017 the likelihood maximization can be performed in two steps by parameter transformation:

1. Maximize $L_T$ with respect to the working parameters;
2. Transform the estimates of the working parameters to estimates of the natural parameters.

The working parameters of $\sigma_k$ and $\nu_k$ are defined as $\eta_k = \log \sigma_k$ and $\delta_k = \log \nu_k$. By contrast, for $K = 2$ the working parameter of $\boldsymbol{\Gamma}$ is defined as follows

$$\boldsymbol{T} = \begin{pmatrix} \tau_{11} & \tau_{12} \\ \tau_{21} & \tau_{22} \end{pmatrix} \tag{5.12}$$

where

$$\tau_{ij} = \log \left( \frac{\gamma_{ij}}{1 - \sum_{k \neq i} \gamma_{ik}} \right) \text{ with } \tau_{ij} \in \mathbb{R}. \tag{5.13}$$

The parameter estimation of a stationary two-state GND-HMM via direct optimization is resumed in the following box.

---

**Parameter estimation of a stationary two-state GND-HMM via direct optimization**

1. **require:** data $x_1, x_2, ..., x_N$.
2. **set the initial estimates:**
3. $\mu_1 \leftarrow 0.5, \mu_2 \leftarrow -0.5$
4. $\sigma_1 \leftarrow 1, \sigma_2 \leftarrow 2$
5. $\nu_1 \leftarrow 2, \nu_2 \leftarrow 2$
6.
$$\boldsymbol{\Gamma} \leftarrow \begin{pmatrix} 0.8 & 0.2 \\ 0.2 & 0.8 \end{pmatrix}$$
7. $\boldsymbol{\pi}$ is computed by solving $\boldsymbol{\pi}(\boldsymbol{I}_K - \boldsymbol{\Gamma} + \boldsymbol{U}) = \boldsymbol{1}$
8. **from natural parameters to working parameters**: $\widehat{\eta}_k \leftarrow \widehat{\sigma}_k, \widehat{\delta}_k \leftarrow \widehat{\nu}_k$ and $\widehat{\boldsymbol{T}} \leftarrow \widehat{\boldsymbol{\Gamma}}$
9. **maximize $L_T$ in equation 5.11 with respect to the working parameters via the R function *nlm***
10. **from working parameters to natural parameters**: $\widehat{\sigma}_k \leftarrow \widehat{\eta}_k, \widehat{\nu}_k \leftarrow \widehat{\delta}_k$ and $\widehat{\boldsymbol{\Gamma}} \leftarrow \widehat{\boldsymbol{T}}$
11. **return** parameters $= \left( \widehat{\pi}_k, \widehat{\mu}_k, \widehat{\sigma}_k, \widehat{\nu}_k, \widehat{\boldsymbol{\Gamma}} \right)$

The EM algorithm can be used as an alternative to direct optimization to perform the parameter estimation of HMMs (Hamilton 1990; Hamilton 1994; G. J. McLachlan and Peel 2000; Zucchini, MacDonald, and Langrock 2017; Catania and Di Mari 2021; Catania, Di Mari, and Santucci De Magistris 2022). In this manuscript the direct maximization method is adopted (Zucchini, MacDonald, and Langrock 2017).

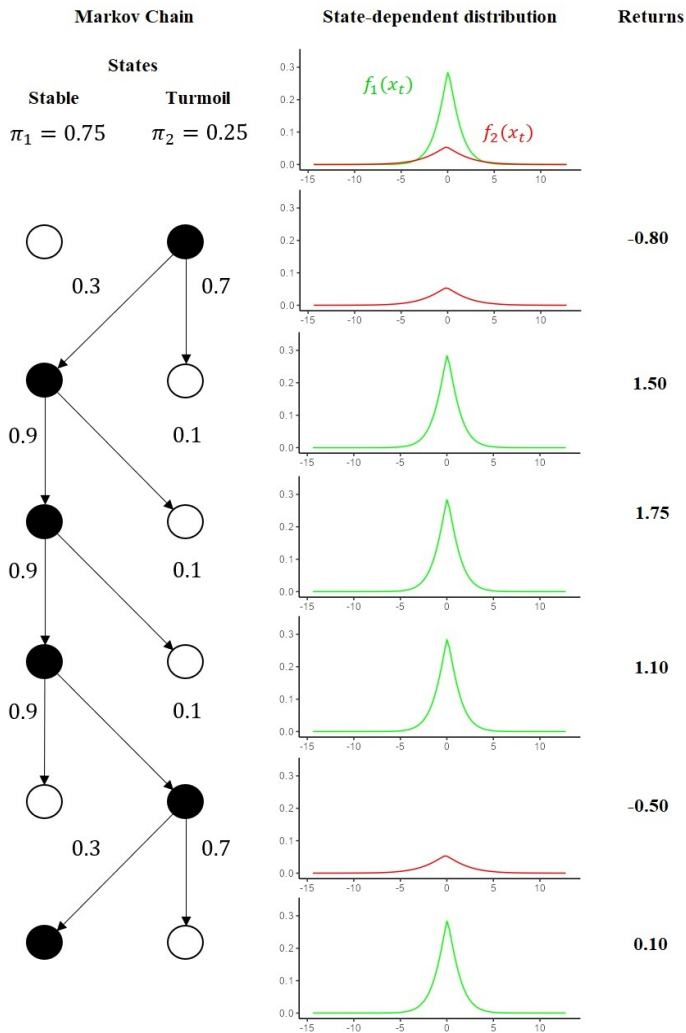

**Figure 5.6:** Example of returns generated by a two-state HMM.

## 5.3 Numerical experiments

### 5.3.1 Simulation study

The simulation study at analysing the performance of two-state GND-HMMs in a simulated financial scenario. Specifically, it is assumed that the financial market is characterized by two different periods[2]: stable and tumultuous (Figure 5.7). The stable period with small price fluctuations has a GND $f_1(x_t) = f_1(x_t|0, 1, 2)$. The turmoil period with large price fluctuations has a GND with negative mean and fat tails $f_2(x_t) = f_2(x_t| - 0.02, 1, 0.8)$. Whereas the stable component assumes a

2: A similar scenario has been simulated for the MGND model in Section 2.7.

normal distribution $\nu_1 = 2$, the tails of the turmoil distribution are thicker than those of a Laplace distribution since $\nu_2 < 1$.

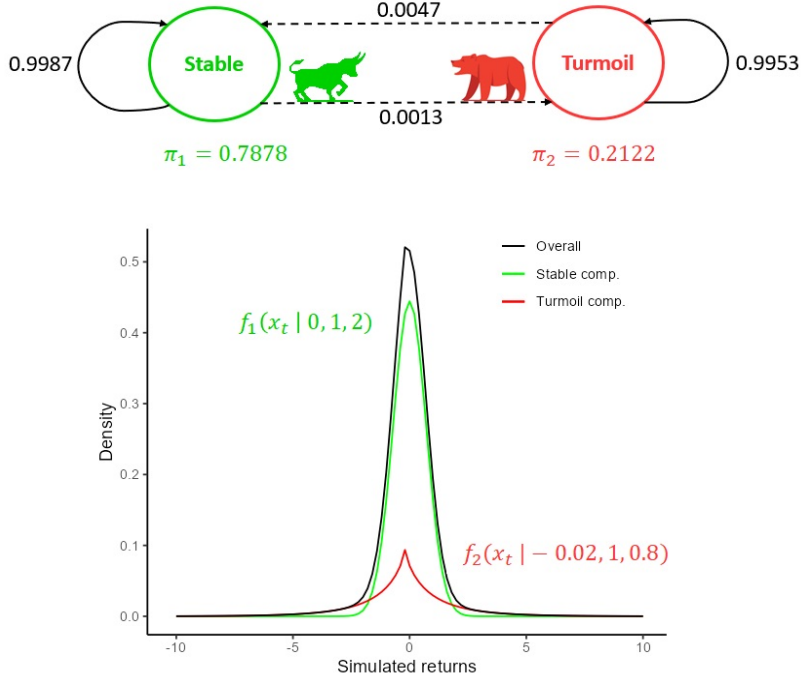

**Figure 5.7:** GND-HMMs simulated financial scenario.

Returns are simulated for $N = 1000, 2000$ trading days with the state sequences below.

**N=1000**

| 160 | 112 | 262 | 100 | 366 | Trading |
|-----|-----|-----|-----|-----|---------|
| Stable | Turmoil | Stable | Turmoil | Stable | days |

**N=2000**

| 320 | 224 | 524 | 200 | 732 | Trading |
|-----|-----|-----|-----|-----|---------|
| Stable | Turmoil | Stable | Turmoil | Stable | days |

The three stable periods are more persistent than the two turmoil periods. Under the homogeneity and stationary assumptions, the t.p.m. $\boldsymbol{\Gamma}$ and $\boldsymbol{\pi}$ are determined by

$$\boldsymbol{\Gamma} = \begin{pmatrix} 0.9987 & 0.0013 \\ 0.0047 & 0.9953 \end{pmatrix}, \quad \boldsymbol{\pi} = (0.7878, 0.2122).$$

For each sample sizes ($N$), $S = 250$ samples are simulated using the R function *rgnorm* of the package *gnorm*. An example of simulated returns is provided in Figure 5.8. The two-state HMMs (Table 5.2) are estimated for each sample. As a measure of quality estimation, for each parameter the mean squared error (MSE) is computed as follows

$$\text{MSE}(\widehat{\theta}) = \frac{1}{S} \sum_{s=1}^{S} (\widehat{\theta}_s - \theta)^2, \tag{5.14}$$

where $\theta$ is the true parameter value and $\widehat{\theta}_s$ is the estimate of $\theta$ for the $s$-th simulated data. The average of the estimates is also computed for

each parameter

$$\text{AVG}(\widehat{\theta}) = \frac{1}{S} \sum_{s=1}^{S} \widehat{\theta}. \tag{5.15}$$

As a measure of parametric efficiency the BIC defined in equation 2.59 is computed for each estimated model. Furthermore, the adjusted Rand index (ARI) is employed to evaluate the similarity between the true clustering of states and the clustering predicted with GND-HMMs via the Viterbi algorithm[3] . The ARI (Hubert and Arabie 1985) is defined as follows

$$\text{ARI} = \frac{\text{RI} - \text{Expexted RI}}{\text{Max RI} - \text{Expected RI}}. \tag{5.16}$$

RI is the Rand index (Rand 1971) which calculates a similarity between the true clustering of states and the predicted clustering of states (Sinnott, Duan, and Sun 2016; Scrucca et al. 2016). Expected RI is the expected value of the RI under random clustering. Max RI is the maximum possible value of the Rand Index. The ARI yields a score in the range $[-1, 1]$. A score of 1 indicates that true and predicted clusterings are the same, while a score of 0 suggests random clustering. Negative scores indicate worse than random clustering.

Tables B.30-B.31 in Appendix B report the average of the estimates and their MSE. Figure 5.9 shows the MSE dotplot for $N = 1000$ (panel a) and $N = 2000$ (panel b). All GND-HMMs show consistent MSE across different sample sizes. The GND-HMM-CCU yields the lowest MSE for all parameters. Specifically, MSE of the turmoil component parameters are significantly lower than those of the unconstrained model. For example, for $N = 1000$ the GND-HMM-CCU yields MSE($\widehat{\mu}_2$) = 0.03, MSE($\widehat{\sigma}_2$) = 0.05 and MSE($\widehat{v}_2$) = 0.04, while the GND-HMM-UUU yields MSE($\widehat{\mu}_2$) = 0.09, MSE($\widehat{\sigma}_2$) = 0.27 and MSE($\widehat{v}_2$) = 0.11. The GND-HMM-UCU also gets low MSE except for $\mu_2$ whose MSE is similar to that of the unconstrained model (0.09 for $N = 1000$ and 0.06 for $N = 2000$). The GND-HMM-UCC and GND-HMM-CCC have the highest MSE, while the GND-HMM-UUC and GND-HMM-CUC show high MSE for $v_1$, $\sigma_2$ and $v_2$.

Based on the average BIC values in Table 5.3, the GND-HMM-CCU is the best model for both sample sizes, while the GND-HMM-CUU and GND-HMM-UCU also report low BICs compared to the unconstrained model. Moreover, all GND-HMMs present an average ARI close to 1. The only two exceptions are the GND-HMM-UCC and GND-HMM-CCC with an average ARI close to 0.

In summary, the GND-HMM-CCU stands out for its efficiency in parametrization, estimation and clustering. On the contrary, the GND-HMM-UCC and GND-HMM-CCC are not suitable to capture financial returns dynamics since they imply homoscedasticity[4] .

3: The Viterbi algorithm (Viterbi 1967) allows to obtain a global decoding of the hidden states, i.e. the state sequence $(s_1, s_2, ..., s_T)$ which maximizes the conditional probability

$$\Pr(\boldsymbol{S}^{(T)} = \boldsymbol{s}^{(T)} | \boldsymbol{X}^{(T)} = \boldsymbol{x}^{(T)}),$$

where $\boldsymbol{S}^{(T)} \equiv (S_1, S_2, ..., S_T)$, and similarly $\boldsymbol{s}^{(T)}$, $\boldsymbol{X}^{(T)}$ and $\boldsymbol{x}^{(T)}$. See the next section for an application on real data.

4: $\sigma_1 = \sigma_2$, $v_1 = v_2$.

**Table 5.3:** Average BIC and ARI by GND-HMM and sample size. In bold the best model.

|            | UUU     | CUU     | UCU     | UUC     | CCU        | CUC     | UCC     | CCC     |
|------------|---------|---------|---------|---------|------------|---------|---------|---------|
| **Average BIC** |         |         |         |         |            |         |         |         |
| 1000       | 2654.47 | 2649.23 | 2648.91 | 2686.60 | **2643.07** | 2680.23 | 2840.96 | 2859.16 |
| 2000       | 5235.61 | 5229.58 | 5229.02 | 5307.12 | **5223.06** | 5297.71 | 5647.80 | 5686.41 |
| **Average ARI** |         |         |         |         |            |         |         |         |
| 1000       | 0.9460  | 0.9457  | 0.9456  | 0.9083  | **0.9465** | 0.9101  | 0       | 0       |
| 2000       | 0.9736  | 0.9737  | 0.9736  | 0.9586  | **0.9738** | 0.9592  | 0       | 0       |

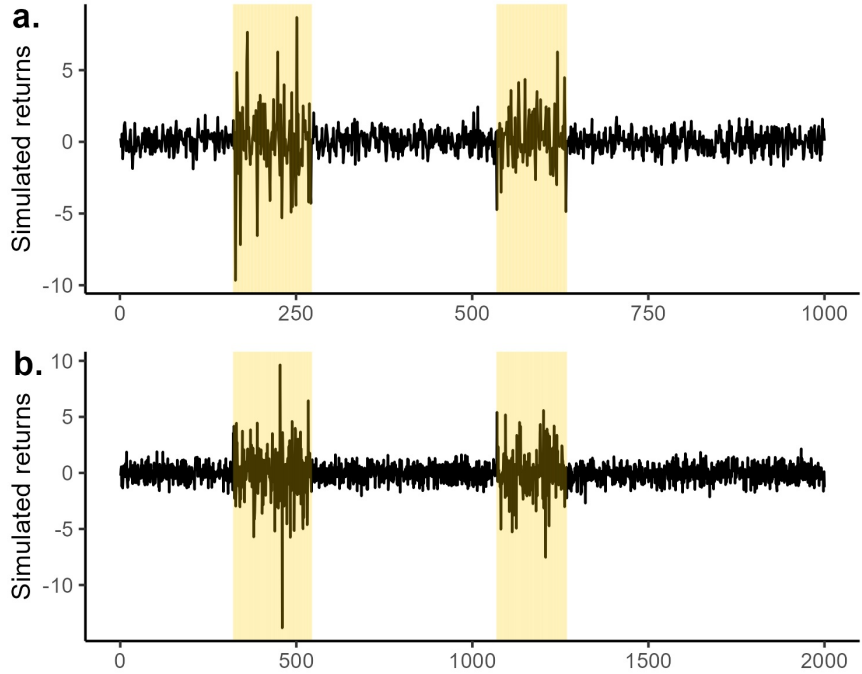

**Figure 5.8:** Example of simulated financial returns with GND-HMMs for $N = 1000$ (panel a) and $N = 2000$ (panel b). The yellow areas highlight turmoil periods.

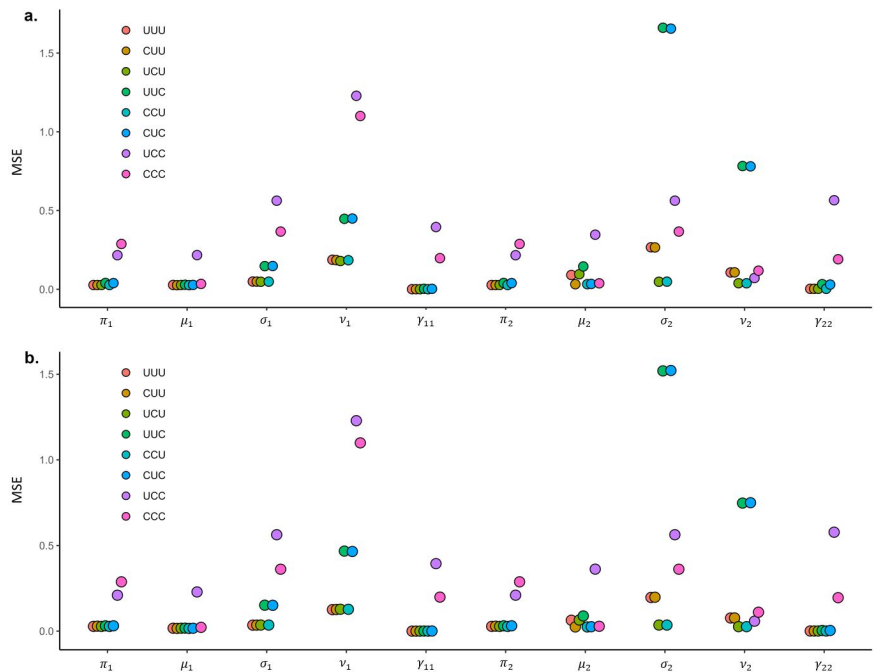

**Figure 5.9:** GND-HMMs simulations, MSE dotplot for $N = 1000$ (panel a) and $N = 2000$ (panel b).

## 5.3.2  An empirical application

An empirical application of the two-state GND-HMMs is presented in this section. As proven before, returns on STOXX50E are characterized by leptokurtosis (Table 4.7) and heteroskedasticity (Figure 5.2).

Table 5.4 shows the estimates of the unconstrained and constrained two-state GND-HMMs for daily returns on STOXX50E. The BIC is used to select the best model and compare the goodness-of-fit performance (G. J. McLachlan and S. Rathnayake 2014; Wen et al. 2022). According to the BIC the GND-HMM-CCU is the best model. Figure 5.10 shows the estimated marginal distribution with the two-state GND-HMM-CCU. It can be seen that the stable component (dotted line) describes the central and intermediate values of the data, while the turmoil component (dashed line) describes the more extreme tail behaviours.

**Table 5.4:** Estimates of the unconstrained and constrained two-state GND-HMM for daily returns on STOXX50E.

| $\theta$ | GND-HMM | | | | | | | |
| --- | --- | --- | --- | --- | --- | --- | --- | --- |
| | UUU | CUU | UCU | UUC | CCU | CUC | UCC | CCC |
| $\pi_1$ | 0.7150 | 0.7197 | 0.7251 | 0.7180 | 0.7255 | 0.7226 | 0.6618 | 0.5000 |
| $\mu_1$ | 0.0665 | 0.0508 | 0.0629 | 0.0662 | 0.0447 | 0.0510 | -0.1589 | 0.0426 |
| $\sigma_1$ | 0.9669 | 0.9710 | 1.0847 | 0.9604 | 1.0886 | 0.9643 | 0.6965 | 0.9185 |
| $\nu_1$ | 1.3224 | 1.3202 | 1.4344 | 1.3105 | 1.4346 | 1.3083 | 0.8406 | 0.9574 |
| $\pi_2$ | 0.2850 | 0.2803 | 0.2749 | 0.2820 | 0.2745 | 0.2774 | 0.3382 | 0.5000 |
| $\mu_2$ | -0.1216 | 0.0508 | -0.1113 | -0.1239 | 0.0447 | 0.0510 | 0.5033 | 0.0426 |
| $\sigma_2$ | 2.2066 | 2.2184 | 1.0847 | 2.2570 | 1.0886 | 2.2699 | 0.6965 | 0.9185 |
| $\nu_2$ | 1.2856 | 1.2830 | 0.8073 | 1.3105 | 0.8089 | 1.3083 | 0.8406 | 0.9574 |
| $\log L(\hat{\theta})$ | -6598.23 | -6602.24 | -6636.82 | -6598.31 | -6639.97 | -6602.33 | -6847.29 | -6863.80 |
| BIC | 13271.33 | 13271.05 | 13340.21 | 13263.18 | **13262.90** | 13338.17 | 13752.82 | 13777.51 |

*Notes.* In bold the best model according to the BIC.

The stable component is predominant compared to the turmoil component ($\pi_1 > \pi_2$). The shape parameter of the turmoil component ($\nu_2 = 0.8089$) is lower than the shape parameter of the stable component ($\nu_1 = 1.4346$). In addition the turmoil component has heavier tails than a Laplace distribution ($0.8089 < 1$). As shown in Table 5.5 the stable component has a low standard deviation ($\text{Std}_1 = 0.9736$), while the turmoil component has a high standard deviation ($\text{Std}_2 = 2.3431$). The kurtosis of turmoil component ($\text{Kur}_2 = 8.3980$) is higher than the kurtosis of the stable component ($\text{Kur}_1 = 3.9232$). The estimated overall standard deviation fit well the respective empirical measure, while the estimated overall kurtosis slightly overestimates the empirical one.

| Stable component | | Turmoil component | | Overall | |
| --- | --- | --- | --- | --- | --- |
| $\text{Std}_1$ | $\text{Kur}_1$ | $\text{Std}_2$ | $\text{Kur}_2$ | Std | Kur |
| 0.9736 | 3.9232 | 2.3431 | 8.3980 | 1.4815 | 14.9559 |

**Table 5.5:** Estimated standard deviation and kurtosis of the two-state GND-HMM-CCU on STOXX50E. The empirical standard deviation and kurtosis are equal to 1.436 and 10.326, respectively.

The estimated t.p.m. of the two-state GND-HMM-CCU is

$$\widehat{\Gamma} = \begin{pmatrix} \widehat{\gamma_{11}} & \widehat{\gamma_{12}} \\ \widehat{\gamma_{21}} & \widehat{\gamma_{22}} \end{pmatrix} = \begin{pmatrix} 0.9934 & 0.0066 \\ 0.0174 & 0.9826 \end{pmatrix}.$$

The probabilities of remaining in stable ($\widehat{\gamma_{11}}$) or turmoil ($\widehat{\gamma_{22}}$) periods are close to 1; these results reflect the stock market inertia, i.e. the tendency of the market to remain in a given state (Salhi et al. 2016). Moreover, the probability from turmoil to stable state ($\widehat{\gamma_{21}}$) is higher than from stable to turmoil ($\widehat{\gamma_{12}}$), suggesting a tendency to have shorter turmoil periods than stable periods in the market (Salhi et al. 2016).

Figure 5.11 shows the turmoil periods detection with the two-state GND-HMM for the STOXX50E. Panel (a) shows the local decoding of turmoil periods, while panel (b) shows the global decoding with the Viterbi algorithm (Viterbi 1967; Zucchini, MacDonald, and Langrock 2017).

Local decoding determines the state which is most probable at a given time. Specifically, for each $t \in \{1, ..., T\}$ the most probable state $k_t^*$ is defined as follows

$$k_t^* = \underset{k=1,...,K}{\mathrm{argmax}} \Pr(S_t = k | \boldsymbol{X}^{(T)} = \boldsymbol{x}^{(T)}), \qquad (5.17)$$

where $\boldsymbol{x}^{(T)} \equiv (x_1, x_2, ..., x_T)$, and similarly $\boldsymbol{X}^{(T)}$. Hence, the conditional distribution of $S_t$ given the observations is defined as follows

$$\Pr(S_t = k | \boldsymbol{X}^{(T)} = \boldsymbol{x}^{(T)}) = \frac{\alpha_t(k) \beta_t(k)}{L_T} \qquad (5.18)$$

where:

► $L_T$ is the likelihood function defined in equation 5.11;
► $\alpha_t(k) = \Pr(\boldsymbol{X}^{(t)} = \boldsymbol{x}^{(t)}, S_t = k)$ for $t = 1, ..., T$ and $k = 1, ..., K$ are the forward probabilities;
► $\beta_t(k) = \Pr(X_{t+1} = x_{t+1}, ..., X_T = x_T | S_t = k)$ for $t = 1, ..., T-1$ and $k = 1, ..., K$ are the backward probabilities.

By contrast, global decoding determines the most likely sequence of hidden states, i.e. the state sequence $(s_1, s_2, ..., s_T)$ which maximize the conditional probability

$$\Pr(\boldsymbol{S}^{(T)} = \boldsymbol{s}^{(T)} | \boldsymbol{X}^{(T)} = \boldsymbol{x}^{(T)}) \qquad (5.19)$$

In this context, global decoding provides more accurate results because considers the entire sequence of observations and states, capturing the global context and dependencies between states. This is crucial for understanding how market conditions persist or change over time. During the 2015-2016 stock market sell-off, global decoding reveals more persistent turmoil periods compared to local decoding. Most notably, the comparison between figures 5.3 and 5.11 confirm that GND-HMMs are able to capture turmoil periods instead of turmoil days by implying temporal dependence.

Figure 5.12 shows the state probabilities (equation 5.18) over time for the STOXX50E. In this context, state probabilities represent the probability of the market being in a specific state at a given time. Turmoil state probabilities highlight important volatility periods like: the 2008 global financial crisis, the 2015-2016 stock market sell-off, the US-China trade war, the COVID-19 pandemic, and the Russian-Ukrainian conflict.

By learning the underlying patterns from observed data, HMMs can be used to forecast future states or observations in a probabilistic way. The state prediction is given by a generalization of equation 5.12 for the forecast horizon $h$:

$$\Pr(S_{T+h} = k | \boldsymbol{X}^{(T)} = \boldsymbol{x}^{(T)}) = \frac{\boldsymbol{\alpha}_T \boldsymbol{\Gamma}^h \boldsymbol{e}_i'}{L_T}, \qquad (5.20)$$

where $\boldsymbol{\alpha_T}$ is the row vector of forward probabilities (equation 5.18) and $\boldsymbol{e}_i'$ is a row vector with 1 in the $i$-th position $(0, ..., 0, 1, 0, ..., 0)$. Table 5.6 reports state prediction for 5 days ahead using a two-state GND-HMM-CCU estimated on STOXX50E returns.

| Day | Aug. 17, 2023 | Aug. 18, 2023 | Aug. 19, 2023 | Aug. 20, 2023 | Aug. 21, 2023 |
|---|---|---|---|---|---|
| Stable | 0.9421 | 0.9369 | 0.9318 | 0.9269 | 0.9221 |
| Turmoil | 0.0579 | 0.0631 | 0.0682 | 0.0731 | 0.0779 |

**Table 5.6:** State prediction for 5 days ahead using a two-state GND-HMM-CCU estimated on STOXX50E returns.

The forecast distribution is given by

$$\Pr(X_{T+h} = x | \boldsymbol{X}^{(T)} = \boldsymbol{x}^{(T)}) = \frac{\boldsymbol{\alpha}_T \boldsymbol{\Gamma}^h \boldsymbol{P}(x) \boldsymbol{1}'}{\boldsymbol{\alpha}_T \boldsymbol{1}'} \qquad (5.21)$$

where $\boldsymbol{1}'$ is a vector of ones and $\boldsymbol{P}(x)$ is the $K \times K$ diagonal matrix with $i$-th diagonal element the state-dependent distribution $f_k(x)$. Figure 5.13 shows the one-step ahead forecast distribution using a two-state GND-HMM-CCU estimated on STOXX50E returns.

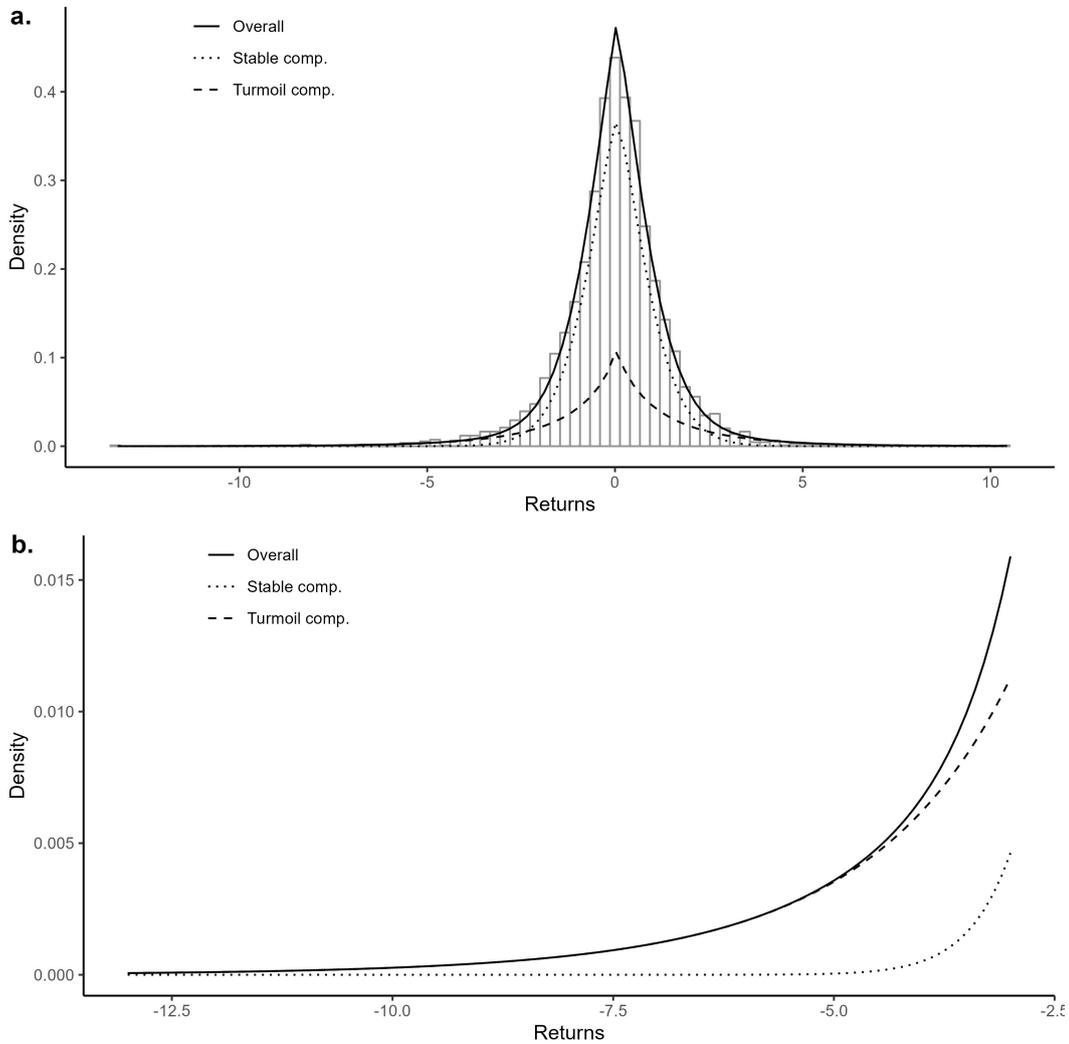

**Figure 5.10:** Estimated marginal distribution with the two-state GND-HMM-CCU. Panel (a) shows the overall distribution, while panel (b) shows the tails.

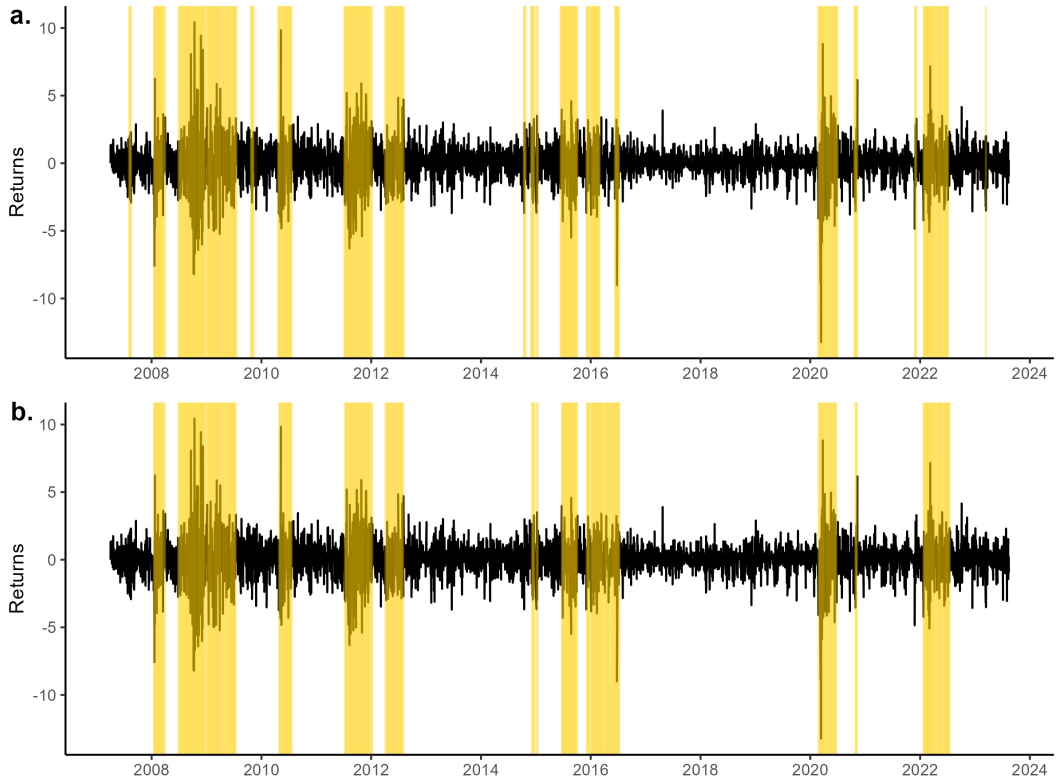

**Figure 5.11:** Turmoil periods detection with the two-state GND-HMM-CCU for the STOXX50E. Panel (a) shows the local decoding, while panel (b) shows the global decoding (Viterbi algorithm).

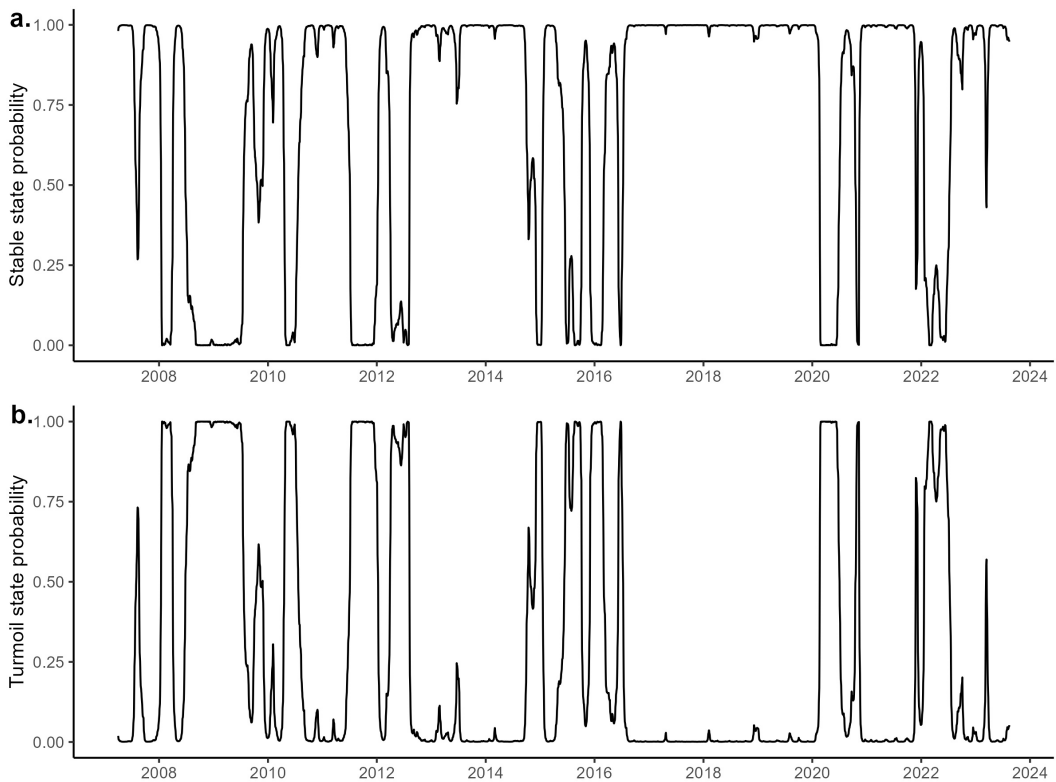

**Figure 5.12:** State probabilities over time for the STOXX50E. Panel (a) shows the stable state probabilities, while panel (b) shows the turmoil state probabilities.

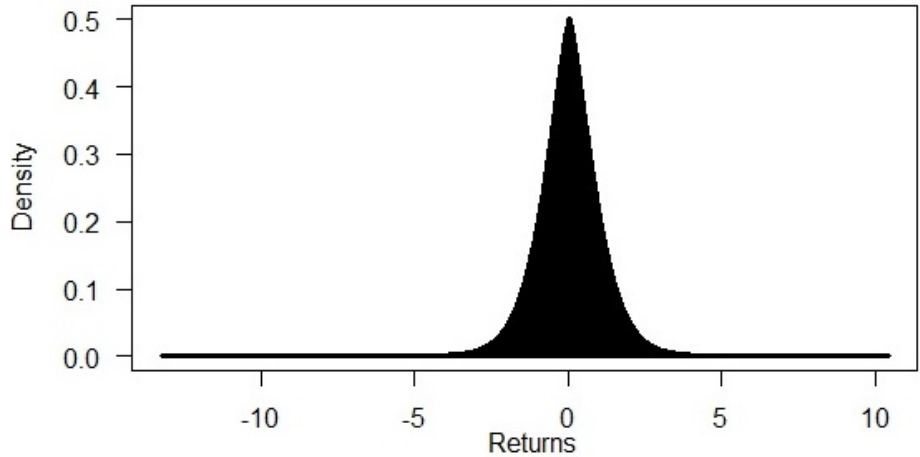

**Figure 5.13:** One-step ahead forecast distribution using a two-state GND-HMM-CCU estimated on STOXX50E returns.

### 5.3.3 Comparative analysis part III

This section aims at answering the following questions: who best fits financial returns among unconstrained and constrained GND-HMMs? and then who best fits financial returns independent GND mixture models or GND-HMMs? A comparative analysis for the 52 stocks that belong to the STOXX50E is performed in order to answer these questions. Firstly, unconstrained and constrained GND-HMMs are compared. Secondly, the best unconstrained and constrained GND-HMMs are compared with independent mixture models, i.e. the best MGND and CMGND models (Section 4.5.4). The goodness-of-fit measures are those used in the comparative analysis part I and II, namely AIC, BIC, KS and AD (Section 2.7.2).

**Data**

Data on daily closing prices of the 52 stocks that belong to the STOXX50E are collected from Yahoo Finance for the period from 01 January 2010 to 30 September 2023 (shorter for stocks entering the STOXX50E later than January 2010). Daily log returns are computed for each stock according to equation 2.4. All descriptive statistics and the JB test are computed for daily log returns for the full period.

Table B.9 in Appendix B shows for each stock (alphabetically ordered) the number of observations ($N$), mean, median, standard deviation, skewness, kurtosis, JB statistics, and minimal and maximal values. All stocks have a mean and median close to 0. The standard deviation range from 1.2624 to 2.7371. Returns are positively and negatively skewed with an empirical kurtosis greater than three, indicating fat-tails (leptokurtic distribution). For all indices the JB test rejects the null hypothesis which states that data are normally distributed.

Table B.29 in Appendix B confirms the presence of serial dependence ($Q_x(M)$) for 57.69% of the stocks that belong to the STOXX50E. By contrast, the Ljung-Box on squared returns ($Q_{xx}(M)$) shows that there is an extreme strong serial dependence of volatility for all 52 stocks.

**Results**

Tables B.32-B.35 in Appendix B present the AIC, BIC, KS and AD values for daily log-returns over the past 13 years. Unconstrained and constrained GND-HMMs are compared for each stock, and the minimum statistic is highlighted in bold to signify the best fit. Table 5.7 summarizes the percentage of times (%) that unconstrained and constrained GND-HMM are selected by a goodness-of-fit measure. AIC prefers both the UUC and CCU, while BIC prefers the CCU as it is more parsimonious than AIC. According to the KS the best models are the UUU and UCC, while AD prefers UCU, UUC, and CCU models.

|     | GND-HMM | | | | | | | |
| --- | UUU | CUU | UCU | UUC | CCU | CUC | UCC | CCC |
| AIC | 13.46 | 13.46 | | 38.46 | 34.62 | | | |
| BIC | 1.92 | 5.77 | | 5.77 | 86.54 | | | |
| KS | 34.62 | 1.92 | 9.62 | 17.31 | 1.92 | 1.92 | 25.00 | 7.69 |
| AD | 9.62 | 5.77 | 46.15 | 19.23 | 17.31 | | 1.92 | |

**Table 5.7:** Percentage of times (%) that unconstrained and constrained GND-HMMs are selected by a goodness-of-fit measure.

Tables B.36-B.39 in Appendix B present the AIC, BIC, KS and AD values. Independent mixture models and HMMs are compared for each stock, and the minimum statistic is highlighted in bold to signify the best fit.

**AIC**  AIC mainly selects (86.54%) constrained GND-HMMs. The unconstrained GND-HMM is selected 7 times. Panel (a) of Figure 5.14 shows the AIC boxplot for all models ordered by median. Constrained GND-HMMs have the lowest mean and median AIC values.

**BIC**  BIC mainly selects (98.08%) constrained GND-HMMs. The unconstrained GND-HMM model is selected only 1 times. Panel (b) of Figure 5.14 shows the BIC boxplot for all models ordered by median. Constrained GND-HMMs have the lowest mean and median AIC values.

**KS**  KS mainly selects (53.84%) CMGND models. GND-HMMs are selected 8 times. Panel (c) of Figure 5.14 shows the KS boxplot for all models ordered by median. The CMGND models have the lowest mean and median KS values. However, according to mean and median KS values, constrained GND-HMMs represent a good alternative to CMGND models.

**AD**  AD mainly selects (59.62%) CMGND models. GND-HMMs are selected 10 times. Panel (d) of Figure 5.14 shows the AD boxplot for all models ordered by median. The CMGND models have the lowest mean and median AD values. However, according to mean and median AD values, constrained GND-HMMs represent a good alternative to CMGND models.

GND-HMMs are selected by AIC and BIC, while independent mixture models are selected by KS and AD. The fact that AD and KS prefer independent mixture models is not that surprising because these goodness-of-fit measures assume i.i.d. ordered observations and do not take into account temporal dependence. To overcome this critical issue, Janczura and Weron 2012 proposed goodness-of-fit measures for the marginal distribution of HMMs based on the weighted empirical distribution function and a generalization of the KS. Costa and De Angelis 2010 provided a review of the model selection procedures for HMMs. It was pointed out that the number of observations, the state-dependent probabilities, and t.p.m. are the main factors that affect information criteria and LRTS. For example, the AIC performs better than BIC for shorter univariate time series (Costa and De Angelis 2010).

To conclude, a significant modelling flexibility is achieved by GND-HMMs. The proposed models are good candidate for modelling financial returns. The empirical application and the comparative analysis part III suggest that returns can be generated by constrained GND-HMMs as the unconstrained model is too complex. Compared to independent mixture models, GND-HMMs obtains better results in terms of AIC and BIC. In addition, it is worth noting that GND-HMMs offer a practical interpretability of financial market dynamics considering both leptokurtosis and heteroskedasticity. Hence, GND-HMMs combine flexibility, dynamicity, parsimony, and interpretability showing a clear advantage over independent mixture models.

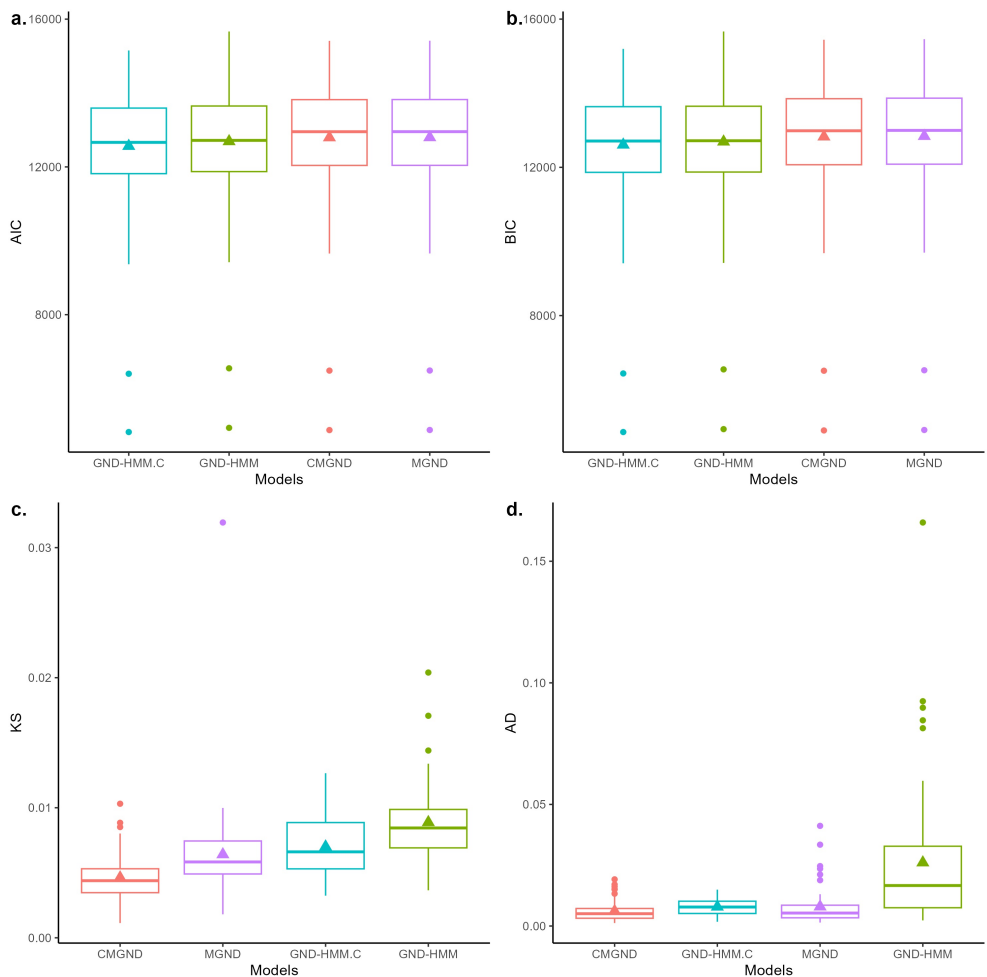

**Figure 5.14:** Comparative analysis part III, AIC, BIC, KS and AD boxplots. Models are ordered by median.

# Final Remarks | 6

This manuscript presents a comprehensive investigation into the analysis of financial returns using mixture models, with a particular focus on mixtures of generalized normal distributions (MGND). Some final remarks are resumed below.

**Chapter 2** focuses on the MGND model and the maximum likelihood estimation via expectation conditional maximization (ECM) and generalized expectation maximization (GEM) algorithms. The study highlights a degeneracy issue when considering the maximum likelihood estimator of the shape parameter $\nu_k$. The ECM algorithm (algorithm 1 in Chapter 2) may have some convergence issues when the shape parameter ($\nu_k$) is higher than 2, since variance and kurtosis are nearly constant and vary slowly for high $\nu_k$. Several solutions are provided to overcome this critical issue with algorithms 2-6 in Appendix A.

Some interesting considerations arise from the simulation study:

- ▶ The degeneracy problem can be controlled by increasing the sample size and using algorithms 2, 3, 5 and 6;
- ▶ The degeneracy issue is not observed if $\nu_k \leq 2$ in the population;
- ▶ Parameters' estimation is heavily affected when the sample size is small;
- ▶ Algorithm 1 does not prevent the degeneracy issue for $\nu_k > 2$ and small sample sizes;
- ▶ Algorithm 2 can be used as an alternative but sometimes its estimation underperforms that of algorithms 3-5-6;
- ▶ Algorithm 3 require experimenting different step-size $\alpha$ to identify the optimal value that depends on the specific scenario and sample size;
- ▶ Algorithm 4 automates the choice of the step-size but does not prevent the degeneracy issue, its estimation accuracy is better than algorithm 1 when $\nu$ is not very high;
- ▶ Algorithm 5 automates the choice of the step-size and prevent the degeneracy issue providing accurate estimates for all parameters in multiple scenarios and sample sizes;
- ▶ Algorithm 6 prevents the degeneracy issue but for some scenarios and sample sizes algorithm 5 get more accurate estimates.

The best solution is represented by algorithm 5 in Appendix A which performs a parameter estimation via GEM algorithm with the backtracking line-search and an additional $\nu_k$ step-size.

The comparative analysis highlights that the GH model is more often selected according to the AIC, KS and AD, while the MGND and MSTD models are often selected as second choice. The student-t model is the most selected according to the BIC. However, if a model is chosen several times as first according to the median value, the second and third model are still competitive.

**Chapter 3** focuses on the theoretical perspective of the MGND model through two empirical studies. The first is focused on ESG[1] indices, while the second on the G7 stock markets[2] . A two step approach is proposed. Firstly, turmoil days are objectively detected employing a two-component MGND model fitted on financial asset returns. Secondly, returns and volatility of several market indices are estimated with generalized autoregressive conditional heteroskedasticity (GARCH) models including the detected turmoil days as exogenous dummy variables. The use of the MGND model introduces an objective and data-driven way to identify financial market turmoil days. Its flexibility to accommodate different shapes of data distributions improves the accuracy in identifying turmoil days. This departure from the subjective approach provides a more rigorous way for understanding periods of market stress, minimizing the risk of subjective bias in the analysis. The inclusion of these dummy variables improves the estimates as well as the interpretation of the results.

**Chapter 4** introduces constrained mixture of generalized normal distributions (CMGND). Constraints are imposed on the location and/or scale and/or shape parameters. In the two-component case taking all possible combinations of these constraints into consideration would result to a 7-model family. The implementation of such constraints not only solves the numerical degeneracy of the log-likelihood but also enhances the interpretability and the parametric efficiency of the final solution. The MLE of the parameters is obtained via the GEM algorithm. The proposed methodology is tested on simulated data across different scenarios such as symmetric models (mixtures with common location parameters) and models with the same kurtosis (i.e. mixtures with common shape parameter).

Simulations show that higher PMRSE gains are observed when estimating the constrained parameters, i.e. the parameters with the same value across the components. However, for the unconstrained MGND model, the presence of constraints also deteriorates the estimates of the 'free' remaining parameters, i.e. parameters not constrained to be equal across the components.

In order to test the null hypothesis '*constraints hold*' (Chauveau and Hunter 2013), an appropriate likelihood ratio test (LRTS) is implemented. The LRTS performance is evaluated by a power analysis. Results show that the power reaches 100% for all sample sizes and models except for the UUC model which displays a low power for small sample sizes but reaching 100% at $N = 2000$. The empirical level is around or lower the nominal value of 5% for all the models but the CUU reporting an empirical level always greater than 5% and sometime more close to 10%.

The comparative analysis part II shows that the CMGND model is more often selected according to the AIC, KS and AD, while the GH, MGND and CMSTD models are often selected as second choice. The student-t model is the most selected according to the BIC.

In summary, a significant modelling flexibility is achieved by allowing certain parametric equalities among the different GND components. The proposed 7-model family of univariate two-component CMGND is a good candidate for modelling financial data. The empirical application and the comparative analysis part II suggest that returns can be modelled

1: Environmental, Social and Governance.

2: Canada, France, Germany, Italy, Japan, UK and US.

by the CMGND model as the MGND model is too complex. By adding constraints the CMGND model obtains better results in terms of goodness-of-fit performance. In addition, it is worth noting that mixture models offer a practical interpretability of financial returns by distinguishing periods of high and low volatility as shown in Chapter 3. To conclude, the CMGND and CMSTD models combine flexibility, parsimony, and interpretability showing a clear advantage over non-mixture models.

**Chapter 5** introduces GND-hidden Markov models (GND-HMMs) with unconstrained and constrained parameters. GND-HMMs improve the MGND and CMGND models by allowing temporal dependence with a Markov chain process for the state variable. Specifically, a first-order Markov chain governs the draws of the states from the mixture components. In this way, it is possible to improve the model ability to capture the dynamic nature of the state variable distinguishing stable and turmoil periods. The simulation study shows that the GND-HMM-CCU stands out for its efficiency in parametrization, estimation and clustering, while the GND-HMM-UCC and GND-HMM-CCC are not suitable to capture financial returns dynamics since they imply homoscedasticity[3] . The comparative analysis part III shows that GND-HMMs are selected by AIC and BIC, while independent mixture models are selected by KS and AD. The fact that AD and KS prefer independent mixture models is not that surprising because these goodness-of-fit measures assume i.i.d. ordered observations and do not take into account temporal dependence. Overall, a significant modelling flexibility is achieved by GND-HMMs. The proposed models are good candidate for modelling financial returns. The empirical application and the comparative analysis part III suggest that returns can be generated by constrained GND-HMMs as the unconstrained model is too complex. Compared to independent mixture models, GND-HMMs obtain better results in terms of AIC and BIC. In addition, it is worth noting that GND-HMMs offer a practical interpretability of financial market dynamics considering both leptokurtosis and heteroskedasticity. Hence, GND-HMMs combine flexibility, dynamicity, parsimony, and interpretability showing a clear advantage over independent mixture models.

The investigation process of this study, as well as the upgrades obtained for each proposed model can be summarized in Figure 6.1.

3: $\sigma_1 = \sigma_2, \nu_1 = \nu_2$.

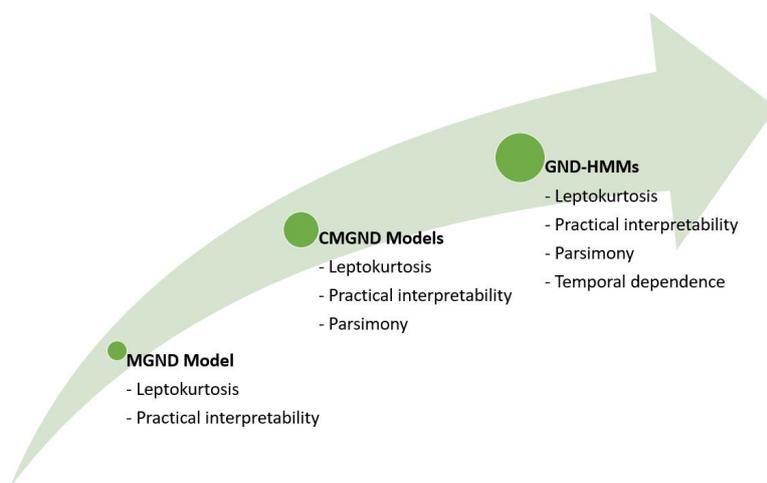

**Figure 6.1:** Investigation process and improvement of the models.

# APPENDIX

# A

# Algorithms

---

**Algorithm 2:** Parameter estimation of the MGND model via the ECM algorithm and *optim* (Brent) updates.

---

1. **require:** data $x_1, x_2, ..., x_N$.

2. **set the initial estimates:** *k-means* initialization
   minimize $\sum_{k=1}^{K} W(P_k)$, where $P_k$ denotes the set of units belonging to the $k$th cluster and $W(P_k)$ is the within cluster variation.
   $\mu_k^{(m-1)} \leftarrow \text{mean}(P_k)$, $\sigma_k^{(m-1)} \leftarrow \text{std}(P_k)$,
   $\nu_k^{(m-1)} \leftarrow$ randomly generated in $[0.5, 3]$
   $\pi_k^{(m-1)} \leftarrow$ randomly generated in $[0, 1]$, $z_{kn}^{(m-1)} \leftarrow$ Eq. 2.41,
   $\epsilon \leftarrow 10^{-5}$

3. **while** $\|\theta^{(m-1)} - \theta^{(m)}\| \leq \epsilon$ not convergence **do**
   $\mu_k^{(m)} \leftarrow \underset{\mu_k}{\text{argmax}} \, Q(\theta, \theta^{(m-1)})$, $\sigma_k^{(m)} \leftarrow \underset{\sigma_k}{\text{argmax}} \, Q(\theta, \theta^{(m-1)})$,
   $\nu_k^{(m)} \leftarrow \underset{\nu_k}{\text{argmax}} \, Q(\theta, \theta^{(m-1)})$, $\pi_k^{(m)} \leftarrow$ Eq. 2.42,
   $z_{kn}^{(m)} \leftarrow$ Eq. 2.41, $\log L(\theta^{(m)}) \leftarrow$ Eq. 2.39
   **evaluate** $\|\theta^{(m-1)} - \theta^{(m)}\| \leq \epsilon$
   $\pi_k^{(m-1)} \leftarrow \pi_k^{(m)}$, $\mu_k^{(m-1)} \leftarrow \mu_k^{(m)}$, $\sigma_k^{(m-1)} \leftarrow \sigma_k^{(m)}$, $\nu_k^{(m-1)} \leftarrow \nu_k^{(m)}$,
   $\log L(\theta^{(m-1)}) \leftarrow \log L(\theta^{(m)})$
   **end while**

4. **return** $\theta^{(m)}$, $\log L(\theta^{(m)})$

---

---

**Algorithm 3:** Parameter estimation of the MGND model via the ECM algorithm and Newton-Raphson updates with a step-size $\alpha$.

---

1. **require:** data $x_1, x_2, ..., x_N$.

2. **set the initial estimates:** *k-means* initialization

   minimize $\sum_{k=1}^{K} W(P_k)$, where $P_k$ denotes the set of units belonging to the $k$th cluster and $W(P_k)$ is the within cluster variation.

   $\mu_k^{(m-1)} \leftarrow \text{mean}(P_k)$, $\sigma_k^{(m-1)} \leftarrow \text{std}(P_k)$,

   $\nu_k^{(m-1)} \leftarrow$ randomly generated in $[0.5, 3]$

   $\pi_k^{(m-1)} \leftarrow$ randomly generated in $[0, 1]$, $z_{kn}^{(m-1)} \leftarrow$ Eq. 2.41,

   $\epsilon \leftarrow 10^{-5}$

3. **while** $\|\theta^{(m-1)} - \theta^{(m)}\| \leq \epsilon$ not convergence **do**

   $\mu_k^{(m)} \leftarrow \mu_k^{(m-1)} - \alpha \frac{g\left(\mu_k^{(m-1)}\right)}{g'\left(\mu_k^{(m-1)}\right)}$   with $\alpha \in [1, 0]$,

   $\sigma_k^{(m)} \leftarrow$ Eq. 2.47,

   $\nu_k^{(m)} \leftarrow \nu_k^{(m-1)} - \alpha \frac{g\left(\nu_k^{(m-1)}\right)}{g'\left(\nu_k^{(m-1)}\right)}$   with $\alpha \in [1, 0]$,

   $\pi_k^{(m)} \leftarrow$ Eq. 2.42, $z_{kn}^{(m)} \leftarrow$ Eq. 2.41,

   $\log L(\theta^{(m)}) \leftarrow$ Eq. 2.39

   **evaluate** $\|\theta^{(m-1)} - \theta^{(m)}\| \leq \epsilon$

   $\pi_k^{(m-1)} \leftarrow \pi_k^{(m)}$, $\mu_k^{(m-1)} \leftarrow \mu_k^{(m)}$, $\sigma_k^{(m-1)} \leftarrow \sigma_k^{(m)}$, $\nu_k^{(m-1)} \leftarrow \nu_k^{(m)}$,

   $\log L(\theta^{(m-1)}) \leftarrow \log L(\theta^{(m)})$

   **end while**

4. **return** $\theta^{(m)}$, $\log L(\theta^{(m)})$

---

---

**Algorithm 4:** Parameter estimation of the MGND model via the GEM algorithm and Newton-Raphson updates with the backtracking line-search

---

1. **require:** data $x_1, x_2, ..., x_N$.

2. **set the initial estimates:** *k-means* initialization
   minimize $\sum_{k=1}^{K} W(P_k)$, where $P_k$ denotes the set of units belonging to the $k$th cluster and $W(P_k)$ is the within cluster variation.
   $\mu_k^{(m-1)} \leftarrow$ mean$(P_k)$, $\sigma_k^{(m-1)} \leftarrow$ std$(P_k)$,
   $\nu_k^{(m-1)} \leftarrow$ randomly generated in $[0.5, 3]$
   $\pi_k^{(m-1)} \leftarrow$ randomly generated in $[0, 1]$, $z_{kn}^{(m-1)} \leftarrow$ Eq. 2.41,
   $\epsilon \leftarrow 10^{-5}$

3. **while** $\|\theta^{(m-1)} - \theta^{(m)}\| \leq \epsilon$ not convergence **do**
   $\alpha^{(i)} \leftarrow 1$, $\log L(\theta^{(m-1)}) \leftarrow$ Eq. 2.39
   **while** $ctrl = $ TRUE **do**
   $\mu_k^{(i)} \leftarrow \mu_k^{(m-1)} - \alpha^{(i)} \frac{g\left(\mu_k^{(m-1)}\right)}{g'\left(\mu_k^{(m-1)}\right)}$     with $\alpha^{(i)} \in [1, 0]$.

   $\log L(\theta^{(i)}) \leftarrow$ Eq. 2.39
   **if** $\log L(\theta^{(i)}) > \log L(\theta^{(m-1)})$ **then**
   $\mu_k^{(m)} \leftarrow \mu_k^{(i)}$ and $ctrl = $ FALSE
   **else** $\alpha^{(i)} \leftarrow 0.8\alpha^{(i)}$
   **if** $\alpha^{(i)} < 0.01$ **then**
   $\mu_k^{(m)} \leftarrow \mu_k^{(m-1)}$ and $ctrl = $ FALSE
   **end while**
   $\sigma_k^{(m)} \leftarrow$ Eq. 2.47
   $\alpha^{(i)} \leftarrow 1$, $\log L(\theta^{(m-1)}) \leftarrow$ Eq. 2.39
   **while** $ctrl = $ TRUE **do**
   $\nu_k^{(i)} \leftarrow \nu_k^{(m-1)} - \alpha^{(i)} \frac{g\left(\nu_k^{(m-1)}\right)}{g'\left(\nu_k^{(m-1)}\right)}$     with $\alpha^{(i)} \in [1, 0]$.

   $\log L(\theta^{(i)}) \leftarrow$ Eq. 2.39
   **if** $\log L(\theta^{(i)}) > \log L(\theta^{(m-1)})$ **then**
   $\nu_k^{(m)} \leftarrow \nu_k^{(i)}$ and $ctrl = $ FALSE
   **else** $\alpha^{(i)} \leftarrow 0.8\alpha^{(i)}$
   **if** $\alpha^{(i)} < 0.01$ **then**
   $\nu_k^{(m)} \leftarrow \nu_k^{(m-1)}$ and $ctrl = $ FALSE
   **end while**
   $\pi_k^{(m)} \leftarrow$ Eq. 2.42, $z_{kn}^{(m)} \leftarrow$ Eq. 2.41, $\log L(\theta^{(m)}) \leftarrow$ Eq. 2.39
   **evaluate** $\|\theta^{(m-1)} - \theta^{(m)}\| \leq \epsilon$
   $\pi_k^{(m-1)} \leftarrow \pi_k^{(m)}$, $\mu_k^{(m-1)} \leftarrow \mu_k^{(m)}$, $\sigma_k^{(m-1)} \leftarrow \sigma_k^{(m)}$, $\nu_k^{(m-1)} \leftarrow \nu_k^{(m)}$,
   $\log L(\theta^{(m-1)}) \leftarrow \log L(\theta^{(m)})$
   **end while**

4. **return** $\theta^{(m)}$, $\log L(\theta^{(m)})$

---

**Algorithm 5:** Parameter estimation of the MGND model via the GEM algorithm and Newton-Raphson updates with the backtracking line-search and adaptive step-size $1/\nu$

---

1. **require:** data $x_1, x_2, ..., x_N$.
2. **set the initial estimates:** *k-means* initialization
   minimize $\sum_{k=1}^{K} W(P_k)$, where $P_k$ denotes the set of units belonging to the $k$th cluster and $W(P_k)$ is the within cluster variation.
   $\mu_k^{(m-1)} \leftarrow \text{mean}(P_k), \sigma_k^{(m-1)} \leftarrow \text{std}(P_k),$
   $\nu_k^{(m-1)} \leftarrow$ randomly generated in $[0.5, 3]$
   $\pi_k^{(m-1)} \leftarrow$ randomly generated in $[0, 1], z_{kn}^{(m-1)} \leftarrow$ Eq. 2.41,
   $\epsilon \leftarrow 10^{-5}$
3. **while** $\|\theta^{(m-1)} - \theta^{(m)}\| \leq \epsilon$ not convergence **do**
   $\alpha^{(i)} \leftarrow 1, \log L(\theta^{(m-1)}) \leftarrow$ Eq. 2.39
   **while** $ctrl = \text{TRUE}$ **do**
   $\mu_k^{(i)} \leftarrow \mu_k^{(m-1)} - \alpha^{(i)} \frac{g\left(\mu_k^{(m-1)}\right)}{g'\left(\mu_k^{(m-1)}\right)}$    with $\alpha^{(i)} \in [1, 0]$.
   $\log L(\theta^{(i)}) \leftarrow$ Eq. 2.39
   **if** $\log L(\theta^{(i)}) > \log L(\theta^{(m-1)})$ **then**
   $\mu_k^{(m)} \leftarrow \mu_k^{(i)}$ and $ctrl = \text{FALSE}$
   **else** $\alpha^{(i)} \leftarrow 0.8\alpha^{(i)}$
   **if** $\alpha^{(i)} < 0.01$ **then**
   $\mu_k^{(m)} \leftarrow \mu_k^{(m-1)}$ and $ctrl = \text{FALSE}$
   **end while**
   $\sigma_k^{(m)} \leftarrow$ Eq. 2.47
   $\alpha^{(i)} \leftarrow 1$
   $\log L(\theta^{(m-1)}) \leftarrow$ Eq. 2.39
   **while** $ctrl = \text{TRUE}$ **do**
   $\nu_k^{(i)} \leftarrow \nu_k^{(m-1)} - \alpha^{(i)} \frac{1}{\nu_k^{(m-1)}} \frac{g\left(\nu_k^{(m-1)}\right)}{g'\left(\nu_k^{(m-1)}\right)}$    with $\alpha^{(i)} \in [1, 0]$.
   $\log L(\theta^{(i)}) \leftarrow$ Eq. 2.39
   **if** $\log L(\theta^{(i)}) > \log L(\theta^{(m-1)})$ **then**
   $\nu_k^{(m)} \leftarrow \nu_k^{(i)}$ and $ctrl = \text{FALSE}$
   **else** $\alpha^{(i)} \leftarrow 0.8\alpha^{(i)}$
   **if** $\alpha^{(i)} < 0.01$ **then**
   $\nu_k^{(m)} \leftarrow \nu_k^{(m-1)}$ and $ctrl = \text{FALSE}$
   **end while**
   $\pi_k^{(m)} \leftarrow$ Eq. 2.42, $z_{kn}^{(m)} \leftarrow$ Eq. 2.41, $\log L(\theta^{(m)}) \leftarrow$ Eq. 2.39
   **evaluate** $\|\theta^{(m-1)} - \theta^{(m)}\| \leq \epsilon$
   $\pi_k^{(m-1)} \leftarrow \pi_k^{(m)}, \mu_k^{(m-1)} \leftarrow \mu_k^{(m)}, \sigma_k^{(m-1)} \leftarrow \sigma_k^{(m)}, \nu_k^{(m-1)} \leftarrow \nu_k^{(m)},$
   $\log L(\theta^{(m-1)}) \leftarrow \log L(\theta^{(m)})$
   **end while**
4. **return** $\theta^{(m)}, \log L(\theta^{(m)})$

---

---

**Algorithm 6:** Parameter estimation of the MGND model via the ECM algorithm and Newton-Raphson updates with the adaptive step-size $1/\nu$

---

1. **require:** data $x_1, x_2, ..., x_N$.
2. **set the initial estimates:** *k-means* initialization

   minimize $\sum_{k=1}^{K} W(P_k)$, where $P_k$ denotes the set of units belonging to the $k$th cluster and $W(P_k)$ is the within cluster variation.

   $\mu_k^{(m-1)} \leftarrow \text{mean}(P_k), \sigma_k^{(m-1)} \leftarrow \text{std}(P_k),$

   $\nu_k^{(m-1)} \leftarrow$ randomly generated in $[0.5, 3]$

   $\pi_k^{(m-1)} \leftarrow$ randomly generated in $[0, 1], z_{kn}^{(m-1)} \leftarrow$ Eq. 2.41,

   $\epsilon \leftarrow 10^{-5}$

3. **while** $\|\theta^{(m-1)} - \theta^{(m)}\| \le \epsilon$ not convergence **do**

   $\mu_k^{(m)} \leftarrow$ Eq. 2.46, $\sigma_k^{(m)} \leftarrow$ Eq. 2.47,

   $\nu_k^{(m)} \leftarrow \nu_k^{(m)} = \nu_k^{(m-1)} - \frac{1}{\nu_k^{(m-1)}} \frac{g\left(\nu_k^{(m-1)}\right)}{g'\left(\nu_k^{(m-1)}\right)}, \pi_k^{(m)} \leftarrow$ Eq. 2.42,

   $z_{kn}^{(m)} \leftarrow$ Eq. 2.41, $\log L(\theta^{(m)}) \leftarrow$ Eq. 2.39

   **evaluate** $\|\theta^{(m-1)} - \theta^{(m)}\| \le \epsilon$

   $\pi_k^{(m-1)} \leftarrow \pi_k^{(m)}, \mu_k^{(m-1)} \leftarrow \mu_k^{(m)}, \sigma_k^{(m-1)} \leftarrow \sigma_k^{(m)}, \nu_k^{(m-1)} \leftarrow \nu_k^{(m)},$

   $\log L(\theta^{(m-1)}) \leftarrow \log L(\theta^{(m)})$

   **end while**

4. **return** $\theta^{(m)}, \log L(\theta^{(m)})$

---

---

**Algorithm 7:** Parameter estimation of the CMGND model via the GEM algorithm and Newton-Raphson updates with the backtracking line-search and adaptive step-size $1/\nu$

---

1. **require:** data $x_1, x_2, ..., x_N$.

2. **set the initial estimates:** *k-means* initialization
   minimize $\sum_{k=1}^{K} W(P_k)$, where $P_k$ denotes the set of units belonging to the $k$th cluster and $W(P_k)$ is the within cluster variation.
   $\mu_k^{(m-1)} \leftarrow \text{mean}(P_k)$, $\sigma_k^{(m-1)} \leftarrow \text{std}(P_k)$,
   $\nu_k^{(m-1)} \leftarrow$ randomly generated in $[0.5, 3]$
   $\pi_k^{(m-1)} \leftarrow$ randomly generated in $[0, 1]$, $z_{kn}^{(m-1)} \leftarrow$ Eq. 2.41,
   $\epsilon \leftarrow 10^{-5}$

3. **while** $\|\theta^{(m-1)} - \theta^{(m)}\| \leq \epsilon$ not convergence **do**
   $\alpha^{(i)} \leftarrow 1$, $\log L(\theta^{(m-1)}) \leftarrow$ Eq. 4.2
   **while** $ctrl = \text{TRUE}$ **do**
   $\mu_k^{(i)} \leftarrow$ Eq. 4.6 (if $\mu_k$ is constrained) or Eq. 2.54 (if $\mu_k$ is unconstrained)
   $\log L(\theta^{(i)}) \leftarrow$ Eq. 4.2
   **if** $\log L(\theta^{(i)}) > \log L(\theta^{(m-1)})$ **then**
   $\mu_k^{(m)} \leftarrow \mu_k^{(i)}$ and $ctrl = \text{FALSE}$
   **else** $\alpha^{(i)} \leftarrow 0.8\alpha^{(i)}$
   **if** $\alpha^{(i)} < 0.01$ **then**
   $\mu_k^{(m)} \leftarrow \mu_k^{(m-1)}$ and $ctrl = \text{FALSE}$
   **end while**
   $\alpha^{(i)} \leftarrow 1$, $\log L(\theta^{(m-1)}) \leftarrow$ Eq. 4.2
   **while** $ctrl = \text{TRUE}$ **do**
   $\sigma_k^{(i)} \leftarrow$ Eq. 4.10 (if $\sigma_k$ is constrained) or Eq. 2.47 (if $\sigma_k$ is unconstrained)
   $\log L(\theta^{(i)}) \leftarrow$ Eq. 4.2
   **if** $\log L(\theta^{(i)}) > \log L(\theta^{(m-1)})$ **then**
   $\sigma_k^{(m)} \leftarrow \sigma_k^{(i)}$ and $ctrl = \text{FALSE}$
   **else** $\alpha^{(i)} \leftarrow 0.8\alpha^{(i)}$
   **if** $\alpha^{(i)} < 0.01$ **then**
   $\sigma_k^{(m)} \leftarrow \sigma_k^{(m-1)}$ and $ctrl = \text{FALSE}$
   **end while**
   $\alpha^{(i)} \leftarrow 1$
   $\log L(\theta^{(m-1)}) \leftarrow$ Eq. 4.2
   **while** $ctrl = \text{TRUE}$ **do**
   $\nu_k^{(i)} \leftarrow$ Eq. 4.13 (if $\nu_k$ is constrained) or Eq. 2.55 (if $\nu_k$ is unconstrained)
   $\log L(\theta^{(i)}) \leftarrow$ Eq. 4.2
   **if** $\log L(\theta^{(i)}) > \log L(\theta^{(m-1)})$ **then**
   $\nu_k^{(m)} \leftarrow \nu_k^{(i)}$ and $ctrl = \text{FALSE}$
   **else** $\alpha^{(i)} \leftarrow 0.8\alpha^{(i)}$
   **if** $\alpha^{(i)} < 0.01$ **then**
   $\nu_k^{(m)} \leftarrow \nu_k^{(m-1)}$ and $ctrl = \text{FALSE}$
   **end while**
   $\pi_k^{(m)} \leftarrow$ Eq. 4.5, $z_{kn}^{(m)} \leftarrow$ Eq. 4.4, $\log L(\theta^{(m)}) \leftarrow$ Eq. 4.2
   **evaluate** $\|\theta^{(m-1)} - \theta^{(m)}\| \leq \epsilon$
   $\pi_k^{(m-1)} \leftarrow \pi_k^{(m)}, \mu_k^{(m-1)} \leftarrow \mu_k^{(m)}, \sigma_k^{(m-1)} \leftarrow \sigma_k^{(m)}, \nu_k^{(m-1)} \leftarrow \nu_k^{(m)},$
   $\log L(\theta^{(m-1)}) \leftarrow \log L(\theta^{(m)})$
   **end while**

4. **return** $\theta^{(m)}$, $\log L(\theta^{(m)})$

---

# B

## Tables

| N | $\pi_1$ 0.7 | $\mu_1$ 0 | $\sigma_1$ 2 | $\nu_1$ 5 | $\pi_2$ 0.3 | $\mu_2$ 0 | $\sigma_2$ 5 | $\nu_2$ 3 |
|---|---|---|---|---|---|---|---|---|
| Algorithm 1 | | | | | | | | |
| 500 | 0.65 | 0.00 | 2.11 | 18.43 | 0.35 | -0.03 | 4.67 | 3.54 |
| 2000 | 0.69 | -0.01 | 1.99 | 5.30 | 0.31 | 0.00 | 4.91 | 3.05 |
| Algorithm 2 | | | | | | | | |
| 500 | 0.61 | -0.02 | 1.86 | 5.89 | 0.39 | -0.01 | 4.20 | 3.02 |
| 2000 | 0.65 | -0.01 | 1.98 | 5.70 | 0.35 | 0.00 | 4.45 | 2.63 |
| Algorithm 3, $\alpha = 0.10$ | | | | | | | | |
| 500 | 0.58 | 0.01 | 1.86 | 4.85 | 0.42 | -0.05 | 3.84 | 2.51 |
| 2000 | 0.57 | -0.01 | 1.84 | 3.91 | 0.43 | 0.00 | 3.30 | 1.78 |
| Algorithm 3, $\alpha = 0.20$ | | | | | | | | |
| 500 | 0.62 | 0.02 | 1.99 | 6.97 | 0.38 | -0.05 | 4.23 | 2.93 |
| 2000 | 0.64 | 0.00 | 1.98 | 5.40 | 0.36 | 0.00 | 4.22 | 2.39 |
| Algorithm 3, $\alpha = 0.50$ | | | | | | | | |
| 500 | 0.64 | 0.00 | 2.11 | 13.88 | 0.36 | -0.03 | 4.51 | 3.36 |
| 2000 | 0.67 | -0.01 | 1.99 | 5.41 | 0.33 | 0.00 | 4.66 | 2.78 |
| Algorithm 3, $\alpha = 0.80$ | | | | | | | | |
| 500 | 0.65 | -0.01 | 2.09 | 15.78 | 0.35 | -0.03 | 4.62 | 3.50 |
| 2000 | 0.68 | -0.01 | 1.99 | 5.35 | 0.32 | 0.00 | 4.84 | 2.97 |
| Algorithm 4 | | | | | | | | |
| 500 | 0.55 | 0.00 | 1.80 | 18.89 | 0.45 | -0.03 | 3.50 | 2.62 |
| 2000 | 0.61 | 0.05 | 1.84 | 5.28 | 0.39 | -0.05 | 4.03 | 2.48 |
| Algorithm 5 | | | | | | | | |
| 500 | 0.65 | 0.01 | 2.01 | 6.21 | 0.35 | -0.04 | 4.62 | 3.28 |
| 2000 | 0.67 | -0.01 | 1.99 | 5.40 | 0.33 | 0.00 | 4.63 | 2.72 |
| Algorithm 6 | | | | | | | | |
| 500 | 0.66 | 0.01 | 2.07 | 6.21 | 0.34 | -0.04 | 4.70 | 3.33 |
| 2000 | 0.67 | 0.00 | 1.99 | 5.27 | 0.33 | 0.00 | 4.67 | 2.76 |

**Table B.1:** MGND simulations, estimated parameters, scenario 1.

| N | $\pi_1$ 0.7 | $\mu_1$ 0 | $\sigma_1$ 2 | $\nu_1$ 5 | $\pi_2$ 0.3 | $\mu_2$ 0 | $\sigma_2$ 5 | $\nu_2$ 3 |
|---|---|---|---|---|---|---|---|---|
| **Algorithm 1** | | | | | | | | |
| 500 | 0.14 | 0.13 | 0.73 | 52.67 | 0.14 | 0.25 | 1.29 | 1.98 |
| 2000 | 0.04 | 0.03 | 0.04 | 1.59 | 0.04 | 0.11 | 0.52 | 0.63 |
| **Algorithm 2** | | | | | | | | |
| 500 | 0.18 | 0.41 | 0.52 | 2.54 | 0.18 | 0.33 | 1.72 | 1.83 |
| 2000 | 0.09 | 0.10 | 0.10 | 1.35 | 0.09 | 0.12 | 0.98 | 0.85 |
| **Algorithm 3, $\alpha = 0.10$** | | | | | | | | |
| 500 | 0.23 | 0.47 | 0.69 | 5.90 | 0.23 | 0.34 | 1.87 | 1.61 |
| 2000 | 0.20 | 0.33 | 0.34 | 1.63 | 0.20 | 0.21 | 2.02 | 1.45 |
| **Algorithm 3, $\alpha = 0.20$** | | | | | | | | |
| 500 | 0.18 | 0.31 | 0.68 | 10.35 | 0.18 | 0.28 | 1.57 | 1.65 |
| 2000 | 0.09 | 0.07 | 0.07 | 1.08 | 0.09 | 0.12 | 1.14 | 0.93 |
| **Algorithm 3, $\alpha = 0.50$** | | | | | | | | |
| 500 | 0.16 | 0.21 | 0.80 | 38.85 | 0.16 | 0.25 | 1.39 | 1.91 |
| 2000 | 0.06 | 0.03 | 0.05 | 1.07 | 0.06 | 0.11 | 0.72 | 0.70 |
| **Algorithm 3, $\alpha = 0.80$** | | | | | | | | |
| 500 | 0.15 | 0.14 | 0.67 | 49.01 | 0.15 | 0.25 | 1.32 | 2.02 |
| 2000 | 0.05 | 0.03 | 0.04 | 1.43 | 0.05 | 0.11 | 0.60 | 0.64 |
| **Algorithm 4** | | | | | | | | |
| 500 | 0.23 | 0.56 | 0.90 | 58.15 | 0.23 | 0.45 | 2.38 | 2.05 |
| 2000 | 0.15 | 0.39 | 0.40 | 2.92 | 0.15 | 0.32 | 1.75 | 1.11 |
| **Algorithm 5** | | | | | | | | |
| 500 | 0.13 | 0.14 | 0.64 | 3.05 | 0.13 | 0.24 | 1.14 | 1.53 |
| 2000 | 0.06 | 0.03 | 0.05 | 0.95 | 0.06 | 0.11 | 0.70 | 0.67 |
| **Algorithm 6** | | | | | | | | |
| 500 | 0.13 | 0.20 | 0.52 | 3.06 | 0.13 | 0.26 | 1.25 | 1.61 |
| 2000 | 0.05 | 0.03 | 0.05 | 0.89 | 0.05 | 0.11 | 0.65 | 0.64 |

**Table B.2:** MGND simulations, MSE, scenario 1.

| N | $\pi_1$ | $\mu_1$ | $\sigma_1$ | $\nu_1$ | $\pi_2$ | $\mu_2$ | $\sigma_2$ | $\nu_2$ |
|---|---|---|---|---|---|---|---|---|
|  | 0.7 | 1 | 1 | 2 | 0.3 | 5 | 3 | 4 |
| Algorithm 1 | | | | | | | | |
| 500 | 0.66 | 1.00 | 0.99 | 2.14 | 0.34 | 4.62 | 3.35 | 10.30 |
| 2000 | 0.68 | 1.00 | 1.00 | 2.05 | 0.32 | 4.80 | 3.19 | 4.39 |
| Algorithm 2 | | | | | | | | |
| 500 | 0.69 | 1.00 | 1.00 | 2.10 | 0.31 | 4.88 | 3.09 | 4.98 |
| 2000 | 0.70 | 1.00 | 1.00 | 2.01 | 0.30 | 4.98 | 3.02 | 4.26 |
| Algorithm 3, $\alpha = 0.10$ | | | | | | | | |
| 500 | 0.72 | 1.05 | 1.05 | 2.10 | 0.28 | 5.22 | 2.70 | 3.88 |
| 2000 | 0.72 | 1.04 | 1.05 | 2.06 | 0.28 | 5.26 | 2.70 | 3.52 |
| Algorithm 3, $\alpha = 0.20$ | | | | | | | | |
| 500 | 0.70 | 1.03 | 1.03 | 2.10 | 0.30 | 5.06 | 2.87 | 4.38 |
| 2000 | 0.71 | 1.03 | 1.03 | 2.04 | 0.29 | 5.17 | 2.80 | 3.80 |
| Algorithm 3, $\alpha = 0.50$ | | | | | | | | |
| 500 | 0.69 | 1.01 | 1.01 | 2.12 | 0.31 | 4.89 | 3.05 | 5.57 |
| 2000 | 0.70 | 1.01 | 1.01 | 2.03 | 0.30 | 4.98 | 2.99 | 4.02 |
| Algorithm 3, $\alpha = 0.80$ | | | | | | | | |
| 500 | 0.67 | 1.00 | 1.00 | 2.13 | 0.33 | 4.73 | 3.23 | 6.81 |
| 2000 | 0.68 | 1.00 | 1.00 | 2.05 | 0.32 | 4.86 | 3.13 | 4.25 |
| Algorithm 4 | | | | | | | | |
| 500 | 0.66 | 1.00 | 0.99 | 2.14 | 0.34 | 4.66 | 3.31 | 8.39 |
| 2000 | 0.68 | 1.00 | 1.00 | 2.05 | 0.32 | 4.78 | 3.21 | 4.44 |
| Algorithm 5 | | | | | | | | |
| 500 | 0.68 | 1.01 | 1.00 | 2.15 | 0.32 | 4.85 | 3.10 | 4.65 |
| 2000 | 0.69 | 1.01 | 1.01 | 2.06 | 0.31 | 4.95 | 3.05 | 4.18 |
| Algorithm 6 | | | | | | | | |
| 500 | 0.68 | 1.01 | 1.00 | 2.15 | 0.32 | 4.81 | 3.13 | 4.74 |
| 2000 | 0.70 | 1.01 | 1.01 | 2.04 | 0.30 | 4.98 | 3.02 | 4.16 |

**Table B.3:** MGND simulations, estimated parameters, scenario 2.

| N | $\pi_1$ 0.7 | $\mu_1$ 1 | $\sigma_1$ 1 | $\nu_1$ 2 | $\pi_2$ 0.3 | $\mu_2$ 5 | $\sigma_2$ 3 | $\nu_2$ 4 |
|---|---|---|---|---|---|---|---|---|
| Algorithm 1 | | | | | | | | |
| 500 | 0.07 | 0.05 | 0.10 | 0.48 | 0.07 | 0.73 | 0.75 | 36.41 |
| 2000 | 0.05 | 0.03 | 0.04 | 0.18 | 0.05 | 0.50 | 0.52 | 1.04 |
| Algorithm 2 | | | | | | | | |
| 500 | 0.05 | 0.05 | 0.08 | 0.41 | 0.05 | 0.52 | 0.49 | 2.16 |
| 2000 | 0.03 | 0.03 | 0.04 | 0.14 | 0.03 | 0.29 | 0.30 | 0.77 |
| Algorithm 3, $\alpha = 0.10$ | | | | | | | | |
| 500 | 0.04 | 0.07 | 0.10 | 0.31 | 0.04 | 0.41 | 0.43 | 1.29 |
| 2000 | 0.02 | 0.05 | 0.06 | 0.16 | 0.02 | 0.28 | 0.32 | 0.68 |
| Algorithm 3, $\alpha = 0.20$ | | | | | | | | |
| 500 | 0.04 | 0.06 | 0.09 | 0.35 | 0.04 | 0.45 | 0.44 | 2.18 |
| 2000 | 0.02 | 0.04 | 0.05 | 0.14 | 0.02 | 0.23 | 0.24 | 0.56 |
| Algorithm 3, $\alpha = 0.50$ | | | | | | | | |
| 500 | 0.05 | 0.05 | 0.09 | 0.43 | 0.05 | 0.52 | 0.50 | 7.39 |
| 2000 | 0.03 | 0.03 | 0.04 | 0.15 | 0.03 | 0.33 | 0.33 | 0.62 |
| Algorithm 3, $\alpha = 0.80$ | | | | | | | | |
| 500 | 0.06 | 0.05 | 0.09 | 0.46 | 0.06 | 0.65 | 0.65 | 13.51 |
| 2000 | 0.05 | 0.03 | 0.04 | 0.16 | 0.05 | 0.44 | 0.45 | 0.88 |
| Algorithm 4 | | | | | | | | |
| 500 | 0.07 | 0.05 | 0.09 | 0.47 | 0.07 | 0.69 | 0.71 | 21.65 |
| 2000 | 0.05 | 0.03 | 0.05 | 0.17 | 0.05 | 0.53 | 0.55 | 1.11 |
| Algorithm 5 | | | | | | | | |
| 500 | 0.06 | 0.05 | 0.09 | 0.51 | 0.06 | 0.58 | 0.55 | 1.71 |
| 2000 | 0.04 | 0.03 | 0.04 | 0.20 | 0.04 | 0.42 | 0.41 | 0.70 |
| Algorithm 6 | | | | | | | | |
| 500 | 0.06 | 0.05 | 0.09 | 0.50 | 0.06 | 0.61 | 0.59 | 1.84 |
| 2000 | 0.04 | 0.03 | 0.04 | 0.18 | 0.04 | 0.38 | 0.37 | 0.70 |

**Table B.4:** MGND simulations estimated, MSE, scenario 2.

| N | $\pi_1$ 0.7 | $\mu_1$ 0 | $\sigma_1$ 1 | $\nu_1$ 2 | $\pi_2$ 0.3 | $\mu_2$ 0 | $\sigma_2$ 3 | $\nu_2$ 0.8 |
|---|---|---|---|---|---|---|---|---|
| Algorithm 1 | | | | | | | | |
| 500 | 0.72 | 0.00 | 0.98 | 1.95 | 0.28 | -0.06 | 4.78 | 1.05 |
| 2000 | 0.71 | 0.00 | 0.99 | 1.95 | 0.29 | -0.03 | 3.89 | 0.89 |
| Algorithm 2 | | | | | | | | |
| 500 | 0.69 | 0.00 | 0.99 | 2.10 | 0.31 | -0.06 | 3.96 | 0.96 |
| 2000 | 0.70 | 0.00 | 1.00 | 2.01 | 0.30 | -0.02 | 3.45 | 0.85 |
| Algorithm 3, $\alpha = 0.10$ | | | | | | | | |
| 500 | 0.74 | 0.00 | 0.98 | 1.87 | 0.26 | -0.09 | 5.57 | 1.13 |
| 2000 | 0.73 | 0.00 | 0.99 | 1.89 | 0.27 | -0.01 | 5.00 | 1.02 |
| Algorithm 3, $\alpha = 0.20$ | | | | | | | | |
| 500 | 0.74 | 0.00 | 0.98 | 1.88 | 0.26 | -0.09 | 5.36 | 1.11 |
| 2000 | 0.73 | 0.00 | 0.99 | 1.86 | 0.27 | -0.01 | 4.90 | 1.01 |
| Algorithm 3, $\alpha = 0.50$ | | | | | | | | |
| 500 | 0.73 | 0.00 | 0.98 | 1.91 | 0.27 | -0.09 | 5.03 | 1.07 |
| 2000 | 0.72 | 0.00 | 0.98 | 1.86 | 0.28 | -0.01 | 4.43 | 0.95 |
| Algorithm 3, $\alpha = 0.80$ | | | | | | | | |
| 500 | 0.72 | 0.00 | 0.98 | 1.93 | 0.28 | -0.10 | 4.94 | 1.06 |
| 2000 | 0.71 | 0.00 | 0.99 | 1.93 | 0.29 | -0.02 | 3.98 | 0.90 |
| Algorithm 4 | | | | | | | | |
| 500 | 0.72 | 0.00 | 0.99 | 2.01 | 0.28 | -0.08 | 4.36 | 1.01 |
| 2000 | 0.71 | 0.00 | 1.00 | 1.99 | 0.29 | -0.01 | 3.60 | 0.86 |
| Algorithm 5 | | | | | | | | |
| 500 | 0.72 | 0.00 | 0.99 | 1.99 | 0.28 | -0.09 | 4.36 | 1.00 |
| 2000 | 0.71 | 0.00 | 1.00 | 1.98 | 0.29 | -0.01 | 3.60 | 0.86 |
| Algorithm 6 | | | | | | | | |
| 500 | 0.72 | 0.00 | 0.98 | 1.91 | 0.28 | -0.07 | 4.93 | 1.06 |
| 2000 | 0.70 | 0.00 | 0.99 | 1.96 | 0.30 | -0.01 | 3.77 | 0.88 |

**Table B.5:** MGND simulations, estimated parameters, scenario 3.

| N | $\pi_1$ | $\mu_1$ | $\sigma_1$ | $\nu_1$ | $\pi_2$ | $\mu_2$ | $\sigma_2$ | $\nu_2$ |
|---|---|---|---|---|---|---|---|---|
|   | 0.7 | 0 | 1 | 2 | 0.3 | 0 | 3 | 0.8 |
| Algorithm 1 | | | | | | | | |
| 500 | 0.07 | 0.05 | 0.10 | 0.45 | 0.07 | 0.58 | 3.06 | 0.60 |
| 2000 | 0.05 | 0.03 | 0.04 | 0.22 | 0.05 | 0.25 | 1.91 | 0.21 |
| Algorithm 2 | | | | | | | | |
| 500 | 0.08 | 0.06 | 0.10 | 0.61 | 0.08 | 0.55 | 2.89 | 0.59 |
| 2000 | 0.03 | 0.03 | 0.04 | 0.21 | 0.03 | 0.22 | 1.29 | 0.14 |
| Algorithm 3, $\alpha = 0.10$ | | | | | | | | |
| 500 | 0.06 | 0.05 | 0.09 | 0.37 | 0.06 | 0.60 | 3.18 | 0.55 |
| 2000 | 0.04 | 0.02 | 0.05 | 0.21 | 0.04 | 0.25 | 2.12 | 0.23 |
| Algorithm 3, $\alpha = 0.20$ | | | | | | | | |
| 500 | 0.05 | 0.05 | 0.10 | 0.41 | 0.05 | 0.61 | 3.05 | 0.57 |
| 2000 | 0.04 | 0.03 | 0.05 | 0.22 | 0.04 | 0.25 | 2.07 | 0.22 |
| Algorithm 3, $\alpha = 0.50$ | | | | | | | | |
| 500 | 0.06 | 0.05 | 0.10 | 0.43 | 0.06 | 0.60 | 3.04 | 0.59 |
| 2000 | 0.04 | 0.02 | 0.05 | 0.22 | 0.04 | 0.25 | 2.02 | 0.22 |
| Algorithm 3, $\alpha = 0.80$ | | | | | | | | |
| 500 | 0.07 | 0.05 | 0.10 | 0.44 | 0.07 | 0.60 | 3.07 | 0.59 |
| 2000 | 0.05 | 0.03 | 0.05 | 0.24 | 0.05 | 0.26 | 1.92 | 0.21 |
| Algorithm 4 | | | | | | | | |
| 500 | 0.06 | 0.05 | 0.09 | 0.45 | 0.06 | 0.60 | 2.76 | 0.58 |
| 2000 | 0.03 | 0.03 | 0.04 | 0.18 | 0.03 | 0.24 | 1.21 | 0.13 |
| Algorithm 5 | | | | | | | | |
| 500 | 0.06 | 0.05 | 0.09 | 0.43 | 0.06 | 0.60 | 2.74 | 0.56 |
| 2000 | 0.03 | 0.03 | 0.04 | 0.18 | 0.03 | 0.24 | 1.22 | 0.13 |
| Algorithm 6 | | | | | | | | |
| 500 | 0.07 | 0.05 | 0.12 | 0.44 | 0.07 | 0.59 | 3.11 | 0.58 |
| 2000 | 0.05 | 0.03 | 0.04 | 0.23 | 0.05 | 0.24 | 1.89 | 0.21 |

**Table B.6:** MGND simulations estimated, MSE, scenario 3.

| N | $\pi_1$ | $\mu_1$ | $\sigma_1$ | $\nu_1$ | $\pi_2$ | $\mu_2$ | $\sigma_2$ | $\nu_2$ |
|---|---------|---------|------------|---------|---------|---------|------------|---------|
|   | 0.7     | 1       | 1          | 2       | 0.3     | 5       | 3          | 0.8     |
| Algorithm 1 | | | | | | | | |
| 500  | 0.7 | 1.00 | 0.98 | 2.01 | 0.3 | 4.92 | 3.82 | 0.91 |
| 2000 | 0.7 | 1.00 | 0.98 | 1.94 | 0.3 | 4.97 | 3.91 | 0.91 |
| Algorithm 2 | | | | | | | | |
| 500  | 0.7 | 1.00 | 1.00 | 2.06 | 0.3 | 4.96 | 3.08 | 0.82 |
| 2000 | 0.7 | 1.00 | 1.00 | 2.02 | 0.3 | 4.98 | 3.06 | 0.81 |
| Algorithm 3, $\alpha = 0.10$ | | | | | | | | |
| 500  | 0.7 | 1.00 | 0.98 | 2.02 | 0.3 | 4.91 | 3.84 | 0.93 |
| 2000 | 0.7 | 1.00 | 0.99 | 1.99 | 0.3 | 4.93 | 3.85 | 0.92 |
| Algorithm 3, $\alpha = 0.20$ | | | | | | | | |
| 500  | 0.7 | 1.00 | 0.99 | 2.03 | 0.3 | 4.93 | 3.67 | 0.90 |
| 2000 | 0.7 | 1.00 | 0.99 | 2.01 | 0.3 | 4.93 | 3.73 | 0.90 |
| Algorithm 3, $\alpha = 0.50$ | | | | | | | | |
| 500  | 0.7 | 1.00 | 0.99 | 2.04 | 0.3 | 4.92 | 3.69 | 0.90 |
| 2000 | 0.7 | 1.00 | 0.99 | 2.00 | 0.3 | 4.94 | 3.68 | 0.89 |
| Algorithm 3, $\alpha = 0.80$ | | | | | | | | |
| 500  | 0.7 | 1.00 | 0.99 | 2.04 | 0.3 | 4.95 | 3.71 | 0.90 |
| 2000 | 0.7 | 1.00 | 0.99 | 1.98 | 0.3 | 4.95 | 3.78 | 0.90 |
| Algorithm 4 | | | | | | | | |
| 500  | 0.7 | 1.00 | 1.00 | 2.06 | 0.3 | 4.95 | 3.15 | 0.83 |
| 2000 | 0.7 | 1.00 | 1.00 | 2.03 | 0.3 | 4.98 | 3.10 | 0.81 |
| Algorithm 5 | | | | | | | | |
| 500  | 0.7 | 1.00 | 0.99 | 2.04 | 0.3 | 4.94 | 3.14 | 0.83 |
| 2000 | 0.7 | 1.00 | 1.00 | 2.01 | 0.3 | 4.96 | 3.09 | 0.81 |
| Algorithm 6 | | | | | | | | |
| 500  | 0.7 | 1.00 | 0.99 | 2.02 | 0.3 | 4.93 | 3.68 | 0.90 |
| 2000 | 0.7 | 1.00 | 0.98 | 1.96 | 0.3 | 4.95 | 3.72 | 0.89 |

**Table B.7:** MGND simulations, estimated parameters, scenario 4.

| N | $\pi_1$ 0.7 | $\mu_1$ 1 | $\sigma_1$ 1 | $\nu_1$ 2 | $\pi_2$ 0.3 | $\mu_2$ 5 | $\sigma_2$ 3 | $\nu_2$ 0.8 |
|---|---|---|---|---|---|---|---|---|
| Algorithm 1 | | | | | | | | |
| 500 | 0.03 | 0.06 | 0.09 | 0.42 | 0.03 | 0.69 | 1.58 | 0.20 |
| 2000 | 0.02 | 0.03 | 0.05 | 0.21 | 0.02 | 0.33 | 1.37 | 0.17 |
| Algorithm 2 | | | | | | | | |
| 500 | 0.03 | 0.04 | 0.08 | 0.38 | 0.03 | 0.38 | 1.06 | 0.15 |
| 2000 | 0.01 | 0.02 | 0.04 | 0.16 | 0.01 | 0.18 | 0.50 | 0.06 |
| Algorithm 3, $\alpha = 0.10$ | | | | | | | | |
| 500 | 0.03 | 0.04 | 0.08 | 0.36 | 0.03 | 0.40 | 1.23 | 0.18 |
| 2000 | 0.01 | 0.02 | 0.04 | 0.17 | 0.01 | 0.19 | 0.98 | 0.14 |
| Algorithm 3, $\alpha = 0.20$ | | | | | | | | |
| 500 | 0.03 | 0.04 | 0.08 | 0.37 | 0.03 | 0.40 | 1.20 | 0.18 |
| 2000 | 0.01 | 0.02 | 0.04 | 0.16 | 0.01 | 0.20 | 0.99 | 0.14 |
| Algorithm 3, $\alpha = 0.50$ | | | | | | | | |
| 500 | 0.03 | 0.04 | 0.09 | 0.39 | 0.03 | 0.40 | 1.30 | 0.18 |
| 2000 | 0.01 | 0.02 | 0.04 | 0.17 | 0.01 | 0.22 | 1.05 | 0.14 |
| Algorithm 3, $\alpha = 0.80$ | | | | | | | | |
| 500 | 0.03 | 0.04 | 0.08 | 0.39 | 0.03 | 0.44 | 1.38 | 0.19 |
| 2000 | 0.01 | 0.02 | 0.05 | 0.18 | 0.01 | 0.22 | 1.19 | 0.15 |
| Algorithm 4 | | | | | | | | |
| 500 | 0.03 | 0.04 | 0.08 | 0.38 | 0.03 | 0.37 | 1.05 | 0.15 |
| 2000 | 0.01 | 0.02 | 0.04 | 0.16 | 0.01 | 0.18 | 0.50 | 0.06 |
| Algorithm 5 | | | | | | | | |
| 500 | 0.03 | 0.04 | 0.08 | 0.35 | 0.03 | 0.37 | 1.07 | 0.15 |
| 2000 | 0.01 | 0.02 | 0.04 | 0.16 | 0.01 | 0.19 | 0.51 | 0.07 |
| Algorithm 6 | | | | | | | | |
| 500 | 0.03 | 0.04 | 0.08 | 0.37 | 0.03 | 0.42 | 1.41 | 0.19 |
| 2000 | 0.01 | 0.02 | 0.05 | 0.20 | 0.01 | 0.22 | 1.17 | 0.15 |

**Table B.8:** MGND simulations, MSE, scenario 4.

**Table B.9:** Stocks of the Euro Stoxx 50 Index, number of observations from 01 January 2010 to 30 September 2023, mean, median, standard deviation, skewness, kurtosis, JB statistics, min and max.

| Stock | N | Mean | Median | Std | Sk | Kur | JB test | Min | Max |
|---|---|---|---|---|---|---|---|---|---|
| ADS.DE | 3,487 | 0.0412 | 0 | 1.8356 | 0.0627 | 8.1216 | 3,820.386 | -14.3944 | 12.779 |
| ADYEN.AS | 1,359 | 0.0817 | 0.1786 | 2.7371 | -0.0713 | 5.5399 | 262.2945 | -12.7242 | 12.5983 |
| AD.AS | 3,521 | 0.0274 | 0.0612 | 1.2834 | -0.3597 | 8.5637 | 4,625.226 | -10.0197 | 7.7183 |
| AI.PA | 3,518 | 0.0344 | 0.0493 | 1.3078 | -0.2723 | 7.2998 | 2,758.759 | -11.8337 | 6.4528 |
| AIR.PA | 3,513 | 0.0601 | 0.0753 | 2.0468 | -0.2312 | 7.9231 | 3,585.51 | -14.8106 | 11.7814 |
| ALV.DE | 3,488 | 0.0316 | 0.0467 | 1.5747 | -0.0094 | 11.0718 | 9,483.951 | -10.9689 | 14.6728 |
| AMS.MC | 3,437 | 0.0508 | 0.0814 | 1.7454 | 0.1465 | 9.5268 | 6,123.306 | -10.5846 | 13.9789 |
| ABI.BR | 3,518 | 0.0151 | 0.0129 | 1.6302 | -0.1836 | 11.8109 | 11,416.7 | -11.8715 | 13.4532 |
| ASML.AS | 3,520 | 0.0883 | 0.1177 | 1.941 | 0.0482 | 6.1129 | 1,425.731 | -13.1142 | 13.0633 |
| CS.PA | 3,515 | 0.0196 | 0.066 | 1.9262 | -0.1651 | 9.3814 | 5,990.093 | -13.341 | 13.7201 |
| BAS.DE | 3,489 | -0.0012 | 0.0304 | 1.7012 | -0.2655 | 6.8781 | 2,231.817 | -12.5494 | 10.1917 |
| BAYN.DE | 3,488 | -0.0014 | 0 | 1.7416 | -0.3232 | 7.2172 | 2,650.582 | -14.0118 | 8.3761 |
| BMW.DE | 3,489 | 0.0316 | 0.02 | 1.8387 | -0.2901 | 7.8276 | 3,443.385 | -13.8933 | 13.5163 |
| BNP.PA | 3,513 | -0.0023 | 0.0183 | 2.1813 | -0.2656 | 7.5572 | 3,086.976 | -14.5261 | 13.2344 |
| CRG.IR | 3,489 | 0.0368 | 0 | 1.9773 | 0.0102 | 6.5138 | 1,798.841 | -13.6124 | 12.6868 |
| DAI.SG | 3,479 | 0.0309 | 0.023 | 1.8545 | -0.2641 | 6.9038 | 2,254.085 | -14.4782 | 10.6972 |
| BN.PA | 3,518 | 0.0054 | 0 | 1.2624 | -0.1605 | 6.8822 | 2,228.857 | -8.8931 | 7.4267 |
| DB1.DE | 3,489 | 0.0294 | 0.034 | 1.5596 | -0.3884 | 9.6774 | 6,580.463 | -12.5985 | 12.3141 |
| DPWA.F | 3,487 | 0.0302 | 0 | 2.463 | -0.0133 | 5.4431 | 869.4945 | -11.6724 | 12.1751 |
| DTE.DE | 3,489 | 0.018 | 0 | 1.3983 | -0.3666 | 9.2537 | 5,773.302 | -11.2673 | 10.6715 |
| ENEL.MI | 3,495 | 0.0165 | 0.0204 | 1.645 | -0.3725 | 5.9107 | 1,317.549 | -10.586 | 7.6674 |
| ENGI.PA | 3,517 | -0.0155 | 0 | 1.6017 | -0.5145 | 8.2303 | 4,171.239 | -14.1452 | 7.9217 |
| ENI.MI | 3,494 | 0.0075 | 0.061 | 1.6479 | -0.0223 | 7.6029 | 3,090.503 | -9.6408 | 13.9159 |
| EL.PA | 3,518 | 0.0392 | 0.0479 | 1.4575 | 0.0194 | 7.7644 | 3,333.752 | -10.9702 | 11.1999 |
| FLTR.IR | 3,486 | 0.047 | 0 | 1.9416 | 0.1216 | 9.7218 | 6,582.256 | -13.7653 | 13.7996 |
| IBE.MC | 3,518 | 0.0172 | 0.0228 | 1.4759 | -0.0469 | 9.0267 | 5,334.397 | -10.7585 | 13.3703 |
| ITX.MC | 3,519 | 0.0395 | 0 | 1.6414 | 0.24 | 7.4655 | 2,963.142 | -11.1276 | 13.1323 |
| IFX.DE | 3,487 | 0.0676 | 0.0603 | 2.2427 | -0.0951 | 5.1992 | 709.8873 | -12.5866 | 13.0759 |
| INGA.AS | 3,515 | 0.0235 | 0.0256 | 2.2649 | -0.1142 | 6.6016 | 1,911.385 | -11.8762 | 13.4696 |
| ISP.MI | 3,492 | 0.0073 | 0 | 2.3659 | -0.2043 | 6.4481 | 1,757.928 | -14.7567 | 13.4928 |
| KER.PA | 3,518 | 0.0476 | 0.0413 | 1.822 | -0.0915 | 7.0884 | 2,459.926 | -13.1416 | 10.0708 |
| KNEBV.HE | 3,458 | 0.0279 | 0.024 | 1.5269 | 0.067 | 5.6651 | 1,028.53 | -8.4035 | 9.0786 |
| OR.PA | 3,518 | 0.0453 | 0.0418 | 1.3689 | 0.1024 | 5.7614 | 1,126.561 | -7.5482 | 8.1003 |
| MC.PA | 3,518 | 0.0628 | 0.0738 | 1.7079 | -0.0403 | 5.7204 | 1,088.316 | -9.0777 | 9.6099 |
| MUV2.DE | 3,487 | 0.0355 | 0.0806 | 1.4411 | -0.1344 | 9.4888 | 6,138.242 | -9.4495 | 11.897 |
| NOKIA.HE | 3,447 | 0.0124 | 0.0402 | 2.2465 | -0.1071 | 7.9201 | 3,489.897 | -14.214 | 12.9643 |
| RI.PA | 3,518 | 0.0271 | 0.0396 | 1.3053 | -0.3043 | 6.7443 | 2,113.654 | -10.3501 | 7.1272 |
| ORA.PA | 3,518 | -0.014 | 0 | 1.4413 | 0.0246 | 7.8195 | 3,411.423 | -10.9307 | 10.6636 |
| PHIA.AS | 3,520 | 0.0018 | 0.0615 | 1.7364 | -0.3872 | 8.3065 | 4,225.415 | -13.0951 | 12.9037 |
| PRX.AS | 1,040 | -0.0214 | -0.036 | 2.5995 | 0.1027 | 4.9001 | 159.8456 | -10.923 | 14.5961 |
| SAF.PA | 3,512 | 0.07 | 0.0436 | 1.8873 | -0.1346 | 8.4552 | 4,373.111 | -14.8928 | 13.0855 |
| SAN.PA | 3,518 | 0.0165 | 0.0343 | 1.3901 | -0.3596 | 6.5236 | 1,899.704 | -11.2447 | 6.1157 |
| SAN.MC | 3,514 | -0.0265 | 0 | 2.105 | -0.0683 | 5.4611 | 891.8522 | -12.752 | 11.3413 |
| SAP.DE | 3,488 | 0.0448 | 0.0669 | 1.4172 | -0.1243 | 7.9306 | 3,548.644 | -10.179 | 11.8219 |
| SU.PA | 3,517 | 0.0423 | 0.0821 | 1.8165 | -0.0236 | 5.7904 | 1,144.052 | -11.3596 | 11.3445 |
| SIE.DE | 3,489 | 0.0218 | 0.0306 | 1.6202 | -0.1187 | 7.9388 | 3,560.692 | -13.5774 | 10.9387 |
| TTE.PA | 3,516 | 0.0185 | 0.0578 | 1.605 | 0.0713 | 10.3436 | 7,916.186 | -14.3177 | 14.0407 |
| TEF.MC | 3,516 | -0.0415 | 0 | 1.6281 | -0.0046 | 7.8482 | 3,449.834 | -10.3225 | 12.8148 |
| DG.PA | 3,516 | 0.0279 | 0.0452 | 1.6324 | -0.359 | 9.4892 | 6,254.896 | -13.2133 | 10.7943 |
| VIV.PA | 3,515 | 0.0004 | 0 | 1.5176 | -0.3975 | 8.4716 | 4,485.044 | -13.532 | 8.6089 |
| VOW.DE | 3,485 | 0.0351 | 0 | 2.0608 | 0.3302 | 7.1296 | 2,544.688 | -10.2645 | 14.6944 |
| VNA.DE | 2,592 | 0.0169 | 0.0457 | 1.7071 | -0.0861 | 6.8932 | 1,644.669 | -8.8811 | 10.6258 |

**Note.** All JB test statistics have a *p*-value ≤ 0.05.

**Table B.10:** Comparative analysis part I, AIC of fitted distributions for daily log-returns.

| Stock | MGND | Laplace | Cauchy | GND | Student-t | GH | MN | MNL | MNC | MSTD |
|---|---|---|---|---|---|---|---|---|---|---|
| ADS.DE | **13,587.6** | 13,628.38 | 14,254.89 | 13,625.64 | 13,590.54 | 13,592.02 | 13,613.51 | 13,587.83 | 13,663.5 | 13,596.05 |
| ADYEN.AS | 6,487.87 | 6,499.08 | 6,772.5 | 6,489.02 | 6,488.2 | **6,485.17** | 6,486.59 | 6,489.07 | 6,522.49 | 6,488.75 |
| AD.AS | 11,108.64 | 11,159.12 | 11,746.98 | 11,159.41 | 11,110.91 | **11,108.1** | 11,130.29 | 11,110.21 | 11,201.55 | 11,111.16 |
| AI.PA | 11,526.42 | 11,576.74 | 12,280.3 | 11,545.39 | **11,517.17** | 11,518.6 | 11,545.63 | 11,522.9 | 11,594.69 | 11,521.37 |
| AIR.PA | 14,464.38 | 14,497.28 | 15,113 | 14,493.57 | **14,461.35** | 14,461.71 | 14,507.51 | 14,466.52 | 14,538 | 14,464.34 |
| ALV.DE | 12,169.38 | 12,217.66 | 12,674.51 | 12,213.97 | 12,171.38 | **12,162.98** | 12,228.83 | 12,171.53 | 12,272.22 | 12,168.06 |
| AMS.MC | 12,954.11 | 12,993.34 | 13,556.96 | 12,993.73 | 12,953.82 | 12,950.55 | 12,997.15 | 12,955.86 | 13,038.21 | **12,949.06** |
| ABI.BR | 12,615.54 | 12,672.69 | 13,219.27 | 12,674.68 | **12,607.38** | 12,610.91 | 12,669.83 | 12,625.14 | 12,682 | 12,614.33 |
| ASML.AS | 14,336.54 | 14,367.76 | 15,068.57 | 14,342.52 | 14,339.79 | **14,334.52** | 14,361.96 | 14,336.3 | 14,406.64 | 14,336.56 |
| CS.PA | 13,772.19 | 13,810.87 | 14,295.49 | 13,810.01 | 13,780.18 | **13,767.67** | 13,818.06 | 13,773.67 | 13,884.85 | 13,772.11 |
| BAS.DE | 13,185.91 | 13,222.42 | 13,868.99 | 13,209.51 | 13,190.93 | **13,181.73** | 13,204.37 | 13,188.44 | 13,290.93 | 13,187.88 |
| BAYN.DE | 13,321.16 | 13,334.48 | 13,980.46 | 13,329.56 | 13,327.4 | 13,322.58 | 13,356.69 | **13,317.58** | 13,399.56 | 13,320.83 |
| BMW.DE | 13,645.54 | 13,659.78 | 14,265.94 | 13,657.66 | 13,649.95 | **13,640.36** | 13,685.76 | 13,646.31 | 13,741.66 | 13,641.35 |
| BNP.PA | 14,910.16 | 14,933.09 | 15,542.29 | 14,931.86 | 14,916.37 | 14,911.75 | 14,939.23 | **14,906.9** | 15,000.53 | 14,914.32 |
| CRG.IR | 14,291.35 | 14,361.18 | 15,054.07 | 14,330.31 | **14,286.25** | 14,288.76 | 14,294.21 | 14,288.44 | 14,379.71 | 14,290.48 |
| DAI.SG | **13,706.43** | 13,717.86 | 14,371.88 | 13,718.14 | 13,756.23 | 13,722.7 | 13,791.44 | 13,713.1 | 13,773.03 | 13,711.95 |
| BN.PA | 11,235.4 | 11,280.6 | 11,963.39 | 11,258.8 | **11,229.43** | 11,230.6 | 11,255.22 | 11,233.76 | 11,312.02 | 11,235.46 |
| DB1.DE | 12,475.73 | 12,512.71 | 13,152.92 | 12,506 | **12,466.98** | 12,476.36 | 12,519.14 | 12,475.52 | 12,540.4 | 12,471.69 |
| DPWA.F | 15,287.13 | 15,619.36 | 16,242.53 | 15,618.28 | 15,757.13 | **9,334.09** | 15,704.19 | 15,624.66 | 15,178.66 | |
| DTE.DE | 11,623.82 | 11,646.67 | 12,228.42 | 11,647.94 | 11,619.54 | 11,618.3 | 11,668.69 | 11,621.46 | 11,708.8 | **11,617.41** |
| ENEL.MI | 13,087.3 | 13,126.94 | 13,840.02 | 13,099.58 | 13,093.38 | 13,088.05 | 13,109.44 | **13,084.09** | 13,160.46 | 13,095.44 |
| ENGI.PA | 12,791.14 | 12,833.98 | 13,468.57 | 12,826.12 | 12,792.08 | **12,788.1** | 12,815.81 | 12,788.42 | 12,882.64 | 12,793.82 |
| ENI.MI | 12,960.72 | 13,014.6 | 13,659.1 | 12,998.78 | 12,962.25 | **12,958.72** | 12,984.25 | 12,969.79 | 13,059.5 | 12,962.86 |
| EL.PA | 12,154.42 | 12,177.93 | 12,811.79 | 12,171.41 | 12,153.51 | **12,150.38** | 12,188.23 | 12,154.01 | 12,235.95 | 12,151.3 |
| FLTR.IR | **13,832.53** | 13,864.76 | 14,441.13 | 13,866.75 | 13,842.14 | 13,842.79 | 13,899.26 | 13,845.31 | 13,891.76 | 13,834.11 |
| IBE.MC | 12,058.94 | 12,078.03 | 12,626.64 | 12,080.03 | 12,061.63 | **12,052.96** | 12,099.07 | 12,057.21 | 12,164.5 | 12,058.04 |
| ITX.MC | 13,105.9 | 13,154.86 | 13,854.04 | 13,127.98 | **13,098.54** | 13,101.36 | 13,133.01 | 13,103.49 | 13,164.11 | 13,102.47 |
| IFX.DE | 15,297.79 | 15,344.19 | 16,095.95 | 15,302.48 | 15,308.05 | 15,304.42 | 15,326.44 | **15,293.22** | 15,350.73 | 15,297.28 |
| INGA.AS | 15,137.45 | 15,139.26 | 15,703.15 | 15,141.13 | 15,166.85 | 15,138.87 | 15,173.28 | **15,134.6** | 15,243.55 | 15,144.5 |
| ISP.MI | 15,415.48 | 15,424.42 | 16,018.21 | 15,424.74 | 15,429.17 | **15,412.46** | 15,438.66 | 15,412.48 | 15,527.43 | 15,420.44 |
| KER.PA | 13,728.68 | 13,762.04 | 14,402.55 | 13,754.35 | 13,726.28 | 13,726.25 | 13,746.65 | **13,725.02** | 13,819.72 | 13,731.39 |
| KNEBV.HE | 12,377.71 | 12,384.03 | 13,036.88 | 12,375.07 | 12,392.87 | **12,374.53** | 12,396.51 | 12,384.2 | 12,470.58 | 12,389.43 |
| OR.PA | 11,904.43 | 11,963.03 | 12,679.53 | 11,920.59 | **11,896.1** | 11,896.89 | 11,912.8 | 11,904.69 | 11,984.65 | 11,901.36 |
| MC.PA | 13,407.35 | 13,442.97 | 14,123.31 | 13,419.68 | 13,407.19 | **13,401.8** | 13,413.81 | 13,408.26 | 13,507.25 | 13,406.2 |
| MUV2.DE | **11,690.9** | 11,742.64 | 12,270.25 | 11,744.44 | 11,697.45 | 11,691.49 | 11,722.03 | 11,695.77 | 11,791.79 | 11,693.35 |
| NOKIA.HE | 14,693.66 | 14,730.34 | 15,267.74 | 14,732.34 | 14,696.45 | 14,693.69 | 14,716.33 | **14,691.24** | 14,797.66 | 14,696.47 |
| RI.PA | 11,491.3 | 11,538.96 | 12,230.76 | 11,514.29 | 11,488.48 | **11,486.45** | 11,507.28 | 11,490.78 | 11,580.45 | 11,492.08 |
| ORA.PA | 12,024.56 | 12,041.36 | 12,661.04 | 12,040.83 | 12,031.19 | 12,026.77 | 12,055.05 | **12,020.64** | 12,127.71 | 12,030.91 |
| PHIA.AS | 13,367.94 | 13,419.03 | 14,056.98 | 13,408.78 | 13,361.41 | **13,359.65** | 13,389.28 | 13,369.73 | 13,462.98 | 13,364.47 |
| PRX.AS | 4,878.32 | 4,894.64 | 5,107.3 | 4,879.27 | 4,879.34 | 4,878.46 | **4,877** | 4,887.67 | 4,910.54 | 4,880.2 |
| SAF.PA | 13,814.6 | 13,832.89 | 14,417.62 | 13,834.55 | 13,819.45 | 13,816.34 | 13,857.11 | 13,815 | 13,894.56 | **13,812.48** |
| SAN.PA | 12,008.92 | 12,094.79 | 12,835.2 | 12,041.15 | **11,999.38** | 12,000.66 | 12,014.46 | 12,007.15 | 12,082.47 | 12,005.84 |
| SAN.MC | 14,929.96 | 14,966.78 | 15,684.7 | 14,932.93 | 14,933.25 | 14,927.86 | 14,951.05 | **14,927.33** | 14,999 | 14,930.5 |
| SAP.DE | 11,896.08 | 11,950.36 | 12,624.43 | 11,932.81 | 11,895.22 | 11,893.68 | 11,917.37 | **11,892.89** | 11,974.07 | 11,897.55 |
| SU.PA | 13,848.53 | 13,863.61 | 14,556.16 | 13,847.65 | 13,863.88 | 13,845.91 | 13,878.67 | **13,844.61** | 13,911.73 | 13,847.52 |
| SIE.DE | 12,826.09 | 12,860.48 | 13,526.21 | 12,848.65 | 12,823.41 | 12,825.2 | 12,857.57 | **12,823.05** | 12,894.17 | 12,825.32 |
| TTE.PA | 12,785.79 | 12,821.75 | 13,486.72 | 12,815.56 | 12,788 | 12,789.45 | 12,850.11 | 12,792.67 | 12,832.23 | **12,783.54** |
| TEF.MC | 12,917.25 | 12,962.52 | 13,598.62 | 12,952.28 | 12,913.16 | **12,911.04** | 12,940.98 | 12,918.68 | 13,014.27 | 12,915.01 |
| DG.PA | **12,775.87** | 12,818.47 | 13,408.27 | 12,818.92 | 12,776.78 | 12,778.36 | 12,817.57 | 12,784.1 | 12,838.73 | 12,777.62 |
| VIV.PA | 12,371.66 | 12,390.26 | 12,997.48 | 12,388.76 | 12,376.99 | 12,371.13 | 12,419.75 | 12,375.96 | 12,459.31 | **12,370.84** |
| VOW.DE | 14,477.53 | 14,497.34 | 15,140.77 | 14,491.13 | 14,480.4 | **14,476.04** | 14,514.18 | 14,476.75 | 14,552.08 | 14,479.28 |
| VNA.DE | **9,657.33** | 9,675.49 | 10,071.56 | 9,677.45 | 9,670.3 | 9,658.32 | 9,667.66 | 9,659.34 | 9,762.9 | 9,666.45 |
| Mean | 12,767.34 | 12,802.05 | 13,413.99 | 12,791.14 | 12,770.36 | **12,765.75** | 12,797.32 | 12,768.01 | 12,849.02 | 12,768.06 |
| Median | 12,954.11 | 12,993.34 | 13,598.62 | 12,993.73 | 12,953.82 | 12949.45 | 12,984.25 | 12,955.86 | 13,038.21 | **12,949.06** |

**Note.** In bold the best model.

**Table B.11:** Comparative analysis part I, BIC of fitted distributions for daily log-returns.

| Stock | MGND | Laplace | Cauchy | GND | Student-t | GH | MN | MNL | MNC | MSTD |
|---|---|---|---|---|---|---|---|---|---|---|
| ADS.DE | 13,630.7 | 13,640.69 | 14,267.2 | 13,644.11 | **13,609.01** | 13,622.8 | 13,644.3 | 13,618.62 | 13,694.28 | 13,639.15 |
| ADYEN.AS | 6,524.38 | 6,509.51 | 6,782.93 | 6,504.67 | **6,503.84** | 6,511.25 | 6,512.66 | 6,515.14 | 6,548.56 | 6,525.25 |
| AD.AS | 11,151.8 | 11,171.45 | 11,759.32 | 11,177.91 | **11,129.41** | 11,138.93 | 11,161.12 | 11,141.05 | 11,232.38 | 11,154.33 |
| AI.PA | 11,569.58 | 11,589.08 | 12,292.64 | 11,563.88 | **11,535.66** | 11,549.42 | 11,576.46 | 11,553.73 | 11,625.52 | 11,564.53 |
| AIR.PA | 14,507.53 | 14,509.61 | 15,125.33 | 14,512.07 | **14,479.85** | 14,492.53 | 14,538.34 | 14,497.35 | 14,568.82 | 14,507.49 |
| ALV.DE | 12,212.48 | 12,229.98 | 12,686.82 | 12,232.44 | **12,189.85** | 12,193.76 | 12,259.61 | 12,202.31 | 12,303 | 12,211.16 |
| AMS.MC | 12,997.1 | 13,005.63 | 13,569.25 | 13,012.16 | **12,972.25** | 12,981.26 | 13,027.86 | 12,986.57 | 13,068.92 | 12,992.06 |
| ABI.BR | 12,658.7 | 12,685.02 | 13,231.6 | 12,693.18 | **12,625.88** | 12,641.74 | 12,700.66 | 12,655.96 | 12,712.83 | 12,657.49 |
| ASML.AS | 14,379.71 | 14,380.1 | 15,080.91 | 14,361.02 | **14,358.28** | 14,365.36 | 14,392.79 | 14,367.13 | 14,437.47 | 14,379.72 |
| CS.PA | 13,815.34 | 13,823.2 | 14,307.82 | 13,828.51 | 13,798.67 | **13,798.5** | 13,848.88 | 13,804.49 | 13,915.68 | 13,815.26 |
| BAS.DE | 13,229.01 | 13,234.74 | 13,881.31 | 13,227.98 | **13,209.4** | 13,212.52 | 13,235.16 | 13,219.22 | 13,321.72 | 13,230.98 |
| BAYN.DE | 13,364.26 | 13,346.79 | 13,992.77 | 13,348.03 | **13,345.87** | 13,353.36 | 13,387.48 | 13,348.36 | 13,430.34 | 13,363.93 |
| BMW.DE | 13,688.64 | 13,672.09 | 14,278.25 | 13,676.13 | **13,668.42** | 13,671.14 | 13,716.55 | 13,677.09 | 13,772.44 | 13,684.45 |
| BNP.PA | 14,953.31 | 14,945.42 | 15,554.62 | 14,950.35 | **14,934.86** | 14,942.57 | 14,970.06 | 14,937.72 | 15,031.35 | 14,957.47 |
| CRG.IR | 14,334.45 | 14,373.49 | 15,066.38 | 14,348.79 | **14,304.73** | 14,319.55 | 14,325 | 14,319.22 | 14,410.5 | 14,333.58 |
| DAI.SG | 13,749.52 | **13,730.17** | 14,384.19 | 13,736.6 | 13,774.69 | 13,753.48 | 13,822.22 | 13,743.88 | 13,803.8 | 13,755.03 |
| BN.PA | 11,278.56 | 11,292.93 | 11,975.72 | 11,277.3 | **11,247.93** | 11,261.42 | 11,286.05 | 11,264.59 | 11,342.85 | 11,278.62 |
| DB1.DE | 12,518.83 | 12,525.02 | 13,165.23 | 12,524.47 | **12,485.45** | 12,507.15 | 12,549.93 | 12,506.3 | 12,571.19 | 12,514.79 |
| DPWA.F | 15,330.22 | 15,631.67 | 16,254.85 | 15,636.75 | 15,775.6 | **9,364.87** | 15,734.98 | 15,655.44 | 15,209.44 | |
| DTE.DE | 11,666.92 | 11,658.99 | 12,240.73 | 11,666.41 | **11,638.01** | 11,649.09 | 11,699.48 | 11,652.25 | 11,739.58 | 11,660.52 |
| ENEL.MI | 13,130.42 | 13,139.26 | 13,852.34 | 13,118.06 | **13,111.86** | 13,118.84 | 13,140.23 | 13,114.88 | 13,191.25 | 13,138.55 |
| ENGI.PA | 12,834.3 | 12,846.31 | 13,480.9 | 12,844.61 | **12,810.58** | 12,818.93 | 12,846.63 | 12,819.24 | 12,913.47 | 12,836.98 |
| ENI.MI | 13,003.83 | 13,026.91 | 13,671.42 | 13,017.25 | **12,980.73** | 12,989.52 | 13,015.05 | 13,000.58 | 13,090.29 | 13,005.97 |
| EL.PA | 12,197.58 | 12,190.26 | 12,824.12 | 12,189.49 | **12,172.01** | 12,181.21 | 12,219.06 | 12,184.83 | 12,266.77 | 12,194.46 |
| FLTR.IR | 13,875.63 | 13,877.07 | 14,453.44 | 13,885.22 | **13,860.61** | 13,873.57 | 13,930.04 | 13,876.09 | 13,922.55 | 13,877.2 |
| IBE.MC | 12,102.09 | 12,090.37 | 12,638.97 | 12,098.53 | **12,080.12** | 12,083.78 | 12,129.9 | 12,088.04 | 12,195.33 | 12,101.2 |
| ITX.MC | 13,149.07 | 13,167.2 | 13,866.37 | 13,146.47 | **13,117.04** | 13,132.19 | 13,163.84 | 13,134.32 | 13,194.94 | 13,145.63 |
| IFX.DE | 15,340.89 | 15,356.5 | 16,108.26 | **15,320.95** | 15,326.52 | 15,335.21 | 15,357.22 | 15,324 | 15,381.51 | 15,340.38 |
| INGA.AS | 15,180.6 | **15,151.59** | 15,715.48 | 15,159.63 | 15,185.35 | 15,169.69 | 15,204.1 | 15,165.43 | 15,274.38 | 15,187.65 |
| ISP.MI | 15,458.58 | **15,436.74** | 16,030.53 | 15,443.21 | 15,447.64 | 15,443.26 | 15,469.45 | 15,443.27 | 15,558.23 | 15,463.55 |
| KER.PA | 13,771.84 | 13,774.37 | 14,414.88 | 13,772.85 | **13,744.78** | 13,757.08 | 13,777.48 | 13,755.85 | 13,850.55 | 13,774.55 |
| KNEBV.HE | 12,420.74 | 12,396.32 | 13,049.18 | **12,393.52** | 12,411.31 | 12,405.28 | 12,427.25 | 12,414.94 | 12,501.33 | 12,432.47 |
| OR.PA | 11,947.59 | 11,975.36 | 12,691.86 | 11,939.09 | **11,914.6** | 11,927.72 | 11,943.63 | 11,935.51 | 12,015.48 | 11,944.52 |
| MC.PA | 13,450.51 | 13,455.3 | 14,135.64 | 13,438.18 | **13,425.69** | 13,432.63 | 13,444.64 | 13,439.09 | 13,538.07 | 13,449.36 |
| MUV2.DE | 11,734 | 11,754.95 | 12,282.57 | 11,762.91 | **11,715.92** | 11,722.27 | 11,751.61 | 11,726.55 | 11,822.57 | 11,736.45 |
| NOKIA.HE | 14,736.67 | 14,742.63 | 15,280.03 | 14,750.77 | **14,714.88** | 14,724.42 | 14,747.05 | 14,721.96 | 14,828.38 | 14,739.49 |
| RI.PA | 11,534.46 | 11,551.29 | 12,243.09 | 11,532.78 | **11,506.98** | 11,517.28 | 11,538.11 | 11,521.61 | 11,611.28 | 11,535.24 |
| ORA.PA | 12,067.72 | 12,053.69 | 12,673.37 | 12,059.32 | **12,049.68** | 12,057.6 | 12,085.88 | 12,051.47 | 12,158.53 | 12,074.07 |
| PHIA.AS | 13,411.1 | 13,431.37 | 14,069.31 | 13,427.28 | **13,379.91** | 13,390.48 | 13,401.13 | 13,400.56 | 13,493.81 | 13,407.63 |
| PRX.AS | 4,912.95 | 4,904.53 | 5,117.19 | **4,894.11** | 4,894.18 | 4,903.19 | 4,901.74 | 4,912.4 | 4,935.27 | 4,914.83 |
| SAF.PA | 13,857.75 | 13,845.22 | 14,429.95 | 13,853.05 | **13,837.94** | 13,847.16 | 13,887.93 | 13,845.82 | 13,925.38 | 13,855.63 |
| SAN.PA | 12,052.08 | 12,107.12 | 12,847.53 | 12,059.65 | **12,017.88** | 12,031.48 | 12,045.29 | 12,037.98 | 12,113.3 | 12,049 |
| SAN.MC | 14,973.11 | 14,979.11 | 15,697.02 | **14,951.42** | 14,951.74 | 14,958.68 | 14,981.87 | 14,958.16 | 15,029.82 | 14,973.65 |
| SAP.DE | 11,939.18 | 11,962.67 | 12,636.75 | 11,951.28 | **11,913.7** | 11,924.47 | 11,948.15 | 11,923.68 | 12,004.85 | 11,940.65 |
| SU.PA | 13,891.69 | 13,875.95 | 14,568.49 | **13,866.14** | 13,882.38 | 13,876.74 | 13,909.5 | 13,875.44 | 13,942.55 | 13,890.68 |
| SIE.DE | 12,869.2 | 12,872.79 | 13,538.52 | 12,867.12 | **12,841.88** | 12,855.98 | 12,888.36 | 12,853.84 | 12,924.96 | 12,868.42 |
| TTE.PA | 12,828.94 | 12,834.08 | 13,499.05 | 12,834.06 | **12,806.49** | 12,820.27 | 12,880.94 | 12,823.5 | 12,863.06 | 12,826.7 |
| TEF.MC | 12,960.41 | 12,974.85 | 13,610.95 | 12,970.78 | **12,931.65** | 12,941.87 | 12,971.8 | 12,949.5 | 13,045.1 | 12,958.17 |
| DG.PA | 12,819.02 | 12,830.8 | 13,420.6 | 12,837.4 | **12,795.27** | 12,809.18 | 12,848.4 | 12,814.93 | 12,869.55 | 12,820.77 |
| VIV.PA | 12,414.81 | 12,402.59 | 13,009.81 | 12,407.25 | **12,395.48** | 12,401.96 | 12,450.58 | 12,406.79 | 12,490.13 | 12,413.99 |
| VOW.DE | 14,520.63 | 14,509.66 | 15,153.08 | 14,509.6 | **14,498.87** | 14,506.82 | 14,544.96 | 14,507.53 | 14,582.86 | 14,522.37 |
| VNA.DE | 9,698.35 | **9,687.21** | 10,083.28 | 9,695.03 | 9,687.88 | 9,687.62 | 9,696.96 | 9,688.64 | 9,792.2 | 9,707.48 |
| Mean | 12,810.13 | 12,814.27 | 13,426.22 | 12,809.48 | **12,788.70** | 12,796.32 | 12,827.89 | 12,798.58 | 12,879.59 | 12,810.85 |
| Median | 12997.10 | 13005.63 | 13610.95 | 13012.16 | **12972.25** | 12981.26 | 13015.05 | 12986.57 | 13068.92 | 12992.06 |

**Note.** In bold the best model.

**Table B.12:** Comparative analysis part I, KS distance between the empirical and fitted distributions for daily log-returns.

| Stock | MGND | Laplace | Cauchy | GND | Student-t | GH | MN | MNL | MNC | MSTD |
|---|---|---|---|---|---|---|---|---|---|---|
| ADS.DE | 0.0082 | 0.0122 | 0.0679 | 0.0077 | 0.0065 | **0.0054** | 0.083 | 0.083 | 0.1684 | 0.0061 |
| ADYEN.AS | **0.0098** | 0.0314 | 0.0855 | 0.0171 | 0.0131 | 0.0124 | 0.1684 | 0.2175 | 0.2597 | 0.0112 |
| AD.AS | **0.0033** | 0.0203 | 0.0714 | 0.0169 | 0.0089 | 0.005 | 0.083 | 0.044 | 0.1048 | 0.0058 |
| AI.PA | 0.006 | 0.0232 | 0.0723 | 0.0103 | 0.0089 | 0.0063 | 0.1684 | 0.0437 | 0.0941 | **0.0044** |
| AIR.PA | 0.0055 | 0.0234 | 0.068 | 0.017 | **0.0039** | 0.0041 | 0.083 | 0.102 | 0.1896 | 0.006 |
| ALV.DE | 0.0051 | 0.0111 | 0.0645 | 0.0111 | 0.0125 | 0.0053 | 0.1684 | 0.0752 | 0.1595 | **0.0042** |
| AMS.MC | 0.0086 | 0.0135 | 0.0761 | 0.0125 | 0.0145 | 0.0099 | 0.083 | 0.0809 | 0.1689 | **0.0051** |
| ABI.BR | 0.0072 | 0.0162 | 0.0667 | 0.0163 | **0.0035** | 0.0039 | 0.1684 | 0.0592 | 0.1587 | 0.0054 |
| ASML.AS | 0.0044 | 0.0225 | 0.0767 | 0.0087 | 0.0095 | 0.0047 | 0.083 | 0.1053 | 0.1706 | **0.0034** |
| CS.PA | 0.0075 | 0.0129 | 0.0645 | 0.0173 | 0.0111 | 0.005 | 0.1684 | 0.0975 | 0.1926 | **0.0043** |
| BAS.DE | 0.0049 | 0.0182 | 0.0744 | 0.0098 | 0.0091 | **0.0038** | 0.083 | 0.0793 | 0.1514 | 0.0044 |
| BAYN.DE | 0.0084 | 0.0167 | 0.0664 | 0.0112 | 0.0075 | **0.0036** | 0.1684 | 0.0771 | 0.1541 | 0.006 |
| BMW.DE | 0.0081 | 0.0092 | 0.0667 | 0.0066 | 0.0062 | **0.0036** | 0.083 | 0.093 | 0.1674 | 0.0064 |
| BNP.PA | 0.0073 | 0.0143 | 0.0715 | 0.0103 | 0.007 | **0.003** | 0.1684 | 0.1226 | 0.2017 | 0.007 |
| CRG.IR | 0.0062 | 0.0219 | 0.0744 | 0.0113 | 0.0058 | 0.0061 | 0.083 | 0.1082 | 0.1782 | **0.0039** |
| DAI.SG | 0.0094 | 0.007 | 0.0676 | **0.0046** | 0.0117 | 0.006 | 0.1684 | 0.0974 | 0.1626 | 0.0083 |
| BN.PA | 0.0049 | 0.0167 | 0.0748 | 0.0091 | 0.0048 | **0.0038** | 0.083 | 0.0475 | 0.0887 | 0.0046 |
| DB1.DE | **0.0034** | 0.0172 | 0.0687 | 0.0109 | 0.0054 | 0.0061 | 0.1684 | 0.0547 | 0.1378 | 0.005 |
| DPWA.F | 0.0319 | **0.0051** | 0.0693 | 0.032 | 0.0168 | 0.1733 | 0.083 | 0.1934 | 0.2302 | |
| DTE.DE | **0.0018** | 0.009 | 0.0642 | 0.0073 | 0.0069 | 0.0055 | 0.1684 | 0.0564 | 0.1201 | 0.0075 |
| ENEL.MI | 0.0053 | 0.0138 | 0.0799 | **0.0049** | 0.0091 | 0.006 | 0.083 | 0.0762 | 0.1366 | 0.0057 |
| ENGI.PA | 0.0056 | 0.0128 | 0.071 | 0.0076 | 0.0085 | **0.0043** | 0.1684 | 0.0558 | 0.1411 | 0.005 |
| ENI.MI | 0.0069 | 0.0293 | 0.0778 | 0.018 | 0.0091 | 0.0069 | 0.083 | 0.0795 | 0.1477 | **0.0056** |
| EL.PA | 0.0074 | 0.0141 | 0.0674 | 0.0093 | 0.0101 | 0.0091 | 0.1684 | 0.0603 | 0.1283 | **0.0062** |
| FLTR.IR | 0.0087 | 0.0073 | 0.066 | 0.0075 | 0.0082 | **0.0059** | 0.083 | 0.0891 | 0.183 | 0.0091 |
| IBE.MC | 0.0049 | 0.0108 | 0.0651 | 0.0108 | 0.006 | **0.0039** | 0.1684 | 0.0694 | 0.1366 | 0.0059 |
| ITX.MC | **0.0058** | 0.0139 | 0.0736 | 0.0074 | 0.0061 | 0.0075 | 0.083 | 0.0659 | 0.1393 | 0.0062 |
| IFX.DE | 0.0074 | 0.0159 | 0.0848 | **0.0048** | 0.0089 | 0.0054 | 0.1684 | 0.1518 | 0.1943 | 0.007 |
| INGA.AS | **0.0023** | 0.0062 | 0.0696 | 0.0063 | 0.0104 | 0.0043 | 0.083 | 0.1444 | 0.2139 | 0.0076 |
| ISP.MI | 0.0034 | 0.0058 | 0.0704 | 0.0062 | 0.0088 | **0.0033** | 0.1684 | 0.1356 | 0.215 | 0.0048 |
| KER.PA | 0.005 | 0.0181 | 0.0695 | 0.0115 | **0.0034** | 0.0052 | 0.083 | 0.0778 | 0.167 | 0.005 |
| KNEBV.HE | **0.0039** | 0.0132 | 0.0748 | 0.0064 | 0.0083 | 0.0049 | 0.1684 | 0.085 | 0.1283 | 0.0054 |
| OR.PA | 0.0067 | 0.027 | 0.081 | 0.0129 | **0.0035** | 0.0051 | 0.083 | 0.0614 | 0.1031 | 0.0053 |
| MC.PA | 0.0045 | 0.0217 | 0.0801 | 0.0103 | 0.0062 | 0.0049 | 0.1684 | 0.0966 | 0.1495 | **0.0041** |
| MUV2.DE | 0.0062 | 0.0175 | 0.0722 | 0.0185 | 0.0114 | 0.0063 | 0.083 | 0.0601 | 0.1312 | **0.0057** |
| NOKIA.HE | 0.0059 | 0.0205 | 0.0673 | 0.0204 | 0.0055 | 0.0073 | 0.1684 | 0.1271 | 0.2142 | **0.0052** |
| RI.PA | **0.0039** | 0.0211 | 0.0747 | 0.0102 | 0.0079 | 0.0041 | 0.083 | 0.0486 | 0.0928 | 0.004 |
| ORA.PA | 0.0049 | 0.0176 | 0.0669 | 0.0118 | 0.0067 | 0.005 | 0.1684 | 0.0589 | 0.1236 | **0.0044** |
| PHIA.AS | **0.0054** | 0.0287 | 0.071 | 0.0181 | 0.0081 | 0.007 | 0.083 | 0.0722 | 0.1597 | 0.0069 |
| PRX.AS | 0.01 | 0.0271 | 0.0874 | 0.0109 | **0.0074** | 0.0086 | 0.1684 | 0.1998 | 0.2445 | 0.0076 |
| SAF.PA | **0.0018** | 0.0073 | 0.0626 | 0.0063 | 0.0062 | 0.0071 | 0.083 | 0.0778 | 0.1793 | 0.0054 |
| SAN.PA | 0.0066 | 0.0292 | 0.0773 | 0.0123 | 0.0075 | 0.0063 | 0.1684 | 0.0445 | 0.1004 | **0.006** |
| SAN.MC | **0.005** | 0.0278 | 0.0773 | 0.013 | 0.0052 | 0.0067 | 0.083 | 0.1251 | 0.182 | 0.0072 |
| SAP.DE | 0.0087 | 0.0209 | 0.0781 | 0.0107 | 0.0095 | **0.0053** | 0.1684 | 0.0522 | 0.112 | 0.0061 |
| SU.PA | 0.0075 | 0.02 | 0.0759 | 0.0096 | 0.0138 | 0.0064 | 0.083 | 0.1014 | 0.1624 | **0.0047** |
| SIE.DE | 0.0058 | 0.0189 | 0.0709 | 0.0106 | 0.0062 | 0.0049 | 0.1684 | 0.0592 | 0.141 | **0.0048** |
| TTE.PA | **0.004** | 0.0239 | 0.0708 | 0.0167 | 0.0101 | 0.0078 | 0.083 | 0.0573 | 0.1459 | 0.0055 |
| TEF.MC | 0.006 | 0.0252 | 0.0775 | 0.0141 | 0.0081 | 0.0057 | 0.1684 | 0.0779 | 0.1479 | **0.0049** |
| DG.PA | 0.0064 | 0.0168 | 0.0696 | 0.0139 | 0.0051 | **0.0042** | 0.083 | 0.0652 | 0.1479 | 0.0061 |
| VIV.PA | 0.0055 | 0.0142 | 0.0624 | 0.0104 | 0.0063 | **0.0039** | 0.1684 | 0.0613 | 0.1317 | 0.0047 |
| VOW.DE | **0.0039** | 0.0161 | 0.0763 | 0.0096 | 0.005 | 0.0057 | 0.083 | 0.1226 | 0.1902 | 0.0061 |
| VNA.DE | 0.0086 | 0.0124 | 0.0715 | 0.013 | 0.0124 | 0.0094 | 0.1684 | 0.0961 | 0.1667 | **0.0083** |
| Mean | 0.0060 | 0.0175 | 0.0722 | 0.0112 | 0.0080 | **0.0057** | 0.0863 | 0.0863 | 0.1566 | 0.0058 |
| Median | 0.0058 | 0.0168 | 0.0714 | 0.0107 | 0.0079 | **0.0054** | 0.0778 | 0.0778 | 0.1541 | 0.0056 |

**Note.** In bold the best model.

**Table B.13:** Comparative analysis part I, AD distance between the empirical and fitted distributions for daily log-returns.

| Stock | MGND | Laplace | Cauchy | GND | Student-t | GH | MN | MNL | MNC | MSTD |
|---|---|---|---|---|---|---|---|---|---|---|
| ADS.DE | 0.0053 | 0.0346 | 0.7054 | 0.0691 | 0.0053 | 0.0052 | 0.8972 | 0.8972 | 1.0865 | **0.0049** |
| ADYEN.AS | 0.0053 | 0.0265 | 0.9084 | 0.018 | 0.0189 | **0.0052** | 2.876 | 2.876 | 3.6355 | 0.0057 |
| AD.AS | 0.0021 | 0.0371 | 0.4347 | 0.054 | 0.0074 | **0.0018** | 0.3052 | 0.3052 | 55.8945 | 0.0018 |
| AI.PA | 0.0412 | 0.0594 | 0.4427 | 1.2692 | **0.0062** | 0.0094 | 0.2859 | 0.2859 | 18.5082 | 0.01 |
| AIR.PA | 0.0112 | 0.0391 | 0.6755 | 0.0738 | **0.0092** | 0.0109 | 1.2386 | 1.2386 | 1.5966 | 0.0114 |
| ALV.DE | **0.0023** | 0.0783 | 0.657 | 0.0396 | 0.0164 | 0.0027 | 0.8007 | 0.8007 | 0.768 | 0.0053 |
| AMS.MC | 0.0036 | 0.0414 | 0.7524 | 0.0602 | 0.011 | 0.0032 | 0.9384 | 0.9384 | 0.9264 | **0.0031** |
| ABI.BR | 0.0188 | 0.2067 | 0.6666 | 0.2001 | 0.0167 | **0.0164** | 0.6892 | 0.6892 | 1.0024 | 0.0188 |
| ASML.AS | 0.0055 | 0.0144 | 0.8 | 0.0286 | 0.0073 | 0.0067 | 1.2586 | 1.2586 | 1.2373 | **0.004** |
| CS.PA | 0.0037 | 0.0738 | 0.6666 | 0.0462 | 0.0172 | **0.0028** | 1.2321 | 1.2321 | 1.3658 | 0.0046 |
| BAS.DE | 0.0059 | 0.0207 | 0.6205 | 0.0574 | 0.013 | **0.0049** | 0.7936 | 0.7936 | 0.7683 | 0.0084 |
| BAYN.DE | 0.0212 | 0.0316 | 0.5107 | 0.0803 | **0.0068** | 0.0127 | 0.8021 | 0.8021 | 0.8863 | 0.008 |
| BMW.DE | 0.0127 | 0.033 | 0.7253 | 0.0635 | 0.0137 | **0.012** | 1.1251 | 1.1251 | 1.0789 | 0.0128 |
| BNP.PA | 0.0082 | 0.0369 | 0.7465 | 0.0596 | 0.0146 | **0.0081** | 1.5176 | 1.5176 | 1.9737 | 0.0135 |
| CRG.IR | **0.0036** | 0.0162 | 0.7938 | 0.0446 | 0.0042 | 0.0041 | 1.157 | 1.157 | 1.3483 | 0.0039 |
| DAI.SG | 0.0073 | 0.0225 | 0.6154 | 0.0339 | 0.0104 | 0.0279 | 1.102 | 1.102 | 1.0814 | **0.0071** |
| BN.PA | 0.0063 | 0.0161 | 0.4648 | 0.0637 | 0.0047 | **0.004** | 0.2742 | 0.2742 | 9.3079 | 0.0055 |
| DB1.DE | 0.0246 | 0.1283 | 0.6758 | 0.3828 | **0.0162** | 0.0486 | 0.6011 | 0.6011 | 2.6608 | 0.0194 |
| DPWA.F | 0.0334 | **0.0055** | 0.6495 | 0.066 | 0.0323 | 0.743 | 2.4472 | 2.4472 | 2.7756 | |
| DTE.DE | **0.0071** | 0.0655 | 0.5498 | 0.0872 | 0.0096 | 0.0089 | 0.4702 | 0.4702 | 13.4383 | 0.0101 |
| ENEL.MI | 0.0095 | 0.0207 | 0.5142 | 0.0521 | 0.0125 | **0.007** | 0.6227 | 0.6227 | 0.6256 | 0.0116 |
| ENGI.PA | 0.0119 | 0.0872 | 0.4992 | 0.3394 | 0.0099 | **0.0057** | 0.5489 | 0.5489 | 3.9975 | 0.0081 |
| ENI.MI | **0.0019** | 0.0143 | 0.8133 | 0.0313 | 0.0116 | 0.0021 | 0.8143 | 0.8143 | 0.6764 | 0.0032 |
| EL.PA | 0.0051 | 0.0161 | 0.6149 | 0.0437 | 0.0053 | 0.0036 | 0.5142 | 0.5142 | 1.1467 | **0.0025** |
| FLTR.IR | 0.0066 | 0.0874 | 0.7082 | 0.0856 | **0.0064** | 0.0096 | 1.1012 | 1.1012 | 1.3797 | 0.0073 |
| IBE.MC | 0.0029 | 0.0237 | 0.6411 | 0.0233 | 0.0125 | **0.0027** | 0.6082 | 0.6082 | 1.3948 | 0.0047 |
| ITX.MC | 0.0062 | 0.0207 | 0.77 | 0.1003 | 0.0025 | 0.0029 | 0.7034 | 0.7034 | 0.6758 | **0.0024** |
| IFX.DE | 0.0034 | 0.0223 | 0.8842 | 0.0107 | 0.007 | **0.0025** | 1.8955 | 1.8955 | 1.9827 | 0.0054 |
| INGA.AS | **0.0049** | 0.012 | 0.7213 | 0.011 | 0.0258 | 0.0072 | 1.9188 | 1.9188 | 2.2264 | 0.0152 |
| ISP.MI | 0.0037 | 0.0108 | 0.7569 | 0.0144 | 0.0188 | 0.004 | 1.9873 | 1.9873 | 2.5184 | 0.0083 |
| KER.PA | **0.0023** | 0.0141 | 0.5975 | 0.035 | 0.0033 | 0.0026 | 0.8673 | 0.8673 | 1.0216 | 0.0029 |
| KNEBV.HE | **0.0014** | 0.0039 | 0.5463 | 0.0029 | 0.0079 | 0.0015 | 0.5853 | 0.5853 | 0.4449 | 0.0041 |
| OR.PA | 0.0024 | 0.0097 | 0.5371 | 0.0137 | 0.0019 | **0.0013** | 0.3537 | 0.3537 | 0.2497 | 0.0014 |
| MC.PA | 0.0022 | 0.0073 | 0.6147 | 0.0082 | 0.0066 | **0.0015** | 0.8061 | 0.8061 | 0.7407 | 0.0017 |
| MUV2.DE | **0.002** | 0.0427 | 0.5988 | 0.0381 | 0.0163 | 0.0032 | 0.5381 | 0.5381 | 0.8458 | 0.002 |
| NOKIA.HE | 0.0065 | 0.0372 | 0.6941 | 0.0376 | 0.0096 | 0.0053 | 1.4845 | 1.4845 | 2.2198 | **0.0039** |
| RI.PA | 0.0039 | 0.0192 | 0.4609 | 0.1145 | 0.0058 | **0.0031** | 0.303 | 0.303 | 65.4832 | 0.0063 |
| ORA.PA | 0.0053 | 0.0169 | 0.5732 | 0.0303 | 0.0058 | 0.0027 | 0.501 | 0.501 | 1.2918 | **0.0021** |
| PHIA.AS | **0.0054** | 0.0501 | 0.7383 | 0.1362 | 0.0139 | 0.0059 | 0.8664 | 0.8664 | 0.8812 | 0.0128 |
| PRX.AS | 0.0086 | 0.0291 | 1.0647 | 0.0078 | 0.0196 | 0.0087 | 3.0516 | 3.0516 | 3.2178 | **0.0076** |
| SAF.PA | **0.0057** | 0.0422 | 0.6791 | 0.0513 | 0.0079 | 0.0073 | 1.0604 | 1.0604 | 1.2197 | 0.0059 |
| SAN.PA | 0.0236 | 0.0275 | 0.4521 | 0.4713 | 0.0069 | **0.0036** | 0.3201 | 0.3201 | 16.7582 | 0.0122 |
| SAN.MC | **0.0028** | 0.0148 | 0.7475 | 0.0132 | 0.0056 | 0.004 | 1.526 | 1.526 | 1.626 | 0.0034 |
| SAP.DE | 0.0041 | 0.0201 | 0.6761 | 0.0741 | 0.0082 | **0.0022** | 0.4397 | 0.4397 | 1.9351 | 0.0029 |
| SU.PA | 0.0033 | 0.0106 | 0.7005 | 0.0097 | 0.0083 | 0.0047 | 1.0336 | 1.0336 | 0.9548 | **0.0024** |
| SIE.DE | 0.0087 | 0.0411 | 0.644 | 0.1874 | 0.0049 | **0.0047** | 0.6531 | 0.6531 | 1.1775 | 0.0049 |
| TTE.PA | 0.0075 | 0.1445 | 0.7544 | 0.5136 | 0.0078 | **0.0069** | 0.6805 | 0.6805 | 8.6142 | 0.0088 |
| TEF.MC | 0.0034 | 0.0166 | 0.7362 | 0.0395 | 0.009 | **0.0024** | 0.7562 | 0.7562 | 0.6612 | 0.0031 |
| DG.PA | 0.0085 | 0.0887 | 0.5872 | 0.1335 | 0.0088 | **0.0057** | 0.5769 | 0.5769 | 0.9634 | 0.0093 |
| VIV.PA | 0.0131 | 0.0944 | 0.4919 | 0.2117 | **0.0081** | 0.0227 | 0.564 | 0.564 | 23.7771 | 0.0108 |
| VOW.DE | **0.0026** | 0.0103 | 0.8663 | 0.0193 | 0.0067 | 0.003 | 1.564 | 1.564 | 1.6107 | 0.0028 |
| VNA.DE | 0.004 | 0.0162 | 0.5819 | 0.0153 | 0.0224 | 0.0061 | 0.8662 | 0.8662 | 0.8138 | **0.0039** |
| Mean | 0.0074 | 0.0403 | 0.6604 | 0.1100 | 0.0101 | 0.0069 | 0.9505 | 0.9505 | 5.2783 | **0.0067** |
| Median | 0.0053 | 0.0265 | 0.6666 | 0.0513 | 0.0083 | **0.0047** | 0.8021 | 0.8021 | 1.2918 | 0.0054 |

**Note.** In bold the best model.

| Index | JB test | ADF test | PP test | ARCH-LM test |
|-------|---------|----------|---------|--------------|
| MSCIW | 24363** | -12** | -1844** | 801** |
| W1DOW | 25064** | -12** | -1814** | 797** |
| W1SGI | 23420** | -12** | -1788** | 1063** |
| DJUS | 38114** | -12** | -2130** | 762** |
| AASGI | 23144** | -12** | -2130** | 782** |
| E1DOW | 20606** | -12** | -1711** | 3406** |
| DJSEUR | 17970** | -12** | -1757** | 2063** |
| W5DOW | 2885** | -12** | -1548** | 816** |
| DJSEMUP | 6412** | -11** | -1533** | 768** |

**Table B.14:** Preliminary statistical hypothesis tests of traditional and ESG indices.

**Note.** Note. ** denotes p-value significance at the 1% level.

| | AIC | BIC | LL |
|-------|-----|-----|-----|
| MGND | **4113.68** | **4151.74** | -2049.84 |
| Mixture of Gaussian | 4185.32 | 4212.51 | -2087.66 |
| Mixture of Gaussian-Laplace | 4123.13 | 4155.32 | -2059.07 |
| Normal distribution | 4764.38 | 4775.25 | -2380.19 |
| GND | 4637.69 | 4654.00 | -2315.85 |

**Table B.15:** AIC, BIC, and log-likelihood (LL) of estimated models on MSCIW index.

**Note.** Values in bold indicate the best model according to AIC and BIC.

| | Weighted Ljung-Box test | | Weighted ARCH-LM test | |
|-------|-----------------|---------|-----------------|---------|
| Index | Test statistic | p-value | Test statistic | p-value |
| W1DOW | 1.8554 | 0.7525 | 0.4569 | 0.4991 |
| W1SGI | 2.3509 | 0.6093 | 0.1420 | 0.7063 |
| DJUS | 0.3955 | 0.9971 | 0.0048 | 0.9455 |
| AASGI | 1.5563 | 0.8327 | 1.0320 | 0.3097 |
| E1DOW | 3.0091 | 0.4290 | 0.1228 | 0.7261 |
| DJSEUR | 2.6158 | 0.5337 | 6e-05 | 0.9934 |
| W5DOW | 2.6954 | 0.5116 | 0.1250 | 0.7237 |
| DJSEMUP | 1.6800 | 0.8006 | 0.7300 | 0.3929 |

**Table B.16:** Diagnostic results of the AR(1)-EGARCH-M model.

**Table B.17:** PRMSE for the 7-model family estimated with CMGND with $\pi_1 = 0.7$.

| $N$ | $\pi_1$ | $\mu_1$ | $\sigma_1$ | $\nu_1$ | $\pi_2$ | $\mu_2$ | $\sigma_2$ | $\nu_2$ |
|---|---|---|---|---|---|---|---|---|
| **CUU** | | | | | | | | |
| 250 | 19.94 | 6.97 | 43.39 | 39.96 | 46.53 | 6.97 | 37.97 | 55.77 |
| 500 | 18.91 | 4.49 | 32.45 | 32.73 | 44.11 | 4.49 | 34.59 | 38.87 |
| 1000 | 18.02 | 3.44 | 18.58 | 25.09 | 42.05 | 3.44 | 24.99 | 34.56 |
| 1500 | 17.44 | 3.06 | 10.84 | 20.79 | 40.69 | 3.06 | 19.65 | 34.51 |
| 2000 | 18.07 | 2.33 | 7.01 | 19.75 | 42.17 | 2.33 | 19.26 | 34.72 |
| **UCU** | | | | | | | | |
| 250 | 16.77 | 44.61 | 15.13 | 61.65 | 39.14 | 13.89 | 15.13 | 56.7 |
| 500 | 11.44 | 30.92 | 11.58 | 31.13 | 26.69 | 10.84 | 11.58 | 39.19 |
| 1000 | 6.47 | 17.88 | 7.03 | 13.81 | 15.09 | 4.99 | 7.03 | 20.87 |
| 1500 | 5.24 | 14.55 | 5.71 | 9.57 | 12.22 | 4.35 | 5.71 | 16.06 |
| 2000 | 4.95 | 13.56 | 5.11 | 9.78 | 11.55 | 4.11 | 5.11 | 16.43 |
| **UUC** | | | | | | | | |
| 250 | 5.29 | 5.23 | 5.39 | 26.34 | 12.35 | 6.5 | 10.79 | 26.34 |
| 500 | 3.85 | 3.16 | 3.8 | 16.73 | 8.98 | 4.89 | 8.68 | 16.73 |
| 1000 | 2.53 | 2.35 | 2.74 | 10.94 | 5.91 | 3.28 | 6.25 | 10.94 |
| 1500 | 2.1 | 1.93 | 2.22 | 8.73 | 4.9 | 2.75 | 4.85 | 8.73 |
| 2000 | 1.95 | 1.63 | 1.96 | 7.71 | 4.56 | 2.28 | 4.07 | 7.71 |
| **CCU** | | | | | | | | |
| 250 | 25.69 | 13.15 | 6.9 | 18.01 | 59.94 | 13.15 | 6.9 | 37.28 |
| 500 | 23.96 | 8 | 5.36 | 11.73 | 55.9 | 8 | 5.36 | 33.04 |
| 1000 | 23.9 | 6.1 | 3.58 | 9.05 | 55.78 | 6.1 | 3.58 | 29.36 |
| 1500 | 23.19 | 5.02 | 2.7 | 7.37 | 54.11 | 5.02 | 2.7 | 25.76 |
| 2000 | 21.07 | 4.05 | 2.5 | 6 | 49.16 | 4.05 | 2.5 | 24.32 |
| **CUC** | | | | | | | | |
| 250 | 7.95 | 6.02 | 7.18 | 20 | 18.56 | 6.02 | 8.74 | 20 |
| 500 | 5.82 | 3.73 | 5.53 | 16.83 | 13.59 | 3.73 | 7.08 | 16.83 |
| 1000 | 3.98 | 2.84 | 3.69 | 11.62 | 9.3 | 2.84 | 4.79 | 11.62 |
| 1500 | 2.97 | 2.54 | 2.99 | 10.08 | 6.93 | 2.54 | 4.03 | 10.08 |
| 2000 | 3.05 | 2.02 | 2.82 | 8.82 | 7.11 | 2.02 | 3.55 | 8.82 |
| **UCC** | | | | | | | | |
| 250 | 7.02 | 19.58 | 7.6 | 26.93 | 16.37 | 5.91 | 7.6 | 26.93 |
| 500 | 4.94 | 13.93 | 5.04 | 15.71 | 11.52 | 3.93 | 5.04 | 15.71 |
| 1000 | 3.73 | 9.79 | 3.57 | 10.99 | 8.69 | 2.69 | 3.57 | 10.99 |
| 1500 | 2.76 | 6.9 | 2.77 | 8.77 | 6.44 | 2.22 | 2.77 | 8.77 |
| 2000 | 2.31 | 6.27 | 2.16 | 7.4 | 5.4 | 1.95 | 2.16 | 7.4 |
| **CCC** | | | | | | | | |
| 250 | 15.74 | 9.67 | 4.2 | 17.84 | 36.74 | 9.67 | 4.2 | 17.84 |
| 500 | 16.01 | 6.22 | 3.23 | 12.18 | 37.36 | 6.22 | 3.23 | 12.18 |
| 1000 | 15.94 | 4.4 | 2.44 | 8.62 | 37.18 | 4.4 | 2.44 | 8.62 |
| 1500 | 16.07 | 3.63 | 2.05 | 7.27 | 37.49 | 3.63 | 2.05 | 7.27 |
| 2000 | 15.97 | 2.98 | 1.85 | 6.34 | 37.27 | 2.98 | 1.85 | 6.34 |

**Table B.18:** PRMSE for the 7-model family estimated with MGND with $\pi_1 = 0.7$.

| $N$ | $\pi_1$ | $\mu_1$ | $\sigma_1$ | $\nu_1$ | $\pi_2$ | $\mu_2$ | $\sigma_2$ | $\nu_2$ |
|---|---|---|---|---|---|---|---|---|
| **CUU** | | | | | | | | |
| 250 | 27.94 | 26.72 | 39.25 | 41.41 | 65.2 | 44.14 | 45.24 | 61.4 |
| 500 | 24.53 | 16.8 | 32.58 | 33.21 | 57.24 | 26.56 | 43.76 | 49.77 |
| 1000 | 22.12 | 11.39 | 28.82 | 25.9 | 51.6 | 15.92 | 42.79 | 38.21 |
| 1500 | 22.32 | 8 | 24.83 | 24.31 | 52.07 | 10.45 | 41.94 | 36.36 |
| 2000 | 20.61 | 3.5 | 23.95 | 21.88 | 48.1 | 7.66 | 41.34 | 34.41 |
| **UCU** | | | | | | | | |
| 250 | 32.99 | 74.81 | 26.76 | 89.36 | 76.98 | 28.54 | 38.87 | 71.55 |
| 500 | 27.04 | 57.75 | 19.96 | 43.79 | 63.1 | 25.02 | 30.87 | 55.39 |
| 1000 | 19.12 | 31.79 | 12.47 | 22 | 44.62 | 18.08 | 27.11 | 32.9 |
| 1500 | 17.64 | 27.01 | 10.61 | 19.78 | 41.17 | 16.74 | 25.48 | 28.28 |
| 2000 | 14.02 | 25.6 | 8.59 | 13.55 | 32.71 | 14.2 | 20.94 | 23.49 |
| **UUC** | | | | | | | | |
| 250 | 5.32 | 5.22 | 5.84 | 32.18 | 12.42 | 7.1 | 13.37 | 56.42 |
| 500 | 3.75 | 3.16 | 4.08 | 19.09 | 8.76 | 4.7 | 9.25 | 37.22 |
| 1000 | 2.54 | 2.36 | 2.97 | 12.95 | 5.93 | 3.41 | 6.71 | 21.74 |
| 1500 | 2.15 | 1.96 | 2.31 | 10.36 | 5.01 | 2.89 | 5.38 | 16.6 |
| 2000 | 1.98 | 1.66 | 2.12 | 8.97 | 4.61 | 2.36 | 4.53 | 15.41 |
| **CCU** | | | | | | | | |
| 250 | 44.19 | 137.19 | 47.32 | 39.86 | 103.1 | 141.01 | 35.1 | 57.41 |
| 500 | 44.35 | 134.92 | 41.46 | 34.65 | 103.49 | 124.78 | 31.76 | 48.25 |
| 1000 | 48 | 141.73 | 39.81 | 27.97 | 111.99 | 104.39 | 25.95 | 46.86 |
| 1500 | 48.11 | 143.95 | 38.35 | 25.91 | 112.25 | 99.83 | 24.61 | 48.4 |
| 2000 | 53.28 | 150.05 | 39.6 | 26.43 | 124.31 | 87.42 | 23.27 | 47.68 |
| **CUC** | | | | | | | | |
| 250 | 14.19 | 8.64 | 11.59 | 56.51 | 33.11 | 20.23 | 20.95 | 60.89 |
| 500 | 11.55 | 5.68 | 7.05 | 39.86 | 26.94 | 12.58 | 16.68 | 39.47 |
| 1000 | 8.58 | 2.84 | 4.04 | 23.39 | 20.03 | 8.1 | 13.03 | 27.04 |
| 1500 | 7.05 | 1.99 | 3.17 | 19.04 | 16.46 | 6.87 | 11.29 | 23.28 |
| 2000 | 6.28 | 1.74 | 2.87 | 16.94 | 14.65 | 5.6 | 10 | 22.08 |
| **UCC** | | | | | | | | |
| 250 | 18.74 | 39.54 | 13.53 | 56.85 | 43.73 | 17.68 | 27.39 | 65.75 |
| 500 | 12.87 | 20.8 | 6.67 | 35.2 | 30.04 | 14.17 | 22.57 | 49.12 |
| 1000 | 9.59 | 15.15 | 5.08 | 23.32 | 22.37 | 11.19 | 19.03 | 29.85 |
| 1500 | 7.66 | 10.94 | 3.64 | 18.32 | 17.88 | 8.27 | 13.17 | 20.25 |
| 2000 | 6.44 | 10.47 | 3.26 | 11.6 | 15.03 | 7.25 | 12.14 | 17.51 |
| **CCC** | | | | | | | | |
| 250 | 16.06 | 100.31 | 34.39 | 35.25 | 37.47 | 179.45 | 53.48 | 47.72 |
| 500 | 16.4 | 105.32 | 35.38 | 33.43 | 38.27 | 173.55 | 52.05 | 41.8 |
| 1000 | 15.78 | 103.24 | 34.8 | 31.94 | 36.82 | 175.15 | 53.6 | 42.8 |
| 1500 | 15.31 | 101.72 | 33.88 | 30.76 | 35.73 | 174.1 | 52.94 | 42.09 |
| 2000 | 15.84 | 101.57 | 33.73 | 30.33 | 36.97 | 176.65 | 53.89 | 43.67 |

**Table B.19:** PRMSE for the 7-model family estimated with CMGND with $\pi_1 = 0.5$.

| $N$ | $\pi_1$ | $\mu_1$ | $\sigma_1$ | $\nu_1$ | $\pi_2$ | $\mu_2$ | $\sigma_2$ | $\nu_2$ |
|---|---|---|---|---|---|---|---|---|
| **CUU** | | | | | | | | |
| 250 | 30.24 | 8.09 | 71.13 | 53.09 | 30.24 | 8.09 | 34.31 | 53.55 |
| 500 | 25.34 | 5.27 | 48.23 | 39.67 | 25.34 | 5.27 | 24.64 | 33.79 |
| 1000 | 21.73 | 3.81 | 33.35 | 29.88 | 21.73 | 3.81 | 17.63 | 20.96 |
| 1500 | 18.32 | 3.42 | 14.45 | 24.41 | 18.32 | 3.42 | 9.86 | 17.53 |
| 2000 | 15.37 | 2.71 | 13.73 | 23.54 | 15.37 | 2.71 | 9.19 | 15.71 |
| **UCU** | | | | | | | | |
| 250 | 19.34 | 48.01 | 13.25 | 62.5 | 19.34 | 8.67 | 13.25 | 45.22 |
| 500 | 14.23 | 30.97 | 9.83 | 31.06 | 14.23 | 6.4 | 9.83 | 31.24 |
| 1000 | 9.13 | 21.09 | 5.7 | 12.62 | 9.13 | 3.52 | 5.7 | 19.31 |
| 1500 | 8.08 | 17.81 | 5.79 | 12.19 | 8.08 | 3.71 | 5.79 | 15 |
| 2000 | 6.19 | 16.23 | 3.7 | 7.83 | 6.19 | 2.47 | 3.7 | 12.93 |
| **UUC** | | | | | | | | |
| 250 | 8.9 | 7.66 | 7 | 29.25 | 8.9 | 5.37 | 9.21 | 29.25 |
| 500 | 6.07 | 4.29 | 5 | 17.19 | 6.07 | 3.45 | 6.47 | 17.19 |
| 1000 | 4.08 | 3.28 | 3.37 | 11.1 | 4.08 | 2.43 | 4.91 | 11.1 |
| 1500 | 3.47 | 2.78 | 2.75 | 8.23 | 3.47 | 2.19 | 3.9 | 8.23 |
| 2000 | 3.22 | 2.38 | 2.48 | 7.13 | 3.22 | 1.94 | 3.48 | 7.13 |
| **CCU** | | | | | | | | |
| 250 | 27.46 | 12.37 | 6.13 | 21.23 | 27.46 | 12.37 | 6.13 | 38.3 |
| 500 | 25.49 | 7.53 | 4.76 | 13.1 | 25.49 | 7.53 | 4.76 | 28.54 |
| 1000 | 22.08 | 5.73 | 3.29 | 9.36 | 22.08 | 5.73 | 3.29 | 23.46 |
| 1500 | 22.74 | 4.69 | 2.57 | 7.64 | 22.74 | 4.69 | 2.57 | 19.96 |
| 2000 | 20.95 | 3.86 | 2.28 | 5.57 | 20.95 | 3.86 | 2.28 | 19.46 |
| **CUC** | | | | | | | | |
| 250 | 12.67 | 7.84 | 10.38 | 21.53 | 12.67 | 7.84 | 6.92 | 21.53 |
| 500 | 9.51 | 5.02 | 7.54 | 17.23 | 9.51 | 5.02 | 6.38 | 17.23 |
| 1000 | 6.62 | 3.58 | 4.78 | 11.67 | 6.62 | 3.58 | 3.92 | 11.67 |
| 1500 | 5.55 | 3.2 | 4.33 | 9.72 | 5.55 | 3.2 | 3.34 | 9.72 |
| 2000 | 5.15 | 2.57 | 3.82 | 8.22 | 5.15 | 2.57 | 2.86 | 8.22 |
| **UCC** | | | | | | | | |
| 250 | 9.72 | 24.67 | 9.08 | 31.91 | 9.72 | 4.81 | 9.08 | 31.91 |
| 500 | 7.28 | 18.02 | 6.08 | 18.32 | 7.28 | 3.44 | 6.08 | 18.32 |
| 1000 | 5.14 | 11.46 | 4 | 11.93 | 5.14 | 2.19 | 4 | 11.93 |
| 1500 | 3.87 | 9.34 | 3.23 | 8.51 | 3.87 | 1.84 | 3.23 | 8.51 |
| 2000 | 3.35 | 8.03 | 2.61 | 7.57 | 3.35 | 1.57 | 2.61 | 7.57 |
| **CCC** | | | | | | | | |
| 250 | 25.33 | 9.67 | 4.2 | 17.84 | 25.33 | 9.67 | 4.2 | 17.84 |
| 500 | 26.6 | 6.22 | 3.23 | 12.18 | 26.6 | 6.22 | 3.23 | 12.18 |
| 1000 | 24.63 | 4.4 | 2.44 | 8.62 | 24.63 | 4.4 | 2.44 | 8.62 |
| 1500 | 24.04 | 3.63 | 2.05 | 7.27 | 24.04 | 3.63 | 2.05 | 7.27 |
| 2000 | 24.88 | 2.98 | 1.85 | 6.34 | 24.88 | 2.98 | 1.85 | 6.34 |

**Table B.20:** PRMSE for the 7-model family estimated with MGND with $\pi_1 = 0.5$.

| $N$ | $\pi_1$ | $\mu_1$ | $\sigma_1$ | $\nu_1$ | $\pi_2$ | $\mu_2$ | $\sigma_2$ | $\nu_2$ |
|---|---|---|---|---|---|---|---|---|
| **CUU** | | | | | | | | |
| 250 | 39.03 | 29.61 | 44.03 | 41.48 | 39.03 | 56.36 | 29.2 | 62.57 |
| 500 | 35.34 | 12.89 | 34.45 | 36.12 | 35.34 | 44.96 | 23.51 | 41.86 |
| 1000 | 31.38 | 6.78 | 24.89 | 32.27 | 31.38 | 29.32 | 15.61 | 27.88 |
| 1500 | 31.7 | 5.21 | 17.5 | 31.62 | 31.7 | 26.37 | 12.41 | 23.83 |
| 2000 | 28.85 | 4.3 | 10.37 | 31.23 | 28.85 | 17.42 | 8.71 | 18.78 |
| **UCU** | | | | | | | | |
| 250 | 34.46 | 93.75 | 31.18 | 85.68 | 34.46 | 15.21 | 24.05 | 60.69 |
| 500 | 29.95 | 79.72 | 23.48 | 45.54 | 29.95 | 11.49 | 16.4 | 44.44 |
| 1000 | 22.22 | 57.88 | 17.65 | 18.94 | 22.22 | 7.3 | 9.44 | 26.07 |
| 1500 | 18.07 | 48.09 | 14.24 | 14.78 | 18.07 | 6.03 | 8.42 | 17.69 |
| 2000 | 14.54 | 37.12 | 9.8 | 10.7 | 14.54 | 4.73 | 6.14 | 14.58 |
| **UUC** | | | | | | | | |
| 250 | 8.57 | 7.72 | 7.99 | 37.21 | 8.57 | 5.41 | 9.55 | 48.39 |
| 500 | 6.16 | 4.43 | 5.2 | 20.88 | 6.16 | 3.59 | 6.74 | 29.68 |
| 1000 | 4.17 | 3.45 | 3.65 | 13.97 | 4.17 | 2.57 | 5.07 | 21.19 |
| 1500 | 3.6 | 2.86 | 2.92 | 10.89 | 3.6 | 2.3 | 4.05 | 13.94 |
| 2000 | 3.31 | 2.47 | 2.56 | 9.15 | 3.31 | 2.01 | 3.74 | 11.98 |
| **CCU** | | | | | | | | |
| 250 | 48.59 | 153.72 | 51.58 | 47.27 | 48.59 | 126.61 | 35.75 | 51.39 |
| 500 | 46.89 | 146.7 | 47.4 | 33.38 | 46.89 | 114.23 | 32.27 | 49.21 |
| 1000 | 41.55 | 149.64 | 45.16 | 29.56 | 41.55 | 86.8 | 25.69 | 44.67 |
| 1500 | 40.71 | 149.82 | 44.15 | 28.99 | 40.71 | 84.31 | 25.42 | 44.39 |
| 2000 | 40.51 | 141.95 | 41.35 | 25.33 | 40.51 | 77.88 | 23.41 | 43.05 |
| **CUC** | | | | | | | | |
| 250 | 25.78 | 15.33 | 19.53 | 54.38 | 25.78 | 34.05 | 17.3 | 54.27 |
| 500 | 18.63 | 7.51 | 10.48 | 42.96 | 18.63 | 25.81 | 12.73 | 32.88 |
| 1000 | 16.53 | 4.58 | 8.74 | 34.2 | 16.53 | 20.1 | 9.84 | 23.89 |
| 1500 | 13.23 | 3.24 | 6.46 | 27.61 | 13.23 | 12.7 | 8.45 | 19.6 |
| 2000 | 10.51 | 2.69 | 4.93 | 24.36 | 10.51 | 4.45 | 7.64 | 18.78 |
| **UCC** | | | | | | | | |
| 250 | 26.24 | 65.53 | 21.1 | 60.24 | 26.24 | 13.57 | 22.1 | 57.66 |
| 500 | 17.25 | 40.09 | 13.57 | 37.18 | 17.25 | 7.75 | 11.88 | 36.31 |
| 1000 | 9.39 | 17.58 | 5.63 | 19.67 | 9.39 | 3.94 | 6.12 | 21.82 |
| 1500 | 8.86 | 22.83 | 7.64 | 15.39 | 8.86 | 4.4 | 6.66 | 15.31 |
| 2000 | 6.61 | 17.39 | 5.87 | 13.65 | 6.61 | 3.42 | 4.05 | 12.63 |
| **CCC** | | | | | | | | |
| 250 | 36.35 | 139.99 | 43.7 | 43.18 | 36.35 | 150.55 | 46.19 | 40.68 |
| 500 | 31.17 | 139.64 | 43.38 | 36.45 | 31.17 | 147.34 | 45.6 | 39.19 |
| 1000 | 32.07 | 136.91 | 43.76 | 37.44 | 32.07 | 150.3 | 46.57 | 38.07 |
| 1500 | 32.85 | 135.73 | 43.01 | 36.69 | 32.85 | 149.12 | 45.83 | 37.04 |
| 2000 | 33.63 | 134.34 | 42.68 | 35.96 | 33.63 | 153.22 | 47.12 | 39.16 |

**Table B.21:** Comparative analysis part II, AIC of two-component MGND and CMGND models.

| Stock | MGND | CMGND | | | | | | |
|---|---|---|---|---|---|---|---|---|
| | | CUU | UCU | UUC | CCU | CUC | UCC | CCC |
| ADS.DE | 13,587.6 | 13,586.83 | 13,588.38 | 13,588.44 | 13,587.15 | **13,586.35** | 13,607.3 | 13,627.75 |
| ADYEN.AS | 6,487.87 | 6,488.72 | 6,487.33 | 6,486.14 | 6,487 | 6,486.37 | **6,485.5** | 6,491.03 |
| AD.AS | 11,108.64 | 11,112.1 | 11,108.32 | **11,106.08** | 11,109.57 | 11,109.08 | 11,144.49 | 11,161.43 |
| AI.PA | 11,526.42 | 11,523.43 | 11,523.3 | 11,526.8 | 11,526.37 | **11,521.29** | 11,534.56 | 11,547.43 |
| AIR.PA | 14,464.38 | 14,462.88 | 14,462.25 | 14,464.18 | 14,461.21 | **14,460.75** | 14,477.78 | 14,495.7 |
| ALV.DE | 12,169.38 | 12,173.25 | **12,168.08** | 12,170.73 | 12,171.96 | 12,170.74 | 12,193.76 | 12,216.26 |
| AMS.MC | 12,954.11 | 12,958.43 | **12,952.98** | 12,956.15 | 12,957.14 | 12,956.12 | 12,972.42 | 12,995.75 |
| ABI.BR | 12,615.54 | 12,611.32 | 12,613.76 | 12,613.9 | 12,610.78 | **12,609.38** | 12,656.9 | 12,676.69 |
| ASML.AS | 14,336.54 | 14,336.66 | 14,334.73 | 14,335.98 | 14,334.75 | **14,334.53** | 14,339.75 | 14,344.52 |
| CS.PA | 13,772.19 | 13,777.13 | **13,771.47** | 13,773.2 | 13,776.48 | 13,776.31 | 13,796.77 | 13,812.08 |
| BAS.DE | 13,185.91 | 13,192.56 | **13,184.68** | 13,186.87 | 13,192.89 | 13,190.59 | 13,195.62 | 13,211.69 |
| BAYN.DE | 13,321.16 | 13,320.16 | 13,320.36 | 13,322.15 | 13,319.98 | **13,318.17** | 13,331.23 | 13,331.76 |
| BMW.DE | 13,645.54 | 13,647.22 | 13,645.32 | 13,643.18 | 13,645.92 | 13,644.96 | **13,642.94** | 13,659.84 |
| BNP.PA | 14,910.16 | 14,912.19 | **14,908.63** | 14,910.16 | 14,910.69 | 14,909.67 | 14,926.34 | 14,933.94 |
| CRG.IR | 14,291.35 | 14,289.37 | 14,289.33 | 14,289.77 | 14,288.72 | **14,287.26** | 14,310.32 | 14,332.36 |
| DAI.SG | 13,706.43 | **13,706.12** | 13,720.09 | 13,722.1 | 13,720.3 | 13,718.83 | 13,721.33 | 13,720.33 |
| BN.PA | 11,235.4 | 11,235.7 | **11,232.59** | 11,237.11 | 11,237.27 | 11,233.29 | 11,245.86 | 11,260.9 |
| DB1.DE | 12,475.73 | 12,472.5 | 12,474.12 | 12,476.46 | 12,473.13 | **12,470.14** | 12,497.27 | 12,508.22 |
| DPWA.F | 15,287.13 | **15,267.08** | 15,604.68 | 15,604.55 | 15,602.55 | 15,602.71 | 15,604.57 | 15,600.67 |
| DTE.DE | 11,623.82 | 11,622.28 | 11,622.77 | 11,624.96 | 11,622.41 | **11,620.35** | 11,634.21 | 11,650.1 |
| ENEL.MI | 13,087.3 | 13,091.09 | **13,085.54** | 13,087.91 | 13,091.8 | 13,088.81 | 13,095.51 | 13,101.61 |
| ENGI.PA | 12,791.14 | 12,795.55 | **12,788.61** | 12,793.55 | 12,797.04 | 12,793.46 | 12,806.1 | 12,828.29 |
| ENI.MI | 12,960.72 | 12,969.81 | **12,959.2** | 12,963.59 | 12,971.02 | 12,968.59 | 12,964.29 | 13,000.97 |
| EL.PA | 12,154.42 | 12,153.94 | **12,151.99** | 12,155.72 | 12,154.01 | 12,152.02 | 12,158.7 | 12,173.65 |
| FLTR.IR | 13,832.53 | **13,832.38** | 13,838.22 | 13,835.57 | 13,835.77 | 13,832.81 | 13,868.54 | 13,868.79 |
| IBE.MC | 12,058.94 | 12,057.28 | 12,057.93 | 12,057.9 | 12,056.51 | **12,056.38** | 12,072.58 | 12,082.09 |
| ITX.MC | 13,105.9 | 13,104 | 13,103.71 | 13,109.33 | 13,104.25 | **13,101.41** | 13,112.7 | 13,129.98 |
| IFX.DE | **15,297.79** | 15,298.43 | 15,298.51 | 15,300.79 | 15,300.05 | 15,298.88 | 15,306.36 | 15,304.54 |
| INGA.AS | 15,137.45 | 15,137.81 | 15,135.56 | 15,136.92 | **15,135.37** | 15,135.92 | 15,139.71 | 15,143.24 |
| ISP.MI | 15,415.48 | 15,414.49 | 15,413.93 | 15,412.58 | **15,411.99** | 15,413.59 | 15,418.31 | 15,426.85 |
| KER.PA | 13,728.68 | 13,727.01 | 13,726.45 | 13,727.44 | 13,724.8 | **13,724.61** | 13,744.32 | 13,756.52 |
| KNEBV.HE | 12,377.71 | 12,376.12 | 12,375.6 | 12,375.73 | **12,373.64** | 12,374.02 | 12,374.98 | 12,377.35 |
| OR.PA | 11,904.43 | 11,902.04 | 11,902.74 | 11,902.79 | **11,900.99** | 11,901.22 | 11,901.27 | 11,922.6 |
| MC.PA | 13,407.35 | 13,405.53 | 13,406.37 | 13,405.07 | **13,404.11** | 13,404.27 | 13,409.65 | 13,421.72 |
| MUV2.DE | 11,690.9 | 11,695.95 | 11,692.8 | **11,689.32** | 11,693.49 | 11,693.24 | 11,735.8 | 11,746.76 |
| NOKIA.HE | 14,693.66 | 14,692.84 | 14,692.5 | 14,691.68 | **14,689.6** | 14,690.81 | 14,713.74 | 14,734.46 |
| RI.PA | 11,491.3 | 11,494.42 | **11,489.55** | 11,493.1 | 11,494.28 | 11,492.34 | 11,504.72 | 11,516.29 |
| ORA.PA | 12,024.56 | 12,025.12 | 12,024.29 | 12,027.9 | 12,025.98 | **12,023.28** | 12,041.47 | 12,043.13 |
| PHIA.MC | **13,367.94** | 13,371.14 | 13,368.9 | 13,372.45 | 13,373.97 | 13,368.44 | 13,385.68 | 13,410.96 |
| PRX.AS | 4,878.32 | 4,876.48 | 4,877.44 | 4,877.5 | 4,877.17 | 4,880.46 | **4,875.59** | 4,881.29 |
| SAF.PA | 13,814.6 | 13,811.93 | 13,813.15 | 13,812.96 | 13,810.62 | **13,809.96** | 13,826.27 | 13,836.57 |
| SAN.PA | 12,008.92 | 12,008.2 | 12,006.81 | 12,010.16 | 12,009.22 | **12,005.51** | 12,017.37 | 12,043.39 |
| SAN.MC | 14,929.96 | 14,929.62 | 14,928.31 | 14,928.36 | 14,927.3 | 14,927.49 | **14,926.96** | 14,934.94 |
| SAP.DE | 11,896.08 | 11,902.24 | **11,893.86** | 11,899.53 | 11,903.15 | 11,899.87 | 11,925.95 | 11,935.07 |
| SU.PA | 13,848.53 | 13,847.65 | 13,846.95 | 13,846.89 | 13,846.96 | **13,845.81** | 13,849.11 | 13,849.83 |
| SIE.DE | 12,826.09 | 12,825.11 | 12,823.8 | 12,829.31 | 12,827.43 | **12,822.81** | 12,845.86 | 12,850.86 |
| TTE.PA | 12,785.79 | 12,783.68 | **12,783.19** | 12,790.78 | 12,793.5 | 12,784.23 | 12,809.66 | 12,817.75 |
| TEF.MC | 12,917.25 | 12,922.59 | **12,915.04** | 12,919.87 | 12,923.71 | 12,920.49 | 12,932.99 | 12,954.49 |
| DG.PA | 12,775.87 | 12,774 | 12,778.4 | **12,771.47** | 12,774.27 | 12,772.75 | 12,801.37 | 12,820.98 |
| VIV.PA | 12,371.66 | 12,373.78 | 12,371.98 | 12,375.52 | 12,373.67 | **12,371.25** | 12,375.07 | 12,391.02 |
| VOW.DE | 14,477.53 | 14,475.57 | 14,476.17 | 14,475.49 | 14,474.83 | **14,473.89** | 14,489.8 | 14,493.26 |
| VNA.DE | 9,657.33 | 9,657.53 | 9,659.06 | 9,655.49 | **9,655.14** | 9,657.73 | 9,667.97 | 9,679.45 |

**Note.** In bold the best model.

**Table B.22:** Comparative analysis part II, BIC of two-component MGND and CMGND models.

| Stock | MGND | CMGND | | | | | | |
|---|---|---|---|---|---|---|---|---|
| | | CUU | UCU | UUC | CCU | CUC | UCC | CCC |
| ADS.DE | 13,630.7 | 13,623.77 | 13,625.33 | 13,625.38 | 13,617.93 | **13,617.13** | 13,638.08 | 13,652.37 |
| ADYEN.AS | 6,524.38 | 6,520.01 | 6,518.62 | 6,517.43 | 6,513.07 | 6,512.44 | **6,511.57** | 6,511.89 |
| AD.AS | 11,151.8 | 11,149.09 | 11,145.32 | 11,143.08 | 11,140.4 | **11,139.92** | 11,175.32 | 11,186.09 |
| AI.PA | 11,569.58 | 11,560.42 | 11,560.29 | 11,563.8 | 11,557.2 | **11,552.12** | 11,565.39 | 11,572.09 |
| AIR.PA | 14,507.53 | 14,499.87 | 14,499.23 | 14,501.17 | 14,492.03 | **14,491.57** | 14,508.6 | 14,520.36 |
| ALV.DE | 12,212.48 | 12,210.2 | 12,205.02 | 12,207.67 | 12,202.74 | **12,201.53** | 12,224.55 | 12,240.89 |
| AMS.MC | 12,997.1 | 12,995.28 | 12,989.83 | 12,993.01 | 12,987.85 | **12,986.83** | 13,003.14 | 13,020.32 |
| ABI.BR | 12,658.7 | 12,648.32 | 12,650.75 | 12,650.89 | 12,641.61 | **12,640.21** | 12,687.72 | 12,701.35 |
| ASML.AS | 14,379.71 | 14,373.66 | 14,371.73 | 14,372.98 | 14,365.58 | **14,365.36** | 14,370.58 | 14,369.19 |
| CS.PA | 13,815.34 | 13,814.12 | 13,808.46 | 13,810.19 | 13,807.3 | **13,807.14** | 13,827.6 | 13,836.74 |
| BAS.DE | 13,229.01 | 13,229.51 | 13,221.63 | 13,223.81 | 13,223.68 | **13,221.38** | 13,226.41 | 13,236.32 |
| BAYN.DE | 13,364.26 | 13,357.1 | 13,357.3 | 13,359.09 | 13,350.77 | **13,348.95** | 13,362.02 | 13,356.39 |
| BMW.DE | 13,688.64 | 13,684.17 | 13,682.27 | 13,680.13 | 13,676.7 | 13,675.75 | **13,673.73** | 13,684.47 |
| BNP.PA | 14,953.31 | 14,949.18 | 14,945.61 | 14,947.14 | 14,941.51 | **14,940.49** | 14,957.16 | 14,958.6 |
| CRG.IR | 14,334.45 | 14,326.32 | 14,326.27 | 14,326.71 | 14,319.51 | **14,318.05** | 14,341.11 | 14,356.99 |
| DAI.SG | 13,749.52 | **13,743.04** | 13,757.01 | 13,759.03 | 13,751.08 | 13,749.61 | 13,752.11 | 13,744.94 |
| BN.PA | 11,278.56 | 11,272.7 | 11,269.59 | 11,274.11 | 11,268.1 | **11,264.12** | 11,276.68 | 11,285.56 |
| DB1.DE | 12,518.83 | 12,509.44 | 12,511.06 | 12,513.41 | 12,503.91 | **12,500.93** | 12,528.06 | 12,532.85 |
| DPWA.F | 15,330.22 | **15,304.02** | 15,641.63 | 15,641.49 | 15,633.33 | 15,633.49 | 15,635.35 | 15,625.3 |
| DTE.DE | 11,666.92 | 11,659.22 | 11,659.72 | 11,661.9 | 11,653.2 | **11,651.14** | 11,664.99 | 11,674.73 |
| ENEL.MI | 13,130.42 | 13,128.05 | 13,122.49 | 13,124.86 | 13,122.6 | **13,119.6** | 13,126.3 | 13,126.24 |
| ENGI.PA | 12,834.3 | 12,832.54 | 12,825.6 | 12,830.54 | 12,827.87 | **12,824.29** | 12,836.93 | 12,852.95 |
| ENI.MI | 13,003.83 | 13,006.77 | 12,996.15 | 13,000.54 | 13,001.81 | 12,999.39 | **12,995.08** | 13,025.61 |
| EL.PA | 12,197.58 | 12,190.93 | 12,188.98 | 12,192.71 | 12,184.83 | **12,182.85** | 12,189.32 | 12,198.32 |
| FLTR.IR | 13,875.63 | 13,869.32 | 13,875.16 | 13,872.5 | 13,866.55 | **13,863.6** | 13,899.32 | 13,893.41 |
| IBE.MC | 12,102.09 | 12,094.27 | 12,094.92 | 12,094.89 | 12,087.34 | **12,087.21** | 12,103.4 | 12,106.75 |
| ITX.MC | 13,149.07 | 13,140.99 | 13,140.71 | 13,146.33 | 13,135.08 | **13,132.24** | 13,143.53 | 13,154.65 |
| IFX.DE | 15,340.89 | 15,335.38 | 15,335.45 | 15,337.73 | 15,330.83 | 15,329.67 | 15,337.14 | **15,329.17** |
| INGA.AS | 15,180.6 | 15,174.8 | 15,172.55 | 15,173.9 | **15,166.2** | 15,166.75 | 15,170.54 | 15,167.9 |
| ISP.MI | 15,458.58 | 15,451.44 | 15,450.88 | 15,449.52 | **15,442.78** | 15,444.38 | 15,449.1 | 15,451.48 |
| KER.PA | 13,771.84 | 13,764 | 13,763.44 | 13,764.43 | 13,755.63 | **13,755.44** | 13,775.15 | 13,781.18 |
| KNEBV.HE | 12,420.74 | 12,413.01 | 12,412.5 | 12,412.62 | 12,404.38 | 12,404.76 | 12,405.73 | **12,401.94** |
| OR.PA | 11,947.59 | 11,939.04 | 11,939.73 | 11,939.79 | **11,931.82** | 11,932.05 | 11,932.1 | 11,947.26 |
| MC.PA | 13,450.51 | 13,442.52 | 13,443.37 | 13,442.06 | **13,434.94** | 13,435.1 | 13,440.48 | 13,446.38 |
| MUV2.DE | 11,734 | 11,732.89 | 11,729.74 | 11,726.26 | 11,724.28 | **11,724.02** | 11,766.59 | 11,771.38 |
| NOKIA.HE | 14,736.67 | 14,729.72 | 14,729.37 | 14,728.55 | **14,720.33** | 14,721.54 | 14,744.46 | 14,759.04 |
| RI.PA | 11,534.46 | 11,531.41 | 11,526.54 | 11,530.09 | 11,525.11 | **11,523.17** | 11,535.55 | 11,540.95 |
| ORA.PA | 12,067.72 | 12,062.12 | 12,061.29 | 12,064.89 | 12,056.81 | **12,054.11** | 12,072.3 | 12,067.8 |
| PHIA.AS | 13,411.1 | 13,408.14 | 13,405.9 | 13,409.45 | 13,404.81 | **13,399.27** | 13,416.51 | 13,435.14 |
| PRX.AS | 4,912.95 | 4,906.16 | 4,907.12 | 4,907.19 | 4,901.9 | 4,905.19 | **4,900.32** | 4,901.08 |
| SAF.PA | 13,857.75 | 13,848.92 | 13,850.13 | 13,849.94 | 13,841.44 | **13,840.78** | 13,857.09 | 13,861.23 |
| SAN.PA | 12,052.08 | 12,045.19 | 12,043.81 | 12,047.15 | 12,040.05 | **12,036.34** | 12,048.2 | 12,068.06 |
| SAN.MC | 14,973.11 | 14,966.61 | 14,965.3 | 14,965.35 | 14,958.12 | 14,958.31 | **14,957.79** | 14,958.64 |
| SAP.DE | 11,939.18 | 11,939.18 | 11,930.8 | 11,936.47 | 11,933.94 | **11,930.65** | 11,956.73 | 11,959.7 |
| SU.PA | 13,891.69 | 13,884.64 | 13,883.94 | 13,883.89 | 13,877.78 | 13,876.63 | 13,879.94 | **13,874.49** |
| SIE.DE | 12,869.2 | 12,862.05 | 12,860.74 | 12,866.26 | 12,858.21 | **12,853.6** | 12,876.64 | 12,875.49 |
| TTE.PA | 12,828.94 | 12,820.68 | 12,820.18 | 12,827.77 | 12,824.32 | **12,815.05** | 12,840.49 | 12,842.41 |
| TEF.MC | 12,960.41 | 12,959.58 | 12,952.03 | 12,956.86 | 12,954.53 | **12,951.32** | 12,963.82 | 12,979.15 |
| DG.PA | 12,819.02 | 12,810.99 | 12,815.39 | 12,808.46 | 12,805.1 | **12,803.57** | 12,832.19 | 12,845.64 |
| VIV.PA | 12,414.81 | 12,410.77 | 12,408.97 | 12,412.51 | 12,404.49 | **12,402.07** | 12,405.9 | 12,415.68 |
| VOW.DE | 14,520.63 | 14,512.5 | 14,513.1 | 14,512.42 | 14,505.61 | **14,504.67** | 14,520.58 | 14,517.88 |
| VNA.DE | 9,698.35 | 9,692.69 | 9,694.22 | 9,690.66 | **9,684.44** | 9,687.03 | 9,697.27 | 9,702.89 |

**Note.** In bold the best model.

**Table B.23:** Comparative analysis part II, KS of two-component MGND and CMGND models.

| Stock | MGND | CMGND | | | | | | |
|---|---|---|---|---|---|---|---|---|
| | | CUU | UCU | UUC | CCU | CUC | UCC | CCC |
| ADS.DE | 0.0082 | **0.0012** | 0.0042 | 0.0045 | 0.004 | 0.0049 | 0.0109 | 0.008 |
| ADYEN.AS | 0.0098 | 0.015 | 0.0108 | 0.0115 | 0.0144 | 0.0143 | **0.0088** | 0.0165 |
| AD.AS | **0.0033** | 0.0072 | 0.0059 | 0.006 | 0.0069 | 0.008 | 0.0081 | 0.0178 |
| AI.PA | **0.006** | 0.008 | 0.0076 | 0.0064 | 0.0069 | 0.007 | 0.0118 | 0.0095 |
| AIR.PA | 0.0055 | 0.0067 | **0.005** | 0.0086 | 0.0113 | 0.0085 | 0.0064 | 0.0184 |
| ALV.DE | **0.0051** | 0.0094 | 0.0073 | 0.0052 | 0.0092 | 0.0088 | 0.0103 | 0.0116 |
| AMS.MC | 0.0086 | 0.0111 | 0.011 | **0.0079** | 0.0111 | 0.0098 | 0.0128 | 0.0123 |
| ABI.BR | 0.0072 | 0.0061 | **0.0049** | 0.0089 | 0.0103 | 0.0085 | 0.0074 | 0.0165 |
| ASML.AS | 0.0044 | 0.0064 | 0.0055 | **0.0041** | 0.0061 | 0.0061 | 0.007 | 0.0087 |
| CS.PA | 0.0075 | 0.0077 | 0.0086 | **0.0052** | 0.0087 | 0.0081 | 0.0248 | 0.016 |
| BAS.DE | 0.0049 | 0.0083 | 0.0069 | **0.0044** | 0.0066 | 0.0071 | 0.0074 | 0.0111 |
| BAYN.DE | 0.0084 | 0.0074 | 0.006 | 0.0067 | 0.0074 | **0.0046** | 0.0081 | 0.01 |
| BMW.DE | 0.0081 | 0.0029 | **0.0014** | 0.0107 | 0.0043 | 0.005 | 0.0099 | 0.0056 |
| BNP.PA | 0.0073 | 0.0091 | **0.0031** | 0.0052 | 0.0054 | 0.0047 | 0.009 | 0.0108 |
| CRG.IR | 0.0062 | **0.0047** | 0.0049 | 0.006 | 0.0058 | 0.0062 | 0.0047 | 0.0108 |
| DAI.SG | 0.0094 | **0.0028** | 0.0056 | 0.0048 | 0.0052 | 0.0049 | 0.0065 | 0.0055 |
| BN.PA | 0.0049 | 0.0043 | 0.0043 | 0.0055 | 0.0053 | **0.0032** | 0.0109 | 0.0084 |
| DB1.DE | 0.0034 | 0.0036 | **0.0027** | 0.0057 | 0.0072 | 0.0052 | 0.0073 | 0.0096 |
| DPWA.F | 0.0319 | 0.0263 | 0.0108 | 0.0114 | 0.0107 | 0.0114 | **0.0103** | 0.0108 |
| DTE.DE | **0.0018** | 0.0049 | 0.0033 | 0.005 | 0.0049 | 0.0049 | 0.0118 | 0.0086 |
| ENEL.MI | 0.0053 | 0.0054 | **0.0044** | 0.0095 | 0.005 | 0.0054 | 0.0085 | 0.0052 |
| ENGI.PA | 0.0056 | 0.0062 | 0.0063 | **0.0043** | 0.0055 | 0.005 | 0.0097 | 0.0078 |
| ENI.MI | 0.0069 | 0.0078 | 0.0066 | **0.0046** | 0.0093 | 0.0063 | 0.0053 | 0.0163 |
| EL.PA | **0.0074** | 0.0089 | 0.0075 | 0.0074 | 0.0076 | 0.0077 | 0.0105 | 0.0091 |
| FLTR.IR | 0.0087 | **0.0011** | 0.011 | 0.0052 | 0.0107 | 0.0045 | 0.0082 | 0.0069 |
| IBE.MC | 0.0049 | 0.0046 | 0.0033 | 0.0045 | 0.0051 | 0.0033 | **0.0031** | 0.0097 |
| ITX.MC | 0.0058 | 0.0037 | 0.0052 | 0.0052 | **0.0032** | 0.0044 | 0.0075 | 0.0075 |
| IFX.DE | 0.0074 | 0.0059 | 0.0064 | **0.0043** | 0.0062 | 0.0055 | 0.0062 | 0.0048 |
| INGA.AS | **0.0023** | 0.0033 | 0.0054 | 0.0028 | 0.0053 | 0.0028 | 0.0039 | 0.0068 |
| ISP.MI | **0.0034** | 0.0046 | 0.0068 | 0.0038 | 0.0046 | 0.0042 | 0.0113 | 0.007 |
| KER.PA | 0.005 | **0.0035** | 0.0048 | 0.0036 | 0.004 | 0.004 | 0.0105 | 0.0104 |
| KNEBV.HE | 0.0039 | 0.0042 | 0.0045 | **0.0034** | 0.0043 | 0.0035 | 0.0061 | 0.0053 |
| OR.PA | 0.0067 | 0.0064 | 0.0059 | 0.007 | 0.0068 | **0.005** | 0.0061 | 0.0131 |
| MC.PA | 0.0045 | 0.004 | 0.0052 | 0.0051 | **0.0039** | 0.0042 | 0.0098 | 0.0096 |
| MUV2.DE | **0.0062** | 0.0104 | 0.0062 | 0.0067 | 0.0104 | 0.0096 | 0.0209 | 0.0169 |
| NOKIA.HE | 0.0059 | **0.0055** | 0.0066 | 0.0093 | 0.0095 | 0.0064 | 0.0068 | 0.0222 |
| RI.PA | **0.0039** | 0.007 | 0.0048 | 0.007 | 0.0067 | 0.0055 | 0.0102 | 0.0099 |
| ORA.PA | 0.0049 | 0.0051 | **0.0042** | 0.0062 | 0.0047 | 0.0042 | 0.008 | 0.0097 |
| PHIA.AS | 0.0054 | 0.0088 | **0.0044** | 0.007 | 0.0104 | 0.011 | 0.0077 | 0.0165 |
| PRX.AS | 0.01 | 0.0103 | 0.0125 | 0.0133 | **0.0059** | 0.0103 | 0.0136 | 0.0102 |
| SAF.PA | **0.0018** | 0.002 | 0.0023 | 0.0046 | 0.002 | 0.0031 | 0.0107 | 0.0058 |
| SAN.PA | 0.0066 | **0.0063** | 0.0063 | 0.0071 | 0.0072 | 0.0078 | 0.0101 | 0.0107 |
| SAN.MC | 0.005 | 0.009 | 0.0062 | 0.0053 | 0.008 | 0.006 | **0.0043** | 0.0133 |
| SAP.DE | 0.0087 | 0.0095 | 0.0083 | **0.0054** | 0.0095 | 0.0079 | 0.0124 | 0.0106 |
| SU.PA | **0.0075** | 0.0102 | 0.008 | 0.0109 | 0.0101 | 0.0119 | 0.0085 |
| SIE.DE | 0.0058 | 0.0068 | 0.0061 | 0.0058 | 0.0058 | **0.0042** | 0.0085 | 0.0093 |
| TTE.PA | **0.004** | 0.0094 | 0.0051 | 0.01 | 0.0131 | 0.0078 | 0.006 | 0.0184 |
| TEF.MC | 0.006 | 0.0086 | **0.004** | 0.0074 | 0.0074 | 0.006 | 0.0079 | 0.0127 |
| DG.PA | 0.0064 | 0.0049 | 0.0076 | 0.0065 | **0.0036** | 0.007 | 0.0077 | 0.0153 |
| VIV.PA | 0.0055 | 0.0049 | **0.002** | 0.0042 | 0.0047 | 0.004 | 0.0071 | 0.0092 |
| VOW.DE | **0.0039** | 0.0044 | 0.0048 | 0.0045 | 0.0048 | 0.0053 | 0.0107 | 0.0109 |
| VNA.DE | 0.0086 | 0.0096 | 0.0104 | **0.0085** | 0.0094 | 0.0105 | 0.0116 | 0.0128 |

**Note.** In bold the best model.

**Table B.24:** Comparative analysis part II, AD of two-component MGND and CMGND models.

| Stock | MGND | CMGND | | | | | | |
|-------|------|-------|-------|-------|-------|-------|-------|-------|
| | | CUU | UCU | UUC | CCU | CUC | UCC | CCC |
| ADS.DE | **0.0053** | 0.0063 | 0.0054 | 0.0069 | 0.0055 | 0.0069 | 0.0382 | 0.0653 |
| ADYEN.AS | 0.0053 | 0.011 | 0.0065 | **0.0051** | 0.0114 | 0.0107 | 0.0087 | 0.0182 |
| AD.AS | **0.0021** | 0.0069 | 0.0024 | 0.0029 | 0.0065 | 0.0078 | 0.0345 | 0.0526 |
| AI.PA | 0.0412 | 0.0332 | 0.0184 | 0.16 | **0.0152** | 0.3419 | 0.2938 | 1.4017 |
| AIR.PA | 0.0112 | 0.0088 | 0.0102 | 0.013 | **0.008** | 0.0103 | 0.0426 | 0.0687 |
| ALV.DE | **0.0023** | 0.0056 | 0.0029 | 0.0028 | 0.006 | 0.0054 | 0.0269 | 0.0438 |
| AMS.MC | **0.0036** | 0.0073 | 0.0066 | 0.0048 | 0.0072 | 0.0068 | 0.0361 | 0.0582 |
| ABI.BR | 0.0188 | 0.0134 | 0.0143 | 0.0174 | **0.0133** | 0.0151 | 0.1237 | 0.1971 |
| ASML.AS | 0.0055 | **0.0052** | 0.0056 | 0.0075 | 0.0056 | 0.0061 | 0.0202 | 0.0287 |
| CS.PA | **0.0037** | 0.0097 | 0.0038 | 0.0043 | 0.0094 | 0.009 | 0.0359 | 0.0484 |
| BAS.DE | 0.0059 | 0.0198 | **0.0056** | 0.0086 | 0.0182 | 0.016 | 0.0298 | 0.0516 |
| BAYN.DE | 0.0212 | 0.0116 | 0.0113 | 0.0186 | **0.0099** | 0.0164 | 0.0672 | 0.0929 |
| BMW.DE | 0.0127 | 0.0135 | **0.0065** | 0.0277 | 0.0109 | 0.0164 | 0.0356 | 0.0713 |
| BNP.PA | 0.0082 | 0.0124 | **0.0073** | 0.0108 | 0.0124 | 0.0144 | 0.0455 | 0.0565 |
| CRG.IR | 0.0036 | 0.0034 | **0.0032** | 0.0065 | 0.0035 | 0.0067 | 0.0165 | 0.0471 |
| DAI.SG | 0.0073 | **0.0068** | 0.0075 | 0.0123 | 0.0081 | 0.0133 | 0.0363 | 0.0303 |
| BN.PA | 0.0063 | 0.0065 | **0.0037** | 0.0115 | 0.0054 | 0.0163 | 0.0367 | 0.0695 |
| DB1.DE | 0.0246 | 0.0219 | 0.0204 | 0.0367 | **0.0192** | 0.0319 | 0.1989 | 0.4435 |
| DPWA.F | 0.0334 | 0.0382 | 0.0176 | 0.0185 | 0.0175 | 0.0184 | **0.0162** | 0.0176 |
| DTE.DE | **0.0071** | 0.0102 | 0.0071 | 0.0145 | 0.0113 | 0.0136 | 0.0601 | 0.0827 |
| ENEL.MI | 0.0095 | 0.0127 | **0.007** | 0.0189 | 0.015 | 0.0209 | 0.0397 | 0.0504 |
| ENGI.PA | **0.0119** | 0.0198 | 0.0151 | 0.0491 | 0.0208 | 0.0754 | 0.1235 | 0.3906 |
| ENI.MI | 0.0019 | 0.0067 | **0.0017** | 0.0061 | 0.007 | 0.0076 | 0.0107 | 0.0348 |
| EL.PA | 0.0051 | 0.0037 | 0.0044 | 0.0086 | **0.0035** | 0.008 | 0.0223 | 0.0379 |
| FLTR.IR | **0.0066** | 0.0076 | 0.0222 | 0.0105 | 0.0073 | 0.0094 | 0.0807 | 0.0892 |
| IBE.MC | 0.0029 | 0.0041 | **0.0028** | 0.0034 | 0.0041 | 0.0044 | 0.0179 | 0.024 |
| ITX.MC | 0.0062 | 0.005 | 0.0048 | 0.0185 | **0.0036** | 0.0081 | 0.0435 | 0.0967 |
| IFX.DE | 0.0034 | 0.004 | 0.0037 | **0.0026** | 0.0042 | 0.0039 | 0.0105 | 0.0102 |
| INGA.AS | **0.0049** | 0.0075 | 0.0058 | 0.0054 | 0.0082 | 0.0067 | 0.0101 | 0.0115 |
| ISP.MI | 0.0037 | 0.0069 | 0.0063 | **0.0036** | 0.0072 | 0.0062 | 0.0124 | 0.0136 |
| KER.PA | **0.0023** | 0.0028 | 0.0025 | 0.0054 | 0.0024 | 0.0033 | 0.0182 | 0.0385 |
| KNEBV.HE | **0.0014** | 0.0018 | 0.0015 | 0.0015 | 0.0018 | 0.0014 | 0.0017 | 0.0033 |
| OR.PA | 0.0024 | 0.0015 | 0.0023 | 0.0041 | **0.0013** | 0.0019 | 0.0042 | 0.0135 |
| MC.PA | 0.0022 | 0.002 | 0.0022 | **0.0012** | 0.0021 | 0.002 | 0.0047 | 0.0086 |
| MUV2.DE | 0.002 | 0.0079 | 0.0025 | **0.0019** | 0.0085 | 0.0082 | 0.0326 | 0.0421 |
| NOKIA.HE | 0.0065 | 0.004 | 0.0083 | 0.0041 | 0.0047 | **0.0037** | 0.0236 | 0.0356 |
| RI.PA | **0.0039** | 0.0111 | 0.006 | 0.0167 | 0.0091 | 0.0204 | 0.0398 | 0.1177 |
| ORA.PA | 0.0053 | 0.007 | 0.0061 | 0.0053 | **0.004** | 0.0051 | 0.0222 | 0.0358 |
| PHIA.AS | **0.0054** | 0.0365 | 0.0066 | 0.0167 | 0.0251 | 0.0239 | 0.0673 | 0.1519 |
| PRX.AS | 0.0086 | 0.0078 | 0.0089 | **0.0075** | 0.0104 | 0.0116 | 0.0077 | 0.008 |
| SAF.PA | 0.0057 | 0.0059 | **0.0053** | 0.0092 | 0.0059 | 0.0061 | 0.0365 | 0.0506 |
| SAN.PA | 0.0236 | 0.0246 | **0.0171** | 0.0668 | 0.0209 | 0.0605 | 0.0724 | 0.598 |
| SAN.MC | **0.0028** | 0.0038 | 0.0063 | 0.0081 | 0.0034 | 0.0031 | 0.0072 | 0.013 |
| SAP.DE | 0.0041 | 0.0094 | **0.004** | 0.0085 | 0.0071 | 0.0111 | 0.0367 | 0.0844 |
| SU.PA | 0.0033 | 0.0041 | 0.0033 | **0.0031** | 0.0041 | 0.004 | 0.0073 | 0.0105 |
| SIE.DE | 0.0087 | 0.0086 | **0.0071** | 0.0348 | 0.0086 | 0.0271 | 0.083 | 0.2219 |
| TTE.PA | **0.0075** | 0.0101 | 0.0106 | 0.0587 | 0.014 | 0.0615 | 0.3558 | 0.4092 |
| TEF.MC | 0.0034 | 0.0079 | **0.0033** | 0.0059 | 0.0062 | 0.0087 | 0.0198 | 0.0437 |
| DG.PA | 0.0085 | 0.0085 | 0.0092 | **0.0057** | 0.0082 | 0.0138 | 0.0761 | 0.127 |
| VIV.PA | 0.0131 | **0.0128** | 0.015 | 0.0434 | 0.017 | 0.0436 | 0.1069 | 0.2541 |
| VOW.DE | 0.0026 | 0.003 | 0.0029 | **0.0023** | 0.0029 | 0.0026 | 0.0132 | 0.0181 |
| VNA.DE | **0.004** | 0.0069 | 0.0086 | 0.0041 | 0.0079 | 0.0067 | 0.0114 | 0.0154 |

**Note.** In bold the best model.

**Table B.25:** Comparative analysis part II, AIC of fitted distributions for daily log-returns.

| Stock | MGND | CMGND | STD | GH | MSTD | CMSTD |
|---|---|---|---|---|---|---|
| ADS.DE | 13,587.6 | **13,586.35** | 13,590.54 | 13,592.02 | 13,596.05 | 13,594.7 |
| ADYEN.AS | 6,487.87 | 6,485.5 | 6,488.2 | **6,485.17** | 6,488.75 | 6,486.765 |
| AD.AS | 11,108.64 | **11,106.08** | 11,110.91 | 11,108.1 | 11,111.16 | 11,109.95 |
| AI.PA | 11,526.42 | 11,521.29 | **11,517.17** | 11,518.6 | 11,521.37 | 11,519.37 |
| AIR.PA | 14,464.38 | **14,460.75** | 14,461.35 | 14,461.71 | 14,464.34 | 14,462.43 |
| ALV.DE | 12,169.38 | 12,168.08 | 12,171.38 | **12,162.98** | 12,168.06 | 12,164.5 |
| AMS.MC | 12,954.11 | 12,952.98 | 12,953.82 | 12,950.55 | 12,949.06 | **12,947.82** |
| ABI.BR | 12,615.54 | 12,609.38 | **12,607.38** | 12,610.91 | 12,614.33 | 12,611.52 |
| ASML.AS | 14,336.54 | 14,334.53 | 14,339.79 | **14,334.52** | 14,336.56 | 14,334.6 |
| CS.PA | 13,772.19 | 13,771.47 | 13,780.18 | **13,767.67** | 13,772.11 | 13,770.38 |
| BAS.DE | 13,185.91 | 13,184.68 | 13,190.93 | **13,181.73** | 13,187.88 | 13,184.31 |
| BAYN.DE | 13,321.16 | **13,318.17** | 13,327.4 | 13,322.58 | 13,320.83 | 13,319.01 |
| BMW.DE | 13,645.54 | 13,642.94 | 13,649.95 | 13,640.36 | 13,641.35 | **13,639.99** |
| BNP.PA | 14,910.16 | **14,908.63** | 14,916.37 | 14,911.75 | 14,914.32 | 14,916.22 |
| CRG.IR | 14,291.35 | 14,287.26 | **14,286.25** | 14,288.76 | 14,290.48 | 14,289.38 |
| DAI.SG | 13,706.43 | **13,706.12** | 13,756.23 | 13,722.7 | 13,711.95 | 13,709.18 |
| BN.PA | 11,235.4 | 11,232.59 | **11,229.43** | 11,230.6 | 11,235.46 | 11,233.3 |
| DB1.DE | 12,475.73 | 12,470.14 | **12,466.98** | 12,476.36 | 12,471.69 | 12,471.13 |
| DPWA.F | 15,287.13 | 15,267.08 | 15,757.13 | **9,334.09** | | |
| DTE.DE | 11,623.82 | 11,620.35 | 11,619.54 | 11,618.3 | 11,617.41 | **11,616.04** |
| ENEL.MI | 13,087.3 | **13,085.54** | 13,093.38 | 13,088.05 | 13,095.44 | 13,093.59 |
| ENGI.PA | 12,791.14 | 12,788.61 | 12,792.08 | **12,788.1** | 12,793.82 | 12,792.52 |
| ENI.MI | 12,960.72 | 12,959.2 | 12,962.25 | **12,958.72** | 12,962.86 | 12,960.28 |
| EL.PA | 12,154.42 | 12,151.99 | 12,153.51 | 12,150.38 | 12,151.3 | **12,149.23** |
| FLTR.IR | 13,832.53 | 13,832.38 | 13,842.14 | 13,842.79 | 13,834.11 | **13,832.14** |
| IBE.MC | 12,058.94 | 12,056.38 | 12,061.63 | **12,052.96** | 12,058.04 | 12,055.71 |
| ITX.MC | 13,105.9 | 13,101.41 | **13,098.54** | 13,101.36 | 13,102.47 | 13,102.7 |
| IFX.DE | 15,297.79 | 15,298.43 | 15,308.05 | 15,304.42 | 15,297.28 | **15,295.82** |
| INGA.AS | 15,137.45 | **15,135.37** | 15,166.85 | 15,138.87 | 15,144.5 | 15,142.58 |
| ISP.MI | 15,415.48 | **15,411.99** | 15,429.17 | 15,412.46 | 15,420.44 | 15,418.42 |
| KER.PA | 13,728.68 | **13,724.61** | 13,726.28 | 13,726.25 | 13,731.39 | 13,729.18 |
| KNEBV.HE | 12,377.71 | **12,373.64** | 12,392.87 | 12,374.53 | 12,389.43 | 12,388.21 |
| OR.PA | 11,904.43 | 11,900.99 | **11,896.1** | 11,896.89 | 11,901.36 | 11,899.38 |
| MC.PA | 13,407.35 | 13,404.11 | 13,407.19 | **13,401.8** | 13,406.2 | 13,403.92 |
| MUV2.DE | 11,690.9 | **11,689.32** | 11,697.45 | 11,691.49 | 11,693.35 | 11,693.47 |
| NOKIA.HE | 14,693.66 | **14,689.6** | 14,696.45 | 14,693.69 | 14,696.47 | 14,696.95 |
| RI.PA | 11,491.3 | 11,489.55 | 11,488.48 | **11,486.45** | 11,492.08 | 11,489.12 |
| ORA.PA | 12,024.56 | **12,023.28** | 12,031.19 | 12,026.77 | 12,030.91 | 12,029.2 |
| PHIA.AS | 13,367.94 | 13,368.44 | 13,361.41 | **13,359.65** | 13,364.47 | 13,362.75 |
| PRX.AS | 4,878.32 | **4,875.59** | 4,879.34 | 4,878.46 | 4,880.2 | 4,878.253 |
| SAF.PA | 13,814.6 | **13,809.96** | 13,819.45 | 13,816.34 | 13,812.48 | 13,810.5 |
| SAN.PA | 12,008.92 | 12,005.51 | **11,999.38** | 12,000.66 | 12,005.84 | 12,003.34 |
| SAN.MC | 14,929.96 | **14,926.96** | 14,933.25 | 14,927.86 | 14,930.5 | 14,937.42 |
| SAP.DE | 11,896.08 | 11,893.86 | 11,895.22 | **11,893.68** | 11,897.55 | 11,893.77 |
| SU.PA | 13,848.53 | 13,845.81 | 13,863.88 | 13,845.91 | 13,847.52 | **13,845.1** |
| SIE.DE | 12,826.09 | **12,822.81** | 12,823.41 | 12,825.2 | 12,825.32 | 12,827.39 |
| TTE.PA | 12,785.79 | **12,783.19** | 12,788 | 12,789.45 | 12,783.54 | 12,788.62 |
| TEF.MC | 12,917.25 | 12,915.04 | 12,913.16 | **12,911.04** | 12,915.01 | 12,911.85 |
| DG.PA | 12,775.87 | **12,771.47** | 12,776.78 | 12,778.36 | 12,777.62 | 12,778.07 |
| VIV.PA | 12,371.66 | 12,371.25 | 12,376.99 | 12,371.13 | 12,370.84 | **12,369.02** |
| VOW.DE | 14,477.53 | **14,473.89** | 14,480.4 | 14,476.04 | 14,479.28 | 14,477.16 |
| VNA.DE | 9,657.33 | **9,655.14** | 9,670.3 | 9,658.32 | 9,666.45 | 9,660.714 |
| Mean | 12,767.34 | **12,764.76** | 12,770.36 | 12,765.75 | 12,768.06 | 12,766.61 |
| Median | 12,954.11 | 12,952.98 | 12,953.82 | 12,950.55 | 12,949.06 | **12,947.82** |

**Note.** In bold the best model.

**Table B.26:** Comparative analysis part II, BIC of fitted distributions for daily log-returns.

| Stock | MGND | CMGND | STD | GH | MSTD | CMSTD |
|---|---|---|---|---|---|---|
| ADS.DE | 13,630.7 | 13,617.13 | **13,609.01** | 13,622.8 | 13,639.15 | 13,625.49 |
| ADYEN.AS | 6,524.38 | 6,511.57 | **6,503.84** | 6,511.25 | 6,525.25 | 6,518.052 |
| AD.AS | 11,151.8 | 11,139.92 | **11,129.41** | 11,138.93 | 11,154.33 | 11,145.9 |
| AI.PA | 11,569.58 | 11,552.12 | **11,535.66** | 11,549.42 | 11,564.53 | 11,552.17 |
| AIR.PA | 14,507.53 | 14,491.57 | **14,479.85** | 14,492.53 | 14,507.49 | 14,496.34 |
| ALV.DE | 12,212.48 | 12,201.53 | **12,189.85** | 12,193.76 | 12,211.16 | 12,201.44 |
| AMS.MC | 12,997.1 | 12,986.83 | **12,972.25** | 12,981.26 | 12,992.06 | 12,984.68 |
| ABI.BR | 12,658.7 | 12,640.21 | **12,625.88** | 12,641.74 | 12,657.49 | 12,642.35 |
| ASML.AS | 14,379.71 | 14,365.36 | **14,358.28** | 14,365.36 | 14,379.72 | 14,371.6 |
| CS.PA | 13,815.34 | 13,807.14 | 13,798.67 | **13,798.5** | 13,815.26 | 13,807.37 |
| BAS.DE | 13,229.01 | 13,221.38 | **13,209.4** | 13,212.52 | 13,230.98 | 13,221.26 |
| BAYN.DE | 13,364.26 | 13,348.95 | **13,345.87** | 13,353.36 | 13,363.93 | 13,355.95 |
| BMW.DE | 13,688.64 | 13,673.73 | **13,668.42** | 13,671.14 | 13,684.45 | 13,676.93 |
| BNP.PA | 14,953.31 | 14,940.49 | **14,934.86** | 14,942.57 | 14,957.47 | 14,951.34 |
| CRG.IR | 14,334.45 | 14,318.05 | **14,304.73** | 14,319.55 | 14,333.58 | 14,321.21 |
| DAI.SG | 13,749.52 | **13,743.04** | 13,774.69 | 13,753.48 | 13,755.03 | 13,746.1 |
| BN.PA | 11,278.56 | 11,264.12 | **11,247.93** | 11,261.42 | 11,278.62 | 11,264.42 |
| DB1.DE | 12,518.83 | 12,500.93 | **12,485.45** | 12,507.15 | 12,514.79 | 12,501.91 |
| DPWA.F | 15,330.22 | 15,304.02 | 15,775.6 | **9,364.87** | | |
| DTE.DE | 11,666.92 | 11,651.14 | **11,638.01** | 11,649.09 | 11,660.52 | 11,652.98 |
| ENEL.MI | 13,130.42 | 13,119.6 | **13,111.86** | 13,118.84 | 13,138.55 | 13,128.33 |
| ENGI.PA | 12,834.3 | 12,824.29 | **12,810.58** | 12,818.93 | 12,836.98 | 12,827.07 |
| ENI.MI | 13,003.83 | 12,995.08 | **12,980.73** | 12,989.52 | 13,005.97 | 12,997.2 |
| EL.PA | 12,197.58 | 12,182.85 | **12,172.01** | 12,181.21 | 12,194.46 | 12,186.22 |
| FLTR.IR | 13,875.63 | 13,863.6 | **13,860.61** | 13,873.57 | 13,877.2 | 13,869.08 |
| IBE.MC | 12,102.09 | 12,087.21 | **12,080.12** | 12,083.78 | 12,101.2 | 12,092.7 |
| ITX.MC | 13,149.07 | 13,132.24 | **13,117.04** | 13,132.19 | 13,145.63 | 13,133.53 |
| IFX.DE | 15,340.89 | 15,329.17 | **15,326.52** | 15,335.21 | 15,340.38 | 15,332.76 |
| INGA.AS | 15,180.6 | **15,166.2** | 15,185.35 | 15,169.69 | 15,187.65 | 15,179.57 |
| ISP.MI | 15,458.58 | **15,442.78** | 15,447.64 | 15,443.26 | 15,463.55 | 15,455.37 |
| KER.PA | 13,771.84 | 13,755.44 | **13,744.78** | 13,757.08 | 13,774.55 | 13,761.27 |
| KNEBV.HE | 12,420.74 | **12,401.94** | 12,411.31 | 12,405.28 | 12,432.47 | 12,425.1 |
| OR.PA | 11,947.59 | 11,931.82 | **11,914.6** | 11,927.72 | 11,944.52 | 11,931.08 |
| MC.PA | 13,450.51 | 13,434.94 | **13,425.69** | 13,432.63 | 13,449.36 | 13,440.92 |
| MUV2.DE | 11,734 | 11,724.02 | **11,715.92** | 11,722.27 | 11,736.45 | 11,730.41 |
| NOKIA.HE | 14,736.67 | 14,720.33 | **14,714.88** | 14,724.42 | 14,739.49 | 14,731.32 |
| RI.PA | 11,534.46 | 11,523.17 | **11,506.98** | 11,517.28 | 11,535.24 | 11,523.47 |
| ORA.PA | 12,067.72 | 12,054.11 | **12,049.68** | 12,057.6 | 12,074.07 | 12,066.16 |
| PHIA.AS | 13,411.1 | 13,399.27 | **13,379.91** | 13,390.48 | 13,407.63 | 13,396.4 |
| PRX.AS | 4,912.95 | 4,900.32 | **4,894.18** | 4,903.19 | 4,914.83 | 4,907.935 |
| SAF.PA | 13,857.75 | 13,840.78 | **13,837.94** | 13,847.16 | 13,855.63 | 13,847.49 |
| SAN.PA | 12,052.08 | 12,036.34 | **12,017.88** | 12,031.48 | 12,049 | 12,034.37 |
| SAN.MC | 14,973.11 | 14,957.79 | **14,951.74** | 14,958.68 | 14,973.65 | 14,968.24 |
| SAP.DE | 11,939.18 | 11,930.65 | **11,913.7** | 11,924.47 | 11,940.65 | 11,930.16 |
| SU.PA | 13,891.69 | **13,874.49** | 13,882.38 | 13,876.74 | 13,890.68 | 13,882.09 |
| SIE.DE | 12,869.2 | 12,853.6 | **12,841.88** | 12,855.98 | 12,868.42 | 12,858.35 |
| TTE.PA | 12,828.94 | 12,815.05 | **12,806.49** | 12,820.27 | 12,826.7 | 12,822.99 |
| TEF.MC | 12,960.41 | 12,951.32 | **12,931.65** | 12,941.87 | 12,958.17 | 12,948.12 |
| DG.PA | 12,819.02 | 12,803.57 | **12,795.27** | 12,809.18 | 12,820.77 | 12,811.77 |
| VIV.PA | 12,414.81 | 12,402.07 | **12,395.48** | 12,401.96 | 12,413.99 | 12,406.01 |
| VOW.DE | 14,520.63 | 14,504.67 | **14,498.87** | 14,506.82 | 14,522.37 | 14,514.09 |
| VNA.DE | 9,698.35 | **9,684.44** | 9,687.88 | 9,687.62 | 9,707.48 | 9,695.875 |
| Mean | 12,810.13 | 12,796.44 | **12,788.70** | 12,796.32 | 12,810.85 | 12,801.43 |
| Median | 12,997.10 | 12,986.83 | **12,972.25** | 12,981.26 | 12,992.06 | 12,984.68 |

Note. In bold the best model.

**Table B.27:** Comparative analysis part II, KS distance between the empirical and fitted distributions for daily log-returns.

| Stock | MGND | CMGND | STD | GH | MSTD | CMSTD |
|---|---|---|---|---|---|---|
| ADS.DE | 0.0082 | **0.0012** | 0.0065 | 0.0054 | 0.0061 | 0.0055 |
| ADYEN.AS | 0.0098 | **0.0088** | 0.0131 | 0.0124 | 0.0112 | 0.0112 |
| AD.AS | **0.0033** | 0.0059 | 0.0089 | 0.005 | 0.0058 | 0.0054 |
| AI.PA | 0.006 | 0.0064 | 0.0089 | 0.0063 | **0.0044** | 0.0047 |
| AIR.PA | 0.0055 | 0.005 | 0.0039 | 0.0041 | 0.006 | **0.0038** |
| ALV.DE | 0.0051 | 0.0052 | 0.0125 | 0.0053 | 0.0042 | **0.0036** |
| AMS.MC | 0.0086 | 0.0079 | 0.0145 | 0.0099 | **0.0051** | 0.0062 |
| ABI.BR | 0.0072 | 0.0049 | 0.0035 | 0.0039 | 0.0054 | **0.0032** |
| ASML.AS | 0.0044 | 0.0041 | 0.0095 | 0.0047 | 0.0034 | **0.0033** |
| CS.PA | 0.0075 | 0.0052 | 0.0111 | 0.005 | 0.0043 | **0.0042** |
| BAS.DE | 0.0049 | 0.0044 | 0.0091 | **0.0038** | 0.0044 | 0.0039 |
| BAYN.DE | 0.0084 | 0.0046 | 0.0075 | **0.0036** | 0.006 | 0.0058 |
| BMW.DE | 0.0081 | **0.0014** | 0.0062 | 0.0036 | 0.0064 | 0.0057 |
| BNP.PA | 0.0073 | 0.0031 | 0.007 | **0.003** | 0.007 | 0.0051 |
| CRG.IR | 0.0062 | 0.0047 | 0.0058 | 0.0061 | 0.0039 | **0.0037** |
| DAI.SG | 0.0094 | **0.0028** | 0.0117 | 0.006 | 0.0083 | 0.0082 |
| BN.PA | 0.0049 | **0.0032** | 0.0048 | 0.0038 | 0.0046 | 0.0039 |
| DB1.DE | 0.0034 | **0.0027** | 0.0054 | 0.0061 | 0.005 | 0.0052 |
| DPWA.F | 0.0319 | **0.0103** | 0.0168 | 0.1733 | | |
| DTE.DE | **0.0018** | 0.0033 | 0.0069 | 0.0055 | 0.0075 | 0.007 |
| ENEL.MI | 0.0053 | **0.0044** | 0.0091 | 0.006 | 0.0057 | 0.0066 |
| ENGI.PA | 0.0056 | **0.0043** | 0.0085 | 0.0043 | 0.005 | 0.0048 |
| ENI.MI | 0.0069 | **0.0046** | 0.0091 | 0.0069 | 0.0056 | 0.0057 |
| EL.PA | 0.0074 | 0.0074 | 0.0101 | 0.0091 | **0.0062** | 0.0063 |
| FLTR.IR | 0.0087 | **0.0011** | 0.0082 | 0.0059 | 0.0091 | 0.0077 |
| IBE.MC | 0.0049 | **0.0031** | 0.006 | 0.0039 | 0.0059 | 0.0051 |
| ITX.MC | 0.0058 | **0.0032** | 0.0061 | 0.0075 | 0.0062 | 0.0058 |
| IFX.DE | 0.0074 | **0.0043** | 0.0089 | 0.0054 | 0.007 | 0.0071 |
| INGA.AS | **0.0023** | 0.0028 | 0.0104 | 0.0043 | 0.0076 | 0.0077 |
| ISP.MI | 0.0034 | 0.0038 | 0.0088 | **0.0033** | 0.0048 | 0.0051 |
| KER.PA | 0.005 | 0.0035 | 0.0034 | 0.0052 | 0.005 | **0.0027** |
| KNEBV.HE | 0.0039 | **0.0034** | 0.0083 | 0.0049 | 0.0054 | 0.0052 |
| OR.PA | 0.0067 | 0.005 | 0.0035 | 0.0051 | 0.0053 | **0.0028** |
| MC.PA | 0.0045 | 0.0039 | 0.0062 | 0.0049 | 0.0041 | **0.0038** |
| MUV2.DE | 0.0062 | 0.0062 | 0.0114 | 0.0063 | 0.0057 | **0.0053** |
| NOKIA.HE | 0.0059 | 0.0055 | 0.0055 | 0.0073 | 0.0052 | **0.0049** |
| RI.PA | **0.0039** | 0.0048 | 0.0079 | 0.0041 | 0.004 | 0.0039 |
| ORA.PA | 0.0049 | **0.0042** | 0.0067 | 0.005 | 0.0044 | 0.0045 |
| PHIA.AS | 0.0054 | **0.0044** | 0.0081 | 0.007 | 0.0069 | 0.0071 |
| PRX.AS | 0.01 | **0.0059** | 0.0074 | 0.0086 | 0.0076 | 0.0075 |
| SAF.PA | **0.0018** | 0.002 | 0.0062 | 0.0071 | 0.0054 | 0.0053 |
| SAN.PA | 0.0066 | 0.0063 | 0.0075 | 0.0063 | **0.006** | 0.0061 |
| SAN.MC | 0.005 | **0.0043** | 0.0052 | 0.0067 | 0.0072 | 0.0049 |
| SAP.DE | 0.0087 | 0.0054 | 0.0095 | 0.0053 | 0.0061 | **0.0043** |
| SU.PA | 0.0075 | 0.008 | 0.0138 | 0.0064 | 0.0047 | **0.0046** |
| SIE.DE | 0.0058 | **0.0042** | 0.0062 | 0.0049 | 0.0048 | 0.0043 |
| TTE.PA | **0.004** | 0.0051 | 0.0101 | 0.0078 | 0.0055 | 0.0059 |
| TEF.MC | 0.006 | **0.004** | 0.0081 | 0.0057 | 0.0049 | 0.0048 |
| DG.PA | 0.0064 | **0.0036** | 0.0051 | 0.0042 | 0.0061 | 0.0041 |
| VIV.PA | 0.0055 | **0.002** | 0.0063 | 0.0039 | 0.0047 | 0.0048 |
| VOW.DE | **0.0039** | 0.0044 | 0.005 | 0.0057 | 0.0061 | 0.0049 |
| VNA.DE | 0.0086 | 0.0085 | 0.0124 | 0.0094 | 0.0083 | **0.0078** |
| Mean | 0.0060 | **0.0045** | 0.0080 | 0.0057 | 0.0058 | 0.0053 |
| Median | 0.0058 | **0.0044** | 0.0079 | 0.0054 | 0.0056 | 0.0051 |

**Note.** In bold the best model.

**Table B.28:** Comparative analysis part II, AD distance between the empirical and fitted distributions for daily log-returns.

| Stock | MGND | CMGND | STD | GH | MSTD | CMSTD |
|---|---|---|---|---|---|---|
| ADS.DE | 0.0053 | 0.0054 | 0.0053 | 0.0052 | 0.0049 | **0.0045** |
| ADYEN.AS | 0.0053 | **0.0051** | 0.0189 | 0.0052 | 0.0057 | 0.0058 |
| AD.AS | 0.0021 | 0.0024 | 0.0074 | **0.0018** | 0.0018 | 0.0019 |
| AI.PA | 0.0412 | 0.0152 | 0.0062 | 0.0094 | 0.01 | **0.006** |
| AIR.PA | 0.0112 | **0.008** | 0.0092 | 0.0109 | 0.0114 | 0.0092 |
| ALV.DE | **0.0023** | 0.0028 | 0.0164 | 0.0027 | 0.0053 | 0.0037 |
| AMS.MC | 0.0036 | 0.0048 | 0.011 | 0.0032 | 0.0031 | **0.0026** |
| ABI.BR | 0.0188 | **0.0133** | 0.0167 | 0.0164 | 0.0188 | 0.0161 |
| ASML.AS | 0.0055 | 0.0052 | 0.0073 | 0.0067 | **0.004** | 0.004 |
| CS.PA | 0.0037 | 0.0038 | 0.0172 | **0.0028** | 0.0046 | 0.0044 |
| BAS.DE | 0.0059 | 0.0056 | 0.013 | **0.0049** | 0.0084 | 0.007 |
| BAYN.DE | 0.0212 | 0.0099 | 0.0068 | 0.0127 | 0.008 | **0.0067** |
| BMW.DE | 0.0127 | **0.0065** | 0.0137 | 0.012 | 0.0128 | 0.0128 |
| BNP.PA | 0.0082 | **0.0073** | 0.0146 | 0.0081 | 0.0135 | 0.0097 |
| CRG.IR | 0.0036 | **0.0032** | 0.0042 | 0.0041 | 0.0039 | 0.0043 |
| DAI.SG | 0.0073 | **0.0068** | 0.0104 | 0.0279 | 0.0071 | 0.0071 |
| BN.PA | 0.0063 | **0.0037** | 0.0047 | 0.004 | 0.0055 | 0.0046 |
| DB1.DE | 0.0246 | 0.0192 | 0.0162 | 0.0486 | 0.0194 | **0.016** |
| DPWA.F | 0.0334 | **0.0162** | 0.0323 | 0.743 | | |
| DTE.DE | **0.0071** | 0.0071 | 0.0096 | 0.0089 | 0.0101 | 0.0095 |
| ENEL.MI | 0.0095 | **0.007** | 0.0125 | 0.007 | 0.0116 | 0.009 |
| ENGI.PA | 0.0119 | 0.0151 | 0.0099 | **0.0057** | 0.0081 | 0.0086 |
| ENI.MI | 0.0019 | **0.0017** | 0.0116 | 0.0021 | 0.0032 | 0.0023 |
| EL.PA | 0.0051 | 0.0035 | 0.0053 | 0.0036 | 0.0025 | **0.0024** |
| FLTR.IR | 0.0066 | 0.0073 | 0.0064 | 0.0096 | 0.0073 | **0.0059** |
| IBE.MC | 0.0029 | 0.0028 | 0.0125 | **0.0027** | 0.0047 | 0.0041 |
| ITX.MC | 0.0062 | 0.0036 | 0.0025 | 0.0029 | **0.0024** | 0.0025 |
| IFX.DE | 0.0034 | 0.0026 | 0.007 | **0.0025** | 0.0054 | 0.0054 |
| INGA.AS | **0.0049** | 0.0054 | 0.0258 | 0.0072 | 0.0152 | 0.0152 |
| ISP.MI | 0.0037 | **0.0036** | 0.0188 | 0.004 | 0.0083 | 0.0071 |
| KER.PA | **0.0023** | 0.0024 | 0.0033 | 0.0026 | 0.0029 | 0.003 |
| KNEBV.HE | **0.0014** | 0.0014 | 0.0079 | 0.0015 | 0.0041 | 0.0048 |
| OR.PA | 0.0024 | **0.0013** | 0.0019 | 0.0013 | 0.0014 | 0.0014 |
| MC.PA | 0.0022 | **0.0012** | 0.0066 | 0.0015 | 0.0017 | 0.0015 |
| MUV2.DE | 0.002 | **0.0019** | 0.0163 | 0.0032 | 0.002 | 0.0026 |
| NOKIA.HE | 0.0065 | **0.0037** | 0.0096 | 0.0053 | 0.0039 | 0.0054 |
| RI.PA | 0.0039 | 0.006 | 0.0058 | **0.0031** | 0.0063 | 0.0055 |
| ORA.PA | 0.0053 | 0.004 | 0.0058 | 0.0027 | **0.0021** | 0.0024 |
| PHIA.AS | **0.0054** | 0.0066 | 0.0139 | 0.0059 | 0.0128 | 0.0107 |
| PRX.AS | 0.0086 | **0.0075** | 0.0196 | 0.0087 | 0.0076 | 0.0077 |
| SAF.PA | 0.0057 | **0.0053** | 0.0079 | 0.0073 | 0.0059 | 0.0059 |
| SAN.PA | 0.0236 | 0.0171 | 0.0069 | **0.0036** | 0.0122 | 0.0055 |
| SAN.MC | **0.0028** | 0.0031 | 0.0056 | 0.004 | 0.0034 | 0.0052 |
| SAP.DE | 0.0041 | 0.004 | 0.0082 | **0.0022** | 0.0029 | 0.0022 |
| SU.PA | 0.0033 | 0.0031 | 0.0083 | 0.0047 | **0.0024** | 0.0024 |
| SIE.DE | 0.0087 | 0.0071 | 0.0049 | **0.0047** | 0.0049 | 0.0047 |
| TTE.PA | 0.0075 | 0.0101 | 0.0078 | **0.0069** | 0.0088 | 0.0073 |
| TEF.MC | 0.0034 | 0.0033 | 0.009 | 0.0024 | 0.0031 | **0.0023** |
| DG.PA | 0.0085 | **0.0057** | 0.0088 | 0.0057 | 0.0093 | 0.0066 |
| VIV.PA | 0.0131 | 0.0128 | **0.0081** | 0.0227 | 0.0108 | 0.0081 |
| VOW.DE | 0.0026 | **0.0023** | 0.0067 | 0.003 | 0.0028 | 0.0028 |
| VNA.DE | 0.0074 | **0.0059** | 0.0101 | 0.0069 | 0.0067 | 0.0058 |
| Mean | 0.0053 | 0.0051 | 0.0083 | **0.0047** | 0.0054 | 0.0054 |
| Median | 0.0058 | **0.0044** | 0.0079 | 0.0054 | 0.0056 | 0.0051 |

**Note.** In bold the best model.

**Table B.29:** Ljung-Box test for the 52 stocks of the STOXX50E.

| Stock | $Q_s(20)$ | p-value | $Q_{ss}(20)$ | p-value |
|-------|-----------|---------|--------------|---------|
| ADS.DE | 46.7 | 0 | 46.7 | 0 |
| ADYEN.AS | 26.3 | 0.16 | 26.3 | 0 |
| AD.AS | 26.3 | 0.16 | 26.3 | 0 |
| AI.PA | 46.2 | 0 | 46.2 | 0 |
| AIR.PA | 47.4 | 0 | 47.4 | 0 |
| ALV.DE | 51 | 0 | 51 | 0 |
| AMS.MC | 69.4 | 0 | 69.4 | 0 |
| ABI.BR | 38 | 0.01 | 38 | 0 |
| ASML.AS | 20.4 | 0.43 | 20.4 | 0 |
| CS.PA | 54.3 | 0 | 54.3 | 0 |
| BAS.DE | 27.4 | 0.12 | 27.4 | 0 |
| BAYN.DE | 14.5 | 0.8 | 14.5 | 0 |
| BMW.DE | 39.5 | 0.01 | 39.5 | 0 |
| BNP.PA | 72.4 | 0 | 72.4 | 0 |
| CRG.IR | 26.5 | 0.15 | 26.5 | 0 |
| DAI.SG | 70.3 | 0 | 70.3 | 0 |
| BN.PA | 23.9 | 0.25 | 23.9 | 0 |
| DB1.DE | 39.9 | 0.01 | 39.9 | 0 |
| DPWA.F | 330 | 0 | 330 | 0 |
| DTE.DE | 24.9 | 0.21 | 24.9 | 0 |
| ENEL.MI | 23.5 | 0.26 | 23.5 | 0 |
| ENGI.PA | 40.5 | 0 | 40.5 | 0 |
| ENI.MI | 34.1 | 0.03 | 34.1 | 0 |
| EL.PA | 48.3 | 0 | 48.3 | 0 |
| FLTR.IR | 52.5 | 0 | 52.5 | 0 |
| IBE.MC | 82.9 | 0 | 82.9 | 0 |
| ITX.MC | 39 | 0.01 | 39 | 0 |
| IFX.DE | 31.5 | 0.05 | 31.5 | 0 |
| INGA.AS | 29.9 | 0.07 | 29.9 | 0 |
| ISP.MI | 19.3 | 0.5 | 19.3 | 0 |
| KER.PA | 29 | 0.09 | 29 | 0 |
| KNEBV.HE | 37 | 0.01 | 37 | 0 |
| OR.PA | 53.4 | 0 | 53.4 | 0 |
| MC.PA | 39.6 | 0.01 | 39.6 | 0 |
| MUV2.DE | 43 | 0 | 43 | 0 |
| NOKIA.HE | 18.5 | 0.55 | 18.5 | 0 |
| RI.PA | 21.5 | 0.37 | 21.5 | 0 |
| ORA.PA | 25.2 | 0.19 | 25.2 | 0 |
| PHIA.AS | 23.6 | 0.26 | 23.6 | 0 |
| PRX.AS | 37.8 | 0.01 | 37.8 | 0 |
| SAF.PA | 26.4 | 0.15 | 26.4 | 0 |
| SAN.PA | 32.1 | 0.04 | 32.1 | 0 |
| SAN.MC | 22.4 | 0.32 | 22.4 | 0 |
| SAP.DE | 24.3 | 0.23 | 24.3 | 0 |
| SU.PA | 59 | 0 | 59 | 0 |
| SIE.DE | 19.1 | 0.52 | 19.1 | 0 |
| TTE.PA | 44 | 0 | 44 | 0 |
| TEF.MC | 38.7 | 0.01 | 38.7 | 0 |
| DG.PA | 56.7 | 0 | 56.7 | 0 |
| VIV.PA | 17.4 | 0.63 | 17.4 | 0 |
| VOW.DE | 35.1 | 0.02 | 35.1 | 0 |
| VNA.DE | 21.3 | 0.38 | 21.3 | 0 |

**Table B.30:** GND-HMMs simulations, estimated parameters.

| $N$ | $\pi_1$ | $\mu_1$ | $\sigma_1$ | $\nu_1$ | $\gamma_{11}$ | $\pi_2$ | $\mu_2$ | $\sigma_2$ | $\nu_2$ | $\gamma_{22}$ |
|---|---|---|---|---|---|---|---|---|---|---|
|  | 0.7878 | 0 | 1 | 2 | 0.9987 | 0.2122 | −0.02 | 1 | 0.8 | 0.9953 |
| **UUU** | | | | | | | | | | |
| 1000 | 0.8135 | 0.0012 | 1.0069 | 2.0561 | 0.9974 | 0.1865 | -0.0208 | 1.0082 | 0.8060 | 0.9884 |
| 2000 | 0.8154 | 0.0009 | 1.0034 | 2.0237 | 0.9988 | 0.1846 | -0.0231 | 1.0159 | 0.8078 | 0.9946 |
| **CUU** | | | | | | | | | | |
| 1000 | 0.8128 | -0.0015 | 1.0065 | 2.0528 | 0.9973 | 0.1872 | -0.0015 | 1.0371 | 0.8169 | 0.9884 |
| 2000 | 0.8145 | -0.0015 | 1.0038 | 2.0254 | 0.9988 | 0.1855 | -0.0015 | 1.0314 | 0.8133 | 0.9946 |
| **UCU** | | | | | | | | | | |
| 1000 | 0.8138 | 0.0007 | 1.0058 | 2.0505 | 0.9974 | 0.1862 | -0.0239 | 1.0058 | 0.8030 | 0.9884 |
| 2000 | 0.8151 | 0.0006 | 1.0038 | 2.0257 | 0.9988 | 0.1849 | -0.0223 | 1.0038 | 0.8024 | 0.9946 |
| **UUC** | | | | | | | | | | |
| 1000 | 0.8249 | 0.0006 | 0.8643 | 1.5718 | 0.9964 | 0.1751 | -0.0293 | 2.6183 | 1.5718 | 0.9808 |
| 2000 | 0.8185 | 0.0006 | 0.8541 | 1.5419 | 0.9984 | 0.1815 | -0.0231 | 2.5068 | 1.5419 | 0.9928 |
| **CCU** | | | | | | | | | | |
| 1000 | 0.8135 | -0.0014 | 1.0065 | 2.0533 | 0.9974 | 0.1865 | -0.0014 | 1.0065 | 0.8036 | 0.9885 |
| 2000 | 0.8151 | -0.0015 | 1.0044 | 2.0271 | 0.9988 | 0.1849 | -0.0015 | 1.0044 | 0.8024 | 0.9946 |
| **CUC** | | | | | | | | | | |
| 1000 | 0.8245 | -0.0005 | 0.8634 | 1.5698 | 0.9965 | 0.1755 | -0.0005 | 2.6146 | 1.5698 | 0.9815 |
| 2000 | 0.8186 | -0.0003 | 0.8547 | 1.5442 | 0.9984 | 0.1814 | -0.0003 | 2.5092 | 1.5442 | 0.9928 |
| **UCC** | | | | | | | | | | |
| 1000 | 0.5852 | 0.0380 | 0.4442 | 0.7732 | 0.6243 | 0.4148 | -0.0513 | 0.4442 | 0.7732 | 0.4600 |
| 2000 | 0.5884 | 0.0004 | 0.4414 | 0.7723 | 0.6156 | 0.4116 | 0.0009 | 0.4414 | 0.7723 | 0.4387 |
| **CCC** | | | | | | | | | | |
| 1000 | 0.5000 | -0.0034 | 0.6392 | 0.9015 | 0.8000 | 0.5000 | -0.0034 | 0.6392 | 0.9015 | 0.8000 |
| 2000 | 0.5000 | -0.0021 | 0.6413 | 0.9015 | 0.8000 | 0.5000 | -0.0021 | 0.6413 | 0.9015 | 0.8000 |

**Table B.31:** GND-HMMs simulations, MSE.

| $N$ | $\pi_1$ | $\mu_1$ | $\sigma_1$ | $\nu_1$ | $\gamma_{11}$ | $\pi_2$ | $\mu_2$ | $\sigma_2$ | $\nu_2$ | $\gamma_{22}$ |
|---|---|---|---|---|---|---|---|---|---|---|
|  | 0.7878 | 0 | 1 | 2 | 0.9987 | 0.2122 | −0.02 | 1 | 0.8 | 0.9953 |
| **UUU** | | | | | | | | | | |
| 1000 | 0.0271 | 0.0276 | 0.0492 | 0.1873 | 0.0004 | 0.0271 | 0.0907 | 0.2665 | 0.1069 | 0.0029 |
| 2000 | 0.0279 | 0.0171 | 0.0351 | 0.1256 | 0.0001 | 0.0279 | 0.0636 | 0.1967 | 0.0766 | 0.0008 |
| **CUU** | | | | | | | | | | |
| 1000 | 0.0271 | 0.0261 | 0.0488 | 0.1850 | 0.0004 | 0.0271 | 0.0320 | 0.2656 | 0.1075 | 0.0029 |
| 2000 | 0.0290 | 0.0161 | 0.0352 | 0.1265 | 0.0001 | 0.0290 | 0.0245 | 0.1981 | 0.0771 | 0.0008 |
| **UCU** | | | | | | | | | | |
| 1000 | 0.0275 | 0.0273 | 0.0478 | 0.1795 | 0.0004 | 0.0275 | 0.0957 | 0.0478 | 0.0385 | 0.0029 |
| 2000 | 0.0276 | 0.0171 | 0.0356 | 0.1279 | 0.0001 | 0.0276 | 0.0651 | 0.0356 | 0.0262 | 0.0008 |
| **UUC** | | | | | | | | | | |
| 1000 | 0.0401 | 0.0288 | 0.1470 | 0.4474 | 0.0031 | 0.0401 | 0.1448 | 1.6606 | 0.7826 | 0.0324 |
| 2000 | 0.0310 | 0.0175 | 0.1515 | 0.4674 | 0.0007 | 0.0310 | 0.0888 | 1.5197 | 0.7477 | 0.0038 |
| **CCU** | | | | | | | | | | |
| 1000 | 0.0270 | 0.0263 | 0.0485 | 0.1846 | 0.0003 | 0.0270 | 0.0322 | 0.0485 | 0.0382 | 0.0027 |
| 2000 | 0.0275 | 0.0161 | 0.0355 | 0.1275 | 0.0001 | 0.0275 | 0.0245 | 0.0355 | 0.0262 | 0.0008 |
| **CUC** | | | | | | | | | | |
| 1000 | 0.0393 | 0.0278 | 0.1477 | 0.4491 | 0.0029 | 0.0393 | 0.0339 | 1.6551 | 0.7805 | 0.0299 |
| 2000 | 0.0311 | 0.0169 | 0.1508 | 0.4651 | 0.0007 | 0.0311 | 0.0259 | 1.5220 | 0.7499 | 0.0038 |
| **UCC** | | | | | | | | | | |
| 1000 | 0.2164 | 0.2171 | 0.5626 | 1.2286 | 0.3952 | 0.2164 | 0.3466 | 0.5626 | 0.0715 | 0.5654 |
| 2000 | 0.2099 | 0.2289 | 0.5630 | 1.2288 | 0.3942 | 0.2099 | 0.3620 | 0.5630 | 0.0582 | 0.5777 |
| **CCC** | | | | | | | | | | |
| 1000 | 0.2878 | 0.0341 | 0.3664 | 1.1002 | 0.1975 | 0.2878 | 0.0378 | 0.3664 | 0.1181 | 0.1906 |
| 2000 | 0.2879 | 0.0221 | 0.3616 | 1.0993 | 0.1987 | 0.2879 | 0.0283 | 0.3616 | 0.1105 | 0.1953 |

**Table B.32:** Comparative analysis part III, AIC of unconstrained and constrained two-state GND-HMMs.

| Stock | UUU | CUU | UCU | UUC | CCU | CUC | UCC | CCC |
|---|---|---|---|---|---|---|---|---|
| ADS.DE | 13,472.2 | 13,474.84 | 13,490.1 | **13,470.28** | 13,472.89 | 13,493.1 | 13,633.64 | 13,631.64 |
| ADYEN.AS | 6,501.022 | 6,403.145 | 6,427.46 | **6,400.229** | 6,401.242 | 6,497.022 | 6,485.889 | 6,495.021 |
| AD.AS | **11,011.06** | 11,019.87 | 11,029.13 | 11,013.93 | 11,022.37 | 11,035.2 | 11,147 | 11,165.41 |
| AI.PA | 11,357.32 | 11,358.57 | 11,388.15 | **11,356.1** | 11,357.5 | 11,386.79 | 11,528.08 | 11,551.39 |
| AIR.PA | 14,241.48 | 14,241.43 | 14,277.09 | 14,239.51 | **14,239.45** | 14,278.96 | 14,481.29 | 14,499.57 |
| ALV.DE | 11,859.65 | 11,866.05 | 11,900.81 | **11,858.14** | 11,865.98 | 11,913.67 | 12,191.51 | 12,219.97 |
| AMS.MC | **12,662.55** | 12,663.75 | 12,699.58 | 12,665.72 | 12,667.18 | 12,698.91 | 12,974.48 | 12,999.73 |
| ABI.BR | 12,436.43 | **12,434.59** | 12,445.99 | 12,443.32 | 12,441.33 | 12,444.21 | 12,670 | 12,680.68 |
| ASML.MC | 14,168.52 | 14,170.8 | 14,227.96 | **14,168.36** | 14,170.59 | 14,233.67 | 14,343.25 | 14,348.52 |
| CS.PA | 13,419.9 | 13,424.07 | 13,482.5 | **13,418.22** | 13,422.34 | 13,484.62 | 13,793.4 | 13,816.01 |
| BAS.DE | 12,984.25 | 12,986.44 | 13,021.67 | **12,983.9** | 12,986.65 | 13,026.28 | 13,199.1 | 13,215.51 |
| BAYN.DE | 13,176.85 | 13,176.35 | 13,199.44 | 13,175.04 | **13,174.52** | 13,199.2 | 13,337.55 | 13,335.55 |
| BMW.DE | 13,412.24 | 13,410.4 | 13,455.02 | 13,410.24 | **13,408.41** | 13,453.33 | 13,643.09 | 13,663.66 |
| BNP.PA | 14,943.86 | 14,490.62 | 14,552.35 | **14,488.14** | 14,489.94 | 14,555.35 | 14,929.75 | 14,937.86 |
| CRG.IR | 14,071.44 | 14,069.48 | 14,101.31 | 14,069.53 | **14,067.57** | 14,099.34 | 14,314.03 | 14,336.31 |
| DAI.SG | **11,673.55** | 13,452.61 | 13,535.48 | 13,443.41 | 13,441.66 | 13,533.62 | 13,726.07 | 13,724.14 |
| BN.PA | 11,112.78 | 11,111.21 | 11,133.18 | 11,112.13 | **11,110.62** | 11,131.38 | 11,245.64 | 11,264.8 |
| DB1.DE | 12,274.38 | 12,273.86 | 12,298.94 | 12,273.1 | **12,272.59** | 12,298.46 | 12,497.43 | 12,512 |
| DPWA.F | 15,610.26 | **14,314.6** | 15,432.29 | 15,447.95 | 15,445.95 | 15,430.18 | 15,617.01 | 15,604.26 |
| DTE.DE | 11,382.83 | 11,382.03 | 11,418.04 | 11,381.39 | **11,380.42** | 11,421.04 | 11,646.48 | 11,653.94 |
| ENEL.MI | 12,889.59 | 12,888.13 | 12,927.29 | 12,887.62 | **12,886.38** | 12,930.8 | 13,090.21 | 13,105.58 |
| ENGI.PA | 12,614.07 | 12,613.45 | 12,646.65 | 12,612.21 | **12,611.57** | 12,646.84 | 12,810.11 | 12,832.12 |
| ENI.MI | 12,695.27 | **12,693.94** | 12,720.56 | 12,702.8 | 12,701.91 | 12,719.14 | 12,967.39 | 13,004.78 |
| EL.PA | 11,987.26 | 11,987.07 | 12,022.25 | 11,985.62 | **11,985.35** | 12,021.54 | 12,157.39 | 12,177.41 |
| FLTR.IR | 13,878.76 | 13,667.53 | 13,699.13 | 13,667.36 | **13,666.05** | 13,699.87 | 13,871.17 | 13,872.75 |
| IBE.MC | 11,654.68 | 11,656.43 | 11,717.67 | **11,652.97** | 11,654.93 | 11,725.05 | 12,067.7 | 12,086.03 |
| ITX.MC | 12,955.9 | 12,957.27 | 12,981.23 | **12,954.07** | 12,955.28 | 12,988.25 | 13,112.88 | 13,133.98 |
| IFX.DE | 15,314.48 | 15,153.08 | 15,207.76 | **15,151.61** | 15,153.04 | 15,209.9 | 15,310.45 | 15,308.48 |
| INGA.AS | 15,153.13 | **14,725.16** | 14,803.49 | 14,729.71 | 14,730.48 | 14,806.98 | 15,143.03 | 15,147.13 |
| ISP.MI | 15,436.73 | 15,059.83 | 15,139.07 | **15,052.98** | 15,057.84 | 15,145.11 | 15,423.72 | 15,430.73 |
| KER.PA | **13,582.72** | 13,586.7 | 13,610.63 | 13,582.8 | 13,586.68 | 13,613.09 | 13,747.74 | 13,760.35 |
| KNEBV.HE | 12,189.79 | **12,188.44** | 12,258 | 12,194.08 | 12,192.82 | 12,256.43 | 12,381.8 | 12,381.07 |
| OR.PA | 11,750.08 | 11,748.98 | 11,797.65 | 11,748.09 | **11,746.99** | 11,796.61 | 11,928.59 | 11,926.59 |
| MC.PA | **13,256.66** | 13,259.27 | 13,294.69 | 13,257.13 | 13,260.48 | 13,296.05 | 13,405.65 | 13,425.68 |
| MUV2.DE | **11,429.43** | 11,439.15 | 11,456.57 | 11,431.74 | 11,440.47 | 11,469.44 | 11,734.34 | 11,750.44 |
| NOKIA.HE | 14,744.34 | **14,492.91** | 14,541.99 | 14,495.18 | 14,493.27 | 14,540.68 | 14,716.28 | 14,738.34 |
| RI.PA | 11,344.69 | 11,345.01 | 11,376.37 | **11,343.18** | 11,343.52 | 11,375.56 | 11,505.31 | 11,520.29 |
| ORA.PA | 11,650.62 | 11,648.64 | 11,728.01 | 11,648.85 | **11,646.87** | 11,726.17 | 12,048.57 | 12,046.83 |
| PHIA.AS | 13,186.08 | 13,188.52 | 13,224.55 | **13,184.89** | 13,187.64 | 13,223.4 | 13,388.47 | 13,414.78 |
| PRX.AS | 4,891.268 | 4,823.742 | 4,839.472 | 4,824.369 | **4,822.4** | 4,887.268 | 4,872.999 | 4,885.268 |
| SAF.PA | 13,457.47 | 13,458.68 | 13,525.96 | **13,456.01** | 13,457.01 | 13,529.32 | 13,829.35 | 13,840.55 |
| SAN.PA | 11,895.52 | 11,896.7 | 11,922.42 | **11,894.38** | 11,895.57 | 11,920.9 | 12,009.9 | 12,047.15 |
| SAN.MC | 14,633.98 | 14,635.56 | 14,706.4 | **14,632.89** | 14,634.63 | 14,707.74 | 14,930.65 | 14,938.93 |
| SAP.DE | **11,775.59** | 11,776.78 | 11,788.86 | 11,778.86 | 11,779.16 | 11,791.71 | 11,928.97 | 11,938.81 |
| SU.PA | 13,603.64 | 13,602.9 | 13,656.44 | 13,602.85 | **13,601.83** | 13,654.74 | 13,853.96 | 13,853.64 |
| SIE.DE | 12,657.11 | 12,656.8 | 12,686.54 | 12,655.11 | **12,654.8** | 12,686.79 | 12,854.96 | 12,854.65 |
| TTE.PA | 12,526.59 | 12,524.83 | 12,560.57 | 12,525.82 | **12,524.06** | 12,558.68 | 12,823.56 | 12,821.56 |
| TEF.MC | 12,613.1 | 12,613.11 | 12,665.24 | **12,611.22** | 12,611.27 | 12,664.21 | 12,936.05 | 12,958.28 |
| DG.PA | 12,463.72 | 12,466.9 | 12,514.72 | **12,462.64** | 12,465.66 | 12,518.2 | 12,807.11 | 12,824.92 |
| VIV.PA | 12,189.86 | 12,189.24 | 12,237.88 | 12,188.25 | **12,187.85** | 12,239.81 | 12,382.22 | 12,394.76 |
| VOW.DE | 14,503.13 | **14,238.91** | 14,262.06 | 14,239.49 | 14,239.81 | 14,264.35 | 14,492.2 | 14,497.13 |
| VNA.DE | 9,368.014 | 9,371.945 | 9,405.773 | **9,366.195** | 9,370.104 | 9,409.971 | 9,674.194 | 9,683.446 |

**Note.** In bold the best model.

**Table B.33:** Comparative analysis part III, BIC of unconstrained and constrained two-state GND-HMMs.

| Stock | UUU | CUU | UCU | UUC | CCU | CUC | UCC | CCC |
|---|---|---|---|---|---|---|---|---|
| ADS.DE | 13,527.61 | 13,524.1 | 13,539.35 | 13,519.54 | **13,515.99** | 13,536.2 | 13,676.74 | 13,668.58 |
| ADYEN.AS | 6,547.952 | 6,444.861 | 6,469.176 | 6,441.945 | **6,437.743** | 6,533.523 | 6,522.39 | 6,526.309 |
| AD.AS | 11,066.56 | 11,069.21 | 11,078.46 | **11,063.26** | 11,065.53 | 11,078.36 | 11,190.16 | 11,202.41 |
| AI.PA | 11,412.81 | 11,407.9 | 11,437.47 | 11,405.42 | **11,400.66** | 11,429.95 | 11,571.24 | 11,588.38 |
| AIR.PA | 14,296.95 | 14,290.74 | 14,326.4 | 14,288.82 | **14,282.6** | 14,322.11 | 14,524.44 | 14,536.56 |
| ALV.DE | 11,915.06 | 11,915.31 | 11,950.07 | **11,907.39** | 11,909.08 | 11,956.77 | 12,234.61 | 12,256.91 |
| AMS.MC | 12,717.83 | 12,712.89 | 12,748.72 | 12,714.86 | **12,710.18** | 12,741.91 | 13,017.48 | 13,036.58 |
| ABI.BR | 12,491.92 | **12,483.92** | 12,495.32 | 12,492.65 | 12,484.49 | 12,487.37 | 12,713.16 | 12,717.68 |
| ASML.AS | 14,224.01 | 14,220.13 | 14,277.29 | 14,217.69 | **14,213.75** | 14,276.84 | 14,386.42 | 14,385.52 |
| CS.PA | 13,475.38 | 13,473.39 | 13,531.82 | 13,467.54 | **13,465.5** | 13,527.77 | 13,836.56 | 13,853 |
| BAS.DE | 13,039.67 | 13,035.7 | 13,070.93 | 13,033.15 | **13,029.75** | 13,069.38 | 13,242.2 | 13,252.46 |
| BAYN.DE | 13,232.26 | 13,225.61 | 13,248.69 | 13,224.3 | **13,217.62** | 13,242.3 | 13,380.65 | 13,372.5 |
| BMW.DE | 13,467.65 | 13,459.66 | 13,504.27 | 13,459.75 | **13,451.51** | 13,496.43 | 13,686.19 | 13,700.6 |
| BNP.PA | 14,999.34 | 14,539.93 | 14,601.67 | 14,537.45 | **14,533.09** | 14,598.5 | 14,972.9 | 14,974.85 |
| CRG.IR | 14,126.86 | 14,118.74 | 14,150.57 | 14,118.79 | **14,110.67** | 14,142.44 | 14,357.13 | 14,373.26 |
| DAI.SG | **11,728.95** | 13,501.85 | 13,584.72 | 13,492.65 | 13,484.74 | 13,576.7 | 13,769.15 | 13,761.07 |
| BN.PA | 11,168.27 | 11,160.53 | 11,182.51 | 11,161.45 | **11,153.78** | 11,174.54 | 11,288.79 | 11,301.8 |
| DB1.DE | 12,329.8 | 12,323.12 | 12,348.2 | 12,322.36 | **12,315.7** | 12,341.56 | 12,540.53 | 12,548.94 |
| DPWA.F | 15,665.68 | **14,363.85** | 15,481.54 | 15,497.2 | 15,489.04 | 15,473.27 | 15,660.1 | 15,641.21 |
| DTE.DE | 11,438.25 | 11,431.29 | 11,467.3 | 11,430.65 | **11,423.52** | 11,464.15 | 11,689.58 | 11,690.89 |
| ENEL.MI | 12,945.02 | 12,937.41 | 12,976.57 | 12,936.9 | **12,929.5** | 12,973.91 | 13,133.33 | 13,142.54 |
| ENGI.PA | 12,669.56 | 12,662.77 | 12,695.97 | 12,661.54 | **12,654.73** | 12,689.99 | 12,853.27 | 12,869.11 |
| ENI.MI | 12,750.7 | **12,743.21** | 12,769.83 | 12,752.07 | 12,745.02 | 12,762.26 | 13,010.5 | 13,041.73 |
| EL.PA | 12,042.75 | 12,036.39 | 12,071.58 | 12,034.95 | **12,028.51** | 12,064.7 | 12,200.55 | 12,214.41 |
| FLTR.IR | 13,934.17 | 13,716.78 | 13,748.38 | 13,716.61 | **13,709.15** | 13,742.96 | 13,914.26 | 13,909.69 |
| IBE.MC | 11,710.17 | 11,705.75 | 11,767 | **11,698.09** | 11,768.21 | 12,110.86 | 12,123.03 |
| ITX.MC | 13,011.39 | 13,006.6 | 13,030.56 | 13,003.4 | **12,998.44** | 13,031.41 | 13,156.05 | 13,170.97 |
| IFX.DE | 15,369.89 | 15,202.33 | 15,257.01 | 15,200.87 | **15,196.13** | 15,253 | 15,353.55 | 15,345.42 |
| INGA.AS | 15,208.62 | 14,774.48 | 14,852.81 | 14,779.03 | **14,773.64** | 14,850.13 | 15,186.18 | 15,184.12 |
| ISP.MI | 15,492.16 | 15,109.09 | 15,188.34 | 15,102.24 | **15,100.95** | 15,188.21 | 15,466.83 | 15,467.68 |
| KER.PA | 13,638.21 | 13,636.02 | 13,659.96 | 13,632.13 | **13,629.84** | 13,656.25 | 13,790.9 | 13,797.34 |
| KNEBV.HE | 12,245.12 | 12,237.62 | 12,307.19 | 12,243.26 | **12,235.86** | 12,299.47 | 12,424.84 | 12,417.96 |
| OR.PA | 11,805.57 | 11,798.31 | 11,846.98 | 11,797.42 | **11,790.15** | 11,839.77 | 11,971.75 | 11,963.58 |
| MC.PA | 13,312.15 | 13,308.6 | 13,344.02 | 13,306.45 | **13,303.64** | 13,339.21 | 13,448.81 | 13,462.67 |
| MUV2.DE | 11,484.85 | 11,488.4 | 11,505.82 | **11,480.99** | 11,483.56 | 11,512.54 | 11,777.43 | 11,787.38 |
| NOKIA.HE | 14,799.64 | 14,542.07 | 14,591.15 | 14,544.35 | **14,536.29** | 14,583.7 | 14,759.29 | 14,775.21 |
| RI.PA | 11,400.18 | 11,394.34 | 11,425.7 | 11,392.51 | **11,386.68** | 11,418.72 | 11,548.47 | 11,557.28 |
| ORA.PA | 11,706.11 | 11,697.97 | 11,777.34 | 11,698.17 | **11,690.03** | 11,769.33 | 12,091.73 | 12,083.82 |
| PHIA.AS | 13,241.58 | 13,237.85 | 13,273.88 | 13,234.22 | **13,230.81** | 13,266.56 | 13,431.63 | 13,451.78 |
| PRX.AS | 4,935.791 | 4,863.318 | 4,879.047 | 4,863.944 | **4,857.029** | 4,921.897 | 4,907.628 | 4,914.95 |
| SAF.PA | 13,512.94 | 13,507.99 | 13,575.27 | 13,505.32 | **13,500.15** | 13,572.47 | 13,872.49 | 13,877.54 |
| SAN.PA | 11,951.01 | 11,946.03 | 11,971.75 | 11,943.71 | **11,938.73** | 11,964.06 | 12,053.06 | 12,084.14 |
| SAN.MC | 14,689.47 | 14,684.88 | 14,755.72 | 14,682.2 | **14,677.79** | 14,750.89 | 14,973.81 | 14,975.92 |
| SAP.DE | 11,831.01 | 11,826.04 | 11,838.12 | 11,828.12 | **11,822.26** | 11,834.81 | 11,972.07 | 11,975.75 |
| SU.PA | 13,659.13 | 13,652.23 | 13,705.77 | 13,652.17 | **13,644.99** | 13,697.9 | 13,897.12 | 13,890.64 |
| SIE.DE | 12,712.52 | 12,706.06 | 12,735.8 | 12,704.37 | **12,697.91** | 12,729.89 | 12,898.06 | 12,891.59 |
| TTE.PA | 12,582.08 | 12,574.15 | 12,609.89 | 12,575.14 | **12,567.21** | 12,601.84 | 12,866.72 | 12,858.55 |
| TEF.MC | 12,668.58 | 12,662.43 | 12,714.56 | 12,660.54 | **12,654.42** | 12,707.36 | 12,979.21 | 12,995.28 |
| DG.PA | 12,519.21 | 12,516.22 | 12,564.04 | 12,511.96 | **12,508.81** | 12,561.35 | 12,850.27 | 12,861.91 |
| VIV.PA | 12,245.35 | 12,238.56 | 12,287.2 | 12,237.56 | **12,231** | 12,282.97 | 12,425.38 | 12,431.75 |
| VOW.DE | 14,558.54 | 14,288.16 | 14,311.31 | 14,288.74 | **14,282.9** | 14,307.44 | 14,535.3 | 14,534.07 |
| VNA.DE | 9,420.756 | 9,418.827 | 9,452.655 | 9,413.076 | **9,411.125** | 9,450.992 | 9,715.215 | 9,718.608 |

**Note.** In bold the best model.

**Table B.34:** Comparative analysis part III, KS of unconstrained and constrained two-state GND-HMMs.

| Stock | UUU | CUU | UCU | UUC | CCU | CUC | UCC | CCC |
|-------|-----|-----|-----|-----|-----|-----|-----|-----|
| ADS.DE | 0.0088 | 0.0117 | 0.0076 | 0.009 | **0.0065** | 0.0119 | 0.0077 | 0.0077 |
| ADYEN.AS | 0.0171 | 0.017 | 0.0138 | 0.0101 | 0.0171 | 0.0168 | **0.0093** | 0.0171 |
| AD.AS | **0.0048** | 0.0109 | 0.009 | 0.0073 | 0.0119 | 0.0109 | 0.008 | 0.0171 |
| AI.PA | **0.0065** | 0.0095 | 0.0095 | 0.0069 | 0.0106 | 0.0098 | 0.0131 | 0.0103 |
| AIR.PA | 0.0112 | 0.0133 | 0.0074 | 0.0111 | 0.0136 | 0.0132 | **0.0038** | 0.017 |
| ALV.DE | 0.0063 | 0.0121 | 0.0067 | **0.0054** | 0.0141 | 0.0115 | 0.02 | 0.011 |
| AMS.MC | **0.0081** | 0.0122 | 0.0161 | 0.0083 | 0.0169 | 0.011 | 0.0175 | 0.0125 |
| ABI.BR | 0.0104 | **0.0102** | 0.0132 | 0.011 | 0.0121 | 0.0108 | 0.0218 | 0.0163 |
| ASML.AS | 0.008 | 0.0121 | **0.0056** | 0.0068 | 0.011 | 0.0113 | 0.0067 | 0.0088 |
| CS.PA | **0.0094** | 0.015 | 0.0112 | 0.0096 | 0.0188 | 0.015 | 0.0141 | 0.017 |
| BAS.DE | 0.0068 | 0.0095 | 0.0076 | **0.0058** | 0.0096 | 0.0087 | 0.0086 | 0.0097 |
| BAYN.DE | **0.0084** | 0.01 | 0.0109 | 0.0084 | 0.0097 | 0.0101 | 0.0111 | 0.0111 |
| BMW.DE | 0.0078 | 0.0069 | 0.0099 | 0.0078 | 0.0086 | 0.0069 | 0.0105 | **0.0065** |
| BNP.PA | 0.0103 | 0.0088 | 0.0083 | **0.0051** | 0.012 | 0.0084 | 0.008 | 0.0103 |
| CRG.IR | 0.0093 | 0.0095 | 0.0115 | 0.0093 | 0.0116 | 0.0095 | **0.004** | 0.0112 |
| DAI.SG | 0.051 | 0.0066 | 0.0109 | 0.0118 | 0.0114 | 0.0101 | 0.0047 | **0.0046** |
| BN.PA | **0.0057** | 0.007 | 0.0077 | 0.0062 | 0.008 | 0.0074 | 0.0133 | 0.0091 |
| DB1.DE | **0.006** | 0.008 | 0.0086 | 0.0065 | 0.0085 | 0.0085 | 0.0107 | 0.0109 |
| DPWA.F | 0.0134 | 0.0178 | 0.0182 | **0.0069** | 0.0121 | 0.0069 | 0.0089 | 0.0133 |
| DTE.DE | 0.0081 | 0.0098 | 0.0059 | 0.0088 | 0.0072 | 0.0102 | **0.0058** | 0.0073 |
| ENEL.MI | 0.0037 | 0.008 | 0.01 | **0.0035** | 0.0117 | 0.0077 | 0.0117 | 0.0049 |
| ENGI.PA | **0.0041** | 0.006 | 0.0076 | 0.0043 | 0.0087 | 0.0058 | 0.0155 | 0.0076 |
| ENI.MI | 0.0091 | 0.0099 | 0.0119 | 0.0106 | 0.0126 | 0.0122 | **0.0062** | 0.018 |
| EL.PA | 0.0088 | 0.0111 | 0.0123 | **0.0085** | 0.0131 | 0.0108 | 0.0121 | 0.0093 |
| FLTR.IR | 0.0074 | 0.0104 | **0.0047** | 0.0087 | 0.0112 | 0.0106 | 0.0096 | 0.0074 |
| IBE.MC | 0.0081 | 0.0121 | 0.0075 | 0.0076 | 0.0117 | 0.0116 | **0.0039** | 0.0108 |
| ITX.MC | 0.0077 | 0.01 | **0.0059** | 0.008 | 0.0081 | 0.01 | 0.0099 | 0.0074 |
| IFX.DE | 0.0048 | 0.0074 | 0.0099 | **0.0047** | 0.0106 | 0.0076 | 0.0048 | 0.0048 |
| INGA.AS | 0.0069 | 0.0074 | 0.0088 | **0.0036** | 0.0109 | 0.0066 | 0.0078 | 0.0069 |
| ISP.MI | **0.0066** | 0.0138 | 0.0122 | 0.0068 | 0.0169 | 0.0137 | 0.0082 | 0.0066 |
| KER.PA | **0.0082** | 0.0125 | 0.0113 | 0.0097 | 0.0127 | 0.0137 | 0.011 | 0.0115 |
| KNEBV.HE | 0.0064 | 0.0079 | 0.0101 | 0.0049 | 0.0107 | 0.0068 | **0.0035** | 0.0063 |
| OR.PA | 0.0119 | 0.0136 | **0.0113** | 0.012 | 0.013 | 0.0137 | 0.0129 | 0.0129 |
| MC.PA | **0.008** | 0.0121 | 0.0095 | 0.0089 | 0.0112 | 0.0131 | 0.0102 | 0.0103 |
| MUV2.DE | **0.0083** | 0.0148 | 0.0121 | 0.0094 | 0.0142 | 0.0162 | 0.0181 | 0.0185 |
| NOKIA.HE | 0.0204 | 0.0194 | 0.0199 | 0.0178 | 0.0193 | 0.0184 | **0.0032** | 0.0204 |
| RI.PA | **0.0069** | 0.0099 | 0.0077 | 0.0075 | 0.0106 | 0.0105 | 0.0118 | 0.0101 |
| ORA.PA | 0.0144 | 0.0147 | 0.018 | 0.0143 | 0.0192 | 0.0146 | **0.0084** | 0.0118 |
| PHIA.AS | 0.0121 | 0.0167 | 0.016 | 0.0125 | 0.0183 | 0.0172 | **0.0066** | 0.0182 |
| PRX.AS | 0.0109 | 0.009 | 0.0097 | **0.0085** | 0.0109 | 0.0088 | 0.0164 | 0.011 |
| SAF.PA | 0.0072 | 0.0115 | 0.0105 | 0.0077 | 0.0129 | 0.0117 | 0.0102 | **0.0061** |
| SAN.PA | **0.0094** | 0.0121 | 0.0099 | 0.0104 | 0.0107 | 0.0131 | 0.0106 | 0.0122 |
| SAN.MC | 0.0105 | 0.0141 | 0.0133 | 0.0099 | 0.0162 | 0.0135 | **0.0041** | 0.013 |
| SAP.DE | **0.0063** | 0.0077 | 0.0082 | 0.0071 | 0.0085 | 0.0088 | 0.0126 | 0.0106 |
| SU.PA | 0.01 | 0.0121 | 0.0177 | 0.0112 | 0.0177 | 0.013 | **0.0089** | 0.0096 |
| SIE.DE | **0.0089** | 0.0104 | 0.0122 | 0.0089 | 0.0096 | 0.0104 | 0.012 | 0.0107 |
| TTE.PA | **0.0125** | 0.0131 | 0.0154 | 0.0127 | 0.0154 | 0.0134 | 0.0167 | 0.0167 |
| TEF.MC | **0.0092** | 0.0121 | 0.0127 | 0.0094 | 0.0156 | 0.0123 | 0.0145 | 0.0141 |
| DG.PA | **0.0098** | 0.0144 | 0.011 | 0.0103 | 0.0139 | 0.0148 | 0.0157 | 0.0139 |
| VIV.PA | 0.0089 | 0.0107 | **0.0052** | 0.0082 | 0.0092 | 0.0103 | 0.0085 | 0.0104 |
| VOW.DE | **0.0096** | 0.0114 | 0.0112 | 0.0102 | 0.0103 | 0.0122 | 0.012 | 0.0096 |
| VNA.DE | 0.009 | 0.0143 | 0.0152 | 0.0091 | 0.0156 | 0.0145 | **0.0081** | 0.013 |

**Note.** In bold the best model.

**Table B.35:** Comparative analysis part III, AD of unconstrained and constrained two-state GND-HMMs.

| Stock | UUU | CUU | UCU | UUC | CCU | CUC | UCC | CCC |
|---|---|---|---|---|---|---|---|---|
| ADS.DE | 0.0191 | 0.0179 | 0.0064 | 0.0215 | **0.0056** | 0.0199 | 0.0688 | 0.0689 |
| ADYEN.AS | 0.0181 | 0.0151 | 0.015 | **0.0059** | 0.0181 | 0.0143 | 0.0099 | 0.0181 |
| AD.AS | 0.0042 | 0.0097 | **0.0032** | 0.0071 | 0.0069 | 0.0137 | 0.0279 | 0.0539 |
| AI.PA | 0.1659 | 0.1623 | **0.0114** | 0.3056 | 0.0124 | 0.3255 | 0.2622 | 1.2654 |
| AIR.PA | 0.011 | 0.0127 | **0.0093** | 0.0105 | 0.0099 | 0.0122 | 0.0377 | 0.0739 |
| ALV.DE | 0.0023 | 0.0067 | 0.0163 | **0.0022** | 0.0112 | 0.0065 | 0.0221 | 0.0396 |
| AMS.MC | 0.0138 | 0.0159 | **0.01** | 0.0274 | 0.0114 | 0.0294 | 0.0306 | 0.0605 |
| ABI.BR | 0.0178 | 0.0176 | 0.0134 | 0.0352 | **0.0123** | 0.0349 | 0.1422 | 0.2004 |
| ASML.AS | 0.0289 | 0.0314 | **0.0084** | 0.0187 | 0.0088 | 0.0207 | 0.0173 | 0.0285 |
| CS.PA | **0.0126** | 0.0185 | 0.0137 | 0.0138 | 0.0142 | 0.0192 | 0.0256 | 0.046 |
| BAS.DE | 0.0094 | 0.016 | 0.0109 | **0.0067** | 0.0145 | 0.0122 | 0.0256 | 0.0573 |
| BAYN.DE | 0.0372 | 0.0408 | **0.0097** | 0.0445 | 0.01 | 0.0484 | 0.0806 | 0.0806 |
| BMW.DE | 0.0383 | 0.0374 | 0.0132 | 0.0373 | **0.0128** | 0.036 | 0.0275 | 0.0635 |
| BNP.PA | 0.0597 | 0.0169 | 0.0207 | **0.0089** | 0.0263 | 0.014 | 0.0421 | 0.0597 |
| CRG.IR | **0.0036** | 0.0036 | 0.0085 | 0.0038 | 0.0085 | 0.0038 | 0.0151 | 0.0446 |
| DAI.SG | 0.5181 | 0.2498 | **0.0129** | 0.0537 | 0.0132 | 0.0544 | 0.0334 | 0.0339 |
| BN.PA | 0.0106 | 0.0114 | **0.0049** | 0.0164 | 0.0051 | 0.0178 | 0.0282 | 0.0636 |
| DB1.DE | 0.03 | 0.0337 | **0.0131** | 0.0399 | 0.0136 | 0.045 | 0.1812 | 0.3827 |
| DPWA.F | 0.0216 | 0.0238 | 0.0362 | 0.0127 | 0.0283 | 0.0127 | **0.0109** | 0.0216 |
| DTE.DE | 0.0204 | 0.024 | **0.0104** | 0.0261 | 0.0107 | 0.0293 | 0.0548 | 0.0873 |
| ENEL.MI | 0.0213 | 0.0398 | **0.0109** | 0.0221 | 0.0142 | 0.0451 | 0.035 | 0.0523 |
| ENGI.PA | 0.047 | 0.0535 | **0.0089** | 0.0576 | 0.0108 | 0.0643 | 0.1152 | 0.3389 |
| ENI.MI | **0.0065** | 0.0077 | 0.01 | 0.014 | 0.0107 | 0.0164 | 0.0086 | 0.0312 |
| EL.PA | 0.0166 | 0.0174 | **0.0046** | 0.0203 | 0.0049 | 0.0208 | 0.018 | 0.0437 |
| FLTR.IR | 0.0846 | 0.0329 | 0.0119 | 0.042 | **0.01** | 0.0408 | 0.0705 | 0.0864 |
| IBE.MC | 0.0045 | 0.007 | 0.016 | **0.0039** | 0.0176 | 0.0058 | 0.0132 | 0.0235 |
| ITX.MC | 0.0133 | 0.012 | 0.0068 | 0.0155 | **0.0051** | 0.0122 | 0.0364 | 0.0996 |
| IFX.DE | 0.0106 | 0.0123 | **0.0067** | 0.008 | 0.0077 | 0.0089 | 0.0106 | 0.0107 |
| INGA.AS | 0.011 | **0.0066** | 0.0294 | 0.0072 | 0.0315 | 0.0099 | 0.0096 | 0.011 |
| ISP.MI | 0.0144 | 0.0113 | 0.0185 | **0.0066** | 0.0225 | 0.0111 | 0.0101 | 0.0144 |
| KER.PA | 0.0072 | 0.0075 | **0.0038** | 0.0117 | 0.004 | 0.0123 | 0.0173 | 0.0347 |
| KNEBV.HE | 0.0034 | 0.0033 | 0.0077 | **0.0018** | 0.0079 | 0.0018 | 0.0026 | 0.0029 |
| OR.PA | 0.0068 | 0.0064 | 0.0027 | 0.007 | **0.0026** | 0.0065 | 0.0137 | 0.0137 |
| MC.PA | **0.003** | 0.0046 | 0.0041 | 0.004 | 0.0048 | 0.0058 | 0.0038 | 0.0082 |
| MUV2.DE | **0.0036** | 0.0105 | 0.0085 | 0.0062 | 0.0123 | 0.0132 | 0.0234 | 0.0382 |
| NOKIA.HE | 0.0376 | 0.0151 | **0.0086** | 0.0089 | 0.0087 | 0.0094 | 0.0206 | 0.0376 |
| RI.PA | 0.0411 | 0.0451 | **0.0098** | 0.0527 | 0.011 | 0.0581 | 0.0501 | 0.1144 |
| ORA.PA | 0.0317 | 0.0316 | **0.0096** | 0.0268 | 0.0097 | 0.0268 | 0.0296 | 0.0303 |
| PHIA.AS | 0.0341 | 0.0393 | **0.0134** | 0.0447 | 0.0156 | 0.0535 | 0.0554 | 0.1355 |
| PRX.AS | **0.0078** | 0.0099 | 0.0339 | 0.0133 | 0.0078 | 0.0133 | 0.0086 | 0.0078 |
| SAF.PA | 0.034 | 0.037 | 0.0116 | 0.0425 | **0.0105** | 0.0444 | 0.0294 | 0.0514 |
| SAN.PA | 0.0924 | 0.0966 | **0.015** | 0.1453 | 0.0167 | 0.1521 | 0.0721 | 0.4742 |
| SAN.MC | 0.007 | 0.0085 | 0.0128 | **0.0055** | 0.0136 | 0.0066 | 0.007 | 0.0132 |
| SAP.DE | 0.0104 | 0.0123 | **0.0064** | 0.0233 | 0.0075 | 0.0232 | 0.0376 | 0.0739 |
| SU.PA | 0.0054 | 0.0064 | 0.0135 | **0.0045** | 0.013 | 0.0056 | 0.0083 | 0.0096 |
| SIE.DE | 0.0221 | 0.0247 | 0.0063 | 0.0224 | **0.0062** | 0.0243 | 0.1748 | 0.1869 |
| TTE.PA | 0.0814 | 0.0812 | **0.0145** | 0.1394 | 0.0147 | 0.1376 | 0.5151 | 0.5151 |
| TEF.MC | **0.008** | 0.0101 | 0.0111 | 0.0091 | 0.0133 | 0.0113 | 0.0186 | 0.0394 |
| DG.PA | 0.0251 | 0.0304 | **0.0086** | 0.0345 | 0.0106 | 0.0401 | 0.0703 | 0.1337 |
| VIV.PA | 0.0898 | 0.1044 | **0.0116** | 0.0664 | 0.0117 | 0.0715 | 0.1086 | 0.2118 |
| VOW.DE | 0.0193 | **0.0053** | 0.0077 | 0.0089 | 0.007 | 0.0071 | 0.0126 | 0.0193 |
| VNA.DE | 0.0045 | 0.0089 | 0.0195 | **0.0044** | 0.0218 | 0.0089 | 0.01 | 0.0153 |

**Note.** In bold the best model.

**Table B.36:** Comparative analysis part III, AIC, comparison between independent mixture models and GND-HMMs.

| Stock | MGND | CMGND | U.GND-HMM | C.GND-HMM |
|---|---|---|---|---|
| ADS.DE | 13,587.6 | 13,586.35 | 13,472.2 | **13,470.28** |
| ADYEN.AS | 6,487.87 | 6,485.5 | 6,501.02 | **6,400.23** |
| AD.AS | 11,108.64 | 11,106.08 | **11,011.06** | 11,013.93 |
| AI.PA | 11,526.42 | 11,521.29 | 11,357.32 | **11,356.1** |
| AIR.PA | 14,464.38 | 14,460.75 | 14,241.48 | **14,239.45** |
| ALV.DE | 12,169.38 | 12,168.08 | 11,859.65 | **11,858.14** |
| AMS.MC | 12,954.11 | 12,952.98 | **12,662.65** | 12,663.75 |
| ABI.BR | 12,615.54 | 12,609.38 | 12,436.43 | **12,434.59** |
| ASML.AS | 14,336.54 | 14,334.53 | 14,168.52 | **14,168.36** |
| CS.PA | 13,772.19 | 13,771.47 | 13,419.9 | **13,418.22** |
| BAS.DE | 13,185.91 | 13,184.68 | 12,984.25 | **12,983.9** |
| BAYN.DE | 13,321.16 | 13,318.17 | 13,176.85 | **13,174.52** |
| BMW.DE | 13,645.54 | 13,642.94 | 13,412.24 | **13,408.41** |
| BNP.PA | 14,910.16 | 14,908.63 | 14,943.86 | **14,488.14** |
| CRG.IR | 14,291.35 | 14,287.26 | 14,071.44 | **14,067.57** |
| DAI.SG | 13,706.43 | 13,706.12 | **11,673.55** | 13,441.66 |
| BN.PA | 11,235.4 | 11,232.59 | 11,112.78 | **11,110.62** |
| DB1.DE | 12,475.73 | 12,470.14 | 12,274.38 | **12,272.59** |
| DPWA.F | 15,287.13 | 15,267.08 | 15,610.26 | **14,314.6** |
| DTE.DE | 11,623.82 | 11,620.35 | 11,382.83 | **11,380.42** |
| ENEL.MI | 13,087.3 | 13,085.54 | 12,889.59 | **12,886.38** |
| ENGI.PA | 12,791.14 | 12,788.61 | 12,614.07 | **12,611.57** |
| ENI.MI | 12,960.72 | 12,959.2 | 12,695.27 | **12,693.94** |
| EL.PA | 12,154.42 | 12,151.99 | 11,987.26 | **11,985.35** |
| FLTR.IR | 13,832.53 | 13,832.38 | 13,878.76 | **13,666.05** |
| IBE.MC | 12,058.94 | 12,056.38 | 11,654.68 | **11,652.97** |
| ITX.MC | 13,105.9 | 13,101.41 | 12,955.9 | **12,954.07** |
| IFX.DE | 15,297.79 | 15,298.43 | 15,314.48 | **15,151.61** |
| INGA.AS | 15,137.45 | 15,135.37 | 15,153.13 | **14,725.16** |
| ISP.MI | 15,415.48 | 15,411.99 | 15,436.73 | **15,052.98** |
| KER.PA | 13,728.68 | 13,724.61 | **13,582.72** | 13,582.8 |
| KNEBV.HE | 12,377.71 | 12,373.64 | 12,189.79 | **12,188.44** |
| OR.PA | 11,904.43 | 11,900.99 | 11,750.08 | **11,746.99** |
| MC.PA | 13,407.35 | 13,404.11 | **13,256.66** | 13,257.13 |
| MUV2.DE | 11,690.9 | 11,689.32 | **11,429.43** | 11,431.74 |
| NOKIA.HE | 14,693.66 | 14,689.6 | 14,744.34 | **14,492.91** |
| RI.PA | 11,491.3 | 11,489.55 | 11,344.69 | **11,343.18** |
| ORA.PA | 12,024.56 | 12,023.28 | 11,650.62 | **11,646.87** |
| PHIA.AS | 13,367.94 | 13,368.44 | 13,186.08 | **13,184.89** |
| PRX.AS | 4,878.32 | 4,875.59 | 4,891.27 | **4,822.4** |
| SAF.PA | 13,814.6 | 13,809.96 | 13,457.47 | **13,456.01** |
| SAN.PA | 12,008.92 | 12,005.51 | 11,895.52 | **11,894.38** |
| SAN.MC | 14,929.96 | 14,926.96 | 14,633.98 | **14,632.89** |
| SAP.DE | 11,896.08 | 11,893.86 | **11,775.59** | 11,776.78 |
| SU.PA | 13,848.53 | 13,845.81 | 13,603.64 | **13,601.83** |
| SIE.DE | 12,826.09 | 12,822.81 | 12,657.11 | **12,654.8** |
| TTE.PA | 12,785.79 | 12,783.19 | 12,526.59 | **12,524.06** |
| TEF.MC | 12,917.25 | 12,915.04 | 12,613.1 | **12,611.22** |
| DG.PA | 12,775.87 | 12,771.47 | 12,463.72 | **12,462.64** |
| VIV.PA | 12,371.66 | 12,371.25 | 12,189.86 | **12,187.85** |
| VOW.DE | 14,477.53 | 14,473.89 | 14,503.13 | **14,238.91** |
| VNA.DE | 9,657.33 | 9,655.14 | 9,368.01 | **9,366.19** |
| Mean | 12815.80 | 12812.88 | 12671.67 | **12579.82** |
| Median | 12957.42 | 12956.09 | 12715.17 | **12678.84** |

**Note.** In bold the best model. 'U.' denotes unconstrained, while 'C.' denotes constrained.

**Table B.37:** Comparative analysis part III, BIC, comparison between independent mixture models and GND-HMMs.

| Stock | MGND | CMGND | U.GND-HMM | C.GND-HMM |
|-------|------|-------|-----------|-----------|
| ADS.DE | 13,630.7 | 13,617.13 | 13,527.61 | **13,515.99** |
| ADYEN.AS | 6,524.38 | 6,511.57 | 6,547.95 | **6,437.74** |
| AD.AS | 11,151.8 | 11,139.92 | 11,066.56 | **11,063.26** |
| AI.PA | 11,569.58 | 11,552.12 | 11,412.81 | **11,400.66** |
| AIR.PA | 14,507.53 | 14,491.57 | 14,296.95 | **14,282.6** |
| ALV.DE | 12,212.48 | 12,201.53 | 11,915.06 | **11,907.39** |
| AMS.MC | 12,997.1 | 12,986.83 | 12,717.83 | **12,710.18** |
| ABI.BR | 12,658.7 | 12,640.21 | 12,491.92 | **12,483.92** |
| ASML.AS | 14,379.71 | 14,365.36 | 14,224.01 | **14,213.75** |
| CS.PA | 13,815.34 | 13,807.14 | 13,475.38 | **13,465.5** |
| BAS.DE | 13,229.01 | 13,221.38 | 13,039.67 | **13,029.75** |
| BAYN.DE | 13,364.26 | 13,348.95 | 13,232.26 | **13,217.62** |
| BMW.DE | 13,688.64 | 13,673.73 | 13,467.65 | **13,451.51** |
| BNP.PA | 14,953.31 | 14,940.49 | 14,999.34 | **14,533.09** |
| CRG.IR | 14,334.45 | 14,318.05 | 14,126.86 | **14,110.67** |
| DAI.SG | 13,749.52 | 13,743.04 | **11,728.95** | 13,484.74 |
| BN.PA | 11,278.56 | 11,264.12 | 11,168.27 | **11,153.78** |
| DB1.DE | 12,518.83 | 12,500.93 | 12,329.8 | **12,315.7** |
| DPWA.F | 15,330.22 | 15,304.02 | 15,665.68 | **14,363.85** |
| DTE.DE | 11,666.92 | 11,651.14 | 11,438.25 | **11,423.52** |
| ENEL.MI | 13,130.42 | 13,119.6 | 12,945.02 | **12,929.5** |
| ENGI.PA | 12,834.3 | 12,824.29 | 12,669.56 | **12,654.73** |
| ENI.MI | 13,003.83 | 12,995.08 | 12,750.7 | **12,743.21** |
| EL.PA | 12,197.58 | 12,182.85 | 12,042.75 | **12,028.51** |
| FLTR.IR | 13,875.63 | 13,863.6 | 13,934.17 | **13,709.15** |
| IBE.MC | 12,102.09 | 12,087.21 | 11,710.17 | **11,698.09** |
| ITX.MC | 13,149.07 | 13,132.24 | 13,011.39 | **12,998.44** |
| IFX.DE | 15,340.89 | 15,329.17 | 15,369.89 | **15,196.13** |
| INGA.AS | 15,180.6 | 15,166.2 | 15,208.62 | **14,773.64** |
| ISP.MI | 15,458.58 | 15,442.78 | 15,492.16 | **15,100.95** |
| KER.PA | 13,771.84 | 13,755.44 | 13,638.21 | **13,629.84** |
| KNEBV.HE | 12,420.74 | 12,401.94 | 12,245.12 | **12,235.86** |
| OR.PA | 11,947.59 | 11,931.82 | 11,805.57 | **11,790.15** |
| MC.PA | 13,450.51 | 13,434.94 | 13,312.15 | **13,303.64** |
| MUV2.DE | 11,734 | 11,724.02 | 11,484.85 | **11,480.99** |
| NOKIA.HE | 14,736.67 | 14,720.33 | 14,799.64 | **14,536.29** |
| RI.PA | 11,534.46 | 11,523.17 | 11,400.18 | **11,386.68** |
| ORA.PA | 12,067.72 | 12,054.11 | 11,706.11 | **11,690.03** |
| PHIA.AS | 13,411.1 | 13,399.27 | 13,241.58 | **13,230.81** |
| PRX.AS | 4,912.95 | 4,900.32 | 4,935.79 | **4,857.03** |
| SAF.PA | 13,857.75 | 13,840.78 | 13,512.94 | **13,500.15** |
| SAN.PA | 12,052.08 | 12,036.34 | 11,951.01 | **11,938.73** |
| SAN.MC | 14,973.11 | 14,957.79 | 14,689.47 | **14,677.79** |
| SAP.DE | 11,939.18 | 11,930.65 | 11,831.01 | **11,822.26** |
| SU.PA | 13,891.69 | 13,874.49 | 13,659.13 | **13,644.99** |
| SIE.DE | 12,869.2 | 12,853.6 | 12,712.52 | **12,697.91** |
| TTE.PA | 12,828.94 | 12,815.05 | 12,582.08 | **12,567.21** |
| TEF.MC | 12,960.41 | 12,951.32 | 12,668.58 | **12,654.42** |
| DG.PA | 12,819.02 | 12,803.57 | 12,519.21 | **12,508.81** |
| VIV.PA | 12,414.81 | 12,402.07 | 12,245.35 | **12,231** |
| VOW.DE | 14,520.63 | 14,504.67 | 14,558.54 | **14,282.9** |
| VNA.DE | 9,698.35 | 9,684.44 | 9,420.76 | **9,411.12** |
| Mean | 12858.59 | 12844.66 | 12671.67 | **12624.54** |
| Median | 13000.47 | 12990.95 | 12715.17 | **12726.69** |

**Note.** In bold the best model. 'U.' denotes unconstrained, while 'C.' denotes constrained.

**Table B.38:** Comparative analysis part III, KS, comparison between independent mixture models and GND-HMMs.

| Stock | MGND | CMGND | U.GND-HMM | C.GND-HMM |
|---|---|---|---|---|
| ADS.DE | 0.0082 | **0.0012** | 0.0088 | 0.0065 |
| ADYEN.AS | 0.0098 | **0.0088** | 0.0171 | 0.0093 |
| AD.AS | **0.0033** | 0.0059 | 0.0048 | 0.0073 |
| AI.PA | **0.006** | 0.0064 | 0.0065 | 0.0069 |
| AIR.PA | 0.0055 | 0.005 | 0.0112 | **0.0038** |
| ALV.DE | **0.0051** | 0.0052 | 0.0063 | 0.0054 |
| AMS.MC | 0.0086 | **0.0079** | 0.0081 | 0.0083 |
| ABI.BR | 0.0072 | **0.0049** | 0.0104 | 0.0102 |
| ASML.AS | 0.0044 | **0.0041** | 0.008 | 0.0056 |
| CS.PA | 0.0075 | **0.0052** | 0.0094 | 0.0096 |
| BAS.DE | 0.0049 | **0.0044** | 0.0068 | 0.0058 |
| BAYN.DE | 0.0084 | **0.0046** | 0.0084 | 0.0084 |
| BMW.DE | 0.0081 | **0.0014** | 0.0078 | 0.0065 |
| BNP.PA | 0.0073 | **0.0031** | 0.0103 | 0.0051 |
| CRG.IR | 0.0062 | 0.0047 | 0.0093 | **0.004** |
| DAI.SG | 0.0094 | **0.0028** | 0.051 | 0.0046 |
| BN.PA | 0.0049 | **0.0032** | 0.0057 | 0.0062 |
| DB1.DE | 0.0034 | **0.0027** | 0.006 | 0.0065 |
| DPWA.F | 0.0319 | 0.0103 | 0.0134 | **0.0069** |
| DTE.DE | **0.0018** | 0.0033 | 0.0081 | 0.0058 |
| ENEL.MI | 0.0053 | 0.0044 | 0.0037 | **0.0035** |
| ENGI.PA | 0.0056 | 0.0043 | **0.0041** | 0.0043 |
| ENI.MI | 0.0069 | **0.0046** | 0.0091 | 0.0062 |
| EL.PA | **0.0074** | 0.0074 | 0.0088 | 0.0085 |
| FLTR.IR | 0.0087 | **0.0011** | 0.0074 | 0.0047 |
| IBE.MC | 0.0049 | **0.0031** | 0.0081 | 0.0039 |
| ITX.MC | 0.0058 | **0.0032** | 0.0077 | 0.0059 |
| IFX.DE | 0.0074 | **0.0043** | 0.0048 | 0.0047 |
| INGA.AS | **0.0023** | 0.0028 | 0.0069 | 0.0036 |
| ISP.MI | **0.0034** | 0.0038 | 0.0066 | 0.0066 |
| KER.PA | 0.005 | **0.0035** | 0.0082 | 0.0097 |
| KNEBV.HE | 0.0039 | **0.0034** | 0.0064 | 0.0035 |
| OR.PA | 0.0067 | **0.005** | 0.0119 | 0.0113 |
| MC.PA | 0.0045 | **0.0039** | 0.008 | 0.0089 |
| MUV2.DE | **0.0062** | 0.0062 | 0.0083 | 0.0094 |
| NOKIA.HE | 0.0059 | 0.0055 | 0.0204 | **0.0032** |
| RI.PA | **0.0039** | 0.0048 | 0.0069 | 0.0075 |
| ORA.PA | 0.0049 | **0.0042** | 0.0144 | 0.0084 |
| PHIA.AS | 0.0054 | **0.0044** | 0.0121 | 0.0066 |
| PRX.AS | 0.01 | **0.0059** | 0.0109 | 0.0085 |
| SAF.PA | **0.0018** | 0.002 | 0.0072 | 0.0061 |
| SAN.PA | 0.0066 | **0.0063** | 0.0094 | 0.0099 |
| SAN.MC | 0.005 | 0.0043 | 0.0105 | **0.0041** |
| SAP.DE | 0.0087 | **0.0054** | 0.0063 | 0.0071 |
| SU.PA | **0.0075** | 0.008 | 0.01 | 0.0089 |
| SIE.DE | 0.0058 | **0.0042** | 0.0089 | 0.0089 |
| TTE.PA | **0.004** | 0.0051 | 0.0125 | 0.0127 |
| TEF.MC | 0.006 | **0.004** | 0.0092 | 0.0094 |
| DG.PA | 0.0064 | **0.0036** | 0.0098 | 0.0103 |
| VIV.PA | 0.0055 | **0.002** | 0.0089 | 0.0052 |
| VOW.DE | **0.0039** | 0.0044 | 0.0096 | 0.0096 |
| VNA.DE | 0.0086 | 0.0085 | 0.009 | **0.0081** |
| Mean | 0.0065 | **0.0046** | 0.0097 | 0.0070 |
| Median | 0.0058 | **0.0044** | 0.0086 | 0.0066 |

**Note.** In bold the best model. 'U.' denotes unconstrained, while 'C.' denotes constrained.

**Table B.39:** Comparative analysis part III, AD, comparison between independent mixture models and GND-HMMs.

| Stock | MGND | CMGND | U.GND-HMM | C.GND-HMM |
|---|---|---|---|---|
| ADS.DE | **0.0053** | 0.0054 | 0.0191 | 0.0056 |
| ADYEN.AS | 0.0053 | **0.0051** | 0.0181 | 0.0059 |
| AD.AS | **0.0021** | 0.0024 | 0.0042 | 0.0032 |
| AI.PA | 0.0412 | 0.0152 | 0.1659 | **0.0114** |
| AIR.PA | 0.0112 | **0.008** | 0.011 | 0.0093 |
| ALV.DE | 0.0023 | 0.0028 | 0.0023 | **0.0022** |
| AMS.MC | **0.0036** | 0.0048 | 0.0138 | 0.01 |
| ABI.BR | 0.0188 | 0.0133 | 0.0178 | **0.0123** |
| ASML.AS | 0.0055 | **0.0052** | 0.0289 | 0.0084 |
| CS.PA | **0.0037** | 0.0038 | 0.0126 | 0.0137 |
| BAS.DE | 0.0059 | **0.0056** | 0.0094 | 0.0067 |
| BAYN.DE | 0.0212 | 0.0099 | 0.0372 | **0.0097** |
| BMW.DE | 0.0127 | **0.0065** | 0.0383 | 0.0128 |
| BNP.PA | 0.0082 | **0.0073** | 0.0597 | 0.0089 |
| CRG.IR | 0.0036 | **0.0032** | 0.0036 | 0.0036 |
| DAI.SG | 0.0073 | **0.0068** | 0.5181 | 0.0129 |
| BN.PA | 0.0063 | **0.0037** | 0.0106 | 0.0049 |
| DB1.DE | 0.0246 | 0.0192 | 0.03 | **0.0131** |
| DPWA.F | 0.0334 | 0.0162 | 0.0216 | **0.0109** |
| DTE.DE | **0.0071** | 0.0071 | 0.0204 | 0.0104 |
| ENEL.MI | 0.0095 | **0.007** | 0.0213 | 0.0109 |
| ENGI.PA | 0.0119 | 0.0151 | 0.047 | **0.0089** |
| ENI.MI | 0.0019 | **0.0017** | 0.0065 | 0.0077 |
| EL.PA | 0.0051 | **0.0035** | 0.0166 | 0.0046 |
| FLTR.IR | **0.0066** | 0.0073 | 0.0846 | 0.01 |
| IBE.MC | 0.0029 | **0.0028** | 0.0045 | 0.0039 |
| ITX.MC | 0.0062 | **0.0036** | 0.0133 | 0.0051 |
| IFX.DE | 0.0034 | **0.0026** | 0.0106 | 0.0067 |
| INGA.AS | **0.0049** | 0.0054 | 0.011 | 0.0066 |
| ISP.MI | 0.0037 | **0.0036** | 0.0144 | 0.0066 |
| KER.PA | **0.0023** | 0.0024 | 0.0072 | 0.0038 |
| KNEBV.HE | **0.0014** | 0.0014 | 0.0034 | 0.0018 |
| OR.PA | 0.0024 | **0.0013** | 0.0068 | 0.0026 |
| MC.PA | 0.0022 | **0.0012** | 0.003 | 0.0038 |
| MUV2.DE | 0.002 | **0.0019** | 0.0036 | 0.0062 |
| NOKIA.HE | 0.0065 | **0.0037** | 0.0376 | 0.0086 |
| RI.PA | **0.0039** | 0.006 | 0.0411 | 0.0098 |
| ORA.PA | 0.0053 | **0.004** | 0.0317 | 0.0096 |
| PHIA.AS | **0.0054** | 0.0066 | 0.0341 | 0.0134 |
| PRX.AS | 0.0086 | **0.0075** | 0.0078 | 0.0078 |
| SAF.PA | 0.0057 | **0.0053** | 0.034 | 0.0105 |
| SAN.PA | 0.0236 | 0.0171 | 0.0924 | **0.015** |
| SAN.MC | **0.0028** | 0.0031 | 0.007 | 0.0055 |
| SAP.DE | 0.0041 | **0.004** | 0.0104 | 0.0064 |
| SU.PA | 0.0033 | **0.0031** | 0.0054 | 0.0045 |
| SIE.DE | 0.0087 | 0.0071 | 0.0221 | **0.0062** |
| TTE.PA | **0.0075** | 0.0101 | 0.0814 | 0.0145 |
| TEF.MC | 0.0034 | **0.0033** | 0.008 | 0.0091 |
| DG.PA | 0.0085 | **0.0057** | 0.0251 | 0.0086 |
| VIV.PA | 0.0131 | 0.0128 | 0.0898 | **0.0116** |
| VOW.DE | 0.0026 | **0.0023** | 0.0193 | 0.0053 |
| VNA.DE | **0.004** | 0.0041 | 0.0045 | 0.0044 |
| Mean | 0.0079 | **0.0061** | 0.0355 | 0.0080 |
| Median | 0.0054 | **0.0052** | 0.0172 | 0.0081 |

**Note.** In bold the best model. 'U.' denotes unconstrained, while 'C.' denotes constrained.

# C

# CMSTD model

A constrained mixture of univariate Student-t distributions (CMSTD) with $K$ components is given by the marginal distribution of the random variable $X$

$$f_{CMSTD}(x|\theta) = \sum_{k=1}^{K} \pi_k f_k(x|\mu_k, \sigma_k, \nu_k), \tag{C.1}$$

where $f_k(\cdot) = f_{St}(\cdot)$ and $-\infty < x < \infty$. The set of all mixture parameters is given by $\theta = \{\pi_k, \mu_k, \sigma_k, \nu_k, k = 1, ..., K\}$ belonging to the parameter space $\Theta = \{\theta : 0 < \pi_k < 1, \sum_{k=1}^{K} \pi_k = 1, -\infty < \mu_k < \infty, \sigma_k > 0, \nu_k > 0, k = 1, ..., K\}$ with $\dim(\theta) = p$.

According to Andrews, Wickins, et al. 2018 constraints are imposed on $\sigma_k$ and $\nu_k$ to be equal across the mixture components: $\sigma_k = \sigma$, $\nu_k = \nu$, for all $k = 1, ..., K$. Thus, taking all possible combinations of these constraints into consideration would result in a 3-model family (Table C.1).

| Model | $\mu_k$ | $\sigma_k$ | $\nu_k$ | $p$ |
|---|---|---|---|---|
| MSTD | U | U | U | $4K - 1$ |
| CMSTD-UCU | U | C | U | $3K$ |
| CMSTD-UUC | U | U | C | $3K$ |
| CMSTD-UCC | U | C | C | $2K + 1$ |

**Table C.1:** 3-model family of CMSTD and MSTD models. 'C' denotes constrained, 'U' denotes unconstrained. $K$ is the number of mixture components.

# Bibliography

Here are the references in alphabetic order.

Abate, G., I. Basile, and P. Ferrari (2021). 'The level of sustainability and mutual fund performance in Europe: An empirical analysis using ESG ratings'. In: *Corporate Social Responsibility and Environmental Management* 28.5, pp. 1446–1455.

Agresti, A. and C. Franklin (2009). *Statistics: The Art and Science of Learning from Data*.

Ahsanullah, M., B. M. G. Kibria, and M. Shakil (2014). 'Student's $$t$$Distribution'. In: *Normal and Student 's t Distributions and Their Applications*. Paris: Atlantis Press, pp. 51–62.

Akaike, H. (1974). 'A new look at the statistical model identification'. In: *IEEE Transactions on Automatic Control* 19.6, pp. 716–723.

Allili, M. S., N. Bouguila, and D. Ziou (2008). 'Finite general Gaussian mixture modeling and application to image and video foreground segmentation'. In: *Journal of Electronic Imaging* 17.1, p. 013005.

Anderson, R. M. et al. (2012). 'Sources of stock return autocorrelation'. In: *Working paper*.

Anderson, T. W. and D. A. Darling (1954). 'A Test of Goodness of Fit'. In: *Journal of the American Statistical Association* 49.268, pp. 765–769.

Andrews, J. L., P. D. McNicholas, and S. Subedi (2011). 'Model-based classification via mixtures of multivariate t-distributions'. In: *Computational Statistics & Data Analysis* 55.1, pp. 520–529.

Andrews, J. L., J. R. Wickins, et al. (2018). 'teigen: An R Package for Model-Based Clustering and Classification via the Multivariate t Distribution'. In: *Journal of Statistical Software* 83.7, pp. 1–32.

Ang, W. R. (2015). 'Sustainable investment in Korea does not catch a cold when the United States sneezes'. In: *Journal of Sustainable Finance & Investment* 5.1-2, pp. 16–26.

Aparicio, F. M. and J. Estrada (2001). 'Empirical distributions of stock returns: European securities markets, 1990-95'. In: *The European Journal of Finance* 7.1, pp. 1–21.

Ardia, D. et al. (2019). 'Markov-Switching GARCH Models in R: The MSGARCH Package'. In: *Journal of Statistical Software* 91.4, pp. 1–38.

Arouri, M. and G. Pijourlet (2017). 'CSR Performance and the Value of Cash Holdings: International Evidence'. In: *Journal of Business Ethics* 140.2, pp. 263–284.

Awan, T. et al. (2021). 'Oil and stock markets volatility during pandemic times: a review of G7 countries'. In: *Green Finance* 3, pp. 15–27.

Bachelier, L. (1900). 'Théorie de la spéculation'. fr. In: *Annales scientifiques de l'École Normale Supérieure* 3e série, 17, pp. 21–86.

Ball, C. A. and W. N. Torous (1983). 'A Simplified Jump Process for Common Stock Returns'. In: *The Journal of Financial and Quantitative Analysis* 18.1, pp. 53–65.

Banfield, J. D. and A. E. Raftery (1993). 'Model-Based Gaussian and Non-Gaussian Clustering'. In: *Biometrics* 49.3, pp. 803–821.

Barndorff-Nielsen, O. E. (1997). 'Normal Inverse Gaussian Distributions and Stochastic Volatility Modelling'. In: *Scandinavian Journal of Statistics* 24.1, pp. 1–13.

Bartolucci, F. (2011). 'An alternative to the Baum-Welch recursions for hidden Markov models'. In: *arXiv: Statistics Theory*.

Barton, D. E. and F. N. David (1957). 'Multiple Runs'. In: *Biometrika* 44.1-2, pp. 168–178.

Basher, S. A., M. K. Hassan, and A. M. Islam (2007). 'Time-varying volatility and equity returns in Bangladesh stock market'. In: *Applied Financial Economics* 17.17, pp. 1393–1407.

Basuony, M. A. K. et al. (2022). 'The effect of COVID-19 pandemic on global stock markets: Return, volatility, and bad state probability dynamics'. In: *Journal of Public Affairs* 22.S1, e2761.

Bauer, R., J. Derwall, and R. Otten (2007). 'The Ethical Mutual Fund Performance Debate: New Evidence from Canada'. In: *Journal of Economic Behavior & Organization* 70.2, pp. 111–124.

Bauer, R. and P. Smeets (2015). 'Social identification and investment decisions'. In: *Journal of Economic Behavior & Organization* 117.C, pp. 121–134.

Baum, L. E., T. Petrie, et al. (1970). 'A maximization technique occurring in the statistical analysis of probabilistic functions of Markov chains'. In: *The annals of mathematical statistics* 41.1, pp. 164–171.

Baum, L. E. and T. Petrie (1966). 'Statistical Inference for Probabilistic Functions of Finite State Markov Chains'. In: *The Annals of Mathematical Statistics* 37.6, pp. 1554–1563.

Bazi, Y., L. Bruzzone, and F. Melgani (2006). 'Image thresholding based on the EM algorithm and the generalized Gaussian distribution'. In: *Pattern Recognition* 40.2, pp. 619–634.

Behr, A. and U. Pötter (2009). 'Alternatives to the normal model of stock returns: Gaussian mixture, generalised logF and generalised hyperbolic models'. In: *Annals of Finance* 5, pp. 49–68.

Belghitar, Y., E. Clark, and N. Deshmukh (2014). 'Does it pay to be ethical? Evidence from the FTSE4Good'. In: *Journal of Banking & Finance* 47, pp. 54–62.

Bellalah, M. and M. Lavielle (2002). 'A Decomposition of Empirical Distributions with Applications to the Valuation of Derivative Assets'. In: *Multinational Finance Journal* 6.2, pp. 99–130.

Benlemlih, M. and M. Bitar (2018). 'Corporate Social Responsibility and Investment Efficiency'. In: *Journal of Business Ethics* 148, pp. 647–671.

Berk, J. and P. DeMarzo (2018). *Finanza aziendale 1 Fondamenti*. 4th ed. Pearson.

Biernacki, C. and S. Chrétien (2003). 'Degeneracy in the maximum likelihood estimation of univariate Gaussian mixtures with EM'. In: *Statistics & Probability Letters* 61.4, pp. 373–382.

Bishop, C. M. (2006). *Pattern Recognition and Machine Learning (Information Science and Statistics)*. Berlin, Heidelberg: Springer-Verlag.

Black, F. and M. Scholes (1973). 'The pricing of options and corporate liabilities'. In: *Journal of political economy* 81.3, pp. 637–654.

Blattberg, R. C. and N. J. Gonedes (1974). 'A Comparison of the Stable and Student Distributions as Statistical Models for Stock Prices'. In: *The Journal of Business* 47.2, pp. 244–280.

Bollerslev, T. (1986). 'Generalized autoregressive conditional heteroskedasticity'. In: *Journal of Econometrics* 31.3, pp. 307–327.

Bollerslev, T. and V. Todorov (2011). 'Tails, Fears, and Risk Premia'. In: *The Journal of Finance* 66.6, pp. 2165–2211.

Boulatov, A., T. Hendershott, and D. Livdan (2013). 'Informed Trading and Portfolio Returns'. In: *The Review of Economic Studies* 80.1, pp. 35–72.

Box, G. E. P. and G. C. Tiao (1973). *Bayesian Inference in Statistical Analysis*. Addison-Wesley.

Brent, R. P. (2013). *Algorithms for minimization without derivatives*. Courier Corporation.

Breymann, W. and D. Lüthi (2013). 'ghyp: A package on generalized hyperbolic distributions'. In: *Manual for R Package ghyp*.

Bujang, M. A. and F. Sapri (2018). 'An Application of the Runs Test to Test for Randomness of Observations Obtained from a Clinical Survey in an Ordered Population'. In: *Malaysian Journal of Medical Sciences* 25, pp. 146–151.

Campbell, J. Y., A. W. Lo, and A. MacKinlay (1997). *The Econometrics of Financial Markets*. Princeton University Press.

Catania, L. and R. Di Mari (2021). 'Hierarchical Markov-switching models for multivariate integer-valued time-series'. In: *Journal of Econometrics* 221.1, pp. 118–137.

Catania, L., R. Di Mari, and P. Santucci De Magistris (2022). 'Dynamic Discrete Mixtures for High-Frequency Prices'. In: *Journal of Business & Economic Statistics* 40.2, pp. 559–577.

Celeux, G. and G. Govaert (1995). 'Gaussian parsimonious clustering models'. In: *Pattern Recognition* 28.5, pp. 781–793.

Chatzitheodorou, K. et al. (2019). 'Exploring socially responsible investment perspectives: A literature mapping and an investor classification'. In: *Sustainable Production and Consumption* 19, pp. 117–129.

Chaudhary, R., P. Bakhshi, and H. Gupta (2020). 'Volatility in International Stock Markets: An Empirical Study during COVID-19'. In: *Journal of Risk and Financial Management* 13.9.

Chauveau, D. and D. R. Hunter (2013). 'ECM and MM algorithms for normal mixtures with constrained parameters'. working paper or preprint.

Christie, A. (1983). *On Information Arrival and Hypothesis Testing in Event Studies*. Tech. rep. Working paper. University of Rochester.

Chu, J., S. Nadarajah, and S. Chan (2015). 'Statistical analysis of the exchange rate of bitcoin'. In: *PloS one* 10.7, e0133678.

Clark, P. K. (1973). 'A Subordinated Stochastic Process Model with Finite Variance for Speculative Prices'. In: *Econometrica* 41.1, pp. 135–155.

Cohen, J. (1992). 'Statistical Power Analysis'. In: *Current Directions in Psychological Science* 1.3, pp. 98–101.

– (2013). *Statistical power analysis for the behavioral sciences*. Academic press.

Collison, D. J. et al. (2008). 'The financial performance of the FTSE4Good indices'. In: *Corporate Social Responsibility and Environmental Management* 15.1, pp. 14–28.

Consolandi, C. et al. (2009). 'Global Standards and Ethical Stock Indexes: The Case of the Dow Jones Sustainability Stoxx Index'. In: *Journal of Business Ethics* 87.1573-0697, pp. 185–197.

Corlu Gunes, C. and A. Corlu (2014). 'Modelling exchange rate returns: which flexible distribution to use?' In: *Quantitative Finance* 15, pp. 1–14.

Costa, M. and L. De Angelis (2010). *Model selection in hidden Markov models: a simulation study*. Quaderni di Dipartimento 7. Department of Statistics, University of Bologna.

Cunha, F. A. F. d. S. et al. (2020). 'Can sustainable investments outperform traditional benchmarks? Evidence from global stock markets'. In: *Business Strategy and the Environment* 29.2, pp. 682–697.

D'Agostino, R. and E. S. Pearson (1973). 'Tests for Departure from Normality. Empirical Results for the Distributions of $b_2$ and $\sqrt{b_1}$'. In: *Biometrika* 60.3, pp. 613–622.

Dempster, A. P., N. M. Laird, and D. B. Rubin (1977). 'Maximum Likelihood from Incomplete Data Via the EM Algorithm'. In: *Journal of the Royal Statistical Society: Series B (Methodological)* 39.1, pp. 1–22.

Divisia, F. (1927). 'Les nombres indices de la variation des prix'. In: *Revue d'économie politique* 41.1, pp. 122–129.

Duttilo, P., M. Bertolini, et al. (2024). 'High volatility, high emissions? a hidden-Markov model approach'. In: *The 52nd Scientific Meeting of the Italian Statistical Society (SIS 2024)*. SpringerNature, pp. 1–6.

Duttilo, P., S. A. Gattone, and T. Di Battista (2021). 'Volatility Modeling: An Overview of Equity Markets in the Euro Area during COVID-19 Pandemic'. In: *Mathematics* 9.11.

Duttilo, P., S. A. Gattone, and B. Iannone (2023). 'Mixtures of generalized normal distributions and EGARCH models to analyse returns and volatility of ESG and traditional investments'. In: *AStA Adv Stat Anal*, pp. 1–33.

Duttilo, P., S. A. Gattone, and A. Kume (2023). 'Mixtures of generalized normal distributions with constraints'. In: *COMPSTAT 2023 Programme and Abstracts*. ISBN 9789073592414. iasc, p. 21.

Duttilo, P., A. Kume, and S. A. Gattone (2023). 'Constrained Mixtures of Generalized Normal Distributions'. In: *SEAS IN Book of short papers 2023*. ISBN 9788891935618AAVV. SIS. Pearson, pp. 611–616.

Dytso, A. et al. (2018). 'Analytical properties of generalized Gaussian distributions'. In: *Journal of Statistical Distributions and Applications* 5.6, pp. 1–40.

Eberlein, E. and U. Keller (1995). 'Hyperbolic Distributions in Finance'. In: *Bernoulli* 1.3, pp. 281–299.

Eberlein, E. and K. Prause (2002). 'The Generalized Hyperbolic Model: Financial Derivatives and Risk Measures'. In: *Mathematical Finance — Bachelier Congress 2000: Selected Papers from the First World Congress of the Bachelier Finance Society, Paris, June 29–July 1, 2000*. Ed. by H. Geman et al. Berlin, Heidelberg: Springer Berlin Heidelberg, pp. 245–267.

Engel, C. and J. D. Hamilton (1990). 'Long Swings in the Dollar: Are They in the Data and Do Markets Know It?' In: *The American Economic Review* 80.4, pp. 689–713.

Engle, R. F. (1982). 'Autoregressive Conditional Heteroscedasticity with Estimates of the Variance of United Kingdom Inflation'. In: *Econometrica* 50.4, pp. 987–1007.

Engle, R. F., D. M. Lilien, and R. P. Robins (1987). 'Estimating Time Varying Risk Premia in the Term Structure: The Arch-M Model'. In: *Econometrica* 55.2, pp. 391–407.

Engle, R. F. and V. K. Ng (1993). 'Measuring and Testing the Impact of News on Volatility'. In: *The Journal of Finance* 48.5, pp. 1749–1778.

Ephraim, Y. and N. Merhav (2002). 'Hidden Markov processes'. In: *IEEE Transactions on Information Theory* 48.6, pp. 1518–1569.

Fama, E. F. (1963). 'Mandelbrot and the Stable Paretian Hypothesis'. In: *The Journal of Business* 36.4, pp. 420–429.

Fama, E. F. and R. Roll (1968). 'Some Properties of Symmetric Stable Distributions'. In: *Journal of the American Statistical Association* 63.323, pp. 817–836.

Ferguson, T. S. (1978). 'Maximum Likelihood Estimates of the Parameters of the Cauchy Distribution for Samples of Size 3 and 4'. In: *Journal of the American Statistical Association* 73.361, pp. 211–213.

Fich, E. M., J. Harford, and A. L. Tran (2015). 'Motivated monitors: The importance of institutional investors' portfolio weights'. In: *Journal of Financial Economics* 118.1, pp. 21–48.

Fisher, L. (1966). 'Some New Stock-Market Indexes'. In: *The Journal of Business* 39.1, pp. 191–225.

Francq, C. and J.-M. Zakoian (2019). *GARCH Models: Structure, Statistical Inference and Financial Applications*. 2nd ed. Chichester, West Sussex, PO19 8SQ, UK: John Wiley & Sons Ltd.

Friede, G. (2019). 'Why don't we see more action? A metasynthesis of the investor impediments to integrate environmental, social, and governance factors'. In: *Business Strategy and the Environment* 28.6, pp. 1260–1282.

Frühwirth-Schnatter, S. (2006). *Finite Mixture and Markov Switching Models*. New York, USA: Springer Science.

Ghalanos, A. (2022). *rugarch: Univariate GARCH models*. R package version 1.4-7.

Gnedenko, B. V. and A. N. Kolmogorov (1968). *Limit distributions for sums of independent random variables*. Translated from the Russian, annotated, and revised by K. L. Chung. With appendices by J. L. Doob and P. L. Hsu. Revised edition. Addison-Wesley Publishing Co., Reading, Mass.-London-Don Mills., Ont., pp. ix+293.

Gomes, O. (2020). 'Optimal growth under socially responsible investment: a dynamic theoretical model of the trade-off between financial gains and emotional rewards'. In: *International Journal of Corporate Social Responsibility* 5.1, p. 5.

Göncü, A., M. O. Karahan, and T. U. Kuzubaş (2016). 'A comparative goodness-of-fit analysis of distributions of some Lévy processes and Heston model to stock index returns'. In: *The North American Journal of Economics and Finance* 36, pp. 69–83.

Gray, J. B. and D. W. French (1990). 'Empirical comparison of distributional model for stock index'. In: *Journal of Business Finance & Accounting* 17.3, pp. 451–459.

Guidolin, M. and A. Timmermann (2006). 'Term structure of risk under alternative econometric specifications'. In: *Journal of Econometrics* 131.1-2, pp. 285–308.

Haas, M. (2009). 'Value-at-Risk via mixture distributions reconsidered'. In: *Applied Mathematics and Computation* 215.6, pp. 2103–2119.

Haas, M. and C. Pigorsch (2009). 'Financial Economics, Fat-Tailed Distributions'. In: *Encyclopedia of Complexity and Systems Science*. Ed. by R. A. Meyers. New York, NY: Springer New York, pp. 3404–3435.

Hamilton, J. D. (1988). 'Rational-expectations econometric analysis of changes in regime: An investigation of the term structure of interest rates'. In: *Journal of Economic Dynamics and Control* 12.2-3, pp. 385–423.

– (1989). 'A new approach to the economic analysis of nonstationary time series and the business cycle'. In: *Econometrica: Journal of the econometric society*, pp. 357–384.

– (1990). 'Analysis of time series subject to changes in regime'. In: *Journal of econometrics* 45.1-2, pp. 39–70.

– (1991). 'A quasi-Bayesian approach to estimating parameters for mixtures of normal distributions'. In: *Journal of Business & Economic Statistics* 9.1, pp. 27–39.

– (1993). 'Estimation, inference and forecasting of time series subject to changes in regime'. In: *Handbook of statistics* 11, pp. 231–260.

– (1994). *Time series analysis*. Princeton, New Jersey: Princeton university press.

Hathaway, R. J. (1986). 'A constrained em algorithm for univariate normal mixtures'. In: *Journal of Statistical Computation and Simulation* 23.3, pp. 211–230.

Hemerijck, A. (2018). 'Social investment as a policy paradigm'. In: *Journal of European Public Policy* 25.6, pp. 810–827.

Hentschel, L. (1995). 'All in the family Nesting symmetric and asymmetric GARCH models'. In: *Journal of Financial Economics* 39.1, pp. 71–104.

Hol, E. M. J. H. (2003). 'Forecasting the Variability of Stock Index Returns with Stochastic Volatility Models and Implied Volatility'. In: *Empirical Studies on Volatility in International Stock Markets*. Boston, MA: Springer US, pp. 71–97.

Hoti, S., M. McAleer, and L. L. Pauwels (2007). 'Measuring Risk in Environmental Finance'. In: *Journal of Economic Surveys* 21.5, pp. 970–998.

Hsu, D. A. (1982). 'A Bayesian Robust Detection of Shift in the Risk Structure of Stock Market Returns'. In: *Journal of the American Statistical Association* 77.377, pp. 29–39.

Hubert, L. and P. Arabie (1985). 'Comparing partitions'. In: *Journal of Classification* 2.1, pp. 193–218.

Imran, N., M. Sheraz, and S. Dedu (2022). 'Mixture Models and Modelling Volatility of Returns - A Study on Gaussian and Heterogeneous Heavy Tail Mixtures'. In: *ECONOMIC COMPUTATION AND ECONOMIC CYBERNETICS STUDIES AND RESEARCH* 56, pp. 5–20.

Izzeldin, M. et al. (2023). 'The impact of the Russian-Ukrainian war on global financial markets'. In: *International Review of Financial Analysis* 87, p. 102598.

Jain, M., G. D. Sharma, and M. Srivastava (2019). 'Can Sustainable Investment Yield Better Financial Returns: A Comparative Study of ESG Indices and MSCI Indices'. In: *Risks* 7.1.

Janczura, J. and R. Weron (2012). 'Goodness-of-fit testing for the marginal distribution of regime-switching models with an application to electricity spot prices'. In: *AStA Advances in Statistical Analysis* 97.

Jiang, K. et al. (2022). 'Forecasting Value-at-Risk of cryptocurrencies using the time-varying mixture-accelerating generalized autoregressive score model'. In: *Research in International Business and Finance* 61, p. 101634.

Jiang, X., S. Nadarajah, and T. Hitchen (2023). 'A Review of Generalized Hyperbolic Distributions'. In: *Computational Economics*, pp. 1–30.

Jones, P. N. and G. J. Mclachlan (1990). 'Laplace-normal mixtures fitted to wind shear data'. In: *Journal of Applied Statistics* 17.2, pp. 271–276.

Kaehler, J. and V. Marnet (1994). 'Markov-Switching Models for Exchange-Rate Dynamics and the Pricing of Foreign-Currency Options'. In: *Econometric Analysis of Financial Markets*. Ed. by J. Kaehler and P. Kugler. Heidelberg: Physica-Verlag HD, pp. 203–230.

Kanji, G. K. (1985). 'A mixture model for wind shear data'. In: *Journal of Applied Statistics* 12.1, pp. 49–58.

Kartal, M. T., H. M. Ertugrul, and T. Ulussever (2022). 'The impacts of foreign portfolio flows and monetary policy responses on stock markets by considering COVID-19 pandemic: Evidence from Turkey'. In: *Borsa Istanbul Review* 22.1, pp. 12–19.

Kaufer, K. and L. Steponaitis (2019). 'Just Money: From Ego-System to Ecosystem Finance'. In: *Just Money: Mission-Driven Banks and the Future of Finance*. The MIT Press.

Kim, T.-H. and H. White (2004). 'On more robust estimation of skewness and kurtosis'. In: *Finance Research Letters* 1.1, pp. 56–73.

Kolmogorov, A. (1933). 'Sulla determinazione empirica di una legge dididistribuzione'. In: *Giorn Dell'inst Ital Degli Att* 4, pp. 89–91.

Kon, S. J. (1983). 'The Market-Timing Performance of Mutual Fund Managers'. In: *The Journal of Business* 56.3, pp. 323–347.

– (1984). 'Models of Stock Returns–A Comparison'. In: *The Journal of Finance* 39.1, pp. 147–165.

Kon, S. J. and W. P. Lau (1979). 'Specification Tests for Portfolio Regression Parameter Stationarity and the Implications for Empirical Research'. In: *The Journal of Finance* 34.2, pp. 451–465.

Koutmos, G. (1997). 'Feedback trading and the autocorrelation pattern of stock returns: further empirical evidence'. In: *Journal of International Money and Finance* 16.4, pp. 625–636.

Kozubowski, T. and S. Nadarajah (2010). 'Multitude of Laplace distributions'. In: *Statistical Papers* 51, pp. 127–148.

Krosinsky, C. and N. Robins (2012). *Sustainable Investing: The Art of Long-Term Performance*. Environmental Market Insights. Taylor & Francis Group.

Lapanan, N. (2018). 'The investment behavior of socially responsible individual investors'. In: *The Quarterly Review of Economics and Finance* 70, pp. 214–226.

Le, N. D. et al. (1992). 'Exact Likelihood Evaluation in a Markov Mixture Model for Time Series of Seizure Counts'. In: *Biometrics* 48.1, pp. 317–323.

Lean, H. H. and D. K. Nguyen (2014). 'Policy uncertainty and performance characteristics of sustainable investments across regions around the global financial crisis'. In: *Applied Financial Economics* 24.21, pp. 1367–1373.

Lee, S. and G. Mclachlan (2013). 'On Mixtures of Skew Normal and Skew t-Distributions'. In: *Advances in Data Analysis and Classification* 7.

Li, J. (2012). *A Cauchy-Gaussian Mixture Model for Basel-compliant Value-at-Risk Estimation in Financial Risk Management*. Theses and Dissertations, Lehigh University, Paper 1075.

Lilliefors, H. W. (1967). 'On the Kolmogorov-Smirnov Test for Normality with Mean and Variance Unknown'. In: *Journal of the American Statistical Association* 62.318, pp. 399–402.

Lindgren, G. (1978). 'Markov regime models for mixed distributions and switching regressions'. In: *Scandinavian Journal of Statistics*, pp. 81–91.

Lintner, J. (1965). 'The Valuation of Risk Assets and the Selection of Risky Investments in Stock Portfolios and Capital Budgets'. In: *The Review of Economics and Statistics* 47.1, pp. 13–37.

Liyan Han, H. Y. and C. Zheng (2019). 'Normal mixture method for stock daily returns over different sub-periods'. In: *Communications in Statistics - Simulation and Computation* 48.2, pp. 447–457.

Llorente, G. et al. (2002). 'Dynamic Volume-Return Relation of Individual Stocks'. In: *The Review of Financial Studies* 15.4, pp. 1005–1047.

Madan, D. B. and E. Seneta (1990). 'The variance gamma (VG) model for share market returns'. In: *Journal of business*, pp. 511–524.

Mandelbrot, B. (1973). 'A Subordinated Stochastic Process Model with Finite Variance for Speculative Prices: Comment'. In: *Econometrica* 41.1, pp. 157–59.

Mandelbrot, B. (1963). 'The Variation of Certain Speculative Prices'. In: *The Journal of Business* 36.4, pp. 394–419.

Markowitz, H. (1952). 'The Utility of Wealth'. In: *Journal of Political Economy* 60.2, pp. 151–158.

Markus Haas, S. M. and M. S. Paolella (2006). 'Modelling and predicting market risk with Laplace–Gaussian mixture distributions'. In: *Applied Financial Economics* 16.15, pp. 1145–1162.

Massing, T. (2019). 'What is the best Lévy model for stock indices? A comparative study with a view to time consistency'. In: *Financial Markets and Portfolio Management* 33, pp. 277–344.

Massing, T., M. Puente-Ajovín, and A. Ramos (2020). 'On the parametric description of log-growth rates of cities' sizes of four European countries and the USA'. In: *Physica A: Statistical Mechanics and its Applications* 551, p. 124587.

Massing, T. and A. Ramos (2021). 'Student's t mixture models for stock indices. A comparative study'. In: *Physica A: Statistical Mechanics and its Applications* 580, p. 126143.

McLachlan, G. J. and D. Peel (2000). *Finite mixture models*. New York: Wiley Series in Probability and Statistics.

McLachlan, G. and T. Krishnan (2008). *The EM algorithm and extensions*. 2.ed. Wiley series in probability and statistics. Hoboken, NJ: Wiley.

McLachlan, G. J., S. X. Lee, and S. I. Rathnayake (2019). 'Finite Mixture Models'. In: *Annual Review of Statistics and Its Application* 6.1, pp. 355–378.

McLachlan, G. J. and D. Peel (1998). 'Robust cluster analysis via mixtures of multivariate t-distributions'. In: *Advances in Pattern Recognition*. Ed. by A. Amin et al. Berlin, Heidelberg: Springer Berlin Heidelberg, pp. 658–666.

McLachlan, G. J. and S. Rathnayake (2014). 'On the number of components in a Gaussian mixture model'. In: *WIREs Data Mining and Knowledge Discovery* 4.5, pp. 341–355.

McLeod, A. I. and W. K. Li (1983). 'DIAGNOSTIC CHECKING ARMA TIME SERIES MODELS USING SQUARED-RESIDUAL AUTOCORRELATIONS'. In: *Journal of Time Series Analysis* 4.4, pp. 269–273.

Meng, X.-L. and D. B. Rubin (1993). 'Maximum likelihood estimation via the ECM algorithm: A general framework'. In: *Biometrika* 80.2, pp. 267–278.

Mills, F. C. (1927). *The Behavior of Prices*. National Bureau of Economic Research, Inc.

Mitchell, W. (1938). *The Making and Using of Index Numbers*. Bulletin / United States Department of Labor, Bureau of Labor Statistics. U.S. Government Printing Office.

Mobin, M. A. et al. (2022). 'COVID-19 pandemic and risk dynamics of financial markets in G7 countries'. In: *International Journal of Islamic and Middle Eastern Finance and Management* 15.2, pp. 461–478.

Mohamed, O. M. M. and M. Jaïdane-Saïdane (2009). 'Generalized Gaussian mixture model'. In: *2009 17th European Signal Processing Conference*, pp. 2273–2277.

Mossin, J. (1966). 'Equilibrium in a Capital Asset Market'. In: *Econometrica* 34.4, pp. 768–783.

Nadarajah, S. (2005). 'A generalized normal distribution'. In: *Journal of Applied Statistics* 32.7, pp. 685–694.

– (2011). 'Making the Cauchy work'. In: *Brazilian Journal of Probability and Statistics* 25.1, pp. 99–120.

Nasir, I., M. Sheraz, and S. Dedu (2022). 'MIXTURE MODELS AND MODELLING VOLATILITY OF RETURNS–A STUDY ON GAUSSIAN AND HETEROGENEOUS HEAVY TAIL MIXTURES.' In: *Economic Computation & Economic Cybernetics Studies & Research* 56.4.

Nelder, J. A. and R. Mead (1965). 'A Simplex Method for Function Minimization'. In: *The Computer Journal* 7.4, pp. 308–313.

Nelson, D. B. (1991). 'Conditional Heteroskedasticity in Asset Returns: A New Approach'. In: *Econometrica* 59.2, pp. 347–370.

Nguyen, T. M., Q. Jonathan Wu, and H. Zhang (2014). 'Bounded generalized Gaussian mixture model'. In: *Pattern Recognition* 47.9, pp. 3132–3142.

Nilsson, J. (2009). 'Segmenting socially responsible mutual fund investors: The influence of financial return and social responsibility'. In: *International Journal of Bank Marketing* 27, pp. 5–31.

Oikonomou, I., E. Platanakis, and C. Sutcliffe (2018). 'Socially responsible investment portfolios: Does the optimization process matter?' In: *The British Accounting Review* 50.4, pp. 379–401.

Ouchen, A. (2022). 'Is the ESG portfolio less turbulent than a market benchmark portfolio?' In: *Risk Management* 24.1, pp. 1–33.

Pearson, K. (1894). 'Contributions to the Mathematical Theory of Evolution'. In: *Philosophical Transactions of the Royal Society of London. A* 185, pp. 71–110.

Peiró, A. (1994). 'The distribution of stock returns: international evidence'. In: *Applied Financial Economics* 4.6, pp. 431–439.

Praetz, P. D. (1972). 'The Distribution of Share Price Changes'. In: *The Journal of Business* 45.1, pp. 49–55.

Prakash, P., V. Sangwan, and K. Singh (2021). 'Transformational Approach to Analytical Value-at-Risk for near Normal Distributions'. In: *Journal of Risk and Financial Management* 14.2.

Press, S. J. (1967). 'A Compound Events Model for Security Prices'. In: *The Journal of Business* 40.3, pp. 317–335.

Quandt, R. E. and J. B. Ramsey (1978). 'Estimating Mixtures of Normal Distributions and Switching Regressions'. In: *Journal of the American Statistical Association* 73.364, pp. 730–738.

Rand, W. M. (1971). 'Objective Criteria for the Evaluation of Clustering Methods'. In: *Journal of the American Statistical Association* 66.336, pp. 846–850.

Revelli, C. (2017). 'Socially responsible investing (SRI): From mainstream to margin?' In: *Research in International Business and Finance* 39, pp. 711–717.

Rocci, R., S. A. Gattone, and R. Di Mari (2018). 'A data driven equivariant approach to constrained Gaussian mixture modeling'. In: *Advances in Data Analysis and Classification* 12.2, pp. 235–260.

Rossi, M. et al. (2019). 'Household preferences for socially responsible investments'. In: *Journal of Banking & Finance* 105, pp. 107–120.

Ruppert, D. (2011). 'Modeling Univariate Distributions'. In: *Statistics and Data Analysis for Financial Engineering*. New York, NY: Springer New York, pp. 79–130.

S&PGlobal (2022). *DJSI Index Family, S&P Global*. Accessed: January 10, 2022. URL: https://www.spglobal.com/esg/performance/indices/djsi-index-family.

Sabbaghi, O. (2022). 'The impact of news on the volatility of ESG firms'. In: *Global Finance Journal* 51, p. 100570.

Salhi, K. et al. (2016). 'Regime switching model for financial data: Empirical risk analysis'. In: *Physica A: Statistical Mechanics and its Applications* 461, pp. 148–157.

Saralees Nadarajah, E. A. and S. Chan (2015). 'A note on "Modelling exchange rate returns: which flexible distribution to use?"' In: *Quantitative Finance* 15.11, pp. 1777–1785.

Scholes, M. and J. Williams (1977). 'Estimating betas from nonsynchronous data'. In: *Journal of Financial Economics* 5.3, pp. 309–327.

Schoutens, W. (2002). 'The Meixner Process: Theory and Applications in Finance'. In: *2nd MaPhySto Lévy Conference*, p. 237.

Schröder, M. (2007). 'Is there a Difference? The Performance Characteristics of SRI Equity Indices'. In: *Journal of Business Finance & Accounting* 34.1-2, pp. 331–348.

Schwarz, G. (1978). 'Estimating the Dimension of a Model'. In: *The Annals of Statistics* 6.2, pp. 461–464.

Scrucca, L. et al. (2016). 'mclust 5: clustering, classification and density estimation using Gaussian finite mixture models'. In: *The R Journal* 8.1, pp. 289–317.

Sharpe, W. F. (1964). 'CAPITAL ASSET PRICES: A THEORY OF MARKET EQUILIBRIUM UNDER CONDITIONS OF RISK*'. In: *The Journal of Finance* 19.3, pp. 425–442.

Shehzad, K., L. Xiaoxing, and H. Kazouz (2020). 'COVID-19's disasters are perilous than Global Financial Crisis: A rumor or fact?' In: *Finance Research Letters* 36, p. 101669.

Shin, J.-Y., T. B. Ouarda, and T. Lee (2016). 'Heterogeneous mixture distributions for modeling wind speed, application to the UAE'. In: *Renewable Energy* 91, pp. 40–52.

Shunsuke, M., O. Tatsuyoshi, and M. Akimi (2012). 'Do socially responsible investment indexes outperform conventional indexes?' In: *Applied Financial Economics* 22.18, pp. 1511–1527.

Sinnott, R., H. Duan, and Y. Sun (2016). 'Chapter 15 - A Case Study in Big Data Analytics: Exploring Twitter Sentiment Analysis and the Weather'. In: *Big Data*. Ed. by R. Buyya, R. N. Calheiros, and A. V. Dastjerdi. Morgan Kaufmann, pp. 357–388.

So, M., A. Mak, and A. Chu (2022). 'Assessing systemic risk in financial markets using dynamic topic networks'. In: *Scientific Reports* 12, p. 2668.

Sudha, S. (2015). 'Risk-return and Volatility analysis of Sustainability Index in India'. English. In: *Environment, Development and Sustainability* 17.6, pp. 1329–1342.

Szczygielski, J. J. et al. (2021). 'The only certainty is uncertainty: An analysis of the impact of COVID-19 uncertainty on regional stock markets'. In: *Finance Research Letters* 43, p. 101945.

Traut, J. and W. Schadner (2023). *Which is Worse: Heavy Tails or Volatility Clusters?* Swiss Finance Institute Research Paper Series 23-61. Swiss Finance Institute.

Turner, C. M., R. Startz, and C. R. Nelson (1989). 'A Markov model of heteroskedasticity, risk, and learning in the stock market'. In: *Journal of Financial Economics* 25.1, pp. 3–22.

University of Cambridge (2022). *What is responsible investment?* Accessed: March 16, 2022. URL: https://www.cisl.cam.ac.uk/business-action/sustainable-finance/investment-leaders-group/what-is-responsible-investment.

Upton, D. E. and D. S. Shannon (1979). 'The Stable Paretian Distribution, Subordinated Stochastic Processes, and Asymptotic Lognormality: An Empirical Investigation'. In: *The Journal of Finance* 34.4, pp. 1031–1039.

Varanasi, M. K. and B. Aazhang (1989). 'Parametric generalized Gaussian density estimation'. In: *The Journal of the Acoustical Society of America* 86.4, pp. 1404–1415.

Viterbi, A. (1967). 'Error bounds for convolutional codes and an asymptotically optimum decoding algorithm'. In: *IEEE Transactions on Information Theory* 13.2, pp. 260–269.

Wald, A. and J. Wolfowitz (1940). 'On a Test Whether Two Samples are from the Same Population'. In: *The Annals of Mathematical Statistics* 11.2, pp. 147–162.

Wen, L. et al. (2022). 'Numerical characteristics and parameter estimation of finite mixed generalized normal distribution'. In: *Communications in Statistics - Simulation and Computation* 51.7, pp. 3596–3620.

Wilks, D. (2011). 'Chapter 4 - Parametric Probability Distributions'. In: *Statistical Methods in the Atmospheric Sciences*. Ed. by D. S. Wilks. Vol. 100. International Geophysics. Academic Press, pp. 71–131.

Wuertz, D. et al. (2023). 'Rmetrics - Markets and Basic Statistics'. In: *Package fBasic*.

Yousef, I. (2020). 'Spillover of COVID-19: Impact on Stock Market Volatility'. In: *nt. J. Psychosoc. Rehabil.* 24.

Zhang, S. et al. (2020). 'Time-Varying Gaussian-Cauchy Mixture Models for Financial Risk Management'. In: *arXiv: Applications*.

Zhang, Y. et al. (2019). 'The generalised hyperbolic distribution and its subclass in the analysis of a new era of cryptocurrencies: Ethereum and its financial risk'. In: *Physica A: Statistical Mechanics and its Applications* 526, p. 120900.

Zigrand, J.-P. (2014). *Systems and systemic risk in finance and economics*.

Zucchini, W., I. MacDonald, and R. Langrock (2017). *Hidden Markov Models for Time Series: An Introduction Using R, Second Edition*. Chapman & Hall/CRC Monographs on Statistics and Applied Probability. CRC Press.

# Acknowledgements

*First and foremost I am extremely grateful to my PhD supervisor Prof. Stefano Antonio Gattone for his invaluable advice, continuous support, and patience during my PhD study. I would like to express my gratitude to Prof. Tonio Di Battista for his important suggestions during my PhD study. I would also like to thank Prof.ssa Barbara Iannone for contributing to the realization of the empirical study 'ESG vs traditional indices' in Chapter 3. I am very grateful to Prof. Alfred Kume for hosting me at the University of Kent and for contributing to the realization of Chapters 4 and 5.*